\documentclass[pre,amsmath,amssymb,onecolumn, showpacs, superscriptaddress,10pt]{revtex4-2}

\usepackage[english]{babel}
\usepackage[utf8]{inputenc}
\usepackage[T1]{fontenc}
\usepackage{amsmath}
\usepackage[colorlinks=true,citecolor=blue]{hyperref}
\usepackage{graphicx}
\usepackage{amsfonts}
\usepackage{amsthm}

\newcommand{\nn}{\nonumber}
\newcommand{\be}{\begin{equation}}
\newcommand{\ee}{\end{equation}}
\newcommand{\bea}{\begin{eqnarray}}
\newcommand{\eea}{\end{eqnarray}}

\usepackage{color}
\definecolor{Blue}{rgb}{0.00, 0.00, 1.00}
\definecolor{Red}{rgb}{1.00, 0.00, 0.00}
\definecolor{Green}{rgb}{0.00, 0.70, 0.00}

\begin{document}

\title{Replica-symmetry breaking transitions in the large deviations of the ground-state of a spherical spin-glass}
\author{Bertrand Lacroix-A-Chez-Toine}
\affiliation{King's College London, Department of Mathematics, London  WC2R 2LS, United Kingdom}
\author{Yan V. Fyodorov}
\affiliation{King's College London, Department of Mathematics, London  WC2R 2LS, United Kingdom}
\author{Pierre Le Doussal}
 \affiliation{Laboratoire de Physique de l'Ecole Normale Sup\'erieure, ENS, Universit\'e PSL, CNRS, Sorbonne Universit\'e, Universit\'e de Paris, 75005 Paris, France}

\begin{abstract}
      We derive, within the replica formalism, a generalisation of the Crisanti-Sommers formula to describe the large deviation function (LDF) ${\cal L}(e)$ for the speed-$N$ atypical fluctuations of the intensive ground-state energy $e$ of a generic spherical spin-glass in the presence of a random external magnetic field of variance $\Gamma$. We then analyse our exact formula for the LDF in much detail for the Replica symmetric, single step Replica Symmetry Breaking (1-RSB) and Full Replica Symmetry Breaking (FRSB) situations. 
Our main qualitative conclusion is that the level of RSB governing the LDF may be different from that for the typical ground state.
      We find that while the deepest ground-states are always controlled by a LDF of replica symmetric form, beyond a finite threshold $e\geq e_{t}$ a replica-symmetry breaking starts to be operative. These findings 
      resolve the puzzling discrepancy between our earlier replica calculations for the $p=2$ spherical spin-glass \cite{fyodorov2014topology} and the rigorous results by Dembo and Zeitouni \cite{dembo2015matrix} which we are able to reproduce invoking an 1-RSB pattern. 
      Finally at an even larger critical energy $e_{c}\geq e_{t}$, acting as a "wall", the LDF diverges logarithmically, 
      which we interpret as a change in the large deviation speed from $N$ to a faster growth. 
      In addition, we show that in the limit $\Gamma \to 0$ the LDF takes non-trivial scaling forms (i) 
      ${\cal L}(e) \sim G((e-e_c)/\Gamma)$ in the vicinity of the wall (ii) 
      ${\cal L}(e) \sim \Gamma^{\eta \nu} F((e-e_{\rm typ})/\Gamma^{\nu})$ in the vicinity of the typical
      energy, characterised by two new exponents $\eta\geq 1$ and $\nu$ characterising universality classes. Via matching
      the latter allows us to formulate several conjectures concerning the regime of 
      {\it typical fluctuations}, identified as $e-e_{\rm typ} \sim N^{-1/\eta}$
      and $\Gamma \sim N^{-1/(\eta \nu)}$. 

\end{abstract}

\maketitle

\section{Introduction}

\subsection{Model}

The properties of high-dimensional random landscapes attracted a lot of attention  in recent years. One of the main driving forces behind that interest is a wish to understand universal features of search algorithms aiming to solve high-dimensional non-convex optimization problems, where the cost/loss functions to be optimized are highly irregular or corrugated and may trap the search considerably in the maze of many local minima and saddle points.  Examples of such problems are abundant in  constraint satisfaction or inference problems \cite{MM2009book} as well as  loss landscapes in supervised learning \cite{Choromanska2015}, and have intimate relations with many problems in economics,  biology, chemistry, condensed matter physics,  string theory and cosmology, see  \cite{RosFyoReview} and 
 \cite{AufMonSubReview} for a concise introduction to the problematic and further references. 
 
 Among many relevant questions to be studied in this framework, one of the most naturally arising is to characterise statistically the value of the global minimum in some 
 benchmark models of random landscapes. The role of such benchmarks is largely served by the so-called  spin glass models,  one of the most canonical choices being the Sherrington-Kirkpatrick (SK) model
 \cite{SK,Montanari_SK}  asking to minimize a random quadratic form $\sum_{ij}J_{ij} S_iS_j$, interpreted in that context as an energy function and called the  Hamiltonian, over the hypercube of $2^N$ vertices parametrized by Ising spin variables $S_{i}=\pm 1, \, i=1,\ldots,N $,  with parameters $J_{i<j}$ being independent, identically distributed real Gaussian variables. Another paradigmatic choice is the so-called spherical spin glass model, with aim to optimize a generic Hamiltonian $H({\bf x}), \, {\bf x}\in \mathbb{R}^N$  assumed to be a mean-zero random Gaussian field with the covariance structure  
\begin{align}
&\overline{H({\bf x})H({\bf y})}=N\,f\left(\frac{{\bf x}\cdot {\bf y}}{N}\right)\;,\\
&f(q)=\Gamma q+g(q)\;,\;\;g(q)=\sum_{r=2}^{\infty}g_r\, q^r\;,\;\;\Gamma\geq 0\;,\;\;\forall r\geq 2\;,\;\;g_r\geq 0\;,\;\;q\in[0,1]\;\label{cov}
\end{align}
 over the sphere ${\bf x}^2=N$. We suppose that the function $g(q)$ defined as a series above has a radius of convergence larger than unity, and is ${\cal C}_{\infty}$ over the interval $[0,1]$,
 so that $f'(1) < +\infty$. 
 In particular, taking $g_p=\delta_{p,r}/p$, the Hamiltonian corresponds to the spherical pure $p$-spin model subject to a random Gaussian magnetic field with zero mean and variance $\Gamma$.

With functions to be optimized being random, their values at the global minimum are necessarily random as well,
and one of the main goals of the theory is to characterize the whole distribution of such a random variable, traditionally called the ground state energy ${ E}_{\min}$, in the limit of large dimension 
$N\gg 1$. One of the possible approaches to such a problem is inspired by ideas from statistical mechanics and aims at extracting the value of the ground state energy per degree of freedom $e_{\min}=E_{\min}/N$ from the zero temperature limit of the associated free energy.   For the case of spherical spin glasses this method amounts to evaluating
\be
e_{\min}=\frac{1}{N}\min_{{\bf x}\in\mathbb{R}^N\,:\,{\bf x}^2=N}H({\bf x})=-\lim_{\beta\to \infty}\frac{1}{N\beta}\ln Z(\beta)\;,
\ee
where $Z(\beta)$ is the partition function at inverse temperature $\beta$, defined as
\be
Z(\beta)=\int_{\sqrt{N}{\cal S}_{N-1}} d{\bf x}\, e^{-\beta H({\bf x})}=\int_{\mathbb{R}^N} d{\bf x}\,\delta\left({\bf x}^2-N\right)\,e^{-\beta H({\bf x})}\;,\label{PS}
\ee
with ${\cal S}_{N-1}$ being the $N$-dimensional unit sphere. Similar formulae can be written for the ground energy of SK or any other optimization model with a given cost/loss function. One defines the probability distribution function (PDF) of the ground state energy
\be 
{\cal P}(e) = \overline{\delta(e-e_{\rm min})}\;.
\ee 
In the large $N$ limit, it is well known that the ground state energy per degree of freedom $e_{\rm min}$ is a {\it self-averaging} quantity, i.e. tends to a nonrandom limit $\overline{e_{\min}}$ when $N\to \infty$
\be 
\lim_{N \to +\infty} e_{\rm min} = \overline{e_{\rm min}}\;.
\ee 
The aim of this paper is to describe the fluctuations of $e_{\min}$ in particular, focusing on the large deviations tails of ${\cal P}(e)$ which describe atypical fluctuations away from $\overline{e_{\min}}$. Before describing the method and our results, it is worth reviewing in some details the literature on the average $\overline{e_{\min}}$, in particular the phenomenon of replica-symmetry breaking (RSB), which will also turn out to be essential to understand the problem at hand.

\subsection{Average ground-state energy}

The theoretical physics framework allowing one to compute the limiting free energy associated with the SK model, and thus eventually to extract  $\overline{e_{\min}}$,  has been originally developed in a series of pioneering papers by Giorgio Parisi starting from \cite{ParisiFirst}. In those works it has been demonstrated that finding the ground state required introducing the fundamental concept of Replica Symmetry Breaking (RSB), see \cite{MPV1987} for a standard introduction. In such framework finding $\overline{e_{\min}}$  amounts to optimizing the so-called  Parisi functional with respect to a positive, non-descreasing function $x(q), \, q\in[0,1]$, and the optimal solution has been predicted to have a part which is continuous in a finite interval $q\in[a,b]\subset [0,1]$. Such solution is then known to describe the situation of the Full Replica Symmetry Breaking (FRSB), which reflects an intricate picture of minima of the associated random landscape informally described as hierarchically organized system of valleys within valleys within valleys, and so on. In such a situation the saddle-points and 
minima are expected to co-exist at the same level of energy down to the global minimal value, which is predicted to imply, among other things, the gapless spectrum of Hessian around the global minimum, known as 
the property of "marginal stability". The picture also suggests an extreme sensitivity of the ground state to 
small changes in random parameters, known as the "disorder chaos", see  \cite{MPV1987,CastellaniCavagna2005pedestrian,rizzo2009chaos} for various aspects of all the above-mentioned phenomena.

Note that although originally the Parisi functional was derived using the powerful, but non-rigorous replica method, more recently it was re-derived and analyzed by fully controlled, mathematically rigorous techniques starting from seminal papers by Guerra and Talagrand \cite{guerra2003broken,talagrand2006parisi}, see  \cite{Panchenko2013} for a detailed exposition and overview of further developments. In particular, there has been considerable progress towards proving the FRSB nature of the solution at zero temperature \cite{auffinger2020sk}.

The problem of extracting the ground state energy for the spherical spin glass Hamiltonian turns out to be as rich as that for the Ising case, but appears to be more easily amenable for a detailed and explicit analysis. An analogue of Parisi functional for the mean free energy (and hence for the ground state value $\overline{e_{\min}}$) for that model has been derived originally in a seminal work of Crisanti and Sommers \cite{CrisantiSommers1992} using the replica approach. In recent years the Crisanti-Sommers variational problem  has been re-derived rigorously  \cite{Talagrand_spherical} and then thoroughly analyzed in various parameter regimes  in many mathematical papers, see e.g. \cite{AufChen2017,ChenSen2017,JagTab2017_spherical,Subag_spherical_ground,AufZEng2019,AufZhou2022}.
These studies demonstrated that depending on the choice 
of parameters the variational problem for the ground state energy may have piecewise-constant solutions,
associated with the so-called k-step replica symmetry breaking (k-RSB) pattern, with the case $k=0$ corresponding to the simplest Replica Symmetric (RS) solution. In particular, $k=1$ is realized generically for all spherical pure $p$-spin models with $p\ge 3$, provided the magnetic field variance $\Gamma$ is small enough, the picture already predicted in the original Crisanti and Sommers paper \cite{CrisantiSommers1992}. For either pure $p=2$ case or any pure case and strong enough  magnetic field variance $\Gamma$ the RS solution is known to hold. Physically, 1RSB type of replica symmetry breaking 
corresponds to the landscape dominated below certain level by exponentially many isolated minima separated 
by high barriers, whereas RS solution is to be expected for subexponential (frequently, just finite) number of  minima in the landscape \cite{MPV1987,CastellaniCavagna2005pedestrian}.

Allowing for some mixtures of two different powers  in the covariance Eq.(\ref{cov}) may lead to 
more complicated solutions, like 2-RSB for a mixture of two powers $r=2$ and $r=4$ with specially chosen coefficients \cite{crisantileuzzi2004,AufZEng2019}. Moreover, taking generic mixtures of two powers (strictly larger than $2$) may lead to solution mixing FRSB features with 1-RSB and 2-RSB (called therefore 
1-FRSB and 2-FRSB)\cite{crisantileuzzi2004,crisantileuzzi2006,AufZhou2022}, but never to pure FRSB solution. The situation for a generic mixture of infinitely many powers is not yet rigorously established, but expected to be of FRSB type. We see therefore that the spherical model allows a very rich behaviour 
for its ground state problem in the limit $N\to \infty$.

It is necessary to mention that, in the spherical case, the method of Crisanti-Sommers functional for characterizing the ground state value in the limit $N\to \infty$ has certain viable alternatives. The first one is based on exact counting of extrema of the energy functional via the so-called  Kac-Rice formula.  Such counting has been  
 originally performed in \cite{Fyodo2004complexity,FyodWilliams2007,FyodRev2015} for a certain type of spin glass models, and independently rediscovered, further developed and successfully applied to studying the spherical pure $p$-spin glasses, in particular to extracting the ground state value in \cite{AufBenArusCerny2013,AufBenArous2013}. Further work in this direction lead to a detailed characterization of landscape geometry close to the ground state value, see e.g. \cite{subag2017geometry}. 
 This line of research is closely related to developing the so-called Thouless-Anderson-Palmer approach to free energy of spherical pure $p$-spin glasses, see \cite{crisanti1995TAP} for the original physics paper and  \cite{subag2018TAP,belius2022TAP} for discussions of the recent efforts in mathematical literature.
 Another interesting alternative suggested very recently for the spherical pure $p$-spin is based on the cavity approach
 \cite {gradenigo2020cavity}.

 A detailed quantitative discussion of the average (and typical) ground-state energy is given in Appendix \ref{app:typical}.

 \subsection{Fluctuations and large deviations of the ground state energy and aim of the paper} 

While the discussion above shows that the problem of finding equilibrium free energy and then eventually the ground state value $\overline{e_{\min}}$ for a general spherical spin glass model in the limit $N\to \infty$  is by now relatively well-studied, the problem of characterizing the ground state fluctuations for large, but finite $N\gg 1$ remains scarcely researched.  For general spherical mixed even $p$-spin models with $\Gamma=0$, the extensive ground state energy  $E_{\min}=N e_{\min}$ "superconcentrates'', which means that ${\cal V}_{\min}=\lim_{N\to \infty}\mbox{Var}\left(E_{\min}\right)/N= 0$,
whereas for any fixed  magnetic field $\Gamma>0$  the above limit is finite and the ground state energy  satisfies
the Central Limit Theorem \cite{ChenSen2017}.  This last fact implicitly suggests that for $\Gamma>0$ the fluctuations of the ground state energy away from $\overline{e_{\min}}$ may actually satisfy a Large Deviation Principle with the speed $N$, at least in some finite interval of values of $e_{\min}$ around  $\overline{e_{\min}}$. This means that, asymptotically at large $N$, the PDF takes the form
\be  \label{Pe} 
{\cal P}(e) \asymp e^{- N {\cal L}(e)}\;,\;\;N\gg 1\;.
\ee 
where ${\cal L}(e)$ is the large deviation function (LDF). Computing this quantity explicitly, using a variant of the replica trick, together with establishing the range of value
$e$ where ${\cal L}(e)$ is finite is the main goal of this paper. In particular one expects that ${\cal L}(e=\overline{e_{\min}})=0$ such that typical and average value of $e_{\min}$ coincide $e_{\rm typ}=\overline{e_{\min}}$.

Let us first discuss the special case of the spherical $p=2$ spin model in a random magnetic field.
Let us recall that the spherical $p=2$ spin model without external field, i.e. $\Gamma=0$, stands out as very special. Indeed, the Hamiltonian in this case can be equivalently written as a quadratic form $H({\bf x})=-\frac{1}{2}\left({\bf x},J{\bf x}\right)$ , 
with $J$ being a $N\times N$ random matrix from the standard Gaussian Orthogonal Ensemble (GOE) \cite{kosterlitz1976spherical}. Hence, the ground state energy in this case has a simple alternative description: $e_{\min}=-\lambda^{(N)}_{\max}/2$  where $\lambda^{(N)}_{\max}$ is the maximum eigenvalue of a $N\times N$ GOE matrix, which as $N\to \infty$ tends to  $\lambda^{(\infty)}_{\max}=2$, implying $\overline{e_{\min}}=-1$ for this case (with e.g.
$g(q)=q^2/2$). This value can be shown to correspond to the simplest replica-symmetric solution to the Crisanti-Sommers variational problem, corroborating with the fact that 
any quadratic form over the sphere may have only two minima, trivially related by the symmetry ${\bf x}\to -{\bf x}$. 
Further, the classical results in random matrix theory allow to find explicitly the distribution of the ground state value $e_{\min}:=e$ which, in the large $N$ limit, take the following forms in the three different asymptotic regimes
\be
{\cal P}(e)\approx \begin{cases}
e^{-N {\cal L}(e)}&\;,\;\;e+1<0\;,\\
N^{2/3}{\cal F}_1'\left(-2N^{2/3}(e+1)\right)&\;,\;\;e+1=O(N^{-2/3})\;,\\
e^{-N^2 \Psi(e)}&\;,\;\;e+1>0\;.
\end{cases}
\ee
The typical fluctuations are described by ${\cal F}_1'(s)$, i.e. the GOE Tracy–Widom law \cite{TracyWidom1994GOE}. The corresponding
regime has a width $e+1=O(N^{-2/3})$.
The large
deviation regimes describe the fluctuations $e- \overline{e_{\min}}=e+1=O(1)$ with a speed $N$ for $e- \overline{e_{\min}}<0$ and a speed $N^2$ for $e- \overline{e_{\min}}>0$. The rate functions have been obtained explicitly
\cite{arous2001aging,deanmajumdar2006LDF,majumdar2014top,fyodorov2014topology}.

 With added magnetic field term $\Gamma>0$ in the covariance Eq.(\ref{cov}) the relation of the ground state energy $e$ for the $p=2$ spherical model to theory of random matrices becomes less transparent (see \cite{baik2021spherical}), although its limiting value $\overline{e_{\min}}$ is not difficult to find \cite{kosterlitz1976spherical,cugliandolo1995dynamics} as the corresponding optimization problem again allows only a replica-symmetric solution for any $\Gamma\ge 0$.
 As to the ensuing statistics of fluctuations in the large deviation regime for the free energy and for the ground state 
 energy $e$, both were a subject of substantial interest in recent years, starting from the physics paper by two of us \cite{fyodorov2014topology} followed by mathematics papers, first by \cite{dembo2015matrix}, and then by several others \cite{mehta2015spherical,baik2021spherical,landon2022fluctuations,kivimae2019critical,landon2020fluctuations,belius2022triviality}.
 In \cite{fyodorov2014topology} a variant of the replica trick was employed
 to calculate the LDF with the speed $N$, ${\cal L}(e)$, whose validity for $\Gamma>0$ was found to cover not only the range $e<\overline{e_{\min}}$, but extended also to a certain positive range of values $\overline{e_{\min}}<e<e_{c,{\rm RS}}$. For $e> e_{c,{\rm RS}}$ the LDF became complex, and it was suggested that this would correspond to a change in the 
 speed of large deviations for $e>e_{c,{\rm RS}}$. To a considerable surprise, the subsequent rigorous optimization procedure for random quadratic forms developed in \cite{dembo2015matrix} only confirmed  the suggested replica-based formula for the LDF in an interval 
 $-\infty<e<e_L$, with $\overline{e_{\min}}<e_L<e_{c,{\rm RS}}$, whereas a different expression was found to hold for LDF in $e\in[e_L,e_c)$, with $e_c\ne e_{c,{\rm RS}}$. Such an apparent, even if only partial, discrepancy between results of the replica-based calculation and the rigorous optimization is certainly puzzling, and poses a challenge to replica practitioners. Indeed, the collective experience accumulated in the theoretical physics community over several decades makes one to believe that the replica method, being inherently mathematically non-rigorous, is nevertheless trustworthy in results it produces when judiciously applied.  The present work stemmed in part from our wish to vindicate the trustworthiness of the replica trick in the large deviation setting, and to apply it for calculation of the LDF beyond the simplest $p=2$ case. 
 
In order to compute the large deviation function, we will follow the same replica procedure as employed in \cite{fyodorov2014topology} and 
compute first the cumulant generating function $\phi(s)$ of $e_{\min}$, resorting to the identity \footnote{See a nice historical exposition of this and related identities in \cite{pastore2019large}.}
\be \label{phi_def}
\phi(s)=\lim_{N\to \infty}\frac{1}{N}\ln\overline{e^{-N s e_{\min}}}=\lim_{N\to \infty}\frac{1}{N}\ln\lim_{\beta\to \infty}\overline{e^{-Ns\left(-\frac{1}{N\beta}\ln Z\right)}}=\lim_{N\to \infty}\frac{1}{N}\ln\lim_{\beta\to \infty}\overline{Z^{s/\beta}}\;,
\ee
Note that from the definition of $\phi(s)$ in Eq. \eqref{phi_def} and by a simple application of Jensen's inequality, the function $\phi(s)$ must be convex. Inserting the form \eqref{Pe} into \eqref{phi_def} we see that 
\be 
\phi(s) = - \min_{e \in \mathbb{R}}\left[ s e + {\cal L}(e) \right]\label{phi_leg_trans}
\ee 
so that the large deviation function ${\cal L}(e)$ is then obtained by inversion of the Legendre transform which can be written as
\be
{\cal L}(e)=-\min_{s\in \mathbb{R}}\left[s e+\phi(s)\right]\;.
\ee
Note that the average and typical ground-state energy can be obtained easily from $\phi(s)$ in the high-dimensional limit as
\be
e_{\rm typ}=\overline{e_{\min}}=-\phi'(0)\;.
\ee
In order to derive $\phi(s)$ we first compute using saddle point methods in the large $N$ limit of the integer moments of the partition sum $\overline{Z(\beta)^n}$, $n \in \mathbb{N}$ for arbitrary inverse temperature $\beta>0$. Using the framework of replica methods we are able to analytically continue its expression to the regime of interest for large deviations $n= s/\beta$ with fixed $s$. 

The following remark is appropriate here. In a relatively recent paper \cite{pastore2019large} the authors used not dissimilar 
technique to address the LDF for the free energy for the spherical $p$-spin model in a random magnetic field. They did not specifically consider the 
zero temperature limit, neither associated properties of the ground state, and mainly resorted to analyzing the corresponding equations numerically. We in contrast concentrate on zero temperature limit and present a very explicit analytical expressions for LDF in most of cases of interest. When attempting to compare some of the low temperature results and claims in \cite{pastore2019large} to our findings, we reveal some qualitative discrepancies in a few cases. Most importantly, our theory generically predicts the threshold value of the ground state energy $e_c$ such that for $e>e_c$ the associated LDF diverges, signalling of the change in the speed from $N$ to higher powers. We call this threshold a "wall".  The paper \cite{pastore2019large} seems to have no mentioning of such phenomenon. Another point worth mentioning is that although the authors seem to be unaware of the recent activity around $p=2$ Large Deviations with magnetic field that we discussed in detail above, they do provide a very nice overview of early physics paper on some aspects relevant to Large Deviations, especially on convexity issues as discussed by Rammal in \cite{Rammal_PhD}. It also contained useful references to several attempts to study Large Deviations of the spin glass free energy in a more recent physics literature, such as e.g. \cite{andreanov2004large,rivoire2005cavity} and \cite{parisi2008large,parisi2009phase,parisi2010universality} which together with a few other related works, like \cite{monthus2010matching} and \cite{malatesta2019fluctuations} we find to be worth of mentioning here for the benefit of a reader interested in this problematic.

The paper has the following structure. As there are a number of results and we want to discuss them extensively, we provide in the main text only the minimal details for the derivation of our results and relegate the full detailed derivation to the appendices. In Section \ref{sec:main} we display the main results for the large-deviation function (LDF) ${\cal L}(e)$ of speed $N$ of the ground-state energy, first
for the case $p=2$ and then for a general mixture model of spherical spins. 
In Section \ref{sec:derivation} we present the derivation of the optimisation problem allowing to obtain the cumulant generating function (CGF) $\phi(s)$, i.e. the generalisation of the Crisanti-Sommers formula, from the expression of the moments of the partition function. In Section \ref{sec:conclusion} we discuss our results and mention some open problems. In Appendix \ref{app:typical}, we recall the known results for the typical and average ground-state energy $e_{\rm typ}(\Gamma)=\overline{e_{\min}}$. In Appendix \ref{app:CGF_res}, we summarise the main results on the cumulant generating function $\phi(s)$. In Appendix \ref{app:analysis}, we provide the detailed analysis of the CGF and LDF in the replica-symmetric (RS), full replica-symmetry broken (FRSB) and one-step replica symmetry broken (1RSB) ansatz. In Appendix \ref{app:stab}, we provide the criteria for stability of the RS and 1RSB ans\"atze. In Appendix \ref{app:comp_min}, we compute explicitly the annealed complexity of minima at fixed energy and compare its expression to the LDF.

\section{Main results}
\label{sec:main} 

\subsection{Spherical $2$-spin model in a random magnetic field}

\begin{figure}
    \centering
    \includegraphics[width=0.6
    \textwidth]{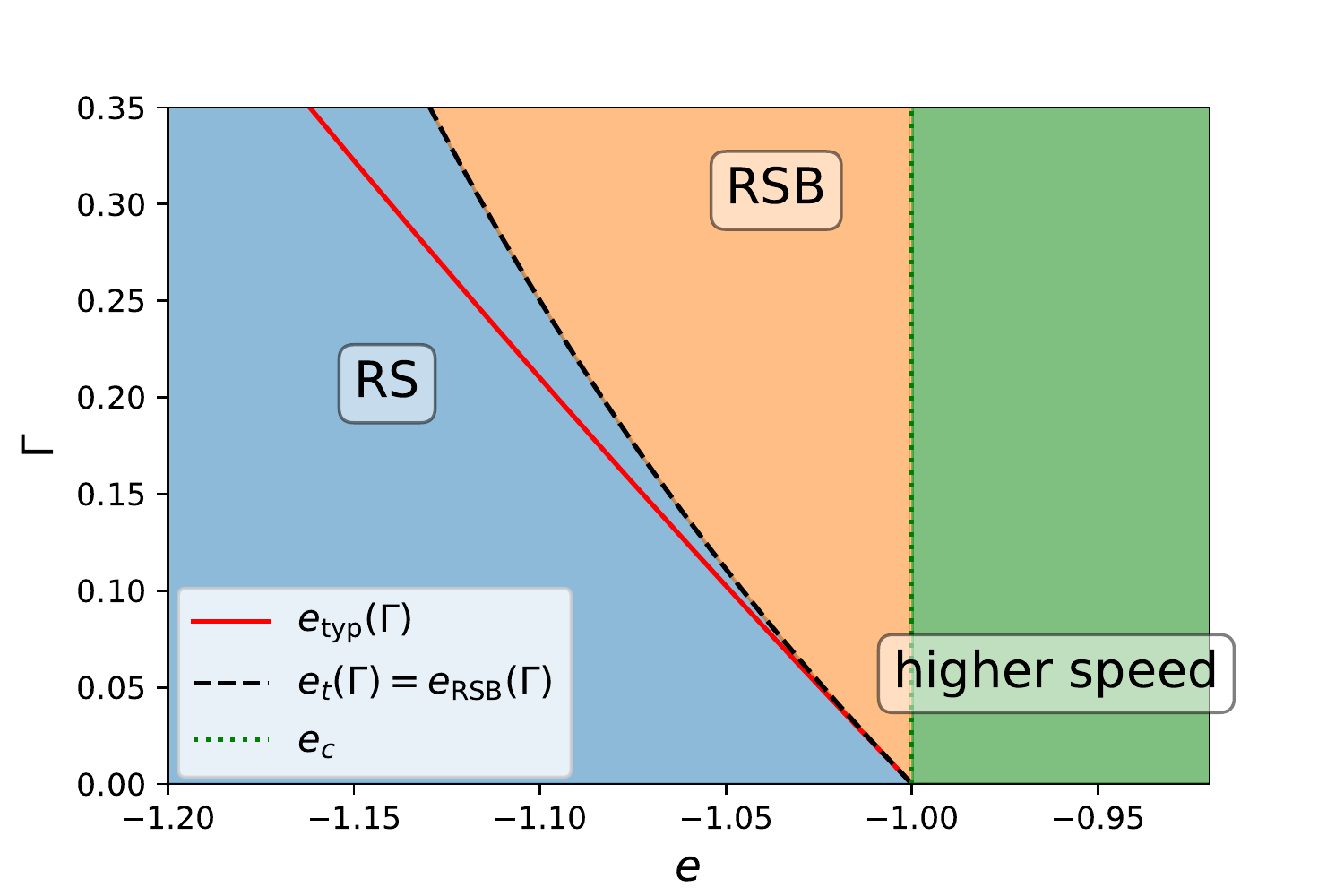}
    \caption{Phase diagram of the LDF ${\cal L}(e)$ in Eq. \eqref{res_2_spin} for the spherical $2$-spin model in the plane $(e,\Gamma)$. The dashed black line $e_t(\Gamma)=e_{\rm RSB}(\Gamma)$ shows the transition separating the FRSB phase (in orange) and the RS phase (in blue) determined by the stability criterion of the RS phase. The red line corresponds to the typical value $e_{\rm typ}(\Gamma)$, which is always smaller than $e_{\rm RSB}(\Gamma)$ for this model. For $e>e_c$ (the energy $e_c$ of the "wall" being marked by the vertical green dotted line), the LDF at speed $N$ is infinite and one expects that a LDF with higher speed will describe the green region.}
    \label{fig:phase_diag_2_spin}
\end{figure}

\begin{figure}
    \centering
    \includegraphics[width=0.45
    \textwidth]{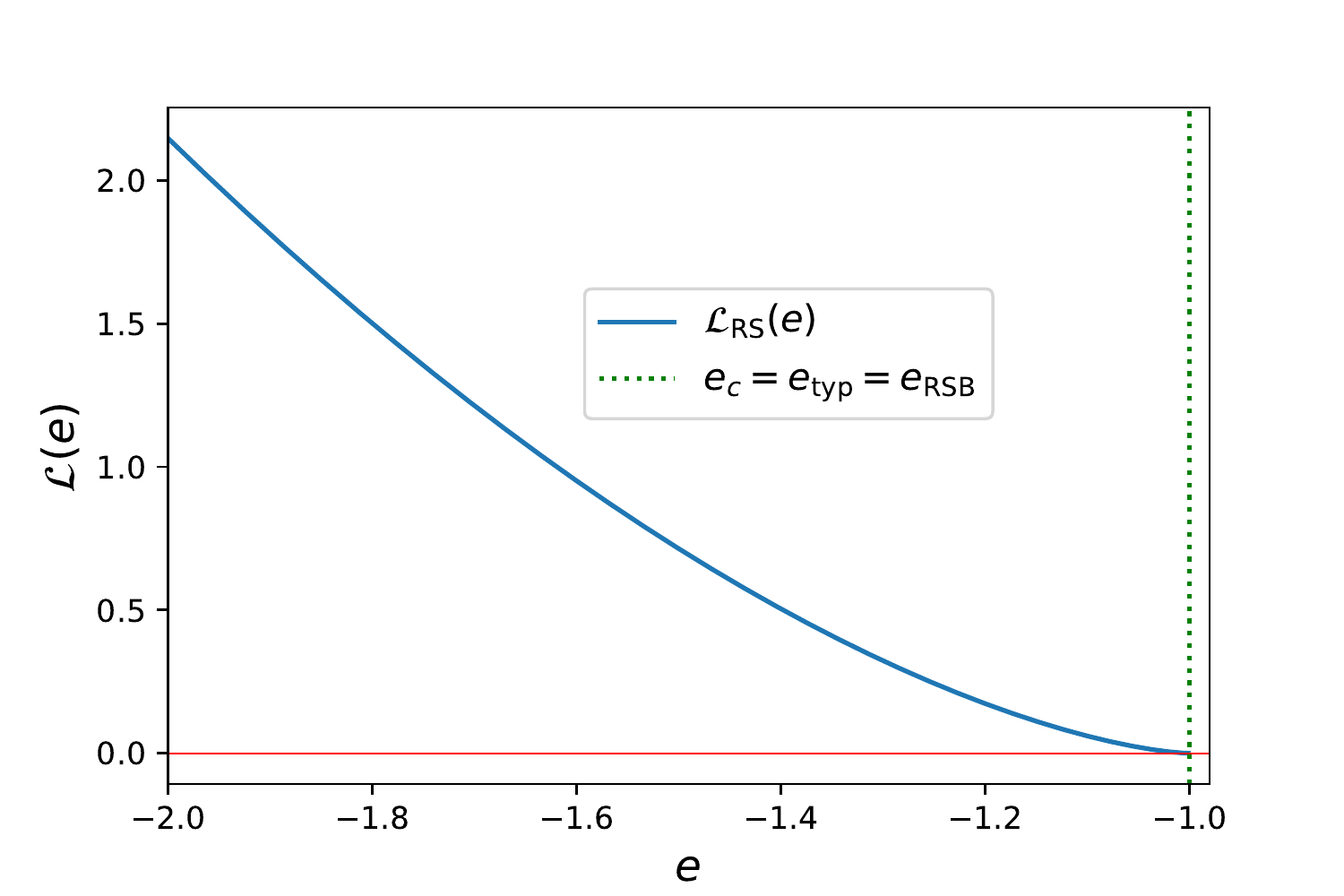}
    \includegraphics[width=0.45
    \textwidth]{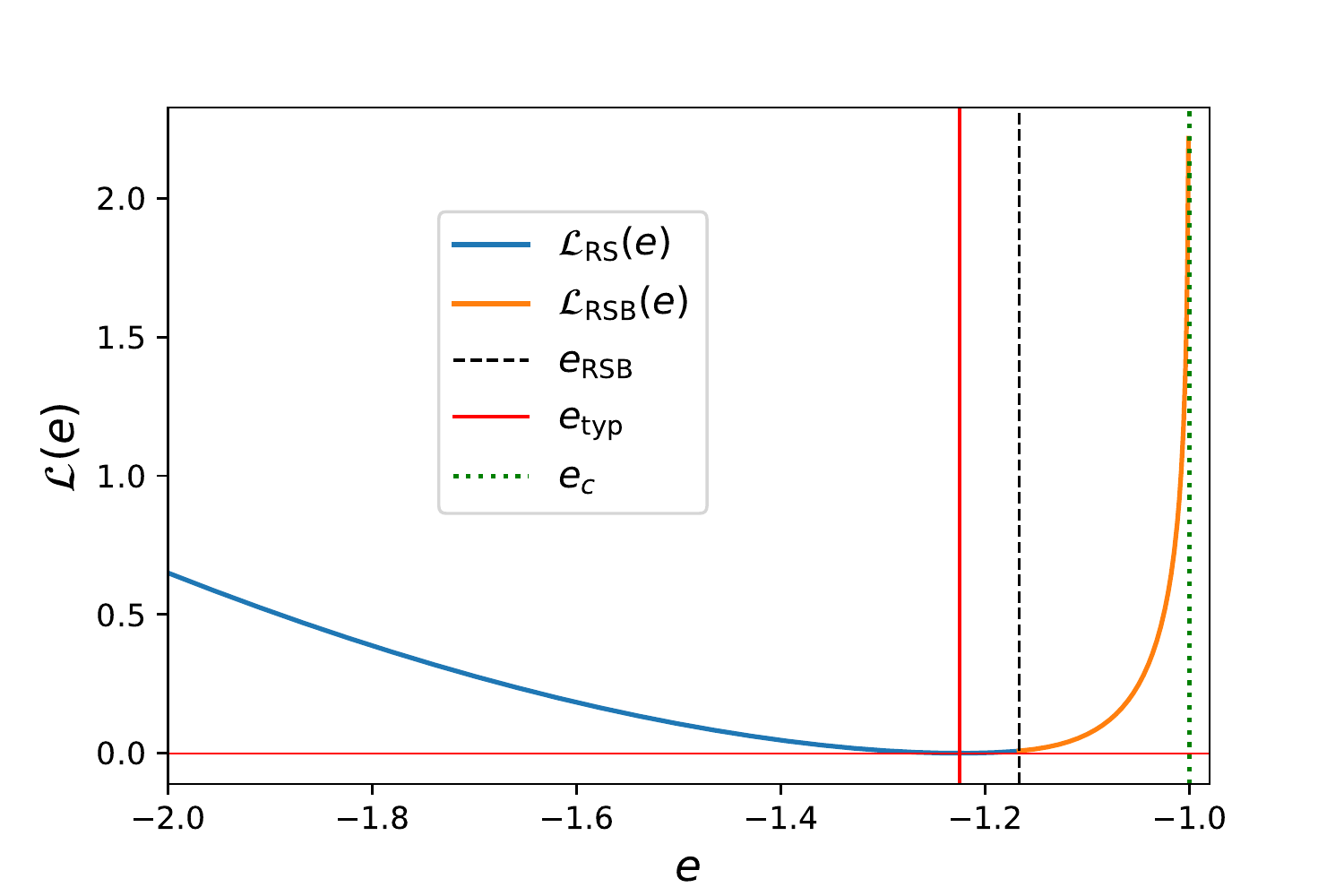}
    \caption{Plot of the large deviation function ${\cal L}(e)$ in Eq. \eqref{res_2_spin} for the spherical $2$-spin model with a random field of variance $\Gamma=0$ (left) and $\Gamma=1/2$ (right).
    For $\Gamma=1/2$, the vertical black dashed line represents the value of $e_{\rm RSB}$ separating on its left the RS regime where ${\cal L}(e)={\cal L}_{\rm RS}(e)$ plotted in blue and on its right the RSB regime where ${\cal L}(e)={\cal L}_{\rm RSB}(e)$ plotted in orange. The solid red line marks the value $e_{\rm typ}$ and the dotted green line, the value of $e_c$ where the LDF diverges logarithmically (see Eq. \eqref{e_quad}). For $\Gamma=0$, all these values coincide and the RSB phase is limited to the single point $e=e_{\rm typ}$.}
    \label{fig:LDF_quad}
\end{figure}

In this case, corresponding to $f(q)=q^2/2+\Gamma\,q$ we obtain the following expression for the LDF 
\begin{align}
    {\cal L}(e)=\begin{cases}\displaystyle
    {\cal L}_{\rm RS}(e)=-\frac{ \displaystyle e \left(e \Gamma+(1+\Gamma)\sqrt{e^2-\frac{1+2\Gamma}{1+\Gamma}}\right)}{1+2\Gamma}-\ln\left(\frac{-e\sqrt{1+\Gamma} +\sqrt{e^2(1+\Gamma)-(1+2\Gamma)}}{1+2\Gamma}\right)&\;,\;\;e\leq e_{t}\;,\\
    \\
    \displaystyle {\cal L}_{\rm RSB}(e)=-\frac{(2(1+e)+\Gamma)(2(1+e)+ 3\Gamma)}{4\Gamma^2} - \frac{1}{2}\ln\left(-\frac{2 (1+e)}{\Gamma}\right)&\;,\;\;e_{t}\leq e\leq e_c\;,
    \end{cases}\label{res_2_spin}
\end{align}
where the typical energy $e_{\rm typ}(\Gamma)$, the transition energy coinciding with the RSB threshold  $e_t(\Gamma)=e_{\rm RSB}(\Gamma)$ and the critical threshold $e_c$ (acting as a "wall") are given by
\be
e_{\rm typ}(\Gamma)=-\sqrt{1+\Gamma}\;,\;\;e_{\rm RSB}(\Gamma)=-\left(1+\frac{\Gamma}{2(1+\Gamma)}\right)\;,\;\;e_{c}=-1\;,\label{e_quad}
\ee
with $e_{\rm typ} < e_{\rm RSB} < e_c$ for $\Gamma>0$. A phase diagram of the LDF is plotted in Fig. \ref{fig:phase_diag_2_spin} and a plot of the LDF ${\cal L}(e)$ is provided in Fig. \ref{fig:LDF_quad} for $\Gamma=0$ and $\Gamma=1/2$. The phase transition is of third order in the sense that the function ${\cal L}(e)$ as well as its first two derivatives ${\cal L}'(e)$ and ${\cal L}''(e)$ are continuous at the transition $e\to e_{\rm RSB}$. As $e\to e_{c}$, the LDF diverges logarithmicaly and one naturally expects that a large deviation function at a larger speed should describe the atypical fluctuations for $e\geq e_{c}$. Note that at zero magnetic field ($\Gamma=0$), the RSB phase of ${\cal L}(e)$ is limited to a single point $e=e_{\rm RSB}$, where it coincides with the RS expression. A plot of the LDF is given in Fig. \ref{fig:LDF_quad} for $\Gamma=0,1/2$. 

The expression of ${\cal L}(e)$ in the first line in Eq. \eqref{res_2_spin} was first obtained from a replica symmetric ansatz in \cite{fyodorov2014topology} and confirmed by a rigorous result in \cite{dembo2015matrix}.
Neither the threshold $e_{\rm RSB}$ nor the second line in Eq. \eqref{res_2_spin} appeared in Ref. \cite{fyodorov2014topology}, but were discovered by Dembo and Zeitouni \cite{dembo2015matrix} using a rigorous approach to the problem of optimisation of random quadratic form over the sphere. It therefore remained a challenge to derive this expression within the framework of replica computations. It turned out that the transition at $e=e_{\rm RSB}$ can be accounted for as the consequence of the instability of the replica-symmetric solution of Ref. \cite{fyodorov2014topology} and can be correctly described using the replica-symmetry breaking ansatz. To the best of our knowledge, this is the first manifestation of a replica-symmetry breaking phenomenon for the spherical pure $p=2$ spin model. Indeed, it is generally believed that RSB phenomena reflect a positive complexity of the topological landscape, i.e. an exponentially large in $N$ number of stationary points, which is not the case for the $2$-spin model: it is strictly less than $2 N$ for any realisation and equal to two for $\Gamma=O(1)$ in a typical realisation
\cite{fyodorov2014topology,cugliandolo1995dynamics}. Note that this model is marginal in the sense of RSB such that the cumulant generating function (the Legendre transform of Eq. \eqref{res_2_spin}) can be retrieved either from a FRSB or a 1RSB ansatz. 

Before analysing in more details the expression above, let us comment in more detail on the nature of the transition observed in the paper by Dembo and Zeitouni and how it can be understood in relation to the RSB transition observed here. In their paper, Dembo and Zeitouni compute two distinct LDF for the ground state energy of the $2$-spin spherical spin model $H({\bf x})=-\sum_{i,j}J_{ij}x_i x_j-\sum_i h_i x_i$: as $N$ increases, they consider both the LDF for a fixed sequence of random matrices $J$ (such that the highest and lowest eigenvalues of $J$ converge to $\pm 2$ as $N\to \infty$) as well as the LDF averaged over the GOE distribution of $J$ \footnote{Note that Dembo and Zeitouni use the notation 'quenched' for a fixed realisation of $J$ and 'annealed' for results averaged over the distribution of $J$ but these notations do not exactly cover the standard use of these terms in the disordered system community.}. In each case, the LDF is computed by averaging with respect to the normal iid components $h_i$'s of the random field and can be expressed as an optimisation problem. Note that the LDF for fixed realisation of $J$ seems out of reach of our replica computation. They show in particular that, as $e$ increases, both LDFs display a freezing transition at $e=e_{\rm RSB}$ in the associated optimisation problem (that we identify here as the RSB transition). The LDF with fixed realisation displays another freezing transition at energy $e=e_U<e_{\rm RSB}$. In addition they show that the two LDF coincide in the range of energy $e\in [e_U,e_{\rm RSB}]$ between the two freezing transitions. If both LDFs can be computed for more involved spherical spin-glass models, we expect that the maximum energy for which the two LDF match identifies with the energy $e_{\rm RSB}$ at which the RSB transition occurs, providing a robust criterion to identify this point without requiring replica methods.

Let us return to the analysis of the LDF around the typical ground-state energy $e_{\rm typ}$. For fixed $\Gamma>0$, the LDF is quadratic in the vicinity of $e_{\rm typ}(\Gamma)=-\sqrt{1+\Gamma}$ with
\be
{\cal L}(e)=\frac{\left(e+\sqrt{1+\Gamma}\right)^2}{\Gamma}+\frac{(1+\Gamma)^{3/2}}{3\Gamma^3}\left(e+\sqrt{1+\Gamma}\right)^3+O\left(e+\sqrt{1+\Gamma}\right)^4\;.
\ee
For $\Gamma=0$, one finds ${\cal L}(e)=-e\sqrt{e^2-1}-\ln(-e+\sqrt{e^2-1})$ for $e<e_{\rm typ}(\Gamma=0)=e_c=-1$, which as noted in \cite{fyodorov2014topology} coincide under the mapping  $e=-\lambda_{\max}/2$, with the LDF of the largest eigenvalue of a GOE matrix. This function can thus be obtained solely from the RS ansatz.

\begin{figure}
    \centering
    \includegraphics[width=0.99\textwidth]{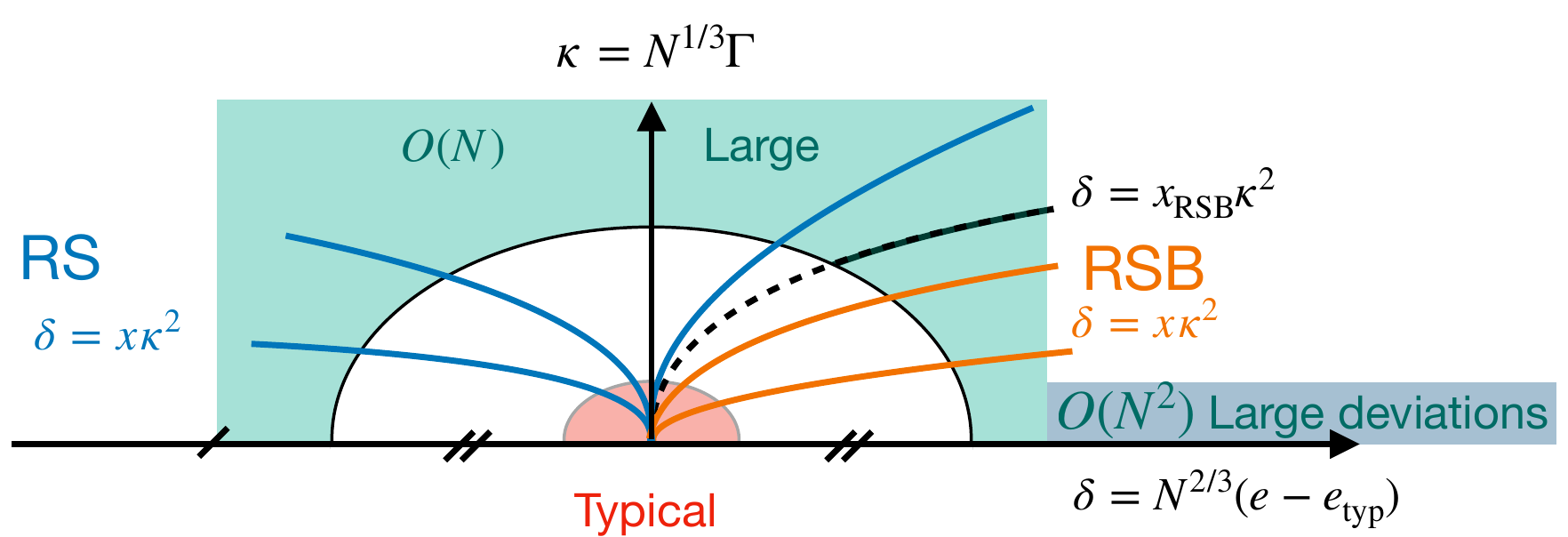}
    \caption{Plot of the different scaling regimes at large $N$ describing our conjectures for the distribution of the typical fluctuations { $p_{\kappa}(\delta)$} for the spherical $2$-spin model with magnetic field in Eq. \eqref{conj_2_spin} and its matching for $\delta, \kappa \to \infty$ with our results for the large deviations. The large deviation region for $\delta =O(N^{2/3})$ and $\kappa = O(N^{1/3})$ is indicated in green, the semi-circle
    represents schematically the matching region.
    The side $\delta>0$ was described in \cite{fyodorov2014topology}. The major novelty is the RSB line, indicated as a solid line
    inside the large deviation regime where it is a true transition, and a dashed line in the typical region where
    it may signal a (smooth) change of behaviour. Finally the region $\kappa=0$ and $\delta<0$ with $|\delta| \sim N^{2/3}$
    is described by the Coulomb gas and has a $N^2$ speed large deviation (it is "on the other side of the wall").
    The vicinity of the wall on the speed $N$ RSB regime is described by yet another scaling behaviour see Eq \eqref{L_2_spin_scaling_G_2}. 
    }
    \label{fig:delta_kappa_2}
\end{figure}

In the limit $\Gamma\to 0$, there is an interesting crossover which involves the RSB branch as well. In this regime, the energy differences scale in a distinct manner as $\Gamma\to 0$, i.e.
\begin{align}
    e_{c}-e_{\rm typ}&=\frac{\Gamma}{2}+O(\Gamma^2)\\
    e_{\rm RSB}-e_{\rm typ}&=\frac{3\Gamma^2}{8}+O(\Gamma^3)\;.
\end{align}
This naturally leads to two different scaling regimes and functions. The first one describes the regime  $e-e_{\rm typ}=O(\Gamma^2)$ and the LDF takes the scaling form
\be \label{firstscaling}
{\cal L}(e)\approx\Gamma^3 F\left(\frac{e-e_{\rm typ}}{\Gamma^2}\right)\;,\;\;F(x)=\begin{cases}
\displaystyle \frac{2}{3}\left(\sqrt{1-2x}-1+x(3-2\sqrt{1-2x})\right)\;,\;\;x\in (-\infty,x_{\rm RSB}]\;,\\
&\\
\displaystyle \frac{(1+8 x)^3}{384}\;,\;\;x\in [x_{\rm RSB},+\infty)\;,
\end{cases}
\ee
with $x_{\rm RSB}=3/8$ and the scaling function has continuous two first derivatives at $x_{\rm RSB}$, which is the mark of third order transition. The first branch, corresponding to the RS phase, was originally derived in \cite{fyodorov2014topology} (see Eq. (47) there). It is quite interesting to note that the full scaling function $F(x)$ in (\ref{firstscaling}) now extends for $x\to + \infty$, while it was thought to be limited to $x<1/2$ in its first derivation. Such a scaling function then has the following asymptotic behaviours
\be
F(x)=\begin{cases}
\displaystyle \frac{2}{3}(-2x)^{3/2}&\;,\;\;x\to -\infty\;,\\
&\\
\displaystyle \frac{1}{6}(2 x)^{3}&\;,\;\;x\to +\infty\;.
\end{cases}
\ee
As was discussed in \cite{fyodorov2014topology}, the asymptotics for $x\to -\infty$ agrees with the behaviour of the right-tail of the $\beta=1$ Tracy-Widom distribution (TW), including the $2/3$ coefficient of the $3/2$ power. It is remarkable that the new RSB branch exhibits a cubic behaviour for $x\to +\infty$, which is reminiscent of the cubic left tail of the Tracy-Widom distribution. Note however that the coefficient of the TW left tail is $1/24$ rather than the $1/6$ found here. We explain below the implications of this remark.
As conjectured in \cite{fyodorov2014topology}, and rigorously shown recently \cite{baik2021spherical,landon2020fluctuations}, in presence of a random field of variance $\Gamma\sim N^{-1/3}$, the distribution of the {\it typical fluctuations} of the ground-state energy of the spherical $p=2$ spin model converges, as $N\to \infty$, to a 
a universal one parameter family of distributions denoted $p_\kappa(\delta)$, i.e. one has
\be
{\cal P}(e)\to N^{2/3}\,p_{\kappa=N^{1/3}\Gamma}\left(\delta=N^{2/3}(e-e_{\rm typ})\right)\;,\;\;N\to \infty\;,\label{conj_2_spin}
\ee
with in particular $p_0(\delta)$ coinciding after a rescaling $\delta=-\zeta/2$ with the TW ${\cal F}_1'(\zeta)$. One expects that there will be a matching of the tails of the distributions $p_{\kappa}(\delta)$ within this family for large $\kappa\to \infty$ and $\delta\to \infty$ with the asymptotic behaviour of the large-deviation form ${\cal P}(e)\asymp e^{-N {\cal L}(e)}$. A sketch of the different scaling regimes discussed below is given in Fig. \ref{fig:delta_kappa_2}. The first scaling regime corresponds to 
\be
p_{\kappa}(\delta)\sim e^{-\kappa^3 F\left(x=\frac{\delta}{\kappa^2}\right)}\;,\;\;\kappa\to \infty\;,\;\;|\delta|\to \infty\;.\label{scal_reg}
\ee
The $\delta>0$ side has been analysed in details in \cite{fyodorov2014topology}. The large $x$ behaviour of the scaling function, $F(x) \simeq \frac{2}{3} x^{3/2}$,
is compatible with the conjecture that 
$p_{\kappa}(\delta)\sim e^{- c_+(\kappa) (-2\delta)^{3/2} }$ 
at any fixed $\kappa$ and that, by matching, $c_+(\kappa) \simeq 2/3=c_+(0)$ at large $\kappa$.
 On the other hand, on the $\delta<0$ side, in the regime $-\delta/\kappa^2\to +\infty$, matching to the $x \to - \infty$ 
cubic behaviour of $F(x)$ on the RSB side leads to 
$
p_{\kappa}(\delta)\sim e^{-\frac{(-2\delta)^3}{6\kappa^3}}
$
which suggests the following conjecture: the left tail of the distribution $p_\kappa(\delta)$ for fixed arbitrary $\kappa=O(1)$ is cubic and takes the form
\be
p_{\kappa}(\delta)\sim e^{- c_-(\kappa)(-2\delta)^3}\;,\;\;\delta\to -\infty\;,\label{Conj_p_2_d}
\ee
with $c_-(0)=1/24$ and  $c_-(\kappa)\sim 1/(6\kappa^3)$ for $\kappa\to +\infty$. This conjecture provides a natural explanation to the mismatch of coefficients of the cubic tail with the TW noted above.

Finally, let us briefly discuss the second scaling regime $e-e_c=O(\Gamma)$ where the Large Deviation Function reads
\be
{\cal L}(e)\approx G\left(\frac{e-e_c}{\Gamma}\right)\;,\;\;G(z)=\begin{cases}
\displaystyle 0\;,\;\;z\in (-\infty,z_{\rm RSB}]\;,\\
&\\
\displaystyle -\frac{1}{4}\left(3+8z+4z^2+2\ln(-2z)\right)\;,\;\;z\in [z_{\rm RSB},0]\;,
\end{cases}\label{L_2_spin_scaling_G_2}
\ee
with $z_{\rm RSB}=z_{\rm typ}=-1/2$. Note that $G(z)>0$ for $z>z_{\rm RSB}$ and vanishes cubically for $z\to z_{\rm RSB}$ as $G(z)\sim \frac{4}{3}(z-z_{\rm RSB})^3$, again a third order transition. Also note that the limit $x \to +\infty$ of the 
scaling function $F(x) \sim \frac{4}{3} x^3$ matches correctly the limit $z\to z_{\rm RSB}=z_{\rm typ}$
of the function $G(z)$. In the case $z<z_{\rm RSB}$, there are sub-leading corrections to the scaling of order $\Gamma^{3/2}$ to Eq. \eqref{L_2_spin_scaling_G_2} described by
\be
{\cal L}(e)\approx \Gamma^{3/2}\, \hat G\left(\frac{e-e_c}{\Gamma}\right)\;,\;\;\hat G(z)=\frac{2}{3}\left(-2z-1\right)^{3/2}\;,\;\;z<z_{\rm RSB}=\frac{1}{2}\;.
\ee

\subsection{Large deviations of the ground state energy for a generic mixture model}
The cumulant generating function (CGF) can be shown to be obtained as 
\be \label{phiscases} 
\phi(s)=\begin{cases}
\displaystyle\max_{z(q),q_0,v}\Phi(s)&\;,\;\;s<0\;,\\
0&\;,\;\;s=0\;,\\
\displaystyle\min_{z(q),q_0,v}\Phi(s)&\;,\;\;s>0\;,
\end{cases}
\ee
where the function $z(q)$ is a real non-decreasing function, satisfying $z(q)=s$ for $q\in[0,q_0)$ and $z(q)\geq s$ for $q\in[q_0,1]$, while the functional $\Phi(s)$ reads
\begin{align}
    \Phi(s)=&\frac{s}{2}\left[s f(q_0)+v f'(1)
    +\int_{q_0}^{1} dq\,z(q)\,f'(q)\right]+\frac{1}{2}\left[\ln\left(s q_0+ v+\int_{q_0}^{1}dq\,z(q)\right)-\ln\left( v+\int_{q_0}^{1}dq\,z(q)\right)\right]\label{phi_z_res}\\
    &+\frac{s}{2}\int_{q_0}^{1}\frac{dq}{\displaystyle v+\int_{q}^{1}dr\,z(r)}\nn\;.
\end{align}
Note that in Eq. \eqref{phiscases}, we have used heuristically that the number of degrees of freedom is $n(n-1)/2$ in the problem,
i.e. $-s/\beta$ in our limit of interest and thus we expect that the correct optimisation problem is solved by taking the maximum of the function $\Phi(s)$ when $s<0$ and the minimum for $s>0$.

From this general expression, one can recover the typical value of the ground state energy (see Appendix \ref{app:typical} for a detailed discussion) as
\be
e_{\rm typ}(\Gamma)=\overline{e_{\min}}=-\phi'(0)\;.
\ee
The rescaled variance can similarly be computed as 
\be
{\cal V}_{\min}(\Gamma)=\lim_{N\to \infty} N\mbox{Var}(e_{\min})=\phi''(0)=g(q_{\rm typ}(\Gamma))+\frac{\Gamma-g'(q_{\rm typ}(\Gamma))}{2}\;,\label{res_gen_var}
\ee
where $q_{\rm typ}(\Gamma)$ is the optimal value of the parameter $q_0$ in the case where $s=0$. For any magnetic field 
\be
\Gamma\geq \Gamma_{\rm RSB}=g''(1)-g'(1)\;,
\ee
the typical energy belongs to the RS phase such that $e_{\rm typ}(\Gamma\geq \Gamma_{\rm RSB})=-\sqrt{g'(1)+\Gamma}$ and $q_{\rm typ}(\Gamma\geq \Gamma_{\rm RSB})=1$.

Finally, as mentioned in the introduction, the large deviation function (LDF) is obtained by taking the Legendre transform of the CGF in Eq. \eqref{phiscases}, i.e.
\be
{\cal L}(e)=-\min_{s\in \mathbb{R}}\left[s e+\phi(s)\right]\;.\label{leg_trans_res}
\ee
Interestingly, one can give a more explicit expression for ${\cal L}(e)$, valid at any level of replica-symmetry breaking:
\be
{\cal L}(e)=\frac{s_*^2}{4}\left( 2f(q_0^*)-q_0^* f'(q_0^*)\right)+\frac{s_* q_0^* f'(q_0^*)}{4}\sqrt{\frac{4}{ q_0^* f'(q_0^*)}+s_*^2}-\ln\left[\frac{\sqrt{q_0^* f'(q_0^*)}}{2}\left(s_*+\sqrt{\frac{4}{ q_0^* f'(q_0^*)}+s_*^2}\right)\right]\;.
\ee
The parameters $q_0^*$ and $s_*$ take values obtained by solving the optimisation problems in Eqs. \eqref{phiscases} and \eqref{leg_trans_res}, and eventually depend on the value of $e$ and on the covariance function $f(q)$ (notably, on the value of $\Gamma$). Expressions for their values  depend on the level of replica-symmetry breaking and  will be given explicitly in the corresponding sections below.

In order to obtain more explicit results, one needs to consider the different regimes of replica-symmetry breaking. There are two relevant scenarios for the emergence of a replica-symmetry broken solution in the optimisation problem for the cumulant generating function above:
\begin{itemize}
    \item If the quadratic form obtained by expanding $\Phi(s)$ around its replica-symmetric saddle-point solution  
    \be
    A_{(ab),(cd)}=\left.\frac{\delta^2 \Phi_n(Q)}{\delta Q_{ab}\delta Q_{cd}}\right|_{Q=Q_{\rm RS}^*}\label{stab_crit_res}
    \ee
    becomes unstable, corresponding to (at least) one of its eigenvalues being positive. This is in particular the mechanism driving the transition to the RSB solution in the large-deviations of the spherical $2$-spin model discussed above.
    \item If the replica-symmetry broken solution is stable, and gives a maximum value of $\Phi(s)$ for $s<0$ (resp. minimum value for $s>0$) larger (resp. smaller) than the value obtained within the replica-symmetric scheme.
\end{itemize}
The criterion determining the type of RSB solution is the same for the cumulant generating function as for the typical value of the ground-state, and involves the Schwarzian derivative of the covariance function:
\be
{\cal S}\left[f'\right](q)={\cal S}\left[g'\right](q)=\frac{f^{(4)}(q)}{f''(q)}-\frac{3}{2}\left(\frac{f^{(3)}(q)}{f''(q)}\right)^2\;.
\ee
Namely, if ${\cal S}\left[f'\right](q)$ is positive for any value $q\in[0,1]$, the solution will be obtained within the Full-RSB scheme, while if it is negative for any value $q\in[0,1]$, the  solution is obtained within the 1RSB scheme. If ${\cal S}\left[f'\right](q)$ changes sign on the interval $q\in [0,1]$, the solution is expected to be given by a combination of FRSB solution with one involving a finite number of replica-symmetry breaking steps. In this article, we will only consider the case where the Schwarzian derivative is of the same sign over the whole interval $q\in[0,1]$.

Let us now discuss the various regimes of replica-symmetry breaking separately.

\subsubsection{Replica symmetric regime} 

In the RS regime, the CGF is obtained by solving the optimisation problem in Eq. \eqref{phiscases} where one imposes that $z(q)=s$ for all $q\in[0,1]$ (or equivalently $q_0=1$). Considering both the saddle-point equation obtained by optimisation of $\Phi(s)$ and the Legendre-transform in Eq. \eqref{leg_trans_res}, the optimal parameters $s_*$ and $v_*$ are determined explicitly as a function of $e$ in that phase as
\be
v_*=-\frac{e+\sqrt{e^2-e_{c,{\rm RS}}^2}}{2(g'(1)-g(1))}\;,\;\;s_*=\frac{1}{2(g'(1)-g(1))}\left[2e+\frac{g'(1)+\Gamma}{g(1)+\Gamma}\left(-e+\sqrt{e^2-e_{c,{\rm RS}}^2}\right)\right]\;,
\ee
where the critical RS energy reads
\be
e_{c,{\rm RS}}(\Gamma)=-2\sqrt{f(1)\left(1-\frac{f(1)}{f'(1)}\right)}=-2\sqrt{(g'(1)-g(1))\frac{g(1)+\Gamma}{g'(1)+\Gamma}}\;.
\ee
Note that $v_*$ and $s_*$ become complex for energies $e>e_{c,{\rm RS}}(\Gamma)$ but as we will now show, such energies cannot be reached within the RS scheme. The criterion of stability obtained from the negativity of the eigenvalues of the quadratic form in Eq. \eqref{stab_crit_res} imposes that the RS expression is stable for (see Appendix \ref{app:stab} for details)
\be
\Lambda_{\rm RS}=g''(1)-\frac{1}{v_*^2}\leq 0\;,\;\;{\rm thus}\;\;e\leq e_{\rm RSB}(\Gamma)=-\frac{1}{\sqrt{g''(1)}}\left[g'(1)-g(1)+g''(1)\frac{g(1)+\Gamma}{g'(1)+\Gamma}\right]\;,\label{e_RSB}
\ee
where $e_{\rm RSB}(0)<0$ and $\partial_{\Gamma}e_{\rm RSB}<0$ (see Eq. \eqref{e_RSB_proof}). Using the identity $e_{\rm RSB}^2-e_{c,{\rm RS}}^2=\frac{1}{g''(1)}\left[g(1)-g'(1)+g''(1)\frac{g(1)+\Gamma}{g'(1)+\Gamma}\right]^2$, one obtains that $e_{\rm RSB}(\Gamma)\leq e_{c,{\rm RS}}(\Gamma)$ for any value of $\Gamma$, ensuring that $v_*$ and $s_*$ are real parameters in the domain of stability of the RS solution.  As described above, this criterion is necessary but not sufficient to ensure that the LDF is indeed given by its RS expression. It ensures nonetheless that the LDF is not correctly reproduced by the RS expression for any $e\geq e_{\rm RSB}(\Gamma)$.

Using the RS scheme to compute the average (and typical) value of the ground-state $e_{\rm typ}(\Gamma)$, one obtains the explicit expression (see Appendix \ref{app:typical})
\be
e_{\rm typ}(\Gamma)=\overline{e_{\min}}=-\sqrt{g'(1)+\Gamma}\;,\;\;\Gamma\geq \Gamma_{\rm RSB}=g''(1)-g'(1)\;,\label{e_typ_RS_res}
\ee
where the value of $\Gamma_{\rm RSB}$ can simply be obtained by ensuring marginality of the stability criterion, i.e. $e_{\rm typ}(\Gamma_{\rm RSB})=e_{\rm RSB}(\Gamma_{\rm RSB})$.
The rescaled variance of the ground state energy is obtained from Eq. \eqref{res_gen_var} using the fact that the parameter $q_0=1$ is frozen in the RS phase. One can thus replace $q_{\rm typ}(\Gamma\geq \Gamma_{\rm RSB})=1$ and obtain
\be
{\cal V}_{\min}(\Gamma)=\lim_{N\to \infty} N\mbox{Var}(e_{\min})=g(1)+\frac{\Gamma-g'(1)}{2}\;,\;\;\Gamma\geq \Gamma_{\rm RSB}=g''(1)-g'(1)\;,\label{V_min_RS_res}
\ee
which is strictly positive for any $\Gamma>g'(1)-2g(1)$ with  $\Gamma_{\rm RSB}\geq g'(1)-2g(1)$.

\begin{figure}
    \centering
    \includegraphics[width=0.58
    \textwidth]{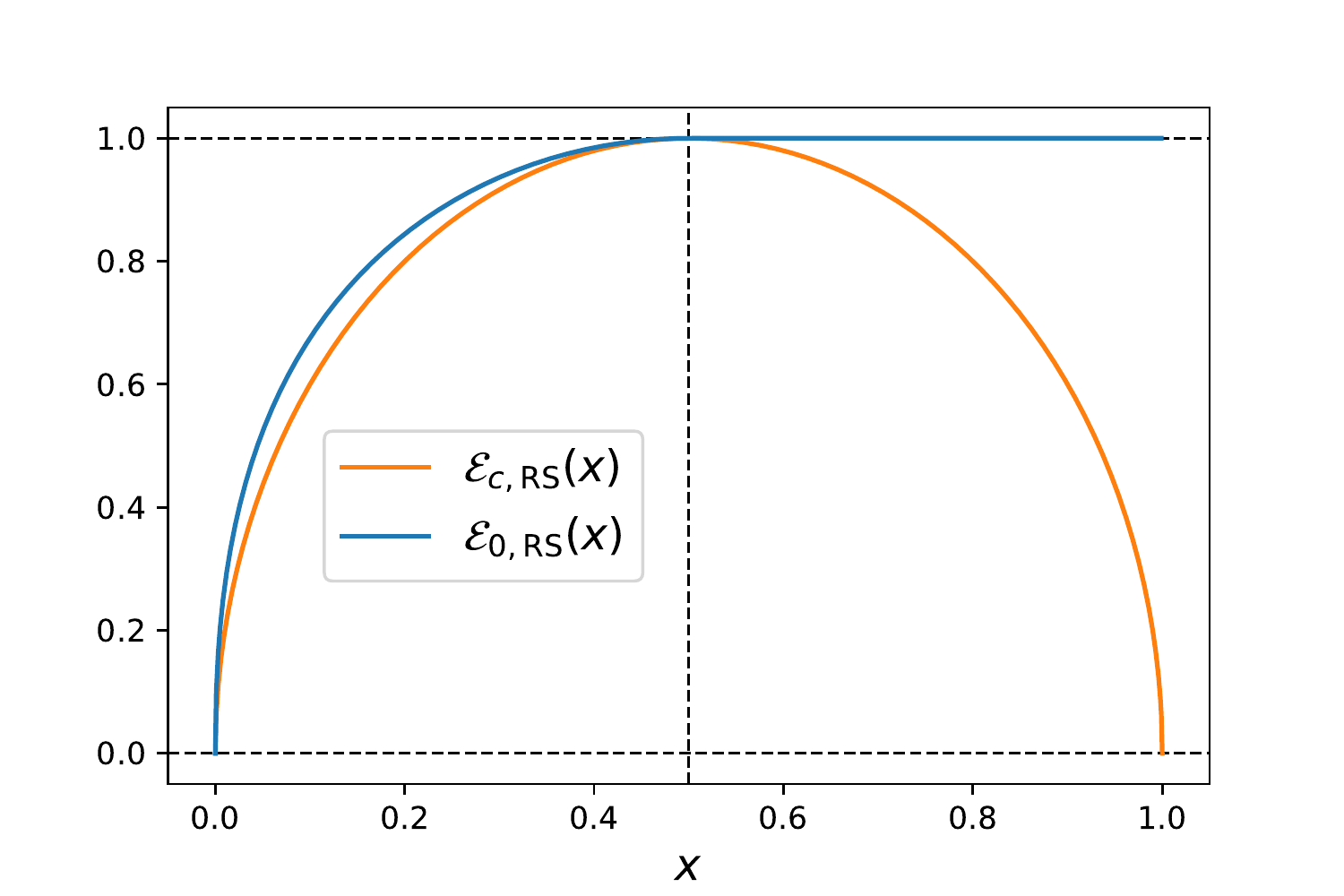}
    \caption{Plot of the rescaled zero ${\cal E}_{0,\rm RS}(x)$ (in blue), solution of Eq. \eqref{e_0}, and the scaling function ${\cal E}_{c,\rm RS}(x)=2\sqrt{x(1-x)}$ (in orange) as a function of $x=(\Gamma+g(1))/(\Gamma+g'(1))$. One always have that ${\cal E}_{0,\rm RS}(x)\geq {\cal E}_{c,\rm RS}(x)$, i.e. $e_{0,\rm RS}(\Gamma)\leq e_{c,{\rm RS}}(\Gamma)$, with the two expressions coinciding only for $x=0$ (non-physical) or $x=1/2$ (the latter corresponding to $\Gamma=g'(1)-2g(1)$). For $x>1/2$, it takes the simple constant value ${\cal E}_{0,\rm RS}(x)=1$. Note that $\partial_\Gamma x=(g'(1)-g(1))/(\Gamma+g'(1))^2>0$ with $x=g(1)/g'(1)$ for $\Gamma=0$, $x=1/2$ for $\Gamma=g'(1)-2g(1)\geq 0$ and $x\to 1$ as $\Gamma\to \infty$.}
    \label{fig:zero_L_RS}
\end{figure}

Finally, the RS expression of the LDF ${\cal L}(e)$ is given by
\begin{align}
{\cal L}_{\rm RS}(e)=&-\frac{2(g(1)+\Gamma)}{g'(1)+\Gamma}\frac{e^2}{e_{c,{\rm RS}}^2}+\frac{e}{e_{c,{\rm RS}}^2}\left(e-\sqrt{e^2-e_{c,{\rm RS}}^2}\right)-\ln\left(\frac{e-\sqrt{e^2-e_{c,{\rm RS}}^2}}{e_{c,{\rm RS}}}\right)+\frac{1}{2}\ln\frac{g(1)+\Gamma}{g'(1)-g(1)}\;.\label{L_RS_res}
\end{align}
This function is well-defined for any $e\leq e_{c,{\rm RS}}(\Gamma)$ but, as previously mentioned, it cannot represent the correct LDF for any $e\geq e_{\rm RSB}(\Gamma)$. The third derivative ${\cal L}_{\rm RS}^{(3)}(e)>0$ for any $e< e_{c,{\rm RS}}(\Gamma)$, hence  ${\cal L}_{\rm RS}''(e)$ is an increasing function, which must be positive  as ${\cal L}_{\rm RS}''(e\to -\infty)=1/(g(1)+\Gamma)>0$. The function ${\cal L}_{\rm RS}(e)$ is thus convex over the interval $e< e_{c,{\rm RS}}(\Gamma)$ and in particular for any $e\leq e_{\rm RSB}(\Gamma)$. The function ${\cal L}_{\rm RS}(e)$ is universal in the sense that it only depends on the values of covariance function $f(q)$ and its first derivative at $q=1$ (the energy $e_{\rm RSB}(\Gamma)$ depends on the second derivative as well). The function ${\cal L}_{\rm RS}(e)$ vanishes at the value $e=e_{0,\rm RS}(\Gamma)$, and this zero is unique in the interval $e\leq e_{c,{\rm RS}}(\Gamma)$. It is given explicitly by
\be
e_{0,\rm RS}(\Gamma)=-\sqrt{g'(1)+\Gamma}\,{\cal E}_{0,\rm RS}\left(\frac{\Gamma+g(1)}{\Gamma+g'(1)}\right)\;,\label{e_0_def}
\ee 
where the scaling function ${\cal E}_{0,\rm RS}(x)$ is obtained as the solution of the equation
 \be
 {\cal E}_{0,\rm RS}(x)\frac{{\cal E}_{0,\rm RS}(x)(1-2x)+\sqrt{{\cal E}_{0,\rm RS}(x)^2-4x(1-x)}}{4x(1-x)}-\ln\left(\frac{{\cal E}_{0,\rm RS}(x)+\sqrt{{\cal E}_{0,\rm RS}(x)^2-4x(1-x)}}{2x}\right)=0\;.\label{e_0}
\ee
Note that the RS critical energy takes the same scaling form 
\be
e_{c,{\rm RS}}(\Gamma)=-\sqrt{g'(1)+\Gamma}\,{\cal E}_{c,{\rm RS}}\left(\frac{\Gamma+g(1)}{\Gamma+g'(1)}\right)\;,\;\;{\cal E}_{c,{\rm RS}}\left(x\right)=-2\sqrt{x(1-x)}\;.
\ee 
The study of Eq. \eqref{e_0}, plotted in Fig. \ref{fig:zero_L_RS}, shows that the scaling function ${\cal E}_{0,\rm RS}(x)=1$ for any value of $x\geq 1/2$. This implies that for any $\Gamma\geq g'(1)-2g(1)$, which coincide with a positive rescaled variance ${\cal V}_{\min}(\Gamma)\geq 0$ (see Eq. \eqref{V_min_RS_res}), the zero takes the simple expression $e_{0,\rm RS}(\Gamma)=-\sqrt{g'(1)+\Gamma}$. For $x>1/2$, the scaling function ${\cal E}_{0,\rm RS}(x)$ does not have a simple analytic expression and thus neither does the zero $e_{0,\rm RS}(\Gamma)$ for $\Gamma<g'(1)-2g(1)$.

{\bf Random field with $\Gamma\geq \Gamma_{\rm RSB}$:} The RS expression of the LDF ${\cal L}_{\rm RS}(e)$ is positive in that case and has a unique minimum $e_{\rm typ}(\Gamma)=e_{0,{\rm RS}}(\Gamma)=-\sqrt{g'(1)+\Gamma}$, where it vanishes ${\cal L}_{\rm RS}(e_{\rm typ})=0$. In this case, the LDF is quadratic around the typical energy $e_{\rm typ}$
\be
{\cal L}_{\rm RS}(e)=\frac{(e-e_{\rm typ})^2}{2{\cal V}_{\min}(\Gamma)}+O(e-e_{\rm typ})^3\;,
\ee
which is consistent with the central limit theorem proved in \cite{ChenSen2017}.\\

\noindent {\bf Random field with $\Gamma\leq \Gamma_{\rm RSB}$:} 
In the regime $\Gamma<\Gamma_{\rm RSB}$, the function ${\cal L}_{\rm RS}(e)$ is decreasing over the interval $e\leq e_{\rm RSB}(\Gamma)$. Clearly, if the zero $e_{0,{\rm RS}}(\Gamma)$ of ${\cal L}_{\rm RS}(e)$ satisfies $e_{0,{\rm RS}}(\Gamma)\leq e_{\rm RSB}(\Gamma)$ then ${\cal L}_{\rm RS}(e_{\rm RSB})\leq 0$, and hence the expression of ${\cal L}_{\rm RS}(e)$ cannot represent a LDF over the whole interval $e\leq e_{\rm RSB}$ and must be limited to energies $e\leq e_{0,\rm RS}(\Gamma)$. 

\begin{figure}
    \centering
    \includegraphics[width=0.48\textwidth]{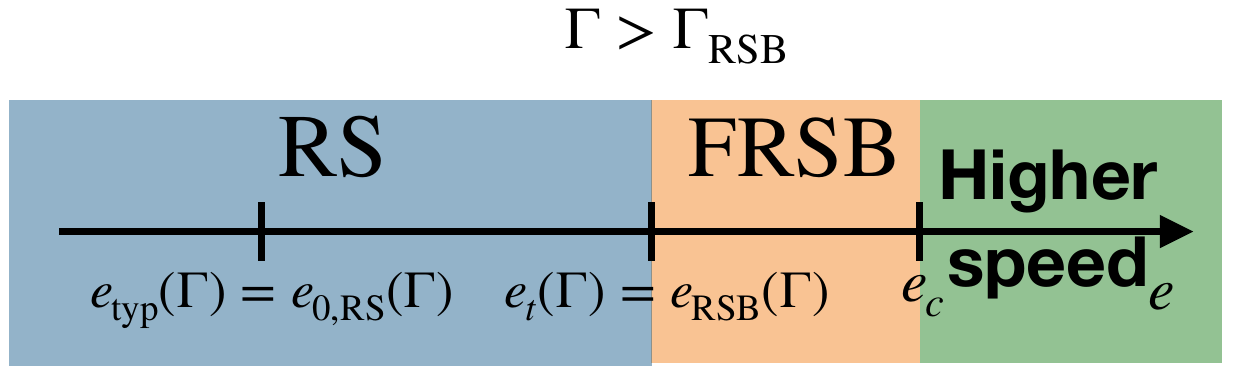}
    \includegraphics[width=0.48\textwidth]{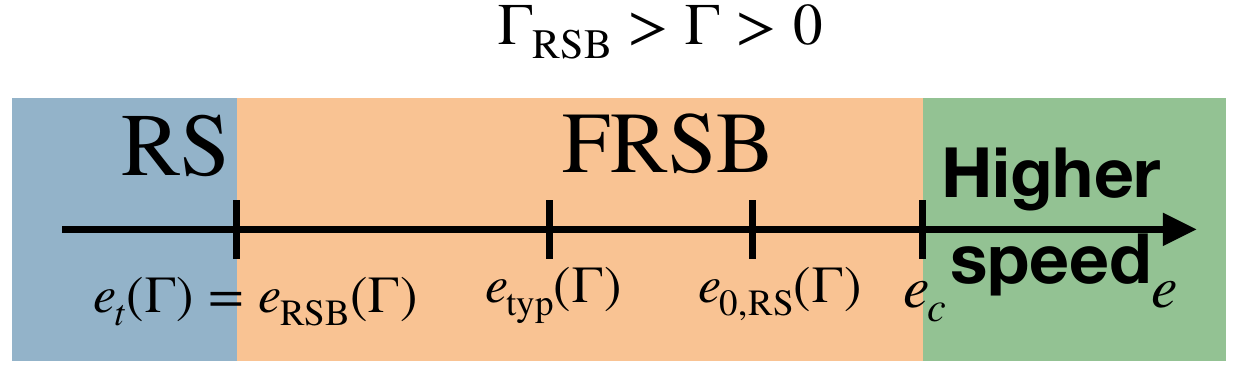}
    \includegraphics[width=0.48\textwidth]{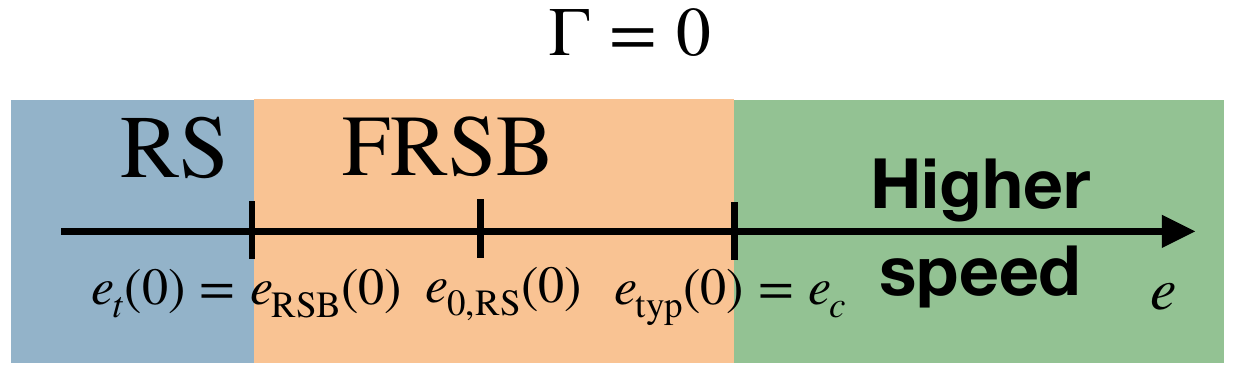}
    \caption{Sketch of the energy scales for different values of $\Gamma$ for models with a FRSB phase. The typical and average energy $e_{\rm typ}(\Gamma)=\overline{e_{\min}}$ is the unique zero of the LDF, i.e. ${\cal L}(e_{\rm typ}(\Gamma))=0$, while $e_{0,{\rm RS}}(\Gamma)$ is the unique zero of the RS expression of the LDF, i.e. ${\cal L}_{\rm RS}(e_{0,{\rm RS}}(\Gamma))=0$. These energy coincide $e_{\rm typ}(\Gamma)=e_{0,{\rm RS}}(\Gamma)$ for $\Gamma>\Gamma_{\rm RSB}$ while $e_{\rm typ}(\Gamma)\neq e_{0,{\rm RS}}(\Gamma)$ for $0<\Gamma<\Gamma_{\rm RSB}$. The critical energy $e_c$ corresponds to the "wall", i.e. the energy beyond which fluctuations of $e_{\min}$ are described by a LDF with speed $\gg N$. The transition between the RS and FRSB phase occurs at the transition energy $e_{t}(\Gamma)$ while the RS expression of the LDF is unstable for energies $e>e_{\rm RSB}(\Gamma)$. These energies coincide $e=e_{\rm SRB}(\Gamma)=e_t(\Gamma)$ for any $\Gamma>0$. Finally for $\Gamma=0$, the typical energy $e_{\rm typ}(0)$ and critical energy $e_c$ coincide.}
    \label{fig:energies_FRSB}
\end{figure}

For models whose replica-symmetry broken phase is of FRSB type, we will show explicitly in the following that ${\cal L}_{\rm RS}(e_{\rm RSB})>0$ in the regime $\Gamma<\Gamma_{\rm RSB}$ (see Fig. \ref{fig:energies_FRSB} for a sketch of the different energy scales for models with FRSB phase). For models whose replica-symmetry broken phase is of 1RSB type instead, ${\cal L}_{\rm RS}(e_{\rm RSB})$ can take either sign. We will show that there exists a threshold magnetic field $0\leq \Gamma_t<\Gamma_{\rm RSB}$, such that for $\Gamma_{\rm RSB}>\Gamma>\Gamma_t$ the value ${\cal L}_{\rm RS}(e_{\rm RSB})$ is positive. For $0<\Gamma<\Gamma_t$, the transition to the RSB regime does not occur at energy $e_{\rm RSB}(\Gamma)$ but rather at an energy $e_t(\Gamma)<e_{\rm RSB}(\Gamma)$ satisfying simultaneously $e_t(\Gamma)<e_{0,\rm RS}(\Gamma)$ which implies ${\cal L}_{\rm RS}(e_t)>0$ (see Fig. \ref{fig:energies_1RSB} for a sketch of the different energy scales for models with 1RSB phase). Note that there exists an additional threshold $\Gamma_{\rm cr}\leq \Gamma_t\leq \Gamma_{\rm RSB}$ such that $e_{\rm RSB}(\Gamma_{\rm cr})=e_{0,{\rm RS}}(\Gamma_{\rm cr})$. Thus for any $\Gamma_{\rm RSB}>\Gamma>\Gamma_{\rm cr}$ one has $e_{0,{\rm RS}}(\Gamma)>e_{\rm RSB}(\Gamma)$ and ${\cal L}_{\rm RS}(e_{\rm RSB})>0$ while for $\Gamma_{\rm cr}>\Gamma$ in contrast $e_{0,{\rm RS}}(\Gamma)<e_{\rm RSB}(\Gamma)$ and ${\cal L}_{\rm RS}(e_{\rm RSB})<0$.

\begin{figure}
    \centering
    \includegraphics[width=0.58
    \textwidth]{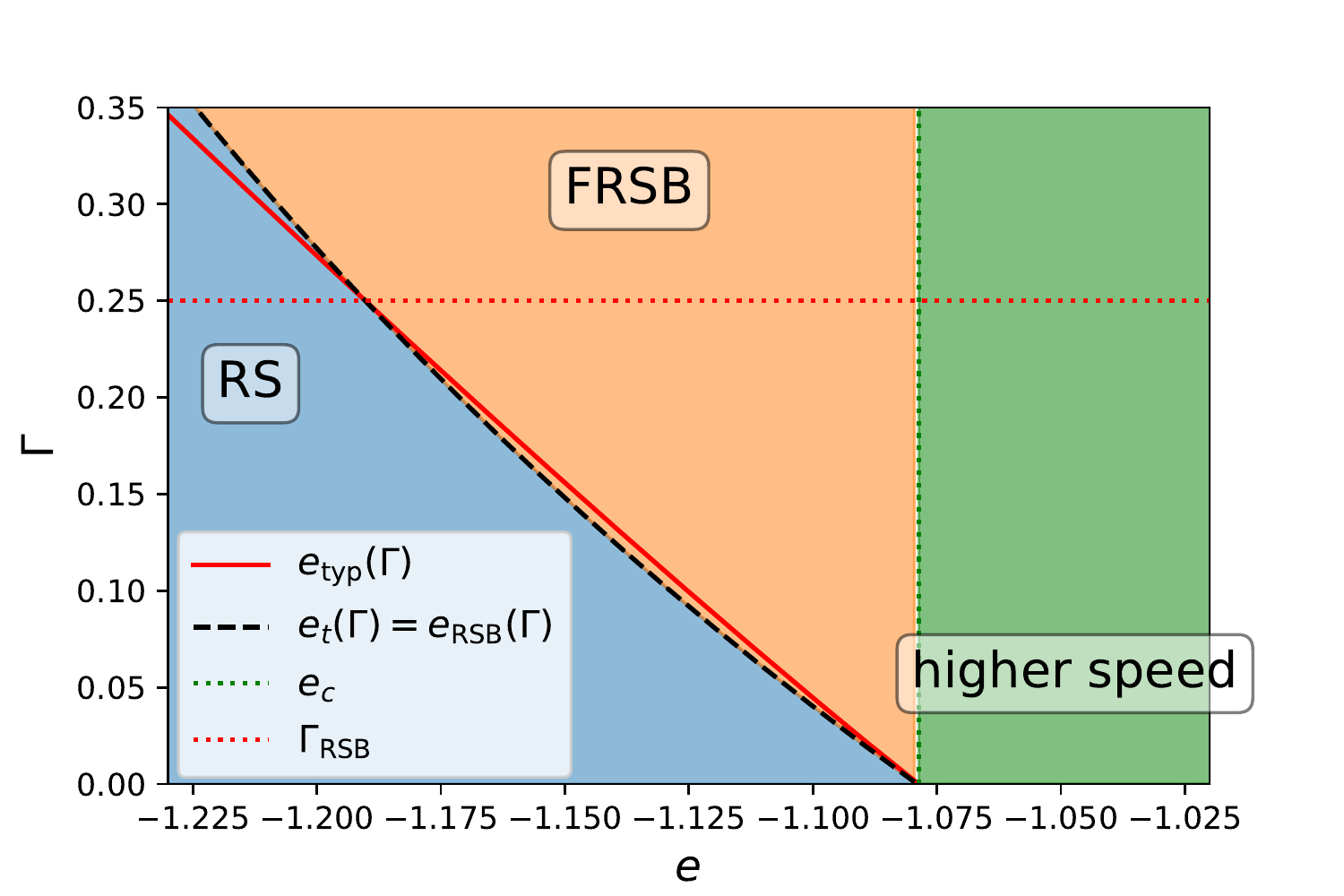}
    \caption{Phase diagram for the LDF for the spherical spin model in Eq. \eqref{f_FRSB} in the plane $e,\Gamma$. The dashed black line $e_t(\Gamma)=e_{\rm RSB}(\Gamma)$ (as the equality always hold for models with FRSB) shows the transition separating the FRSB phase (in orange) and the RS phase (in blue) determined by the stability criterion of the RS phase. The red line corresponds to the typical value $e_{\rm typ}(\Gamma)$, which crosses the line $e_t(\Gamma)$ for $\Gamma=\Gamma_{\rm RSB}$ (marked by a horizontal red dotted line). For $e>e_c$ (the energy $e_c$ of the "wall" being marked by the vertical green dotted line), the LDF at speed $N$ is infinite and one expects that a LDF with higher speed will describe the green region.}
    \label{phase_diag_FRSB}
\end{figure}

\subsubsection{Fully broken replica-symmetry regime}

Let us now consider the RSB phase for models which satisfy
\be
\forall q\in[0,1]\;,\;\;{\cal S}\left[f'\right](q)={\cal S}\left[g'\right](q)=\frac{g^{(4)}(q)}{g''(q)}-\frac{3}{2}\left(\frac{g^{(3)}(q)}{g''(q)}\right)^2\geq 0\;,
\ee
i.e. models for which the replica symmetry breaking  is of FRSB type. Note that $p$-spin models with $g(q)=q^p/p$  for $p>2$ satisfy ${\cal S}\left[g'\right](q)=- p(p-2)/(2 q^2) <0$, hence do not belong to this type. The phase diagram of the LDF for a model with FRSB phase is plotted in Fig. \ref{phase_diag_FRSB} and the different energy scales are provided in Fig. \ref{fig:energies_FRSB}. In the FRSB regime, the CGF is obtained by solving the optimisation problem in Eq. \eqref{phiscases} where one imposes that $z(q)$ is a positive increasing function over the range $q\in[q_0,1]$. We find that the optimisation problems in Eqs. \eqref{phiscases} and \eqref{leg_trans_res} are most easily solved by expressing the optimal parameter $s=s_*$ as
\be
s_*=\frac{1}{\sqrt{g''(q_0^*)}}\left(\frac{g''(q_0^*)}{g'(q_0^*)+\Gamma}-\frac{1}{q_0^*}\right)\,,\label{s_FRSB_res}
\ee
in terms of the optimal value of $q_0=q_0^*$. The parameter $v$ and the function $z(q)$ then take the optimal values (which are actually the same as arise in the optimisation problem for the typical  ground-state energy in the FRSB phase, see Appendix \ref{app:typical})
\be
v_*=\frac{1}{\sqrt{g''(1)}}\;,\;\;z_*(q)=\frac{g^{(3)}(q)}{2g''(q)^{3/2}}\;,\;\;q\in[q_0^*,1]\;.
\ee
Finally, the optimal value $q_0^*$ for fixed energy $e$ is obtained by inverting the following parametric expression
\be
e=-\phi'(s_*)=-\left(\frac{1}{\sqrt{g''(q_0^*)}}\left(g'(q_0^*)-\frac{g(q_0^*)}{q_0^*}\right)+\frac{g(q_0^*)+\Gamma q_0^*}{g'(q_0^*)+\Gamma}\sqrt{g''(q_0^*)}+\int_{q_0^*}^{1} dq\,\sqrt{g''(q)}\right)\;.\label{e_FRSB_res}
\ee
One can show explicitly that $\partial_{q_0^*}e<0$ (see Appendix \ref{FRSB_app} for details). Comparing Eq. \eqref{e_FRSB_res} and Eq. \eqref{e_RSB}, one can simply conclude that the limit $q_0^*\to 1$ corresponds to $e\to e_{\rm RSB}(\Gamma)$. In the opposite limit $q_0^*\to 0$, one defines the critical energy
\be
e_c=-\int_{0}^{1} dq\,\sqrt{g''(q)}\;,
\ee
where we have used that $g(0)=g'(0)=0$ while $g''(0)=2g_2>0$. Indeed, if it were that $g''(0)=0$ the function $z_*(q)$ would diverge to $+\infty$ as $q\to 0$, in contradiction with the FRSB condition that $z_*(q)$ must be an increasing function of $q$.

If $0<\Gamma<\Gamma_{\rm RSB}=g''(1)-g'(1)$, there is a unique value  $0<q_{\rm typ}(\Gamma)<1$ of the parameter $q^*_0$ such that
\be
\Gamma=q_{\rm typ}(\Gamma) g''(q_{\rm typ}(\Gamma))-g'(q_{\rm typ}(\Gamma))\;.\label{q_typ_FRSB_res}
\ee
The associated value of $e$ through the relation \eqref{e_FRSB_res} belongs to the finite interval $(e_{\rm RSB}(\Gamma),e_c)$ and corresponds precisely to the typical ground-state energy (see Appendix \ref{app:typical})
\be
e_{\rm typ}(\Gamma)=\overline{e_{\min}}=-\left(q_{\rm typ}(\Gamma)\sqrt{g''(q_{\rm typ}(\Gamma))}+\int_{q_{\rm typ}(\Gamma)}^{1} dq\,\sqrt{g''(q)}\right)\;,\;\;\Gamma\leq\Gamma_{\rm RSB}\;.\label{e_typ_FRSB_res}
\ee
Note that for $\Gamma=\Gamma_{\rm RSB}=g''(1)-g'(1)$, one can simply check from Eq. \eqref{q_typ_FRSB_res} that $q_{\rm typ}(\Gamma_{\rm RSB})=1$ and Eq. \eqref{e_typ_FRSB_res} coincides with Eq. \eqref{e_typ_RS_res}. The rescaled variance is similarly obtained by simply inserting the value of $q_{\rm typ}(\Gamma)$ obtained from Eq. \eqref{q_typ_FRSB_res} into Eq. \eqref{res_gen_var}. In the limit $\Gamma\to 0$, the value $q_{\rm typ}(\Gamma)\to 0$ such that $e_{\rm typ}(\Gamma)\to e_c$ and the rescaled variance ${\cal V}_{\min}(\Gamma)$ vanishes. This limit will be investigated further in a separate section below.

Let us now give the result for the FRSB expression of the LDF, ${\cal L}_{\rm FRSB}(e)$. The LDF ${\cal L}_{\rm FRSB}(e)$ describes,
through the parametrisation in Eq. \eqref{e_FRSB_res}, the range of energies $e \in [e_{\rm RSB},e_c]$ and reads
\be
{\cal L}_{\rm FRSB}(e)=\frac{1}{2}\left[\frac{f(q_0^*)}{f''(q_0^*)}\left(\frac{f''(q_0^*)}{f'(q_0^*)}-\frac{1}{q_0^*}\right)^2+1-\frac{f'(q_0^*)}{q_0^* f''(q_0^*)}-\ln\left(\frac{q_0^* f''(q_0^*)}{f'(q_0^*)}\right)\right]\;.
\ee
The transition between the RS and FRSB expressions of the LDF occurs at the energy $e=e_t(\Gamma)=e_{\rm RSB}(\Gamma)$, corresponding to $q_0^*\to 1$, for any value of $\Gamma$. Expanding the difference as $e\to e_{\rm RSB}$, one obtains for any $\Gamma\geq 0$
\be
{\cal L}_{\rm FRSB}(e)-{\cal L}_{\rm RS}(e)=-\frac{f'(1)^3 f''(1)^{5/2} \left(\Gamma-\Gamma_{\rm RSB}\right)(e-e_{t})^3}{3 \left(f(1)
   \left(f''(1)+f'(1)\right)-f'(1)^2\right)^3 \left(f'(1) \left(f^{(3)}(1)+2
   f''(1)\right)-2 f''(1)^2\right)}+O(e-e_{t})^4\;.\label{tr_FRSB}
\ee
The transition between the RS and FRSB phases is thus generically of third order in the sense that the third derivative of the LDF ${\cal L}(e)$ is the lowest order discontinuous derivative at the transition. Note however that precisely for $\Gamma=\Gamma_{\rm RSB}$, the transition is of the fourth order
\be
{\cal L}_{\rm FRSB}(e)-{\cal L}_{\rm RS}(e)=-\frac{2g''(1)^3 (e-e_{t})^4}{3g^{(3)}(1)\left(g''(1)-2(g'(1)-g(1))\right)^4}+O(e-e_{t})^5\;,\;\;\Gamma=\Gamma_{\rm RSB}\;.\label{tr_G_rsb_frsb}
\ee

\begin{figure}
    \centering
    \includegraphics[width=0.45\textwidth]{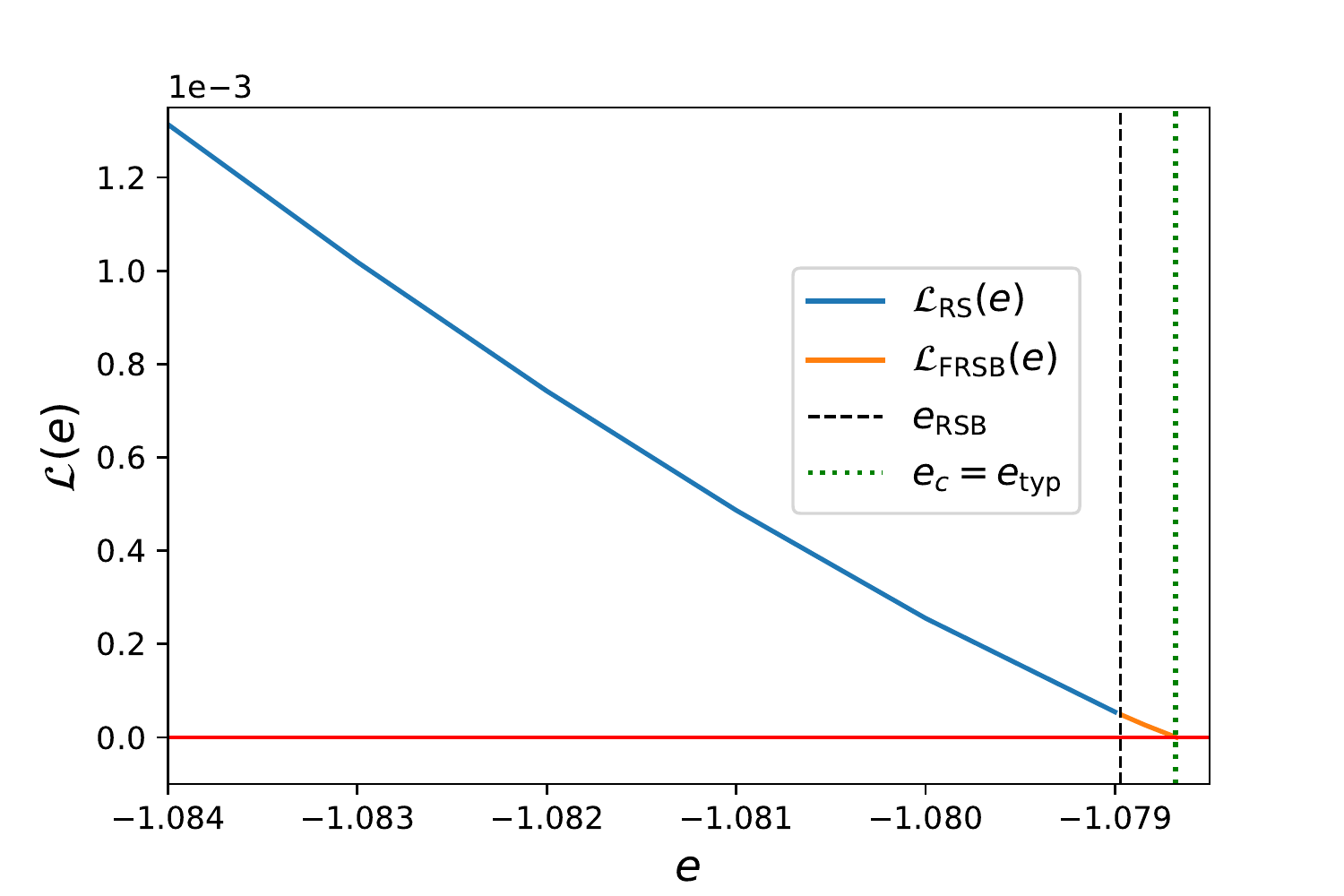}
    \includegraphics[width=0.45\textwidth]{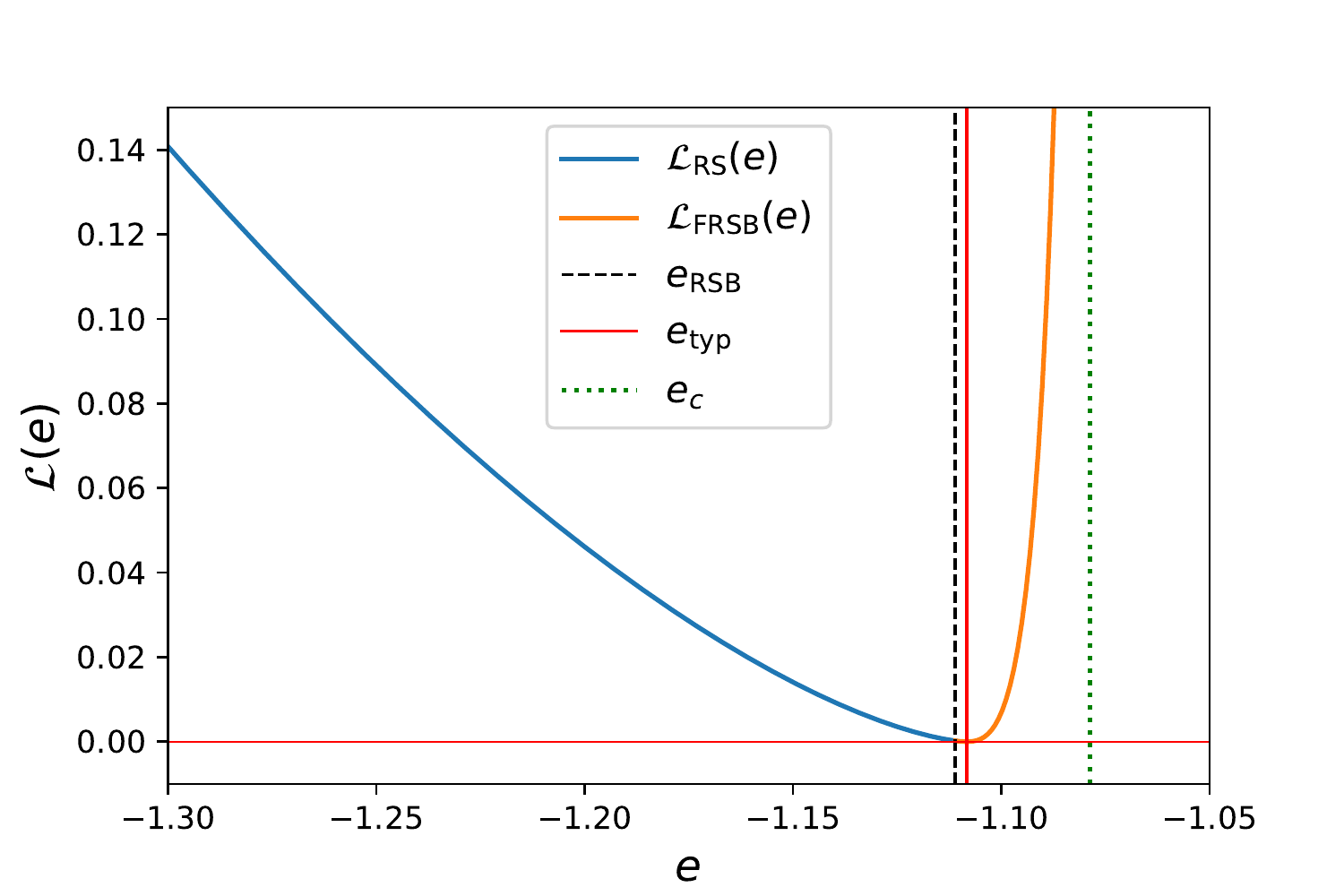}
    \includegraphics[width=0.45\textwidth]{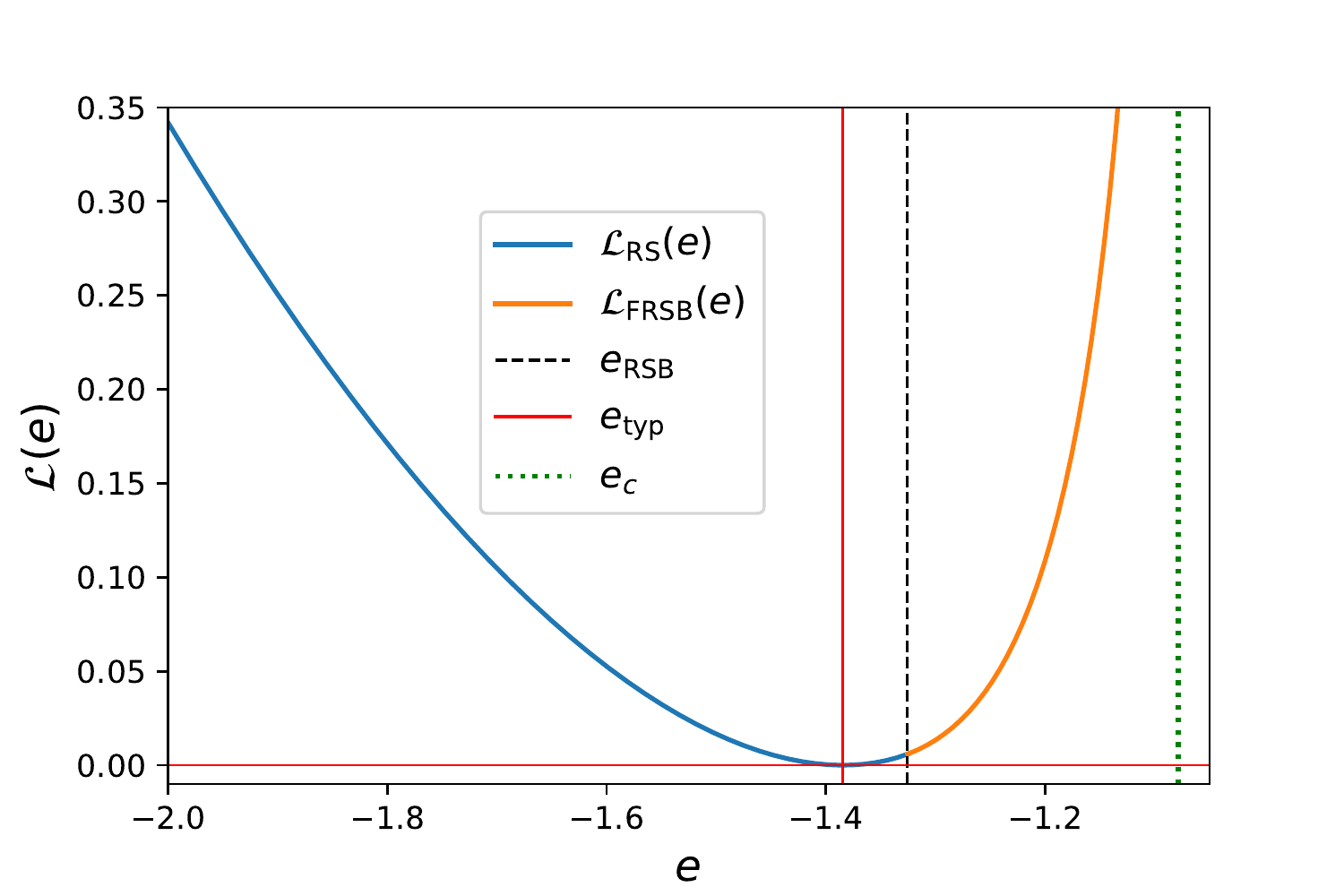}
    \caption{Plot of the large deviation function ${\cal L}(e)$ for the covariance function in Eq. \eqref{f_FRSB} with $g_4=1/48$, $g_3=1/36$ and three different values of $\Gamma=0$ (top left), $\Gamma=1/16$ (top right) and $\Gamma=3/4$ (bottom). The vertical black dashed line represents the value of $e_{\rm RSB}$ separating on its left the RS regime where ${\cal L}(e)={\cal L}_{\rm RS}(e)$ plotted in blue and on its right the FRSB regime where ${\cal L}(e)={\cal L}_{\rm FRSB}(e)$ plotted in orange. The vertical solid red line marks the value $e_{\rm typ}$, where ${\cal L}(e_{\rm typ})=0$ while the vertical green dotted line marks the value $e_c$ where the LDF diverges for $\Gamma>0$ (for $\Gamma=0$, $e_{\rm typ}=e_c$). The typical value of the ground state $e_{\rm typ}$ belongs to the RS phase for $\Gamma>\Gamma_{\rm RSB}=1/4$.}
    \label{fig:LDF_FRSB}
\end{figure}

One can check simply that ${\cal L}_{\rm FRSB}'(e)=-s_*$ and that 
\be
{\cal L}_{\rm FRSB}''(e)=\frac{f'(q_0^*)+q_0^* f''(q_0^*)}{f(q_0^*)(f'(q_0^*)+q_0^* f''(q_0^*))-q_0^* f'(q_0^*)^2}\geq 0\;,
\ee
which can be shown to be positive from Eq. \eqref{nice_id}, such that the LDF ${\cal L}_{\rm FRSB}(e)$ is convex. For a random field $\Gamma\leq \Gamma_{\rm RSB}$, the LDF reaches its global minimum, equal to zero for $e=e_{\rm typ}(\Gamma)$, such that ${\cal L}_{\rm FRSB}(e_{\rm RSB}(\Gamma))={\cal L}_{\rm RS}(e_{\rm RSB}(\Gamma))\geq {\cal L}_{\rm FRSB}(e_{\rm typ}(\Gamma))=0$. This ensures in particular that, for any value of $\Gamma$, the RS expression of the LDF ${\cal L}_{\rm RS}(e)$ is positive for any energy $e\leq e_{\rm RSB}$ if the corresponding phase with broken replica symmetry is of FRSB type. A plot of the LDF ${\cal L}(e)$, both in the RS and FRSB phases, for a representative model is shown in Fig. \ref{fig:LDF_FRSB} for several values of $\Gamma$.

For positive magnetic field, the LDF diverges logarithmically for $e\to e_{\rm c}$ (corresponding to $q_0^*\to 0$) with
\be
{\cal L}_{\rm FRSB}(e)= -\frac{1}{2}\ln\frac{2\sqrt{2g_2}(e_c-e)}{\Gamma}-\frac{3}{4}+O(e_c-e)\;,\;\;\Gamma>0\;,\label{div_wall_FRSB}
\ee
which hints at the fact that a large deviation of speed larger than $N$ should describe the probability that $e>e_c$. In the range of magnetic field $0<\Gamma\leq \Gamma_{\rm RSB}$, the LDF is quadratic around the typical value $e_{\rm typ}$
\be
{\cal L}_{\rm FRSB}(e)=\frac{(e-e_{\rm typ})^2}{2{\cal V}_{\min}(\Gamma)}+O(e-e_{\rm typ})^3\;,\label{quad_typ_FRSB}
\ee
which is again consistent with the central limit theorem proved in \cite{ChenSen2017}. For zero magnetic field $\Gamma=0$ we have instead $e_{\rm typ}(\Gamma=0)=e_c$ and the behaviour of the LDF is quite non-trivial as $e\to e_c$. Namely, supposing that the covariance function has the form
\be
f(q)=\Gamma q+g(q)\;,\;\;g(q)=g_2 q^2+g_r q^r+o(q^r)\;,\label{small_q_cov}
\ee
with $r$ the exponent of the leading correction to the quadratic form, the LDF vanishes as
\be
{\cal L}_{\rm FRSB}(e)=\xi_r (e_c-e)^{\eta}+o(e_c-e)^{\eta}\;,\;\;1\leq \eta=\frac{3(r-2)}{2r-3}\leq \frac{3}{2}\;,\label{powlaw_typ_FRSB}
\ee
where
\be
\xi_r=2^{\frac{r-12}{2(2r-3)}}\left(\frac{r(2r-3)}{r-2}\right)^{\frac{3(r-2)}{2r-3}}\frac{(r-3)^{-\frac{r-3}{2r-3}}\left(r (r-2)g_r\right)^{\frac{3}{2r-3}}}{3 r g_2^{\frac{3r}{2(2r-3)}}}\;.\label{xi_p}
\ee
In particular, for $r=3$, the LDF vanishes linearly as ${\cal L}_{\rm FRSB}(e)=z(0)(e_c-e)+o(e_c-e)$. The opposite limit $r\to\infty$ corresponds to the spherical $2$-spin model, which is marginal ${\cal S}[g'](q)=0$ and for which the RSB branch of the LDF can be obtained either from a FRSB or 1RSB ansatz. In that limit, one can check that ${\cal L}_{\rm FRSB}(e)=2/3(2(e_c-e))^{3/2}+o(e_c-e)^{3/2}$ as mentioned above, which smoothly matches the tail of the Tracy-Widom $\beta=1$ distribution. For zero magnetic field, one does not expect a central limit theorem and the distribution of typical fluctuations becomes non-trivial. It is notably the case in this limit $r\to \infty$ where it corresponds to the TW  \cite{baik2016fluctuations}. From the result in Eq. \eqref{powlaw_typ_FRSB} and assuming the simplest possible matching between the large deviation regime $e_c-e=O(1)$ and the regime of
typical fluctuations, we naturally conjecture that the
{\it typical} fluctuations of the ground state energy will be of order $e_c - e = O(N^{-1/\eta})$
with $\eta = \frac{3(r-2)}{2 r-3}$. More precisely we conjecture that 
the PDF of the ground-state energy ${\cal P}(e)$ 
takes the following scaling form in the regime of typical fluctuations
\be 
{\cal P}(e)=N^{1/\eta}\, p\left(N^{1/\eta}(e-e_c)\right) \;,\;\;{\rm where}\;\; p(\delta) \sim e^{- \xi_r|\delta|^{\eta}}\;,\;\;\delta\to -\infty\;.
\ee

\subsubsection{One-step replica-symmetry breaking regime}

\begin{figure}
    \centering
    \includegraphics[width=0.58
    \textwidth]{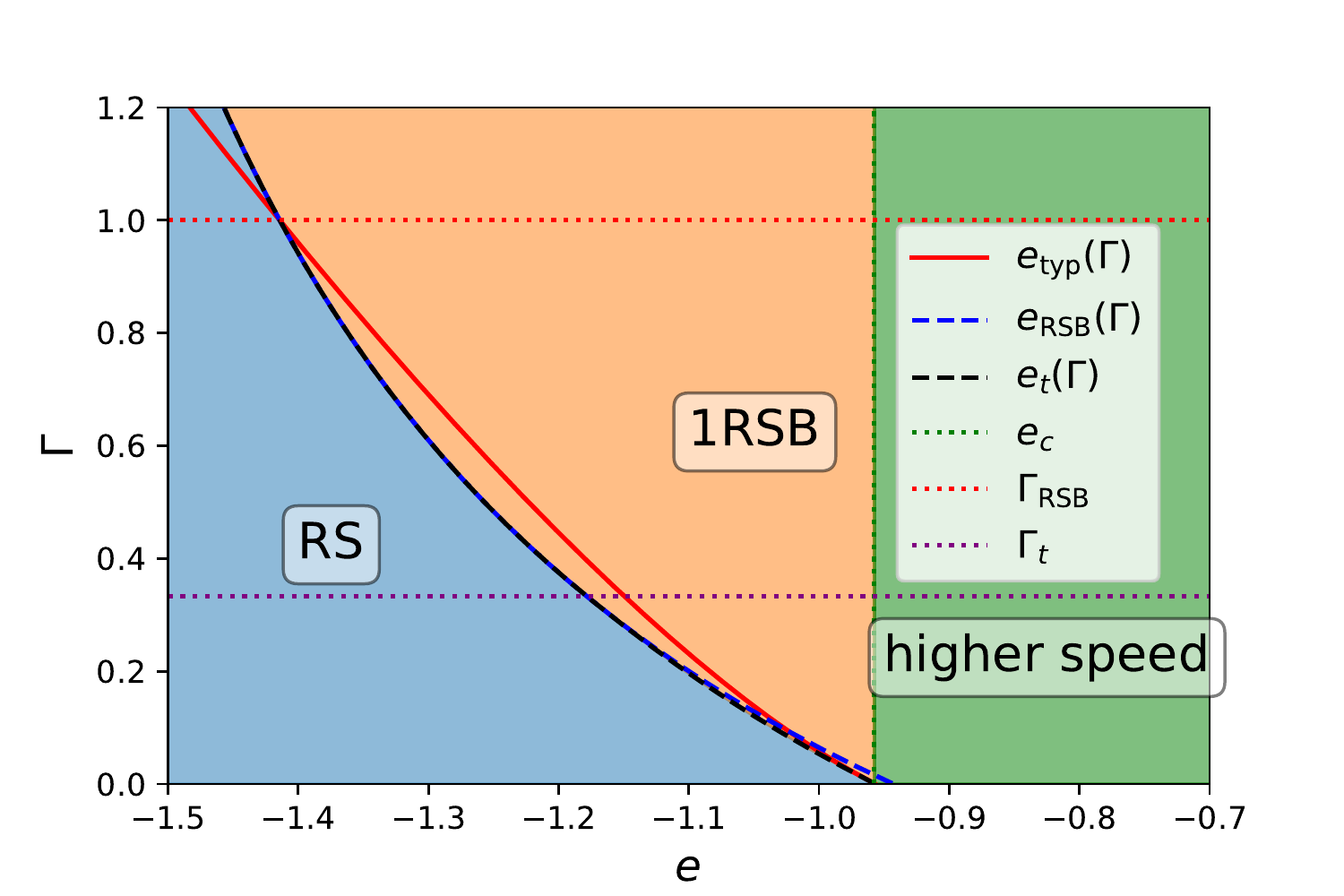}
    \caption{Phase diagram for the LDF for the $3$-spin model in the plane $e,\Gamma$. The black dashed line $e_t(\Gamma)$ shows the transition separating the 1RSB phase (in orange) and the RS phase (in blue). The red line corresponds to the typical value $e_{\rm typ}(\Gamma)$, which crosses the line $e_t(\Gamma)$ for $\Gamma=\Gamma_{\rm RSB}$ (marked by a horizontal red dotted line). The criterion of stability of the RS phase yields the dashed blue line 
    $e_{\rm RSB}(\Gamma)$ but only coincides with the correct transition line $e_t(\Gamma)$ for $\Gamma>\Gamma_t$ (marked by a horizontal purple dotted line) as for $\Gamma<\Gamma_t$ one has $e_t(\Gamma)<e_{\rm RSB}(\Gamma)$. For $e>e_c$ (the energy $e_c$ of the "wall" being marked by the vertical green dotted line), the LDF at speed $N$ is infinite and one expects that a LDF with higher speed will describe the green region. Note that for $\Gamma=0$, the LDF only has a RS phase as the 1RSB phase is reduced to the single point $e_{\rm typ}(0)=e_{t}(0)=e_c$.}
    \label{phase_diag_1RSB}
\end{figure}

\begin{figure}
    \centering
    \includegraphics[width=0.48\textwidth]{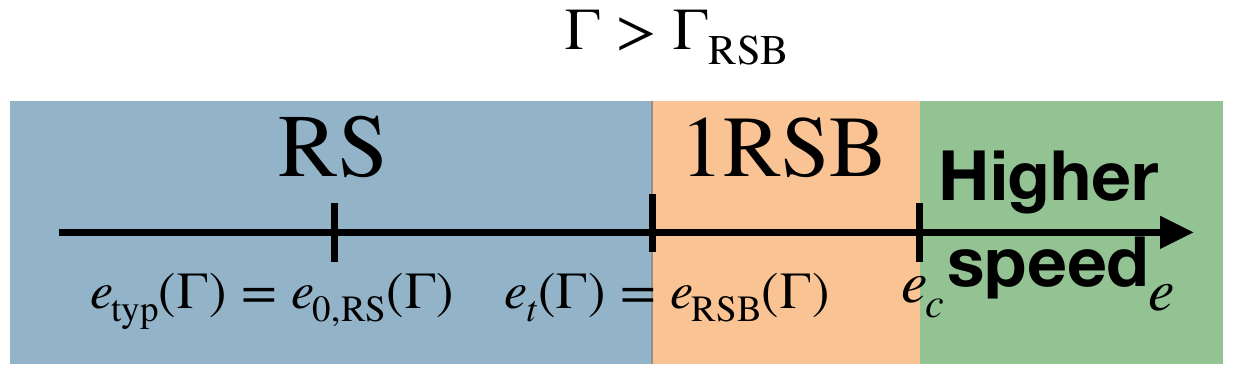}
    \includegraphics[width=0.48\textwidth]{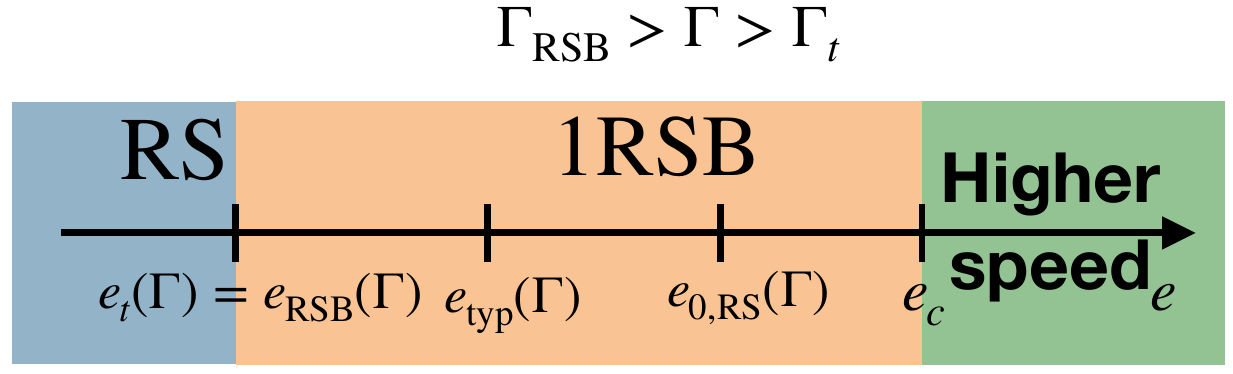}
    \includegraphics[width=0.48\textwidth]{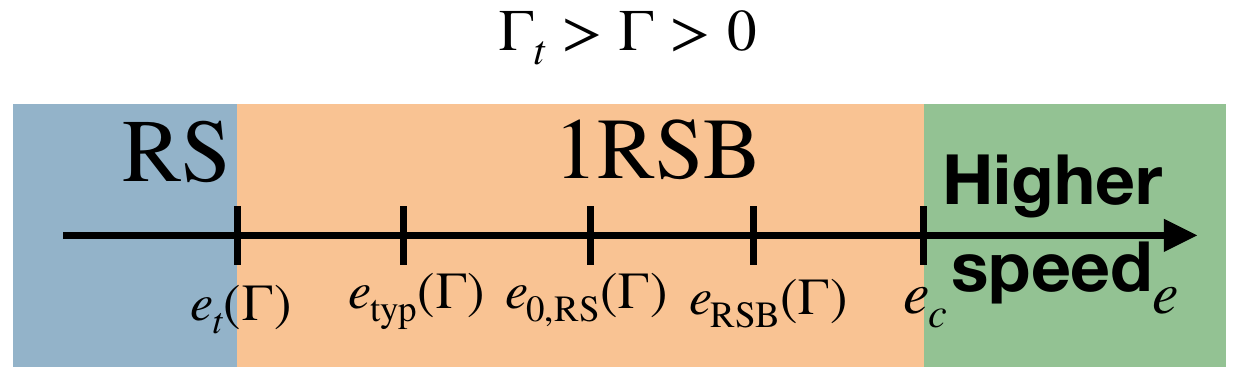}
    \includegraphics[width=0.48\textwidth]{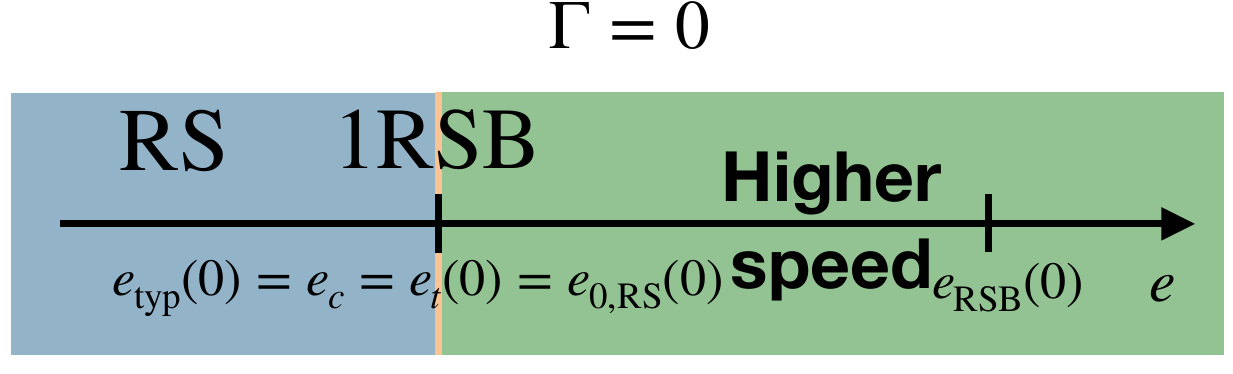}
    \caption{Sketch of the energy scales for different values of $\Gamma$ for models with a 1RSB phase. The typical and average energy $e_{\rm typ}(\Gamma)=\overline{e_{\min}}$ is the unique zero of the LDF, i.e. ${\cal L}(e_{\rm typ}(\Gamma))=0$, while $e_{0,{\rm RS}}(\Gamma)$ is the unique zero of the RS expression of the LDF, i.e. ${\cal L}_{\rm RS}(e_{0,{\rm RS}}(\Gamma))=0$. These energy coincide $e_{\rm typ}(\Gamma)=e_{0,{\rm RS}}(\Gamma)$ for $\Gamma>\Gamma_{\rm RSB}$ while $e_{\rm typ}(\Gamma)\neq e_{0,{\rm RS}}(\Gamma)$ for $0<\Gamma<\Gamma_{\rm RSB}$. The critical energy $e_c$ corresponds to the "wall", i.e. the energy beyond which fluctuations of $e_{\min}$ are described by a LDF with speed $\gg N$. The transition between the RS and 1RSB phase in the large deviation function occurs at the transition energy $e_{t}(\Gamma)$ while the RS expression of the LDF is unstable for energies $e>e_{\rm RSB}(\Gamma)$. These energies coincide $e=e_{\rm SRB}(\Gamma)=e_t(\Gamma)$ for $\Gamma>\Gamma_t$ while $e_t(\Gamma)>e_{\rm RSB}(\Gamma)$ for $\Gamma<\Gamma_t$. Finally for $\Gamma=0$, the transition energy $e_t(0)$, typical energy $e_{\rm typ}(0)$, zero of the RS expression $e_{0,{\rm RS}}(0)$ and critical energy $e_c$ all coincide.}
    \label{fig:energies_1RSB}
\end{figure}

Finally we discuss the RSB phase for models which satisfy
\be
\forall q\in[0,1]\;,\;\;{\cal S}\left[f'\right](q)={\cal S}\left[g'\right](q)=\frac{g^{(4)}(q)}{g''(q)}-\frac{3}{2}\left(\frac{g^{(3)}(q)}{g''(q)}\right)^2\leq 0\;,
\ee
i.e. for which the replica symmetry breaking pattern is of 1RSB type. The $p$-spin models  with $p>2$ for which $g(q)=q^p/p$, satisfy ${\cal S}\left[g'\right](q)=- p(p-2)/(2 q^2) <0$, hence belong to this class and will constitute important examples. The phase diagram of the LDF for a model with 1RSB phase is plotted in Fig. \ref{phase_diag_1RSB} and the sketches of different energy scales in the variable $e$ are provided in Fig. \ref{fig:energies_1RSB}. In the 1RSB regime, the CGF is obtained by solving the optimisation problem in Eq. \eqref{phiscases} where one imposes that $z(q)=s$ for $q\in[0,q_0)$ and $z(q)=m_0\geq s$ for $q\in(q_0,1]$ is piece-wise constant. As in the FRSB phase, it will be most convenient to express all the optimal value of the different parameters in terms of the optimal value $q_0=q_0^*$, i.e.
\begin{align}
  s_*&=\frac{1}{\sqrt{q_0^* f'(q_0^*)}}\left(\frac{1}{\sqrt{\alpha(q_0^*)}}-\sqrt{\alpha(q_0^*)}\right)\;,\label{param_1}\\
  v_*&=\rho(q_0^*)\sqrt{\frac{q_0^*}{f'(q_0^*)\alpha(q_0^*)}}\;,\label{param_2}\\
  m_0^*&=\frac{1}{1-q_0^*}\sqrt{\frac{q_0^* \rho(q_0^*)}{f'(q_0^*)}}\left(\sqrt{\frac{\alpha(q_0^*)}{\rho(q_0^*)}}-\sqrt{\frac{\rho(q_0^*)}{\alpha(q_0^*)}}\right)\;,\label{param_3}
\end{align}
where the function $\rho(q)$  is given by
\be
\rho(q)=\frac{(1-q)(g'(q)+\Gamma)}{(g'(1)-g'(q))q}\;,
\ee
and together with the function $\alpha(q_0^*)$ satisfies the self-consistent equation
\be
\mu(q_0^*)=\rho(q_0^*)\frac{\rho(q_0^*)-\alpha(q_0^*)-\alpha(q_0^*)\ln (\rho(q_0^*)/\alpha(q_0^*))}{(\alpha(q_0^*)-\rho(q_0^*))^2}\;,\;\;\mu(q)=\frac{1}{g'(1)-g'(q)}\left(\frac{g(1)-g(q)}{1-q}-g'(q)\right)\;.\label{mu_alph_res}
\ee
Finally, the value of $q_0^*$ for fixed $e$ is obtained by inverting the parametric expression
\be
e=-\phi'(s_*)=\frac{(1-q_0^*)\mu(q_0^*)f'(1)-f(1)+[f(1)-\mu(q_0^*)f'(1)-(1-\mu(q_0^*))f'(q_0^*)]\alpha(q_0^*)}{\sqrt{q_0^* f'(q_0^*)\alpha(q_0^*)}}\;.\label{e_1rsb_res}
\ee
In the limit $q_0^*\to 0$, the associated critical energy reads
\be
e_c=-\frac{1}{\sqrt{g'(1)}}\left(\sqrt{\rho_{\rm typ}}(g'(1)-g(1))+\frac{g(1)}{\sqrt{\rho_{\rm typ}}}\right)\;,
\ee
where the parameter $\rho_{\rm typ}=\lim_{q_0^*\to 0}[\rho(q_0^*)/\alpha(q_0^*)]$ is independent of $\Gamma$ and satisfies 
\be
\frac{g(1)}{g'(1)}=\rho_{\rm typ}\frac{\rho_{\rm typ}-1-\ln \rho_{\rm typ}}{(1-\rho_{\rm typ})^2}\;.\label{rho_typ_eq_res}
\ee
For a magnetic field 
\be
\Gamma\geq\Gamma_t=\frac{2g''(1)^2}{2g''(1)+g^{(3)}(1)}-g'(1)\;,\label{G_t_res}
\ee
with $\Gamma_t$ satisfying $\Gamma_t\leq \Gamma_{\rm RSB}$, the parametrisation in Eq. \eqref{e_1rsb_res} is operative for any $q_0^*\in[0,1]$ and describes the range $e\in[e_{\rm RSB}(\Gamma),e_c]$. In contrast, for $0<\Gamma<\Gamma_t$, the parametrisation is only operative in a limited range $q_0^*\in[0,q_t(\Gamma)]$, where $q_t(\Gamma)$ is the smallest positive solution of 
\be
\mu(q_t)=(\Gamma+g'(q_t))\frac{\displaystyle\Gamma(1-q_t)+g'(q_t)-q_t g'(1)-q_t (\Gamma+g'(1))\ln \frac{\Gamma+g'(q_t)}{q_t(\Gamma+g'(1))}}{(\Gamma(1-q_t)+g'(q_t)-q_t g'(1))^2}\;.
\ee
Note that $q_t=1$ is always a solution of this equation. The range of energy thus described is $e\in[e_t(\Gamma),e_c]$, with
\be
e_t(\Gamma)=-\frac{q_t f(1)f'(1)+((1-q_t)f'(1)-f(1))f'(q_t)}{\sqrt{(1-q_t)q_t f'(q_t)f'(1)(f'(1)-f'(q_t))}}<e_{\rm RSB}(\Gamma)\;,\;\;\Gamma< \Gamma_{t}\;.
\ee
For convenience, for $\Gamma\geq \Gamma_t$ we define the two energies $e_t(\Gamma)$ and $e_{\rm RSB}(\Gamma)$ to be equal: 
$e_t(\Gamma\geq \Gamma_t)=e_{\rm RSB}(\Gamma\geq \Gamma_t)$. Let us introduce the function 
\be
w(q)=\frac{2q g''(q)^2}{2g''(q)+q g^{(3)}(q)}-g'(q)\,\,{\rm which\;satisfies}\,\,w(0)=0\,,\,\,w(1)=\Gamma_t\,,\,\,w'(q)=-\frac{2q^2 g''(q)^3}{(2g''(q)+q g^{(3)}(q))^2} S[g'](q)\,,
\ee
with $2q^2 g''(q)^3/(2g''(q)+q g^{(3)}(q))^2>0$. One can thus show that for models with 1RSB phase such that $S[g'](q)<0$ for any $q\in[0,1]$ one must have $\Gamma_t\geq 0$. In contrast, $S[g'](q)>0$ for models with FRSB phase, such that $\Gamma_t\leq 0$.

In the regime $\Gamma\leq \Gamma_{\rm RSB}$, the typical energy $e_{\rm typ}(\Gamma)\in [e_t(\Gamma),e_c]$ and is thus obtained within the 1RSB scheme. The associated value of $q_{\rm typ}(\Gamma)$ is obtained ensuring that $\alpha(q_{\rm typ}(\Gamma))=1$ in Eq. \eqref{mu_alph_res}. The associated value of $e_{\rm typ}(\Gamma)$ then reads
\be
e_{\rm typ}(\Gamma)=\overline{e_{\min}}=-\frac{q_{\rm typ}(\Gamma)\mu(q_{\rm typ}(\Gamma))(\Gamma+g'(1))+(1-\mu(q_{\rm typ}(\Gamma)))(\Gamma+g'(q_{\rm typ}(\Gamma)))}{\sqrt{q_{\rm typ}(\Gamma)(\Gamma+g'(q_{\rm typ}(\Gamma)))}}\;,\;\;\Gamma\leq \Gamma_{\rm RSB}\;.
\ee
The variance can similarly be obtained by inserting the value of $q_{\rm typ}(\Gamma)$ obtained from ensuring $\alpha(q_{\rm typ}(\Gamma))=1$ in Eq. \eqref{mu_alph_res} into Eq. \eqref{res_gen_var}.

\begin{figure}
    \centering
    \includegraphics[width=0.45
    \textwidth]{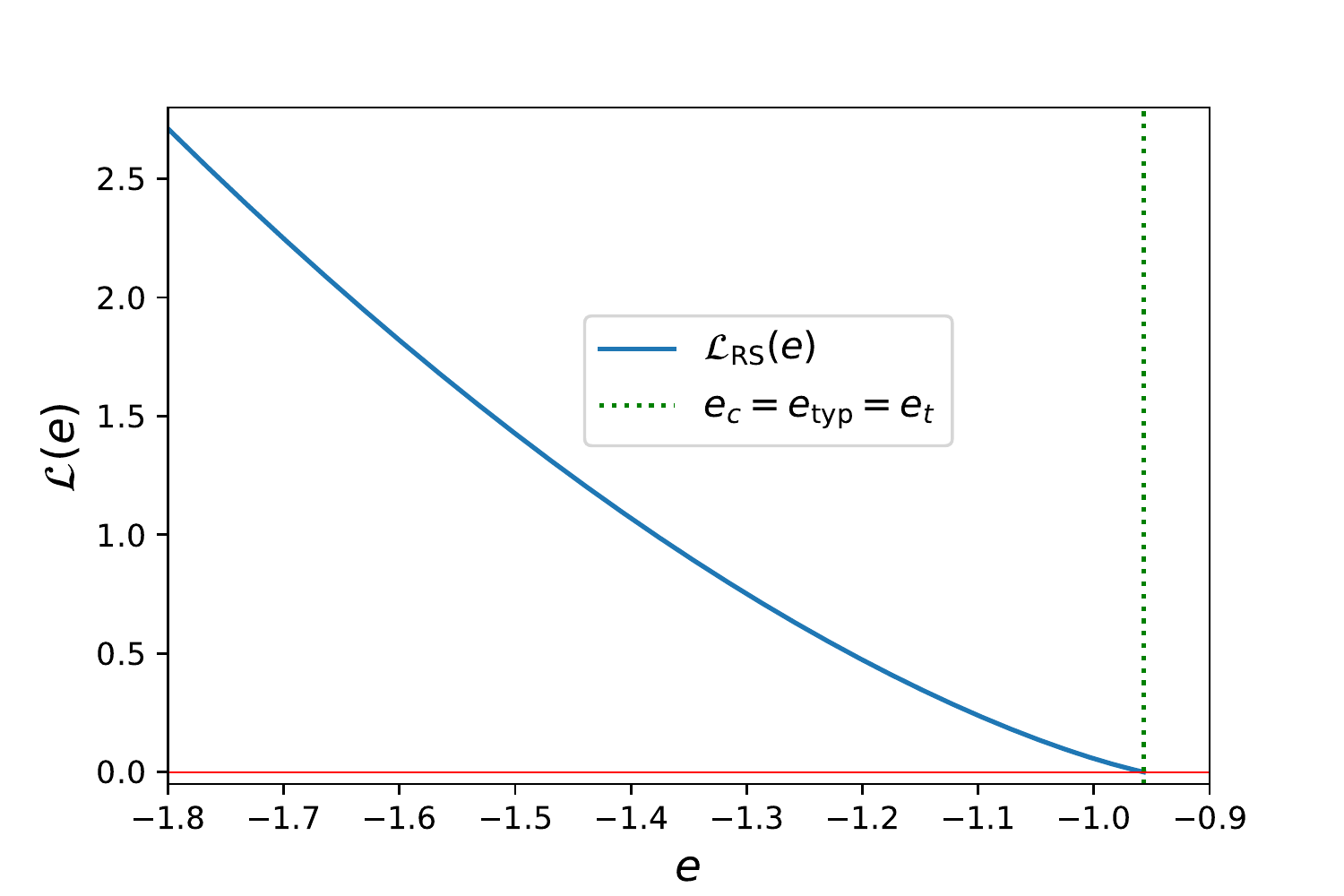}
    \includegraphics[width=0.45
    \textwidth]{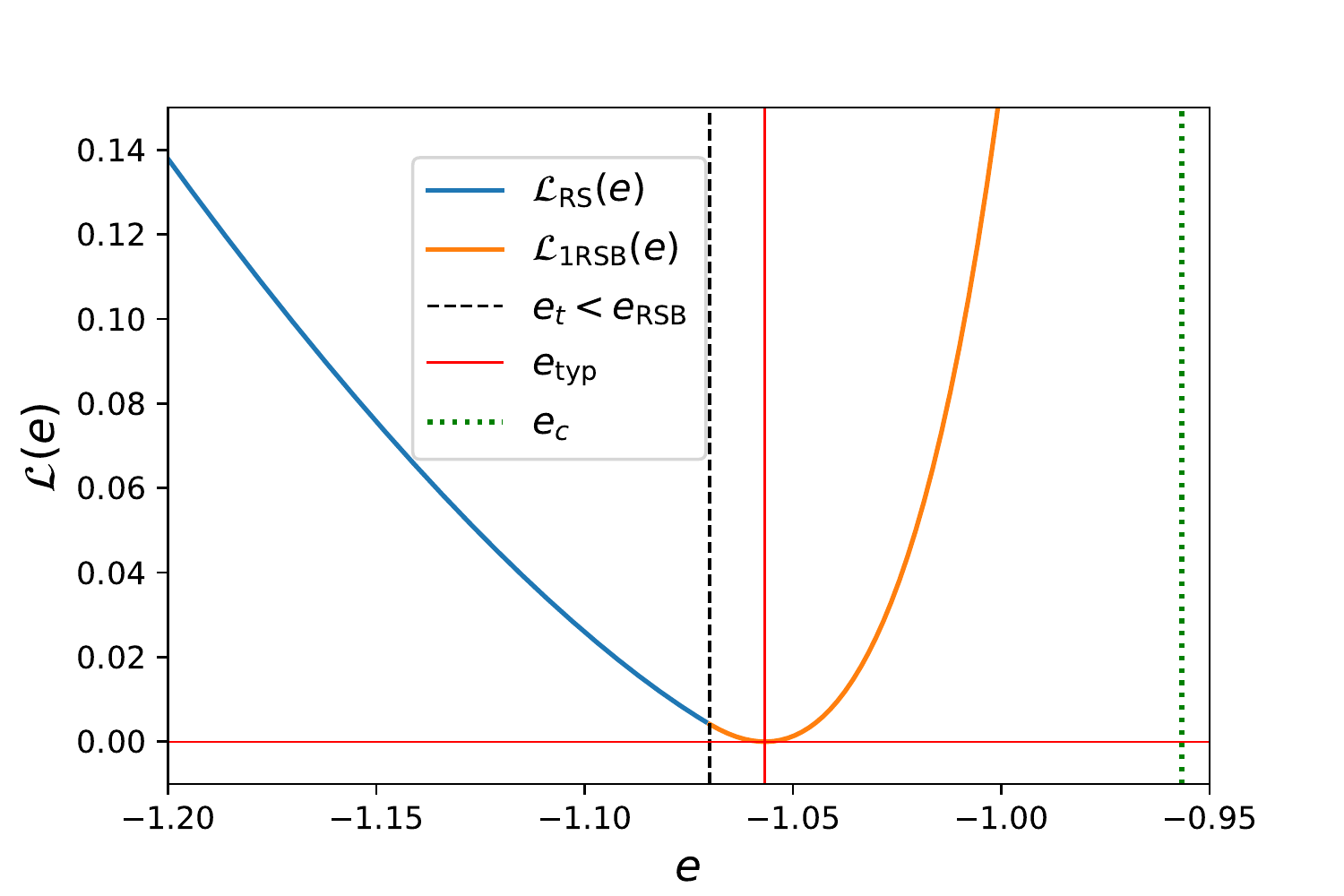}
    \includegraphics[width=0.45
    \textwidth]{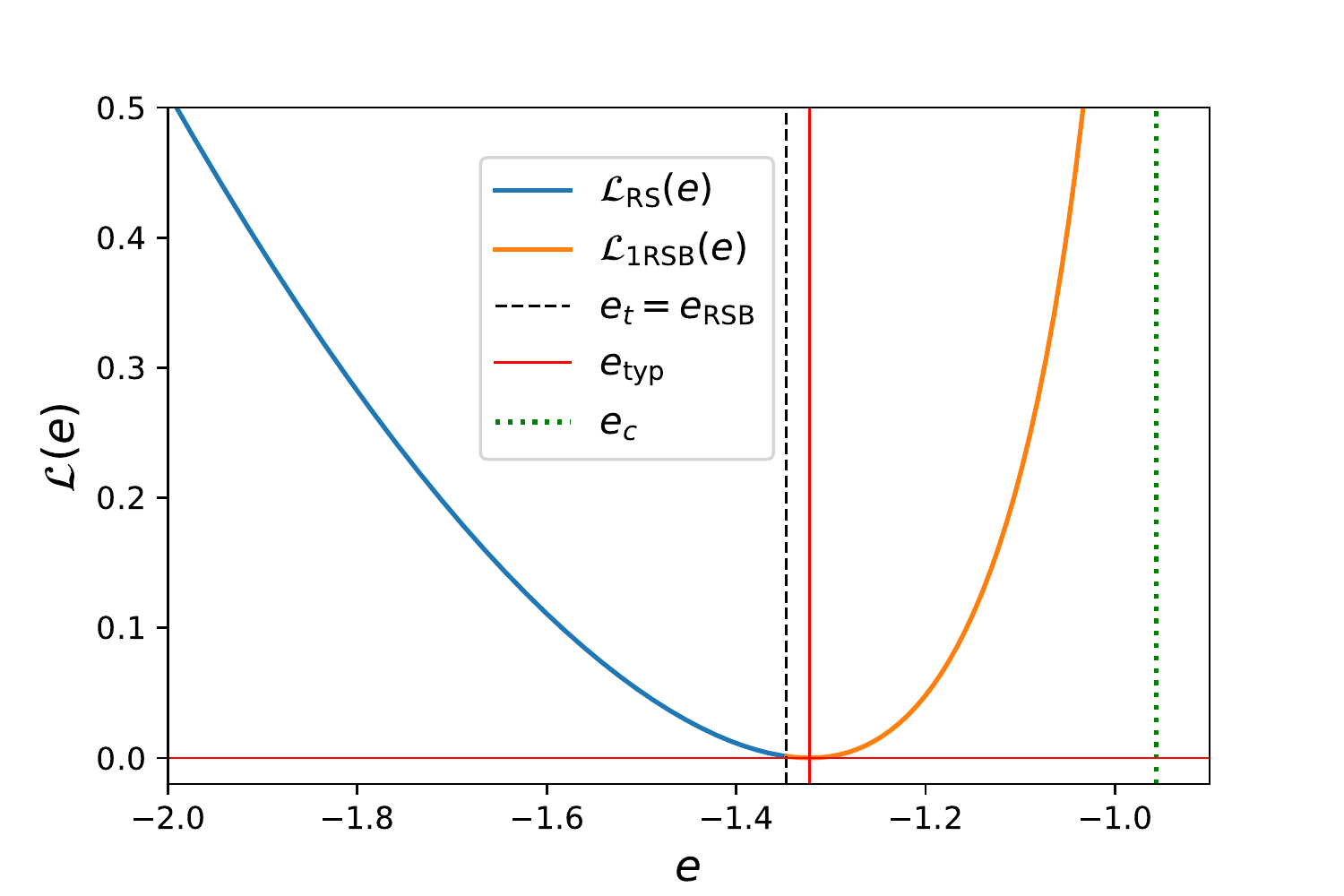}
    \includegraphics[width=0.45
    \textwidth]{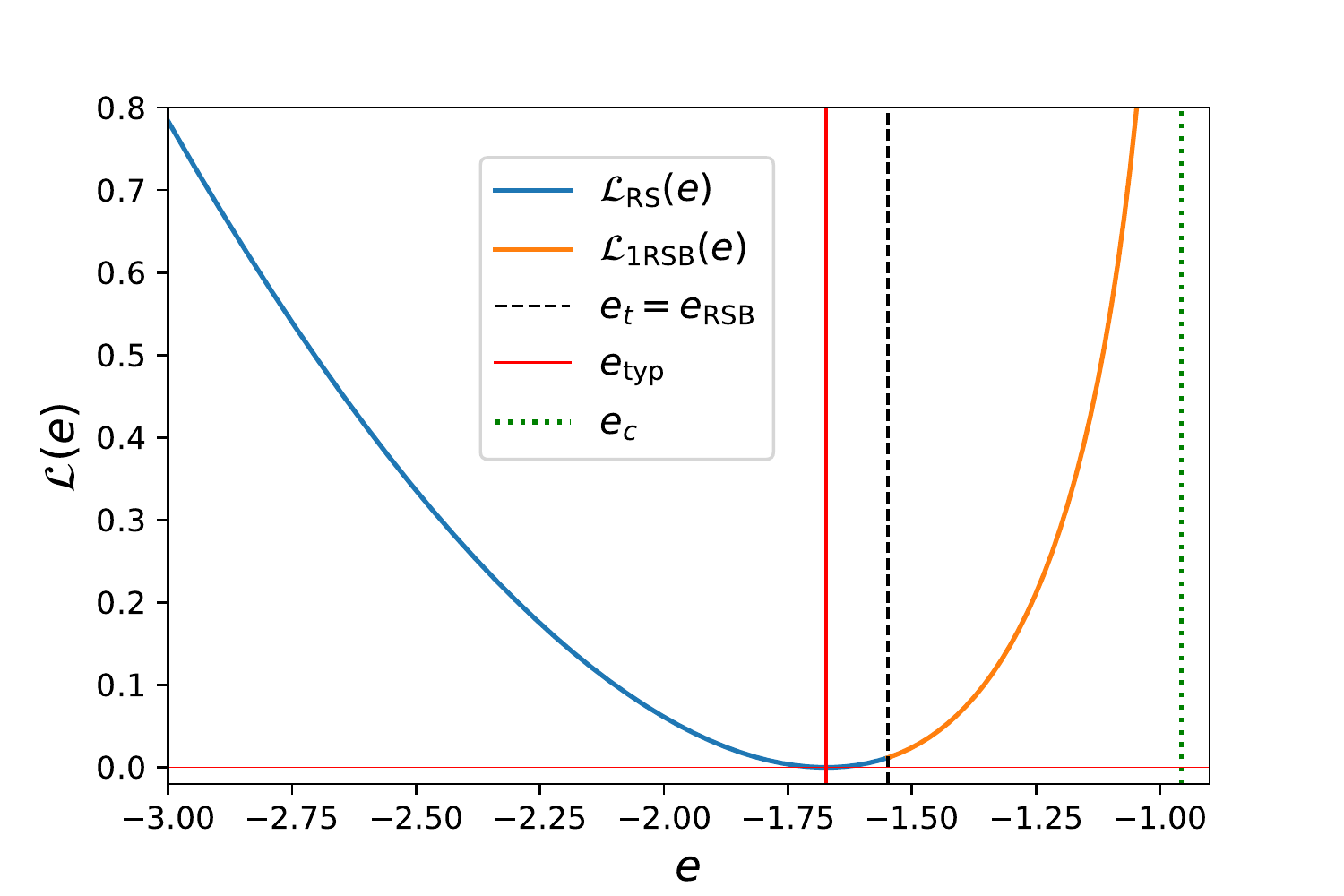}
    \caption{Plot of the large deviation function ${\cal L}(e)$ for the $3$-spin model for  $\Gamma=0$ (top left), $\Gamma=0.15$ (top right), $\Gamma=0.75$ (bottom left) and $\Gamma=1.8$ (bottom right).
    For $\Gamma>0$, the vertical black dashed line represents the value of $e_{t}$ separating on its left the RS regime where ${\cal L}(e)={\cal L}_{\rm RS}(e)$ plotted in blue and on its right the RSB regime where ${\cal L}(e)={\cal L}_{\rm 1RSB}(e)$ plotted in orange. While for $\Gamma>\Gamma_t=1/3$, the values $e_t$ and $e_{\rm RSB}$ coincide, $e_t<e_{\rm RSB}$ in the converse case. The solid red line marks the value $e_{\rm typ}$ and the dotted green line, the value of $e_c$ where the LDF diverges logarithmically. The typical value of the ground state $e_{\rm typ}$ belongs to the RS phase for $\Gamma>\Gamma_{\rm RSB}=1$. For $\Gamma=0$, the values $e_c,e_{\rm typ},e_{t}$ coincide and the 1RSB phase is limited to the single point $e=e_{\rm typ}$.}
    \label{fig:LDF_1rsb}
\end{figure}

Let us now give the result for the 1RSB expression of the LDF, ${\cal L}_{\rm 1RSB}(e)$. The energy $e$ is now fixed and parametrised in terms of $q_0^*$ and $\alpha(q_0^*)$ as in Eq. \eqref{e_1rsb_res}. The LDF ${\cal L}_{\rm 1RSB}(e)$ thus describes parametrically the range of energies $e \in [e_{t}(\Gamma),e_c]$ and reads
\be
{\cal L}_{\rm 1RSB}(e)=\frac{1}{2}\left(1-\alpha(q_0^*)+\ln\alpha(q_0^*)+\frac{g(q_0^*)+\Gamma q_0^*}{q_0^* (g'(q_0^*)+\Gamma)}\frac{(1-\alpha(q_0^*))^2}{\alpha(q_0^*)}\right)\;.
\ee
A plot of the LDF ${\cal L}(e)$, both in the RS and 1RSB phases, for the $3$-spin spherical model is shown in Fig. \ref{fig:LDF_1rsb} for several values of $\Gamma$. The transition from the RS to the 1RSB expression of the LDF occurs for the energy $e=e_t(\Gamma)=e_{\rm RSB}(\Gamma)$ for $\Gamma\geq \Gamma_t$ while it occurs for $e=e_t(\Gamma)<e_{\rm RSB}(\Gamma)$ for $\Gamma<\Gamma_t$.

{\bf Random field $\Gamma\geq \Gamma_{t}$:} 
The transition between the RS and 1RSB expressions of the LDF occurs at energy $e=e_t(\Gamma)=e_{\rm RSB}(\Gamma)$, corresponding to $q_0^*=1$. Expanding the difference as $e\to e_{\rm RSB}$, one obtains
\be
{\cal L}_{\rm 1RSB}(e)-{\cal L}_{\rm RS}(e)=\frac{\sqrt{g''(1)} \left(\Gamma +g'(1)\right)^3
   \left(g'(1)+\Gamma_t\right) \left(\Gamma -\Gamma_{\rm RSB}\right)}{6
   (\Gamma_t-\Gamma ) \left((\Gamma +g(1)) g''(1)+\left(g(1)-g'(1)\right)
   \left(\Gamma +g'(1)\right)\right)^3}(e-e_{t})^3+O(e-e_{t})^4\;.\label{tr_1RSB}
\ee
Quite remarkably, inserting the explicit expression of $\Gamma_t$ from Eq. \eqref{G_t_res}, one obtains, up to third order, the same expansion in Eqs. \eqref{tr_FRSB} and \eqref{tr_1RSB}. The transition between the RS and 1RSB phase is thus generically of third order for $\Gamma\geq \Gamma_{t}$ in the sense that the third derivative of the LDF ${\cal L}(e)$ is the lowest order discontinuous derivative at the transition. Note that at the special value $\Gamma=\Gamma_{\rm RSB}$ the transition is again of the fourth order with
\be
{\cal L}_{\rm 1RSB}(e)-{\cal L}_{\rm RS}(e)=-\frac{2g''(1)^3 (e-e_{t})^4}{3g^{(3)}(1)\left(g''(1)-2(g'(1)-g(1))\right)^4}+O(e-e_{t})^5\;,\;\;\Gamma=\Gamma_{\rm RSB}\;,
\ee
which takes again the same form as in the FRSB case in Eq. \eqref{tr_G_rsb_frsb}.

{\bf Random field $\Gamma<\Gamma_{t}$:} 
The transition between the RS and 1RSB expressions of the LDF occurs at energy $e=e_t(\Gamma)<e_{\rm RSB}(\Gamma)$, corresponding to $q_0^*=q_t<1$. Expanding the difference as $e\to e_{t}$, one obtains
\be
{\cal L}_{\rm 1RSB}(e)-{\cal L}_{\rm RS}(e)=\frac{{\cal L}_{\rm 1RSB}''(e_t)-{\cal L}_{\rm RS}''(e_t)}{2}(e-e_{t})^2+O(e-e_{t})^3\;,
\ee
where the expression of ${\cal L}_{\rm 1RSB}''(e_t)-{\cal L}_{\rm RS}''(e_t)$ is cumbersome but explicit (see Appendix \ref{1RSB_app} for details). The transition between the RS and 1RSB phase is thus of second order for $\Gamma< \Gamma_{t}$ in the sense that the second derivative of the LDF ${\cal L}(e)$ is the lowest order discontinuous derivative at the transition. 

For any $\Gamma>0$, the LDF diverges logarithmically close to the critical energy $e\to e_c$, corresponding to $q_0^*\approx 2(e_c-e)/\sqrt{\rho_{\rm typ}g'(1)}$ with $\alpha(q_0^*)\approx \Gamma/(2\sqrt{\rho_{\rm typ}g'(1)}(e_c-e))$ and where $\rho_{\rm typ}$ is the solution of Eq. \eqref{rho_typ_eq_res}, i.e.
\be
{\cal L}_{\rm 1RSB}(e)\approx\frac{1}{2}\ln\frac{2\sqrt{\rho_{\rm typ}g'(1)}(e_c-e)}{\Gamma}-\frac{1}{2}-\frac{g_2}{2g'(1)\rho_{\rm typ}}\;,\;e\to e_c\;.\label{div_wall_1RSB}
\ee
It is thus natural to expect that the fluctuations in the range of energies $e>e_{c}$ are described by a LDF with speed larger than $N$. In the regime of magnetic field $0<\Gamma\leq\Gamma_{\rm RSB} $, the LDF is quadratic about the typical energy
\be
{\cal L}_{\rm 1RSB}(e)=\frac{(e-e_{\rm typ})^2}{2{\cal V}_{\min}(\Gamma)}+O(e-e_{\rm typ})^3\;,\;\;\Gamma>0\;,\label{quad_typ_1RSB}
\ee
which is consistent with a central limit theorem and a typical distribution taking a Gaussian expression. 

For $\Gamma=0$ instead, one can show explicitly that $e_t(\Gamma=0)=e_{\rm typ}(\Gamma=0)=e_c$ and the 1RSB expression for the LDF is valid only for the single value $e=e_c$. The behaviour of the LDF in the vicinity of $e_{\rm typ}(\Gamma=0)=e_c$ is described by the RS expression instead. The typical energy $e_{\rm typ}(\Gamma=0)=e_c=e_{0,{\rm RS}}(\Gamma=0)$ coincides with the unique zero of ${\cal L}_{\rm RS}(e)$, defined in Eq. \eqref{e_0_def}, and the LDF behaves linearly in the vicinity of the typical energy
\be
{\cal L}(e)={\cal L}_{\rm RS}'(e_{\rm typ})(e-e_{\rm typ})+O(e-e_{\rm typ})^2\;,\;\;\Gamma=0\;.\label{lin_beh}
\ee
This expression is consistent with a typical distribution of energy displaying an exponential tail for $N(e_{\rm typ}-e)\gg 1$.

\subsubsection{Connection between large deviation function and complexities}

Let us now discuss briefly the connection between our results for the LDF and the statistics of the number of minima of the potential landscape associated to the spherical spin-glass. Denoting the empirical density of isolated minima of the random landscape as 
\be
n_{\min}(e)=\sum_{\alpha:{\rm minima}}\delta(e-e_{\alpha})\;,
\ee
where (at least) one of the $e_{\alpha}=e_{\min}$ corresponds to the ground-state energy. Note that an isolated minimum $\alpha$ is defined as a point ${\bf x}_\alpha$ on the $N$-sphere such that
\be
H({\bf x}_{\alpha})=e_{\alpha}\;,\;\;\nabla H({\bf x}_{\alpha})={\bf 0}\;,\;\;\min_{{\bf x}:{\bf x}^2=1}\left[ {\bf x}\cdot\nabla^2 H({\bf x}_{\alpha})\cdot {\bf x}\right]>0\;.
\ee
Taking the average, one obtains that 
\be
\overline{n_{\min}(e)}=\sum_{\alpha:{\rm minima}}\overline{\delta(e-e_{\alpha})}\geq {\cal P}(e)=\overline{\delta(e-e_{\min})}\;.
\ee
Using the definition of the annealed complexity
\be
\Sigma_{\min}(e)=\lim_{N\to \infty}\frac{1}{N}\ln\overline{ n_{\min}(e)}\;,
\ee
as well as the definition of the large-deviation function
\be
{\cal L}(e)=-\lim_{N\to \infty}\frac{1}{N}\ln {\cal P}(e)\;,
\ee
one thus obtains the simple bound
\be
{\cal L}(e)\geq -\Sigma_{\min}(e)\;,\label{bound}
\ee
valid for any energy $e$. Analysing the annealed complexity, one can check that in the regime $\Gamma\geq \Gamma_{\rm RSB}$, it is negative and has a unique maximum $e_{\rm typ}(\Gamma)=-\sqrt{g'(1)+\Gamma}$, where it reaches zero $\Sigma_{\min}(e_{\rm typ}(\Gamma\geq \Gamma_{\rm RSB}))=0$. This bound is thus meaningful for any energy $e$ in this regime. In the regime $\Gamma\leq \Gamma_{\rm RSB}$ instead, the annealed complexity is positive for energies within a finite band of energies $e\in[{\cal E}_{\min},{\cal E}_{\max}]$ which contains the typical energy, i.e. $\Sigma_{\min}(e_{\rm typ}(\Gamma))\geq 0$ (see Appendix \ref{app:comp_min} for details). As ${\cal L}(e)\geq 0$, the bound is meaningless within the range $e\in[{\cal E}_{\min},{\cal E}_{\max}]$ and in particular in the vicinity of $e_{\rm typ}(\Gamma)$.

From our results on the RS scheme, one can further show that (see appendix \ref{app:comp_min} for details) the bound in equation \eqref{bound} becomes an equality in the regime where the LDF is given by its RS expression, i.e.
\be
{\cal L}_{\rm RS}(e)=-\Sigma_{\min}(e)\;,\;\;e\leq e_{t}(\Gamma)\;.\label{comp_LDF}
\ee
\\
The quenched complexity of minima at fixed energy is defined as 
\be
\Xi_{\min}(e)=\lim_{N\to \infty}\frac{1}{N}\overline{\ln n_{\min}(e)}\leq \Sigma_{\min}(e)\;,
\ee
and by simple application of Jensen's inequality gives a lower bound for the annealed complexity as well. In addition, this quenched complexity is expected to vanish exactly for $e=e_{\rm typ}$, similarly to ${\cal L}(e)$. For the pure $p$-spin model without external field (discussed in more details in the next section), the annealed and quenched complexities have been shown to coincide exactly \cite{subag2017complexity} for any energy $e\leq e_{\rm RSB}(0)=-2\sqrt{p-1}/p$.

There are two natural questions that can be asked from our analysis:  Does the LDF ${\cal L}(e)$ and the quenched complexity $\Xi_{\min}(e)$ coincide in some finite range of energy, beyond the case of the pure $p$-spin? Do $e_t(\Gamma)$ or $e_{\rm RSB}(\Gamma)$ provide the threshold energy below which the quenched and annealed complexity coincide $\Sigma_{\min}(e)=\Xi_{\min}(e)$?

\subsubsection{The spherical pure $p$-spin model without magnetic field}

Let us consider in more details the case of the pure $p$-spin for $\Gamma=0$, where
\be
f(q)=g(q)=\frac{q^p}{p}\;.
\ee
The typical ground state energy in this case coincides with the critical energy $e_c$, where the LDF diverges for $\Gamma>0$, while it vanishes for $\Gamma=0$. It is given by
\begin{align}
    e_{\rm typ}&=e_c=\frac{e_{\rm RSB}}{2}\left(\sqrt{(p-1)\rho_{\rm typ}}+\frac{1}{\sqrt{(p-1)\rho_{\rm typ}}}\right)\geq e_{\rm RSB}\;,\\
    e_{c,{\rm RS}}&=e_{\rm RSB}=-\frac{2}{p}\sqrt{p-1}\;,
\end{align}
where the constant $\rho_{\rm typ}$ satisfies the equation
\be
\frac{1}{p}=\rho_{\rm typ}\frac{\rho_{\rm typ}-1-\ln\rho_{\rm typ}}{(1-\rho_{\rm typ})^2}\;.\label{rho_typ_p}
\ee
Since for $\Gamma=0$ the LDF for the range of energies $e\leq e_{\rm typ}$ is  given by the RS expression, the RSB regime is confined to a single point $e=e_c=e_{\rm typ}$. This case of the spherical model is quite specific as can already be seen from the identity $e_{c,{\rm RS}}=e_{\rm RSB}$. Note that for any $p>2$, one has that $e_{\rm RSB}> e_{\rm typ}$ (for $p=2$, one has $e_{\rm RSB}=e_{\rm typ}=-1$), thus we expect the RS ansatz for the large-deviation function to be stable around the typical ground-state $e_{\rm typ}$ given by a 1RSB ansatz.  
As mentioned in the previous sections, the quenched and annealed complexity of minima at fixed energy coincide and match the opposite of the LDF within the RS scheme (see appendix \ref{app:comp_min} for details) for any $e\leq e_{\rm typ}$
\be
{\cal L}_{\rm RS}(e)=-\Sigma_{\min}(e)=-\Xi_{\min}(e)=-\frac{1}{2}\ln(p-1)+\frac{p-2}{p}\frac{e^2}{e_{\rm RSB}^2}-\frac{e}{e_{\rm RSB}^2}\sqrt{e^2-e_{\rm RSB}^2}-\ln\left(\frac{e-\sqrt{e^2-e_{\rm RSB}^2}}{e_{\rm RSB}}\right)\;.
\ee
Finally, as has been shown explicitly in \cite{subag2017extremal} for $p>2$ the distribution of typical fluctuations of the ground-state energy is not given by a Gaussian distribution (as is the case in the presence of a finite external magnetic field $\Gamma$) but rather by a Gumbel distribution:  the ground state energy behaves at large $N$ as 
$E_{\rm min} \simeq N e_{\rm typ} + \frac{\log(c_p N^2)}{2 c_p} - K_0 - \frac{G}{c_p}$, where $G$ is a standard
Gumbel random variable with cumulative distribution $e^{-e^{-G}}$,  $K_0$ is a constant of order unity and 
\be
c_p=\Sigma_{\min}'(e_{\rm typ})=-{\cal L}_{\rm RS}'(e_{\rm typ})=s_*(e_{\rm typ})>0\;.
\ee
Equivalently one can state that the distribution of the ground state energy takes the scaling form
\be
{\cal P}(e)\approx N c_p\, {\cal G}\left(-N c_p(e-e_{\rm typ})+\frac{\ln (N c_p^2)}{2}-K_0 c_p\right)\;,\;\;{\cal G}(\epsilon)= \exp\left( -\epsilon-e^{-\epsilon}\right)\;,
\ee
The tail of this distribution takes the exponential form (for $e<e_{\rm typ}$)
\be
\ln{\cal P}\left(e\right)\approx -N{\cal L}_{\rm RS}'(e_{\rm typ})(e-e_{\rm typ})\;,\;\;(e_{\rm typ}-e)\gg \frac{\ln N}{N}\;,
\ee
which quite obviously matches the behaviour of the large-deviation function ${\cal L}_{\rm RS}(e)$ as $e_{\rm typ}-e\ll 1$. We therefore expect the LDF ${\cal L}_{\rm RS}(e)$ to correctly reproduce the large-deviation function of $e_{\min}$ for values $e\leq e_{\rm typ}$. As mentioned in the section on the RS phase, the fact that ${\cal L}_{\rm RS}(e)$ vanishes linearly around the typical value $e_{\rm typ}$ is not restricted to the pure $p$-spin model. It remains valid for any spherical spin model with zero external magnetic field, for which the typical value $e_{\rm typ}$ corresponds to the 1RSB scheme. It may indicate that the Gumbel distribution for the typical fluctuations of the ground state energy obtained for the pure $p$-spin extends to these more general situations.

Finally, since it was shown that for $\Gamma>0$ the typical fluctuations of the ground state energy of a generic spherical spin-glass model is described by a Gaussian distribution \cite{ChenSen2017}, there should naturally exist a scale of magnetic field, which allows to interpolate between the Gumbel distribution and the Gaussian distribution. More precisely comparing the variance of the ground state energy $N\mbox{Var}(e_{\min})\sim N \Gamma^2$ for $\Gamma>0$ with the Gumbel fluctuations $N\mbox{Var}(e_{\min}) = O(1)$ for $\Gamma=0$, we surmise that, in the regime
\be
N\to \infty\;,\;\;\Gamma\to 0\;,\;\;{\rm with}\;\;N^{1/2}\Gamma= \kappa=O(1)\;,
\ee
the probability distribution takes the scaling form
\be
{\cal P}(e)\to N^{-1}P_{\kappa=N^{1/2}\Gamma,g_p}\left(\delta=N(e-e_{\rm typ}(\Gamma))\right)\;,\;\;N\to \infty\;,
\ee
with the scaling function $P_{\kappa,g_p}(\delta)$ satisfying uniformly for any $\delta$ the asymptotic behaviours
\be
P_{\kappa,g_p}(\delta)\to \begin{cases}
\displaystyle P_{0,g_p}(\delta)=e^{-e^{\delta}+\delta}&\;,\;\;\kappa\to 0\;,\\
\displaystyle \frac{1}{\kappa}P_{\infty,g_p}\left(\frac{\delta}{\kappa}\right)&\;,\;\;\kappa\to \infty\;,
\end{cases}\label{P_g_p}
\ee
where 
\be
P_{\infty,g_p}(x)=\sqrt{\frac{\rho_{\rm typ}}{\pi}}\, e^{-\sqrt{\rho_{\rm typ}}x^2}\;,
\ee
and we remind that $\rho_{\rm typ}$ is the solution of Eq. \eqref{rho_typ_p}. Note that the conjectured form of the function in Eq. \eqref{P_g_p} describing the typical fluctuations of the ground-state of the spherical $p>2$ spin model behaves quite differently for finite $\kappa$ than the function describing the typical fluctuations of the $2$-spin in Eq. \eqref{conj_2_spin}.

\subsubsection{Scaling form for $\Gamma\to 0$ and matching to typical fluctuations}

In the above subsections we have considered the generic mixture model for a fixed value of $\Gamma$, either $\Gamma>0$
or $\Gamma=0$. We found that for $\Gamma>0$ the LDF takes a quadratic form near the typical value $e_{\rm typ}$, which is consistent
with a Gaussian form for the typical ground state energy fluctuations. For $\Gamma=0$ by contrast
we found that the LDF behaves as a power law near $e_{\rm typ}$ with an exponent $\eta$ which depends
on the model and the type of RSB, e.g. $\eta=1$ in the 1RSB case. Two natural questions are thus: (i) how does the
LDF interpolates between these two limits as $\Gamma \to 0$ ? (ii) how does the LDF interpolates with the
typical fluctuations ? In this subsection we present our results for the first question, and
provide some conjectures about the more difficult second question. Note that the presence of
a "wall" at $e=e_c$ must necessarily be addressed when dealing with the limit $\Gamma \to 0$ since
for $\Gamma=0$ the two energies $e_{\rm typ}(\Gamma)$ and $e_c$ coincide. 
We only provide here the main features and a more detailed description can be found in in Appendix \ref{app:analysis}.

{\bf Behaviour near the typical value}
The behaviour of the LDF in the vicinity of the the typical ground-state $e_{\rm typ}(\Gamma)$ is naturally influenced by the scaling of the rescaled variance ${\cal V}_{\min}(\Gamma)$ in the limit $\Gamma\to 0$. This scaling depends on the type of RSB and in the case of FRSB, on the small overlap behaviour of the covariance function
\be
g(q)=g_2 q^2+g_r q^r+o(q^r)\;.\label{p_def_G0}
\ee

For models which are described by a FRSB ansatz, this rescaled variance reads
\be
{\cal V}_{\min}(\Gamma)=\frac{g_r}{2}(r-1)(r-2)\left(\frac{\Gamma}{g_r r(r-2)}\right)^{\frac{r}{r-1}}\sim \Gamma^{\nu(2-\eta)}\;,\;\;\Gamma\to 0\;,
\ee
where we define the exponents
\be
\eta=\frac{3(r-2)}{2r-3}\;,\;\;\nu=\frac{2r - 3}{r-1}\;.
\ee
For these models, the difference $e_{\rm typ}(\Gamma)- e_{t}(\Gamma)=e_{\rm typ}(\Gamma)- e_{\rm RSB}(\Gamma)=O(1)$ and positive as $\Gamma\to 0$. In a vicinity of $e_{\rm typ}(\Gamma)$ of width of order $\Gamma^{\nu}$, the LDF takes the following scaling form
\be
{\cal L}(e)=\Gamma^{\eta\nu}L_{\rm FRSB}\left(\frac{e-e_{\rm typ}(\Gamma)}{\Gamma^{\nu}}\right)\;,\label{FRSB_scal_typ}
\ee
where $p$ is obtained from the leading order behaviours of the covariance function in Eq. \eqref{p_def_G0}. The scaling function $L_{\rm FRSB}$ is universal in the sense that it only depends on the small $q$ behaviour of the covariance function and given parametrically in Eqs. (\ref{scal_FRSB_typ_1}-\ref{scal_FRSB_typ_2}). Its asymptotic behaviours read
\be
L_{\rm FRSB}(u)\approx \begin{cases}
\displaystyle a_- |u|^{\eta}\;,\;\;u\to -\infty\;,\\
&\\
\displaystyle a_0 u^2\;,\;\;u\to 0\;,\\
&\\
\displaystyle a_+ u^3\;,\;\;u\to +\infty\;,
\end{cases}
\ee
where the values of $a_{-},a_0,a_+$ are given in Eq. \eqref{prefactors_scal_typ}. In particular, the first line matches the typical behaviour for $\Gamma=0$ in Eq. \eqref{powlaw_typ_FRSB}, the second line the typical behaviour for $\Gamma>0$ in Eq. \eqref{quad_typ_FRSB} and the third line matches the behaviour in the vicinity of the wall, which we will describe below.

For models which are described by a 1RSB ansatz, the rescaled variance reads
\be
{\cal V}_{\min}(\Gamma)=\frac{\Gamma^2}{2(\rho_{\rm typ}g'(1)-2g_2)}+O(\Gamma^3)\;,
\ee
where we remind that $0\leq \rho_{\rm typ}\leq 1$ is the solution of $g(1)/g'(1)=\rho_{\rm typ}(\rho_{\rm typ}-1-\ln \rho_{\rm typ})/(1-\rho_{\rm typ})^2$. For these models, the difference between the typical energy and the transition energy
is of order $e_{\rm typ}(\Gamma) - e_t(\Gamma)=-u_t\Gamma^2>0$. The LDF takes, in a vicinity of $e_{\rm typ}(\Gamma)$ of order $\Gamma^2$, a scaling form with both an RS and RSB branch
\begin{align}
  {\cal L}(e)&\approx\Gamma^2 L_{\rm 1RSB}\left(\frac{e-e_{\rm typ}(\Gamma)}{\Gamma^2}\right)\;,\;\;L_{\rm 1RSB}\left(u\right)=\begin{cases}
\displaystyle A-B\,u&\;,\;\;u\leq u_t<0\;,\\
&\\
\displaystyle C\,u^2&\;,\;\;u>u_t\;,
\end{cases}  \label{1RSB_scal_typ}
\end{align}
where the value of $u_t$ is given in Eq. \eqref{u_t} and of the amplitudes $A,B,C$ are given in \eqref{scal_1RSB_typ}. In the notations introduced before, this corresponds to $\nu=2$ and in addition one finds $\eta=1$. In particular, the limit $u\to -\infty$ matches smoothly the typical behaviour for $\Gamma=0$ in Eq. \eqref{lin_beh}, while the behaviour for $u\to 0$ matches the typical behaviour for $\Gamma>0$  in Eq. \eqref{quad_typ_1RSB}. Finally, the behaviour for $u\to +\infty$ matches the behaviour in the vicinity of the wall, see below.
\\

{\bf Behaviour near the wall.}
Let us now discuss the behaviour close to the wall as $\Gamma\to 0$. First note that the position of the wall $e_c$ is independent of $\Gamma$. For positive $\Gamma>0$, the LDF diverges close to the wall, i.e. as $e\to e_c$, while for $\Gamma=0$ it vanishes at the wall  since one has in that case $e_c=e_{\rm typ}(\Gamma=0)$. A non-trivial scaling function must therefore describe the interpolation between
these two situations. It is obtained by considering energies within a range $e_c-e\sim \Gamma$ and depends on the nature of the RSB phase. 

In the FRSB phase, after a proper rescaling, the scaling function is universal with respect to the choice of the covariance in the spherical spin-glass model and reads
\be
{\cal L}(e)\approx G\left(\frac{\sqrt{2g_2}(e-e_c)}{\Gamma}\right)\;,\;\;G(z)=\begin{cases}
\displaystyle 0 &\;,\;\;z\in (-\infty,z_{\rm typ})\;,\\
\displaystyle -\frac{1}{4}\left(3+8z+4z^2+2\ln(-2z)\right)&\;,\;\;z\in [z_{\rm typ},0]\;.\label{wall_scal_FRSB_res}
\end{cases}
\ee
As $z\to z_{\rm typ}=-1/2$, corresponding to the position of $e_{\rm typ}(\Gamma)$ for finite $\Gamma$, this scaling function vanishes cubically and matches the asymptotic behaviour $u\to +\infty$ of the scaling function $L_{\rm FRSB}(u)$ describing the vicinity of $e_{\rm typ}(\Gamma)$. Note that the region $z\leq z_{\rm typ}=-1/2$ can alternatively be described from Eq. \eqref{FRSB_scal_typ}, where ${\cal L}(e)$ is parametrically smaller in $\Gamma$. As $z\to 0$ instead, the scaling function $G(z)$ diverges logarithmically, thus reproducing the behaviour close to the wall for $\Gamma>0$ in Eq. \eqref{div_wall_FRSB}.

In the 1RSB phase, after a proper rescaling, the scaling function only depends on the specific model through the parameter $w=g_2/(\rho_{\rm typ}g'(1))$ and reads
\begin{align}
 {\cal L}(e)&\approx G_{\rm 1RSB}\left(\frac{\sqrt{\rho_{\rm typ}g'(1)}(e-e_c)}{\Gamma};\frac{g_2}{\rho_{\rm typ}g'(1)}\right)\;,\label{wall_scal_1RSB_res}\\
 G_{\rm 1RSB}(z;w)&=\begin{cases}
\displaystyle 0 &\;,\;\;z\in (-\infty,z_{\rm typ})\;,\\
\displaystyle -\frac{1}{2}\left((1+2z)\left(1+w(1+2z)\right)+\ln(-2z)\right)&\;,\;\;z\in[ z_{\rm typ},0]\;.
\end{cases}
\end{align}
As $z\to z_{\rm typ}=-1/2$, corresponding to the position of $e_{\rm typ}(\Gamma)$ for finite $\Gamma$, this scaling function vanishes quadratically (cubically for the special case $w=1/2$) and matches the asymptotic behaviour $u\to +\infty$ of the scaling function $L_{\rm 1RSB}(u)$ describing the vicinity of $e_{\rm typ}$. As $z\to 0$ instead, the scaling function $G_{\rm 1RSB}(z;w)$ diverges logarithmically, thus reproducing the behaviour close to the wall for $\Gamma>0$ in Eq. \eqref{div_wall_1RSB}.
\\

\begin{figure}
    \centering
    \includegraphics[width=0.8\textwidth]{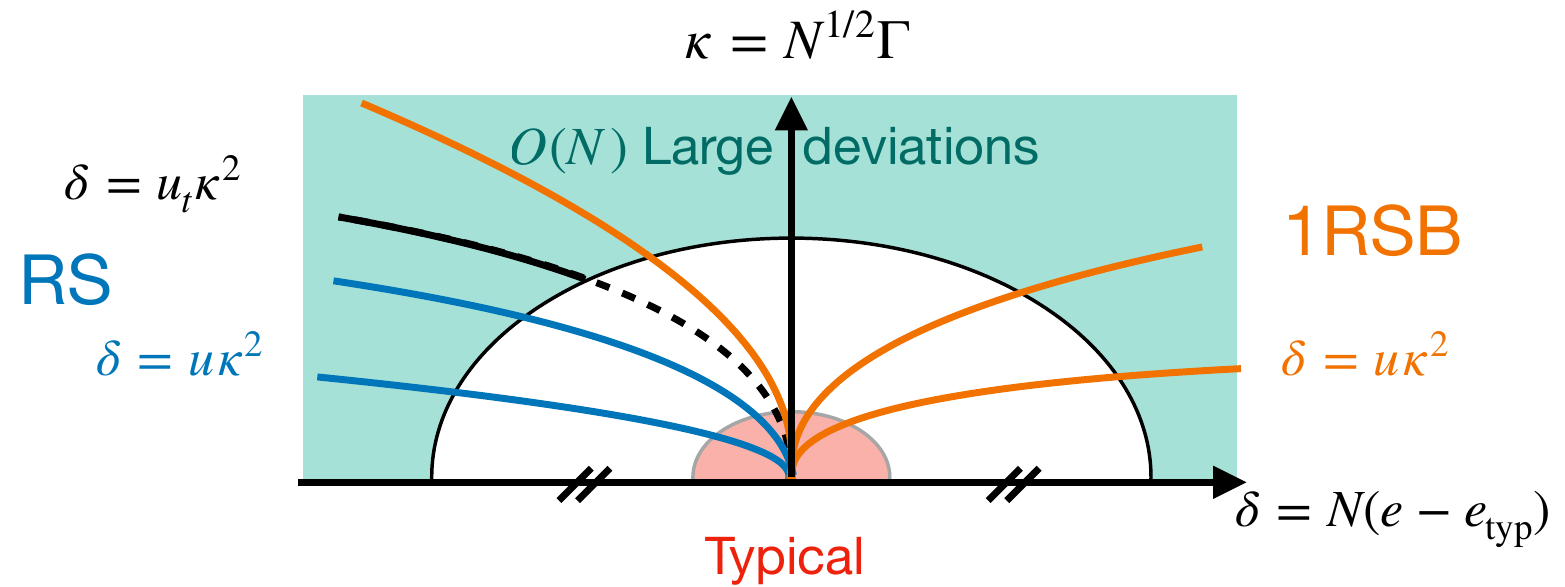}
    \caption{Plot of the different scaling regimes at large $N$ describing our conjectures for the distribution of the typical fluctuations { $P_{\kappa}(\delta)$} for models with a 1RSB phase with magnetic field in Eq. \eqref{conj_1RSB} and its matching for $\delta, \kappa \to \infty$ with our results for the large deviations. The large deviation region for $\delta =O(N)$ and $\kappa = O(N^{1/2})$ is indicated in green, the semi-circle
    represents schematically the matching region. The transition line is indicated as a solid line
    inside the large deviation regime where it is a true transition, and a dashed line in the typical region where
    it may signal a (smooth) change of behaviour. The vicinity of the wall on the speed $N$ RSB regime is described by yet another scaling behaviour see Eq \eqref{wall_scal_1RSB_res}. 
    }
    \label{fig:delta_kappa_1RSB}
\end{figure}

{\bf Matching from large deviations to typical fluctuations}. Let us now finally discuss the matching to the regime of typical fluctuations in the limit $\Gamma\to 0$. For any $\Gamma>0$, the LDF behaves quadratically around its typical value and one naturally expects that the typical distribution of $e_{\min}$ is Gaussian, as predicted in \cite{ChenSen2017}. For $\Gamma\to 0$ instead, one expects a non-trivial typical distribution to emerge. Although 
the methods of the present paper do not allow to obtain exact results in that regime, one can still propose some natural conjectures from simple considerations
of matching.

For models of spherical spin-glasses with a 1RSB phase, a natural conjecture from our analysis is that in the limit $N\to \infty$ with $\Gamma\sim N^{-1/2}$ and for $e-e_{\rm typ}(\Gamma) \sim 1/N$ there exists a family of distributions $P_{\kappa,g}(\delta)$ for the typical fluctuations such that
\be
{\cal P}(e)\to N P_{\kappa=N^{1/2}\Gamma,g}\left(\delta=N(e-e_{\rm typ}(\Gamma))\right)\;,\;\;N\to \infty\;,\label{conj_1RSB}
\ee
where $e_{\rm typ}(\Gamma)$ is the average ground-state energy including its finite $N$ corrections. In Fig. \ref{fig:delta_kappa_1RSB}, we show a scheme summarizing the matching and conjectures that we develop below for models with 1RSB phase. In the joint limit $\delta\to \infty$ and $\kappa\to \infty$, assuming a simple matching between our results for the LDF and the typical distribution, one expects that
\be
P_{\kappa,g}(\delta)\sim e^{-\kappa^2 L_{\rm 1RSB}\left(u=\frac{\delta}{\kappa^2}\right)}\;,\;\;\kappa\to \infty\;,\;\;|\delta|\to \infty\;,\;\;u=\frac{\delta}{\kappa^2}=O(1)\;.
\ee
Within the same joint limit, further considering 
$u\to -\infty$ one expects, in particular, that the function $P_{\kappa,g}(\delta)$ displays an exponential tail:
\be
P_{\kappa,g}(\delta)\sim e^{-B|\delta|}\;,\;\;\kappa\to +\infty\;,\;\;\delta\to -\infty\;,\;\;u=\frac{\delta}{\kappa^2}\to -\infty\;,
\ee
while in the opposite limit $u\to +\infty$ one expects instead that the function $P_{\kappa,g}(\delta)$ displays a Gaussian tail:
\be
P_{\kappa,g}(\delta)\sim e^{-C\frac{\delta^2}{\kappa^2}}\;,\;\;\kappa\to +\infty\;,\;\;\delta\to +\infty\;,\;\;u=\frac{\delta}{\kappa^2}\to +\infty\;.
\ee
The case of the spherical pure $p$-spin model has been mentioned in the previous section.

\begin{figure}
    \centering
    \includegraphics[width=0.8\textwidth]{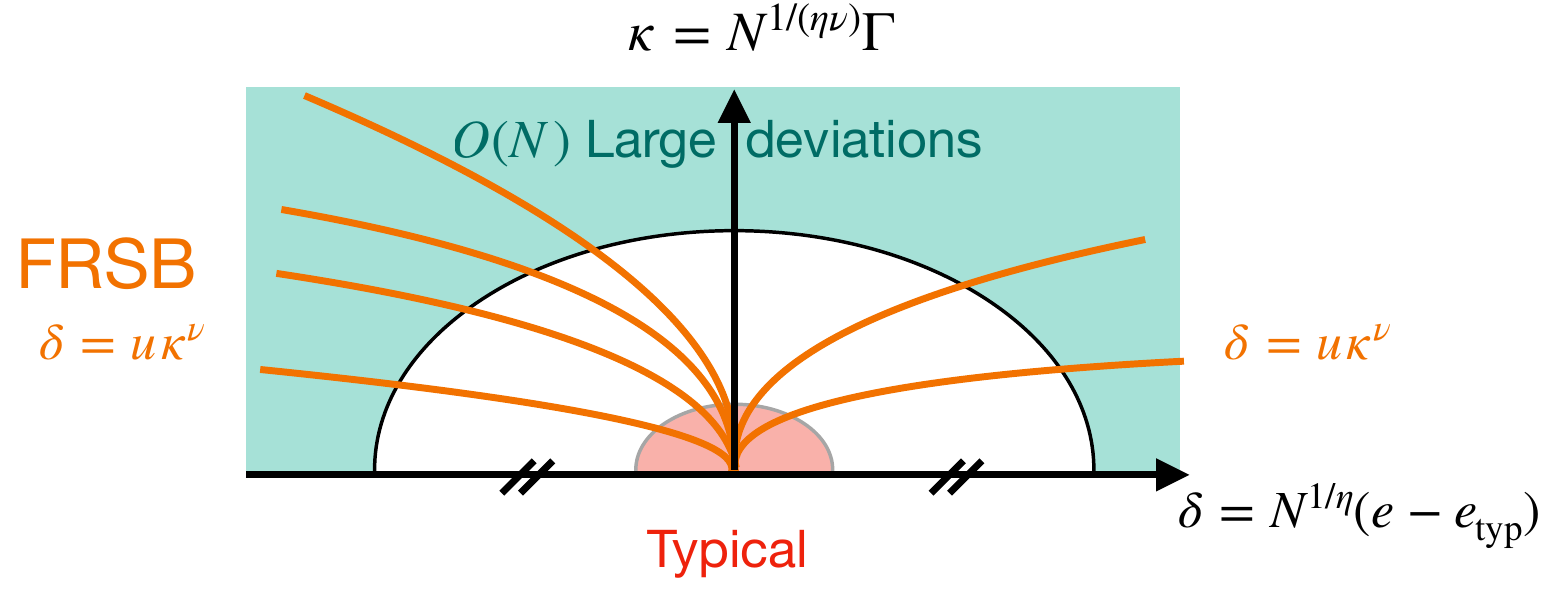}
    \caption{Plot of the different scaling regimes at large $N$ describing our conjectures for the distribution of the typical fluctuations { $P_{\kappa}(\delta)$} for models with a FRSB phase with magnetic field in Eq. \eqref{conj_FRSB} and its matching for $\delta, \kappa \to \infty$ with our results for the large deviations. The large deviation region for $\delta =O(N^{1/\eta})$ and $\kappa = O(N^{1/(\eta\nu)})$ is indicated in green, the semi-circle
    represents schematically the matching region. The transition line is indicated as a solid line
    inside the large deviation regime where it is a true transition, and a dashed line in the typical region where
    it may signal a (smooth) change of behaviour. The vicinity of the wall on the speed $N$ RSB regime is described by yet another scaling behaviour see Eq \eqref{wall_scal_FRSB_res}.  
    }
    \label{fig:delta_kappa_FRSB}
\end{figure}

For models of spherical spin-glasses with a FRSB phase, the natural scaling of $\Gamma$ as a function of $N$ depends on the behaviour of the covariance function $g(q)$ in Eq. \eqref{small_q_cov} and in particular the value of $r$. A natural conjecture from our analysis is that in the limit $N\to \infty$ with $\Gamma\sim N^{-1/\eta\nu}$, and for $e-e_{\rm typ}(\Gamma) \sim 1/N^{1/\eta}$, 
with $\eta=3(r-2)/(2r-3)$ and $\nu=(2r-3)/(r-1)$, there exist a family of distributions 
\be
{\cal P}(e)\to N^{{1/\eta}}P_{\kappa=N^{1/(\eta\nu)}\Gamma,g}\left(\delta=N^{1/\eta}(e-e_{\rm typ})\right)\;,\;\;N\to \infty\;,\label{conj_FRSB}
\ee
where $e_{\rm typ}$ is the average ground-state energy including its finite $N$ corrections. In Fig. \ref{fig:delta_kappa_FRSB}, we outline the matching schemes and conjectures that we develop below for models with FRSB phase. In the joint limit $\delta\to \infty$ and $\kappa\to \infty$, a simple matching between our results for the LDF and the typical distribution leads to the following conjecture:
\be
P_{\kappa,g}(\delta)\sim e^{-\kappa^{\eta\nu} L_{\rm FRSB}\left(u=\frac{\delta}{\kappa^\nu}\right)}\;,\;\;\kappa\to \infty\;,\;\;|\delta|\to \infty\;,\;\;u=\frac{\delta}{\kappa^\nu}=O(1)\;.
\ee
Within the same joint limit and in the further limit $u\to -\infty$, one expects the super-exponential tail of the distribution:
\be
P_{\kappa,g}(\delta)\sim e^{-a_{-}|\delta|^{\eta}}\;,\;\;\kappa\to +\infty\;,\;\;\delta\to -\infty\;,\;\;u=\frac{\delta}{\kappa^\nu}\to -\infty\;,
\ee
while in the opposite limit $u\to +\infty$ one instead expects the distribution to display the tail
\be
P_{\kappa,g}(\delta)\sim e^{- a_{+}\frac{\delta^{3}}{\kappa^3}}\;,\;\;\kappa\to +\infty\;,\;\;\delta\to +\infty\;,\;\;u=\frac{\delta}{\kappa^\nu}\to +\infty\;.
\ee
Note that this conjecture takes the same form as that proposed by us for the $2$-spin in Eq. \eqref{Conj_p_2_d}.

\section{Derivation of the cumulant generating function}\label{sec:derivation}

In this section we compute the cumulant generating function $\phi(s)$. We start by expressing the 
integer moments $\overline{Z(\beta)^n}$ of the partition sum in the large $N$ limit in terms
of the overlap matrix integral. We then evaluate the integral using a saddle point method 
exact for $N \to +\infty$. To perform the analytical continuation $n=s/\beta$ we 
introduce the Parisi parametrization. The cumulant generating function $\phi(s)$ 
is obtained formally from the solution of the saddle point equations, as 
well as the LDF ${\cal L}(e)$. These equations 
will be solved in detail in the Appendix \ref{app:analysis}, providing 
explicit expressions for these functions.

\subsection{Integer moments}

The integer moments of the partition sum \eqref{PS} can be expressed as 
\begin{align}
    \overline{Z(\beta)^n}&= \int_{\sqrt{N}{\cal S}_{N-1}} \prod_{a=1}^n d{\bf x}_a\,\overline{e^{-\beta\sum_{a=1}^n H({\bf x}_a)}}=\int_{\sqrt{N}{\cal S}_{N-1}} \prod_{a=1}^n d{\bf x}_a\,e^{\frac{N\beta^2}{2}\sum_{a,b}f\left(\frac{{\bf x}_a\cdot {\bf x}_b}{N}\right)}\,\\
    &=C_{N,n}N^{-n}\int_{Q>0}dQ\,\prod_{a=1}^n \delta(1-Q_{aa})\,(\det Q)^{-\frac{n+1}{2}}e^{N\Phi_n(Q)}\;,\\
    C_{N,n}&=(\pi N)^{Nn/2}\frac{\pi^{-\frac{n(n-1)}{4}}}{\prod_{k=0}^{n-1}\Gamma\left(\frac{N-k}{2}\right)}\;,
\end{align}
where in the second line we have used a standard replica method together with rotational invariance to re-exrpressed the integration over the replicated ${\cal S}_{N-1}^n$ sphere as an integral over the overlap matrix \cite{fyodorov2007classical,fyodorov2014topology}
\be
Q_{ab}=\frac{{\bf x}_a\cdot {\bf x}_b}{N}\;,\;\;Q_{aa}=1\;,\;\;a,b=1,\cdots,n\;.
\ee
The action $\Phi_n(Q)$ reads
\be
\Phi_n(Q)=\frac{1}{2}\left[\beta^2\sum_{a,b}f(Q_{ab})+\ln \det Q\right]\;.
\ee
In the large $N$ limit, we expect the integral over $Q$ to be dominated by a saddle-point at $Q=Q_*$ leading to the asymptotic behaviour
\begin{align}
 \overline{Z(\beta)^n}&\approx 2^{-n/2}(2\pi)^{\frac{n}{4}(2N-n-1)} N^{\frac{n(n-3)}{4}}e^{\frac{Nn}{2}}\frac{(\det Q_*)^{-\frac{n+1}{2}}}{\sqrt{\det A}}
e^{N\phi_n}\;,\\
\phi_n&=\max_{Q>0,Q_{aa}=1}\Phi_n(Q)=\Phi_n(Q_*)\;,\\
A_{(ab),(cd)}&=\left.\frac{\delta^2 \Phi_n(Q)}{\delta Q_{ab}\delta Q_{cd}}\right|_{Q=Q^*}=f''(Q_{ab}^*)\delta_{(ab),(cd)}-({Q^*}^{-1})_{ac}({Q^*}^{-1})_{bd}-({Q^*}^{-1})_{ad}({Q^*}^{-1})_{bc}\;.\label{quad_form}
\end{align}
The identity above is valid for any positive integer $n\ll N$.

\subsection{Parisi scheme}

We now suppose that the matrix $Q$ maximising the action $\Phi_n(Q)$ takes a Parisi block form with $k$ replica-symmetry breaking and parameters
\begin{align}
    1=q_{k+1}\geq q_k\geq\cdots\geq q_0\geq 0\;,\\
    n=m_0\geq m_1\geq\cdots\geq m_{k+1}=1\;.
\end{align}
The action can be expressed in terms of the positive function
\be
x(q)=n+\sum_{l=0}^k (m_{l+1}-m_l)\Theta(q-q_l)\;.
\ee
Note that one can similarly express \cite{fyodorov2007classical}
\be
\frac{1}{x(q)}=\frac{1}{n}+\sum_{l=0}^k \left(\frac{1}{m_{l+1}}-\frac{1}{m_l}\right)\Theta(q-q_l)\;.
\ee
In particular, this function satisfies
\be
x'(q)=\sum_{l=0}^k (m_{l+1}-m_l)\delta(q-q_l)\;,\;\;-\frac{x'(q)}{x(q)^2}=\sum_{l=0}^k \left(\frac{1}{m_{l+1}}-\frac{1}{m_l}\right)\delta(q-q_l)\;,
\ee
from which one obtains easily
\begin{align}
    \sum_{a,b}f(Q_{ab})&=n\left[f(1)+\sum_{l=0}^{k}(m_l-m_{l+1})f(q_l)\right]=n\left[f(1)-\int_{{q_0}_-}^1 dq\,x'(q)\,f(q)\right]\nn\\
    &=n^2 f(q_0)+n\int_{q_0}^1 dq\,x(q)\,f'(q)\;.
\end{align}
The eigenvalues of the matrix $Q$ are simply expressed as
\begin{align}
 \lambda_i&=1-q_{i-1} m_{i}-\sum_{l=i}^k (m_{l+1}-m_l)q_l=\int_{q_{i-1}}^1 x(q)dq\;,\;\;d_i=n\left(\frac{1}{m_{i}}-\frac{1}{m_{i-1}}\right)\;,\;\;i=1,\cdots,k+1\;,\\
 \lambda_0&=n q_0+\lambda_1=n q_0+\int_{q_0}^1 dq\,x(q)\;,\;\;d_0=1\;.
\end{align}
Finally, the determinant is obtained as \cite{fyodorov2007classical}
\begin{align}
   \ln \det Q =&\ln \lambda_0+n\sum_{l=0}^{k}\left(\frac{1}{m_{l+1}}-\frac{1}{m_l}\right)\ln \lambda_{l+1}=\ln \lambda_0+n\sum_{l=0}^{k}\left(\frac{1}{m_{l+1}}-\frac{1}{m_l}\right)\ln\left(\int_{q_{l}}^1 dq\,x(q)\right)\\
   =&\ln\left(nq_0+\int_{q_{0}}^1 dq\,x(q)\right)-n\int_{{q_0}_-}^{{q_k}_+}\frac{dq\,x'(q)}{x(q)^2}\ln\left(\int_{q}^1 dr\,x(r)\right)\\
   =&\ln\left(n q_0+\int_{q_0}^{1}dq\,x(q)\right)-\ln\left(\int_{q_0}^{1}dq\,x(q)\right)+n\ln(1-q_k)+n\int_{{q_0}}^{q_k}\frac{dq}{\displaystyle \int_{q}^{1}dr\,x(r)}\;,
\end{align}
where we have performed an integration by part and used that $x(q<q_0)=n$ and $x(q>q_k)=1$. Thus, the full expression of the action for arbitrary $n$ and $\beta$ reads
\begin{align}
    \Phi_n=&\frac{n\beta^2}{2}\left[n f(q_0)+f(1)-f(q_k)
    +\int_{q_0}^{q_k} dq\,x(q)\,f'(q)\right]+\frac{n}{2}\ln(1-q_k)\\
    &+\frac{1}{2}\left[\ln\left(n q_0+1-q_k+\int_{q_0}^{q_k}dq\,x(q)\right)-\ln\left(1-q_k+\int_{q_0}^{q_k}dq\,x(q)\right)\right]+\frac{n}{2}\int_{q_0}^{q_k}\frac{dq}{\displaystyle 1-q_k+\int_{q}^{q_k}dr\,x(r)}\;.\nn
\end{align}
Supposing that this function can be continued analytically from positive integer $n$ to arbitrary real value, we take the limit $\beta\to \infty$ of this expression such that
\be
n\beta=s=O(1)\;,\;\; 0\leq q_0\leq 1\;,\;\;\beta(1-q_k)=v=O(1)\;,\;\;\beta x(q)=z(q)=O(1)\;.
\ee
It yields
\begin{align}
    \lim_{\beta\to\infty}\Phi_{s/\beta}=\Phi(s)=&\frac{s}{2}\left[s f(q_0)+v f'(1)
    +\int_{q_0}^{1} dq\,z(q)\,f'(q)\right]\label{phi_z}\\
    &+\frac{1}{2}\left[\ln\left(s q_0+ v+\int_{q_0}^{1}dq\,z(q)\right)-\ln\left( v+\int_{q_0}^{1}dq\,z(q)\right)\right]+\frac{s}{2}\int_{q_0}^{1}\frac{dq}{\displaystyle v+\int_{q}^{1}dr\,z(r)}\nn\;,
\end{align}
where the function $z(q)=s$ for $q\in[0,q_0)$. In that limit, the cumulant generating function is given by the extremum of this action
\be
\phi(s)=\begin{cases}
\displaystyle\max_{z(q),q_0,v}\Phi(s)&\;,\;\;s<0\;,\\
0&\;,\;\;s=0\;,\\
\displaystyle\min_{z(q),q_0,v}\Phi(s)&\;,\;\;s>0\;.
\end{cases}\label{ext_phi}
\ee
The extremisation with respect to $q_0$ and $v$ yield
\begin{align}
\left.\partial_{q_0}\Phi(s)\right|_{\rm s.p.}&=\frac{s}{2}(s-z_*(q_0^*))\left(f'(q_0^*)-\frac{q_0^*}{\left(v_*+\int_{q_0^*}^1 dq\, z_*(q)\right)\left(s q_0^*+v_*+\int_{q_0^*}^1 dq\, z_*(q)\right)}\right)=0\;,\label{sp_eq_q_0}\\
\left.\partial_{v}\Phi(s)\right|_{\rm s.p.}&=\frac{s}{2}\left(f'(1)-\int_{q_0^*}^{1}\frac{dq}{\left(v_*+\int_{q^*}^1 dr\, z_*(r)\right)^2}-\frac{q_0^*}{\left(v_*+\int_{q_0^*}^1 dq\, z_*(q)\right)\left(s q_0^*+v_*+\int_{q_0^*}^1 dq\, z_*(q)\right)}\right)=0\;,\label{sp_eq_v}
\end{align}
where ${\rm s.p.}$ stands for the saddle-point solution $z=z_*$, $q_0=q_0^*$, $v=v_*$. From the first equation, one obtains that $q_0^*=0$ is a solution of the saddle-point equations in the absence of magnetic field $\Gamma=f'(0)=0$. However, this solution yields values of $v_*$ and $z_*(q)$ that are independent of $s$. For such solution, the cumulant generating function $\phi(s)$ is a linear function of $s$. In the general case, there exist two possible solutions of the first equation, namely 
\be
s=z_*(q_0^*)\label{sol_q_not_ok}
\ee
or
\be
v_*+\int_{q_0^*}^1 dq\, z_*(q)=\frac{q_0^*}{2}\left(-s+\sqrt{\frac{4}{q_0^* f'(q_0^*)}+s^2}\right)\;.\label{sol_q}
\ee
where in solving the quadratic equation for the term in the right-hand side obtained from Eq. \eqref{sp_eq_v}, we have kept the positive root.

Using the second solution \eqref{sol_q}, one obtains an additional identity from Eq. \eqref{sp_eq_v} 
\be
\int_{q_0^*}^1\frac{dq}{\left(v_*+\int_{q}^1 dr\, z_*(r)\right)^2}=f'(1)-f'(q_0^*)\;.\label{id_var}
\ee

Inserting the expression in Eq. \eqref{sol_q} into the expression of the CGF in Eq. \eqref{phi_z}, one obtains the following expression
\begin{align}
    \phi(s)=&\frac{s}{2}\left[s f(q_0^*)+\frac{f'(1)q_0^*}{2}\left(-s+\sqrt{\frac{4}{q_0^* f'(q_0^*)}+s^2}\right)
    -\int_{q_0^*}^{1} dq\,z_*(q)\,[f'(1)-f'(q)]\right]\\
    &+\ln\left(\frac{s \sqrt{q_0^* f'(q_0^*)}+\sqrt{4+q_0^* f'(q_0^*)s^2} }{2}\right)+\frac{s}{2}\int_{q_0^*}^{1}\frac{dq}{\displaystyle \frac{q_0^*}{2}\left(-s+\sqrt{\frac{4}{q_0^* f'(q_0^*)}+s^2}\right)-\int_{q_0^*}^{q}dr\,z_*(r)}\;,\nn
\end{align}
as a function of $q_0^*$ and $z_*(q)$ to be determined by extremisation. From this general expression of the CGF, one expects to recover the typical energy by computing
\begin{align}
    e_{\rm typ}&=\overline{e_{\min}}=-\phi'(s=0)\nn\\
    &=-\frac{1}{2}\left(\sqrt{\frac{q_{\rm typ}}{f'(q_{\rm typ})}}[f'(1)+ f'(q_{\rm typ})]-\int_{q_{\rm typ}}^{1} dq\,z_*(q)\,\left[f'(1)-f'(q)\right]+\int_{q_{\rm typ}}^{1}\frac{dq}{\displaystyle \sqrt{\frac{q_{\rm typ}}{f'(q_{\rm typ})}}-\int_{q_{\rm typ}}^{q}dr\,z_*(r)}
\right)\;,
\end{align}
where we have used that for $s=0$, $q_0^*=q_{\rm typ}$. Similarly, the variance can be computed as
\be
{\cal V}_{\min}(\Gamma)=\lim_{N\to \infty} N\mbox{Var}(e_{\min})=\phi''(s=0)=f(q_{\rm typ})-\frac{q_{\rm typ} f'(q_{\rm typ})}{2}=\frac{1}{2}\left[\Gamma q_{\rm typ}-\sum_{r=2}^{\infty}(r-2)\,g_r q_{\rm typ}^r
\right]\;,\label{var_gen}
\ee
where we have used Eq. \eqref{sol_q} as well as Eq. \eqref{id_var}. Importantly, from this expression, one can check that the rescaled variance ${\cal V}_{\min}(\Gamma)$ of the ground-state energy is of order $O(1)$ for any positive magnetic field $\Gamma>0$ but vanishes for zero magnetic field (as $q_{\rm typ}(\Gamma=0)=0$).
\\

We now turn to the LDF ${\cal L}(e)$. Supposing that $\phi(s)$ is strictly convex and using that ${\cal L}(e)$ is the Legendre transform of $\phi(s)$,
\be
{\cal L}(e)=-\min_{s\in \mathbb{R}}\left[s e+\phi(s)\right]\;,
\ee
one may express the value of $s_*$ minimising the action above as \be
e=-\phi'(s_*)=-\frac{1}{2}\left(2 s_* f(q_0^*)+v_* f'(1)+\int_{q_0^*}^{1} dq\,z_*(q)\,f'(q)+\int_{q_0^*}^{1}\frac{dq}{\displaystyle v_*+\int_{q}^{1}dr\,z_*(r)}+\frac{q_0^*}{s_* q_0^*+v_*+\int_{q_0^*}^{1} dq\,z_*(q)}\right)\;,\label{e_s}
\ee
where $v_*$, $z_*(q)$ and $q_0^*$ are the values that maximise the action $\Phi(s)$. A parametric expression for ${\cal L}(e)$ can then be obtained in terms of $s_*$, determined parametrically from $e$ in Eq. \eqref{e_s}, which reads 
\be
{\cal L}(e)=s_* \phi'(s_*)-\phi(s_*)=\frac{1}{2}\left[s_*^2 f(q_0^*)+\frac{q_0^* s_*}{s_* q_0^*+v_*+\int_{q_0^*}^{1} dq\,z_*(q)}-\ln\left(1+\frac{q_0^* s_*}{v_*+\int_{q_0^*}^{1} dq\,z_*(q)}\right)\right]\;.
\ee
Note that if Eq. \eqref{sol_q} holds, this function takes a simpler expression in terms of $q_0^*$ and $s_*$ only, i.e.
\be
{\cal L}(e)=\frac{s_*^2}{4}\left( 2f(q_0^*)-q_0^* f'(q_0^*)\right)+\frac{s_* q_0^* f'(q_0^*)}{4}\sqrt{\frac{4}{ q_0^* f'(q_0^*)}+s_*^2}-\ln\left[\frac{\sqrt{q_0^* f'(q_0^*)}}{2}\left(s_*+\sqrt{\frac{4}{ q_0^* f'(q_0^*)}+s_*^2}\right)\right]\;.
\ee

We have given here some properties of the saddle point equations, and of the
resulting functions $\phi(s)$ and ${\cal L}(e)$ which hold generally. To obtain
explicit expressions for these functions we need now to determine $z^*(q)$ and $q_0^*$ 
and distinguish between several phases, with either replica symmetry or various forms of replica symmetry breaking. This is done in details in Appendix \ref{app:analysis}.

\section{Discussion and conclusion}\label{sec:conclusion}

We have demonstrated that the replica method can be successfully employed  to compute analytically the full expression of the speed $N$ large-deviation function of the ground-state energy for a generic mean-field spherical spin-glass model in a random magnetic field of variance $\Gamma$. While depending on the strength of the field the typical (or, which turns out to be the same, mean) value of the ground-state energy is obtained either from a replica-symmetric or replica-symmetry broken Ansatz, the LDF displays for any value of $\Gamma$ both a branch obtained from a RS ansatz (characterising deviations with lowest ground-state energy) and a branch obtained from a RSB ansatz (which may reduce to a single point for zero magnetic field) characterising deviations with highest ground-state energy. In particular, our results reproduce the expression derived rigorously in \cite{dembo2015matrix} for the spherical $2$-spin model. A derivation of  general LDF results presented in this paper for broader classes of spherical spin-glass model using rigorous methods and identification of the mechanisms responsible for the RSB effects in large deviations constitute an interesting challenge. Beyond a finite critical energy $e_c$, corresponding to the mean value of the ground-state in the absence of external field, the ground state fluctuations are expected instead to be described by a large-deviation function of larger speed. The characterisation of this LDF with higher speed is an interesting open problem. A possible direction to consider is employing replica computations by scaling properly the number of replica with the dimension $N$ as explored in \cite{parisi2010universality}.

While for a variance $\Gamma=O(1)$ the distribution of typical fluctuations takes a simple Gaussian form, one expects a non-trivial distribution to arise by properly rescaling the variance with the dimension $N$. This problem seems currently to be beyond the reach of replica computations and innovative methods will need to be developed to obtain the typical distribution.

Another problem worth exploring is a derivation of the LDF for a fixed external magnetic field. While the results of the present article could be recovered from the results with fixed magnetic field, taking into account the large deviation function for the norm of the magnetic field, the converse is not true.

Finally, our method can be used to investigate the speed $N$ LDF of the free-energy at positive temperature. One expects that the main qualitative features observed for zero temperature, namely the two branches of the LDF and its logarithmic divergence, will be operative at finite temperature as well. Note that one can expect additional transitions to arise in pure models beyond the critical temperature, where the annealed and quenched free-energy coincide. In particular, this is indicated by the fact of the CDF being linear in a finite range of $s$ and that the LDF vanishes linearly in the vicinity of the typical free-energy.\\

{\bf Data availability satement:}
The datasets generated during and/or analysed during the current study are available from the corresponding author on reasonable request.

{\bf Conflict of interest satement:}
The authors have no competing interests to declare that are relevant to the content of this article.

\begin{acknowledgments}
 B. Lacroix-A-Chez-Toine  would like to thank P. Vivo for a  discussion which helped to initiate the present reasearch and V. Ros for useful communications.
  Y. V. Fyodorov and P. Le Doussal aknowledge  O. Zeitouni and J. Baik for stimulating interest in the work and useful discussions at various stages of this project.
The research by B. Lacroix-A-Chez-Toine and Y. V. Fyodorov was supported by the EPSRC Grant EP/V002473/1 {\it Random Hessians and Jacobians: theory and applications}. P. Le Doussal thanks King's College London for hospitality. 
\end{acknowledgments}

\appendix

\section{Typical ground state energy for generic mixture model}
\label{app:typical} 

In this first appendix, let us discuss the behaviour of the typical energy $e_{\rm typ}(\Gamma)=\overline{e_{\min}}$ as a function of the covariance function and the variance $\Gamma$ of the magnetic field.
The typical and mean-value of the ground-state energy is given by
\begin{align}
e_{\rm typ}(\Gamma)=\overline{e_{\min}}=&-\frac{1}{2}\min_{v\in \mathbb{R}^+,q_0\in [0,1],z(q)}\left[v f'(1)
    +\int_{q_0}^{1} dq\,z(q)\,f'(q)+\frac{q_0}{v+\int_{q_0}^{1}dq\,z(q)}+\int_{q_0}^{1}\frac{dq}{\displaystyle v+\int_{q}^{1}dr\,z(r)}\right]\;,\label{e_typ_FRSB}
\end{align}
where $0\leq q_0\leq 1$, $v$ is positive and the function $z(q)$ is a real non-negative non-decreasing function of $q$, satisfying $z(q)=0$ for $q\in[0,q_0)$ and $z(q)\geq 0$ for $q\in [q_0,1]$. There exists a value of the variance
of the random field
\be
\Gamma_{\rm RSB}=g''(1)-g'(1)=\sum_{r=2}^{\infty}r(r-2)\,g_r\,,\label{G_RSB}
\ee
such that any $\Gamma>\Gamma_{\rm RSB}$, the typical energy $e_{\rm typ}(\Gamma)$ is obtained within the RS scheme. In that scheme, the optimal parameter $q_0^*=q_{\rm typ}=1$, and the function $z(q)=0$ for any $q\in[0,1]$. The typical energy in this scheme is given by
\be
e_{\rm typ}(\Gamma)=-\frac{1}{2}\min_{v\in \mathbb{R}^+}\left[v f'(1)+\frac{1}{v}\right]=-\sqrt{f'(1)}=-\sqrt{g'(1)+\Gamma}\;,\;\;\Gamma\geq \Gamma_{\rm RSB}\;.\label{RS_e_min}
\ee
In contrast, for $\Gamma\leq \Gamma_{\rm RSB}$ it is obtained from a RSB solution and its expression depends on the level of replica-symmetry breaking. We consider three cases:
\\

{\bf First case: Full replica symmetry breaking}.
Assuming that the Schwarzian derivative of $f'(q)$ is non-negative, i.e.
\be
{\cal S}\left[f'\right](q)=\frac{f^{(4)}(q)}{f''(q)}-\frac{3}{2}\left(\frac{f^{(3)}(q)}{f''(q)}\right)^2\geq 0\;,\;\;q\in[0,1]\;,\label{cond_FRSB}
\ee
then the solution to the optimisation problem is obtained within the full replica-symmetry breaking (FRSB) framework as
\be
z_*(q)=\frac{f^{(3)}(q)}{2f''(q)^{3/2}}\geq 0\;,\;\;q\in[q_{\rm typ},1]\;.\label{z_eq_res}
\ee
In fact this can be easily understood, since the derivative of $z_*(q)$ has a simple expression in terms of this criterion
\be
z_*'(q)=\frac{{\cal S}\left[f'\right](q)}{2\sqrt{f''(q)}}\geq 0\;,\;\;q\in[q_{\rm typ},1]\;.
\ee
The optimal value of $v$ is then $v_*=f''(1)^{-1/2}$. The expression of the ground-state energy then reads for $\Gamma\leq \Gamma_{\rm RSB}$ as
\be
e_{\rm typ}(\Gamma)=-\frac{1}{2}\min_{q_0\in [0,1]}\left[
    \frac{f'(q_0)+q_0 f''(q_0)}{\sqrt{f''(q_0)}}+2\int_{q_0}^{1} dq\,\sqrt{f''(q)}\right]=-\left[q_{\rm typ}\sqrt{f''(q_{\rm typ})}+\int_{q_{\rm typ}}^{1} dq\,\sqrt{f''(q)}\right]\;,
\ee
where $q_0^*=q_{\rm typ}$ is the solution of the equation
\be
(q_{\rm typ} f''(q_{\rm typ})-f'(q_{\rm typ}))z(q_{\rm typ})=0\;.\label{q_typ_FRSB}
\ee
The solution $q_{\rm typ}(\Gamma)$ is found by inverting the following function
\be
\Gamma=\Gamma_{\rm FRSB}(q_{\rm typ})=q_{\rm typ}g''(q_{\rm typ})-g'(q_{\rm typ})=\sum_{r\geq 2}r(r-2)\,g_r\,q_{\rm typ}^{r-1}\;.
\ee
The function $\Gamma_{\rm FRSB}(q_{\rm typ})$ is a non-decreasing function of $q_{\rm typ}$ with $\Gamma_{\rm FRSB}(0)=0$ and $\Gamma_{\rm FRSB}(1)= \Gamma_{\rm RSB}$. In the case of zero magnetic field, the value of $q_{\rm typ}=0$. In the limit $\Gamma\to \Gamma_{\rm RSB}$, $q_{\rm typ}\to 1$ and one has that $f''(1)=f'(1)=\Gamma+g'(1)$. The replica-symmetry is thus recovered and the value of $e_{\rm typ}$ obtained via the FRSB scheme matches the RS result in Eq. \eqref{RS_e_min}. The transition occurring for $\Gamma=\Gamma_{\rm RSB}$ is of third order as the value of $e_{\rm typ}(\Gamma)$ and its first two derivatives are continuous for $\Gamma=\Gamma_{\rm RSB}$ while
\be
e_{\rm typ}^{(3)}(\Gamma_{\rm RSB}^+)-e_{\rm typ}^{(3)}(\Gamma_{\rm RSB}^-)=\frac{1}{24g''(1)^{3/2} g^{(3)}(1)}\;.\label{e_typ_d_3}
\ee
Note that in the case of the $2$-spin, the typical value of the energy can also be retrieved through the FRSB scheme by optimising Eq. \eqref{e_typ_FRSB}. In that case, $z_*(q)=0$ for any $q\in [0,1]$, which simplifies the computation. The optimal value $v_*=f'(1)^{-1/2}$ and not $f''(1)^{-1/2}$ and the parameter $q_0$ plays no role. 
\\

{\bf Second case: $k$-step replica symmetry breaking}. 
Assuming instead that 
\be
{\cal S}\left[f'\right](q)=\frac{f^{(4)}(q)}{f''(q)}-\frac{3}{2}\left(\frac{f^{(3)}(q)}{f''(q)}\right)^2\leq 0\;,\;\;q\in[0,1]\;,
\ee
the solution is given by taking a finite level $k$ of replica-symmetry breaking. The function $z(q)$ then consists of $k$ number of positive plateaus. In the most important case of one-step replica-symmetry breaking (1RSB), the minimisation can be recast as
\be
e_{\rm typ}(\Gamma)=-\frac{1}{2}\min_{v,m_0\in \mathbb{R}^+,q_0\in [0,1]}\left[v f'(1)
    +m_0(f(1)-f(q_0))+\frac{q_0}{v+m_0(1-q_0)}+\frac{1}{m_0}\ln\left(1+\frac{m_0(1-q_0)}{v}\right)\right]\;.\label{1RSB_e_min}
\ee
In this regime, the optimal values of the parameters $v$ and $m_0$ are given by 
\be
v_*=\rho(q_{\rm typ})\sqrt{\frac{q_{\rm typ}}{f'(q_{\rm typ})}}\;,\;\;
  m_0^*=\frac{1-\rho(q_{\rm typ})}{1-q_{\rm typ}}\sqrt{\frac{q_{\rm typ}}{f'(q_{\rm typ})}}\;,
\ee
where the function 
\be
\rho(q)=\frac{(1-q)f'(q)}{(f'(1)-f'(q))q}\;.\label{rho_exp}
\ee
The remaining saddle-point equation sets the value of $q_{\rm typ}$. It is conveniently expressed by introducing the function
\be
\mu(q)=\frac{1}{f'(1)-f'(q)}\left(\frac{f(1)-f(q)}{1-q}-f'(q)\right)\;,\label{mu_exp}
\ee
and reads
\be
\mu(q_{\rm typ})-\rho(q_{\rm typ})\frac{\rho(q_{\rm typ})-1-\ln\rho(q_{\rm typ})}{(1-\rho(q_{\rm typ}))^2}=0\;.\label{q_typ_1rsb}
\ee
One can show that this equation has at least one solution as the left-hand side is negative and equal to $g(1)/g'(1)-1<0$ for $q\to 0$ (for any $\Gamma>0$) while it is positive and equal to 
\be
\frac{1}{2}-(g'(1)+\Gamma) \frac{g'(1)+\Gamma-g''(1)-\ln[(g'(1)+\Gamma)/g''(1)]}{(g''(1)-g'(1)-\Gamma)^2}\;,
\ee
for any $\Gamma>\Gamma_{\rm RSB}$. Note that it remains to be shown that there is a single solution.

The typical energy can then be expressed for $\Gamma\leq \Gamma_{\rm RSB}$ as
\begin{align}
  e_{\rm typ}(\Gamma)=&-\sqrt{f'(1)\mu(q_{\rm typ})(1-\mu(q_{\rm typ}))}\left(\sqrt{\frac{f'(q_{\rm typ})(1-\mu(q_{\rm typ}))}{q_{\rm typ}f'(1)\mu(q_{\rm typ})}}+\sqrt{\frac{q_{\rm typ}f'(1)\mu(q_{\rm typ})}{f'(q_{\rm typ})(1-\mu(q_{\rm typ}))}}\right)\;.\label{e_typ_1RSB}  
\end{align}
The transition occurring for $\Gamma=\Gamma_{\rm RSB}$ is, as in the case of FRSB, of third order as the value of $e_{\rm typ}(\Gamma)$ and its first two derivatives are continuous for $\Gamma=\Gamma_{\rm RSB}$ while the third derivative satisfies Eq. \eqref{e_typ_d_3} as well.

In the limit of zero magnetic field $\Gamma\to 0$, one finds that $q_{\rm typ}=\nu \Gamma$ also vanishes, where the value of $\nu$ remains to be determined. The function $\rho(q)$ then reaches a finite value in that limit
\be
\rho(q_{\rm typ})=\rho_{\rm typ}=\frac{1+2g_2\nu}{g'(1)\nu}\;,\;\;\Gamma\to 0\;.\label{lim_rho}
\ee
The value of $\nu$ is then determined by inserting this expression into Eq. \eqref{q_typ_1rsb} and using that $\mu(q_{\rm typ})=\mu(0)=g(1)/g'(1)$, i.e.
\be
\frac{g(1)}{g'(1)}=\rho_{\rm typ}\frac{\rho_{\rm typ}-1-\ln\rho_{\rm typ}}{(1-\rho_{\rm typ})^2}\;.\label{rho_typ_eq}
\ee
Similarly, the ground-state typical energy in Eq. \eqref{e_typ_1RSB} is obtained by using that $\mu(q_{\rm typ})=\mu(0)=g(1)/g'(1)$ and that $f'(q_{\rm typ})/(q_{\rm typ}f'(1))$ shares the same limit as $\rho(q_{\rm typ})$, given in Eq. \eqref{lim_rho}, yielding
\be
e_{\rm typ}(\Gamma=0)=-\frac{1}{\sqrt{g'(1)}}\left(\sqrt{\rho_{\rm typ}}(g'(1)-g(1))+\frac{g(1)}{\sqrt{\rho_{\rm typ}}}\right)\;.\label{e_typ_G0}
\ee\\

It is important to note that the ground state of the pure $p$-spin model was first derived in the physics literature and shown to satisfy Eq. \eqref{q_typ_1rsb} (setting $g(q)=q^p/p$) in \cite{CrisantiSommers1992}. In the past ten years, this statement has been shown rigorously by solving the Crisanti-Sommers optimisation problem in \cite{ChenSen2017} and by showing that it corresponds to the unique minimum of the complexity of stable stationary points in \cite{AufBenArusCerny2013}.
\\

{\bf Marginal cases}. Finally, there exist two marginal cases such that the Schwarzian derivative vanishes
\be
{\cal S}\left[f'\right](q)=\frac{f^{(4)}(q)}{f''(q)}-\frac{3}{2}\left(\frac{f^{(3)}(q)}{f''(q)}\right)^2= 0\;,\;\;q\in[0,1]\;.\label{cond_1RSB}
\ee
The $2$-spin, where $f(q)=q^2/2+\Gamma q$ is one of these cases and has been discussed in details previously. The typical energy $e_{\rm typ}$ can be recovered by considering either of the minimisation problems in the RS \eqref{RS_e_min}, FRSB \eqref{e_typ_FRSB} or 1RSB \eqref{1RSB_e_min} schemes. The other marginal model is given by the logarithmic potential
\be
f(q)=-y\ln(1-a q)+(\Gamma-ya)q=\Gamma q+y\sum_{r\geq 2}(a q)^{r}\;,\;\;0<a<1\;.
\ee
The typical value of the ground-state is
\be
e_{\rm typ}(\Gamma)=\begin{cases}
\displaystyle \sqrt{y}\left[\ln(1-a)-\sqrt{\frac{\Gamma}{a y}}+\ln\left(1+\sqrt{\frac{\Gamma}{a y}}\right)\right]&\displaystyle\;,\;\;0\leq \Gamma\leq \Gamma_{\rm RSB}=\frac{ya^3}{(1-a)^2}\;,\\
&\\
\displaystyle -\sqrt{\frac{y a^2}{1-a}+\Gamma}&\displaystyle\;,\;\;\Gamma\geq \Gamma_{\rm RSB}
\end{cases}
\ee
and its expression for $0\leq \Gamma\leq \Gamma_{\rm RSB}$ can then be obtained either from working within the FRSB or 1RSB scheme. \\

Finally, note that as mentioned in the introduction, there exist cases where the Schwarzian changes sign over the interval $q\in [0,1]$, and in that case the typical energy is obtained by a $k$-FRSB, i.e. a combination of full and a finite number of replica-symmetry breaking \cite{crisantileuzzi2004,crisantileuzzi2006,AufZhou2022,zhou2023spherical}.  

\section{Summary of results for the CGF}\label{app:CGF_res}

In this appendix, let us now summarise the results for the RS, FRSB and 1RSB expressions of the CGF. It is given in all these different phases by solving the optimisation problem 
\be \label{phiscases_app} 
\phi(s)=\begin{cases}
\displaystyle\max_{z(q),q_0,v}\Phi(s)&\;,\;\;s<0\;,\\
0&\;,\;\;s=0\;,\\
\displaystyle\min_{z(q),q_0,v}\Phi(s)&\;,\;\;s>0\;,
\end{cases}
\ee
where the function $z(q)$ is a real non-decreasing function, satisfying $z(q)=s$ for $q\in[0,q_0)$ and $z(q)\geq s$ for $q\in[q_0,1]$ such that  and the functional $\Phi(s)$ reads
\begin{align}
    \Phi(s)=&\frac{s}{2}\left[s f(q_0)+v f'(1)
    +\int_{q_0}^{1} dq\,z(q)\,f'(q)\right]+\frac{1}{2}\left[\ln\left(s q_0+ v+\int_{q_0}^{1}dq\,z(q)\right)-\ln\left( v+\int_{q_0}^{1}dq\,z(q)\right)\right]\label{phi_z_res}\\
    &+\frac{s}{2}\int_{q_0}^{1}\frac{dq}{\displaystyle v+\int_{q}^{1}dr\,z(r)}\nn\;.
\end{align}

\subsection{RS phase}

The only parameter to optimise is then $v$, which takes a $s$-dependent optimal value
\be
v_*=\frac{1}{2}\left(-s+\sqrt{\frac{4}{f'(1)}+s^2}\right)\;.
\ee
For any model of spherical spin-glass, this RS solution is unstable for values of the Laplace parameter $s\leq s_{\rm RSB}$, where
\be
s_{\rm RSB}=\frac{1}{\sqrt{f'(1)}}\left(\sqrt{\frac{f''(1)}{f'(1)}}-\sqrt{\frac{f'(1)}{f''(1)}}\right)\;.\label{s_RSB_res}
\ee
The RS expression of the cumulant generating function is given by 
\be
\phi_{\rm RS}(s)=\frac{s^2}{4}(2f(1)-f'(1))+\frac{s f'(1)}{4}\sqrt{\frac{4}{f'(1)}+s^2 }+\ln\left[\frac{\sqrt{f'(1)}}{2}\left(s+\sqrt{\frac{4}{f'(1)}+s^2 }\right)\right]\;,\;\;s\geq s_{\rm RSB}\;.\label{phi_RS_res}
\ee
This function and the bound $s_{\rm RSB}$ are universal in the sense that they only depend on the values of $f(q)$ and its first two derivatives for $q=1$. The CGF is convex over the interval $s\geq s_{\rm RSB}$, as one can show explicitly that
\be
\phi_{\rm RS}''(s)=f(1)-\frac{f'(1)}{2}\left(1-\frac{s\sqrt{f'(1)}}{\sqrt{4+s^2 f'(1)}}\right)\geq 0\;,\;\;s\geq s_{\rm conv, RS}=\frac{2f(1)-f'(1)}{\sqrt{f(1)f'(1)(f'(1)-f(1))}}\;,
\ee
and $s_{\rm conv, RS}\leq s_{\rm RSB}$. The parameter $s_{\rm conv, RS}$ will thus not play any role in the following. Note that only in the case of the pure $p$-spin does $s_{\rm RSB}=s_{\rm conv, RS}$.  The optimal energy associated to a given value of the Laplace parameter $s$ is obtained by computing
\be
e_{*,\rm RS}(s)=-\phi_{\rm RS}'(s)=-\frac{1}{2}\left(s(2 f(1)-f'(1))+\sqrt{f'(1)\left(4+s^2 f'(1)\right)}\right)\;,
\ee
and in particular the energy associated to the value $s_{\rm RSB}$ is given by
\be
e_{\rm RSB}=e_{*,\rm RS}(s_{\rm RSB})=-\frac{1}{\sqrt{f''(1)}}\left[f'(1)+f(1)\left(\frac{f''(1)}{f'(1)}-1\right)\right]\;.
\ee
Note that $e_{*,\rm RS}'(s)=-\phi_{\rm RS}''(s)<0$, which ensures that for any $s> s_{\rm RSB}$, the associated energy $e_{*,\rm RS}(s)< e_{\rm RSB}$. If $s_{\rm RSB}\leq 0$, which can be shown to coincide with the condition $\Gamma\geq  \Gamma_{\rm RSB}$ , where $\Gamma_{\rm RSB}$ is given in \eqref{G_RSB},
the typical value of the ground-state is given within the RS scheme and reads $e_{\rm typ}=e_{*,\rm RS}(0)=-\sqrt{f'(1)}\leq e_{\rm RSB}$. In that case, the rescaled variance can similarly be computed as
\be
{\cal V}_{\min}(\Gamma)=\lim_{N\to \infty} N\mbox{Var}(e_{\min})=\phi_{\rm RS}''(0)=g(1)+\frac{\Gamma-g'(1)}{2}\;,\;\;\Gamma\geq \Gamma_{\rm RSB}\;.
\ee
Let us now describe the different RSB schemes.

\begin{figure}
    \centering
    \includegraphics[width=0.48
    \textwidth]{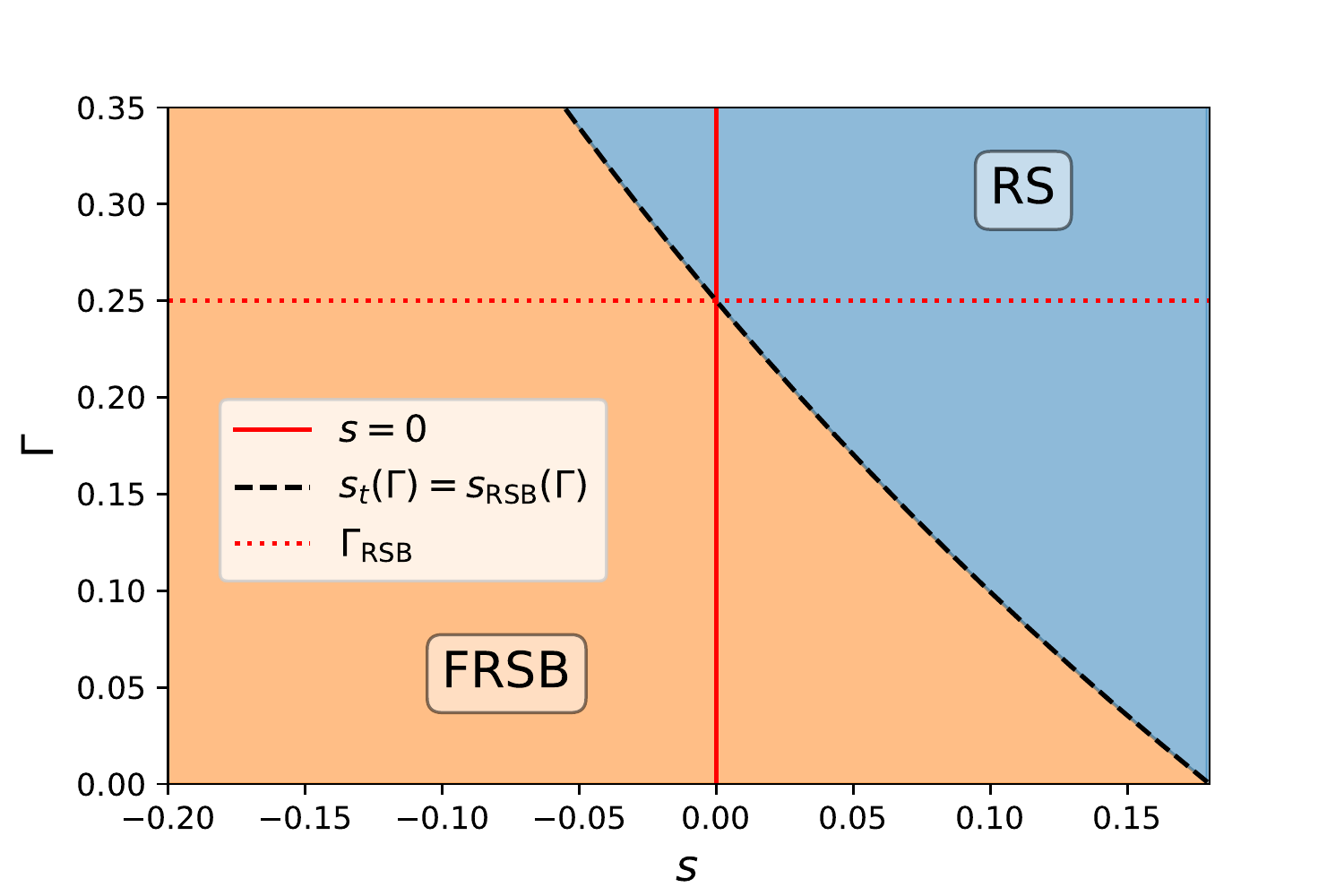}
    \caption{Phase diagram for the CGF for the spherical spin model in Eq. \eqref{f_FRSB} in the plane $s,\Gamma$. The dashed black line $s_t(\Gamma)=s_{\rm RSB}(\Gamma)$ shows the transition separating the FRSB phase (in orange) and the RS phase (in blue) determined by the stability criterion of the RS phase. The red line corresponds to the typical value $s=0$, which crosses the line $s_t(\Gamma)$ for $\Gamma=\Gamma_{\rm RSB}$ (marked by a horizontal red dotted line). Note that there is no change of speed for the CGF, as the "wall" is relegated to $s\to -\infty$.}
    \label{phase_diag_s_FRSB}
\end{figure}

\subsection{FRSB phase}

Supposing that the Schwarzian derivative is positive as given in Eq. \eqref{cond_FRSB}, the RSB expression of the CGF is given within the FRSB scheme. The phase diagram of the CGF for a representative model with a FRSB phase is shown in Fig. \ref{phase_diag_s_FRSB}. In that phase, the function $z(q)$ is continuous and decreasing over the range $[q_0^*,1]$, where $q_0^*$ is the optimal value of $q_0$. As mentioned when discussing the main results for the LDF, in that phase it is more convenient to express parametrically $s$ as a function of the optimal parameter $q_0=q_0^*$ as
\be
s=\frac{1}{\sqrt{g''(q_0^*)}}\left(\frac{g''(q_0^*)}{g'(q_0^*)+\Gamma}-\frac{1}{q_0^*}\right)\,.\label{s_FRSB_res}
\ee
The parameter $v$ and the function $z(q)$ then take the optimal values (which coincide with the values for the typical value of the ground-state in the FRSB phase)
\be
v_*=\frac{1}{\sqrt{g''(1)}}\;,\;\;z_*(q)=\frac{g^{(3)}(q)}{2g''(q)^{3/2}}\;,\;\;q\in[q_0^*,1]\;.
\ee
This parametrisation ensures both that $s\geq z(q_0^*)$ and $\partial_{q_0^*}s\geq 0$ as we show below. In particular, comparing with Eq. \eqref{s_RSB_res}, we see that $s=s_{\rm RSB}$ for $q_0^*=1$. Also if $\Gamma<\Gamma_{\rm RSB}$, $s=0$ for $q_0^*=q_{\rm typ}$ satisfying $q_{\rm typ}f''(q_{\rm typ})=f'(q_{\rm typ})$ as obtained in Eq. \eqref{q_typ_FRSB}. In the limit $q_0^*\to 0$, the parameter $s\to -\infty$ if the magnetic field $\Gamma>0$ and $s\to z_*(0)=f^{(3)}(0)/(2f''(0)^{3/2})\geq 0$ if $\Gamma=0$. Thus, as $q_0^*$ increases from $0$ to $1$, the whole range $s\leq s_{\rm RSB}$ is covered by this parametrisation for positive magnetic field $\Gamma>0$, while only the range $[z_*(0),s_{\rm RSB}]$ is covered for $\Gamma=0$. The implications of this statement are discussed further below.

For models that display this FRSB regime, the transition to the RS regime occurs at the value of the Laplace parameter $s=s_t(\Gamma)=s_{\rm RSB}(\Gamma)$ for any value of the magnetic field $\Gamma$. Hence, the RS regime described in the section above describe exactly the whole range of parameters $s\geq s_{\rm RSB}$. 
Using Eq. \eqref{s_FRSB_res}, we give the result for the CGF $\phi_{\rm FRSB}(s)$ in two equivalent forms
\begin{align}
    \phi_{\rm FRSB}(s)&=\frac{s}{2}\left[s f(q_0^*)+\frac{f'(q_0^*)}{\sqrt{f''(q_0^*)}}
    +2\int_{q_0^*}^{1} dq\,\sqrt{f''(q)}\right]+\frac{1}{2}\ln\left(1+s\,q_0^*\,\sqrt{f''(q_0^*)}\right)\nn\\
    &=\frac{f''(q_0^{*})q_0^{*}-f'(q_0^{*})}{2q_0^{*}f'(q_0^{*})\sqrt{f''(q_0^{*})}}\left[ \frac{f(q_0^{*})(f''(q_0^{*})q_0^{*}-f'(q_0^{*}))+q_0^{*}f'(q_0^{*})^2}{q_0^{*}f'(q_0^{*})\sqrt{f''(q_0^{*})}}
    +2\int_{q_0^{*}}^{1} dq\,\sqrt{f''(q)}\right]+\frac{1}{2}\ln\left(\frac{q_0^{*}\,f''(q_0^{*})}{f'(q_0^{*})}\right)\;.
\end{align}
Computing the second derivative of the CGF, using the parametrisation in Eq. \eqref{s_FRSB_res}, yields
\be \label{phi2rsb} 
\phi_{\rm FRSB}''(s)=f(q_0^*)-\frac{q_0^* f'(q_0^*)^2}{q_0^* f''(q_0^*)+f'(q_0^*)}\;,
\ee
which, using Eq. \eqref{nice_id}, is shown to be strictly positive for any mixed model. Hence this shows that
for any mixture model in the full RSB regime $\phi_{\rm FRSB}(s)$ is strictly convex. The optimal energy associated to a given value of $s$ is obtained by computing
\be
e_{*,\rm FRSB}(s)=-\phi_{\rm FRSB}'(s)=-\left(\frac{ f(q_0^*)}{\sqrt{f''(q_0^*)}}\left(\frac{f''(q_0^*)}{f'(q_0^*)}-\frac{1}{q_0^*}\right)+\frac{f'(q_0^*)}{\sqrt{f''(q_0^*)}}+\int_{q_0^*}^{1} dq\,\sqrt{f''(q)}\right)\;,\;\;s\leq s_{\rm RSB}\;.\label{e_FRSB_app}
\ee
The identity $e_{*,\rm FRSB}'(s)=-\phi_{\rm FRSB}''(s)\leq 0$ ensures that $e_{*,\rm FRSB}(s)\geq e_{\rm RSB}$ for all $s\leq s_{\rm RSB}$. The range of energies thus described by \eqref{e_FRSB_app} belong to a finite interval $[e_{\rm RSB},e_c]$, where 
\be
e_{*,\rm FRSB}(s=s_{\rm RSB})=e_{\rm RSB}=-\frac{1}{\sqrt{f''(1)}}\left[f'(1)+f(1)\left(\frac{f''(1)}{f'(1)}-1\right)\right]\;,
\ee
corresponding to $q_0^*\to 1$, and 
\be
e_{*,\rm FRSB}(s \to s(q_0^*=0))=e_{c}=-\int_0^{1}dq\,\sqrt{f''(q)}\;,
\ee
where $s(q_0^*=0)=-\infty$ if $\Gamma>0$ and $s(q_0^*=0)=z_*(0)$ if $\Gamma=0$. If $\Gamma<\Gamma_{\rm RSB}$, the FRSB expression of the typical ground state $e_{*,\rm FRSB}(0)=e_{\rm typ}=-\left[q_{\rm typ} \sqrt{f''(q_{\rm typ})}+\int_{q_{\rm typ}}^1 dq\,\sqrt{f''(q)}\right]$ is retrieved from Eq. \eqref{e_FRSB_app}, using the parametrisation \eqref{s_FRSB_res} where $\Gamma=q_{\rm typ}g''(q_{\rm typ})-g'(q_{\rm typ})$. In particular, for $\Gamma=0$, one can check that $q_{\rm typ}=0$ and $e_{\rm typ}(\Gamma=0)=e_c$. Finally, the rescaled variance can be computed as well for $0<\Gamma\leq \Gamma_{\rm RSB}$, and one finds
\be
{\cal V}_{\min}(\Gamma)=\lim_{N\to \infty} N\mbox{Var}(e_{\min})=\phi_{\rm FRSB}''(0)=g(q_{\rm typ})+\frac{\Gamma-g'(q_{\rm typ})}{2}>0 \;,\;\;0<\Gamma\leq \Gamma_{\rm RSB}\;,
\ee
where $\Gamma=q_{\rm typ}g''(q_{\rm typ})-g'(q_{\rm typ})$. For $\Gamma=0$, the rescaled variance vanishes. 

Finally, in the special case of $\Gamma=0$, the CGF is given by a linear function for any $s\leq z_*(0)$, i.e
\be
\phi(s)=-s\, e_c\;,\;\;s\leq z_*(0)=\frac{g^{(3)}(0)}{2g''(0)^{3/2}}\;.
\ee

\subsection{1RSB phase}

\begin{figure}
    \centering
    \includegraphics[width=0.48
    \textwidth]{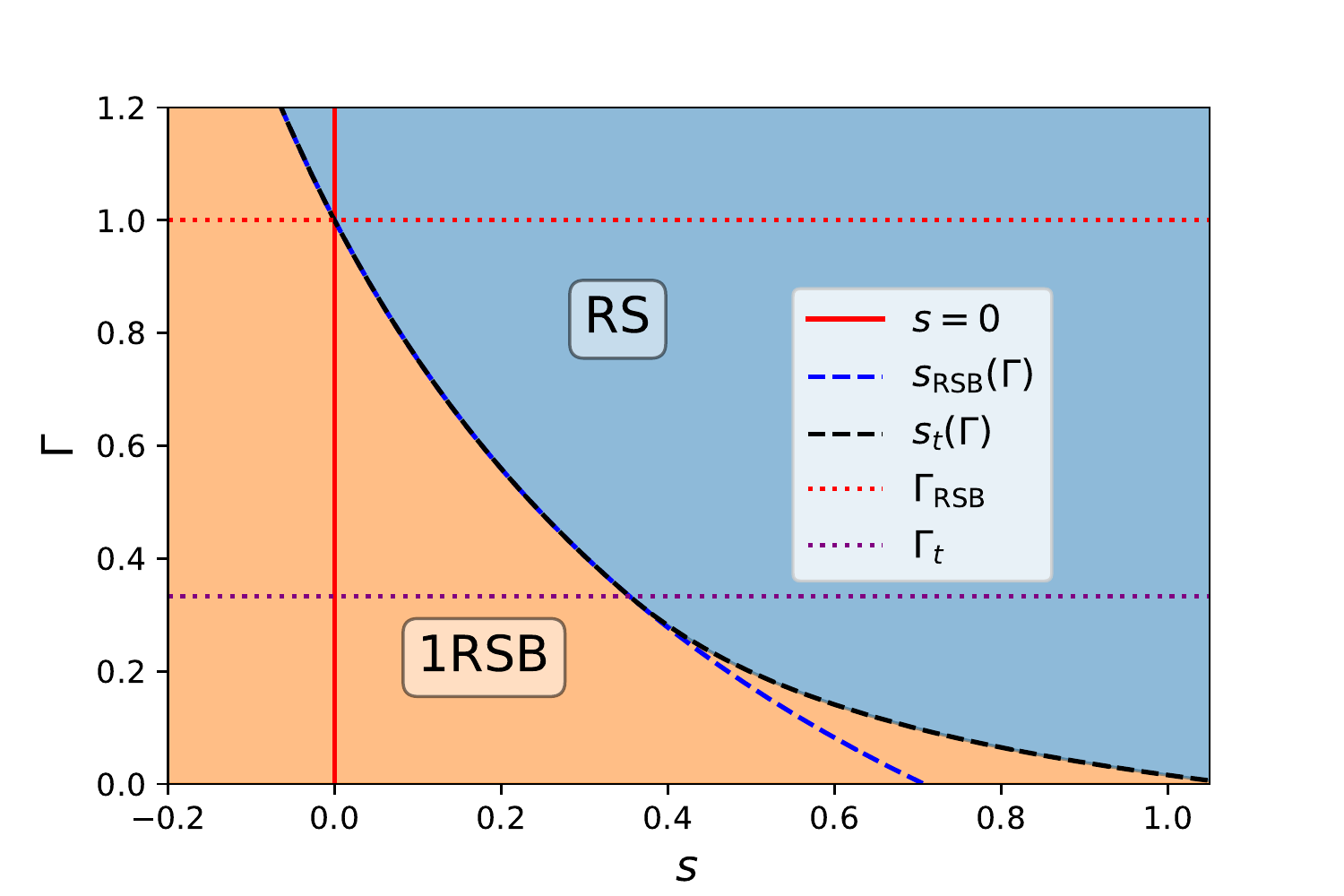}
    \caption{Phase diagram for the CGF for the $3$-spin model in the plane $s,\Gamma$. The black line $s_t(\Gamma)$ shows the transition separating the 1RSB phase (in orange) and the RS phase (in blue). The red line corresponds to the typical value $s=0$, which crosses the line $s_t(\Gamma)$ for $\Gamma=\Gamma_{\rm RSB}$ (marked by a horizontal red dotted line). The criterion of stability of the RS phase yields the dashed blue line $s_{\rm RSB}(\Gamma)$ but only coincides with the correct transition line $s_t(\Gamma)$ for $\Gamma>\Gamma_t$ (marked by a horizontal purple dotted line). Note that there is no change of speed for the CGF, as the "wall" is relegated to $s\to -\infty$.}
    \label{phase_diag_1RSB_s}
\end{figure}

Supposing that the Schwarzian derivative is negative as given in Eq. \eqref{cond_1RSB}, the RSB expression of the CGF is given within the 1RSB scheme. The phase diagram of the CGF for a representative model with a 1RSB phase is shown in Fig. \ref{phase_diag_1RSB_s}. In that phase, the function $z(q)$ is constant and given by $m_0\geq s$ over the range $[q_0^*,1]$, where $q_0^*$ is the optimal value of $q_0$. As mentioned when discussing the main results for the LDF, in that phase it is more convenient to express parametrically $s$ as a function of the optimal parameter $q_0=q_0^*$ as
\be
  s_*=\frac{1}{\sqrt{q_0^* f'(q_0^*)}}\left(\frac{1}{\sqrt{\alpha(q_0^*)}}-\sqrt{\alpha(q_0^*)}\right)\;, \label{param_1_app}
\ee
where the function $\alpha(q_0^*)$ satisfies the self-consistent equation
\be
\mu(q_0^*)=\rho(q_0^*)\frac{\rho(q_0^*)-\alpha(q_0^*)-\alpha(q_0^*)\ln (\rho(q_0^*)/\alpha(q_0^*))}{(\alpha(q_0^*)-\rho(q_0^*))^2}\;,\label{mu_alph_app}
\ee
with 
\be
\mu(q)=\frac{1}{g'(1)-g'(q)}\left(\frac{g(1)-g(q)}{1-q}-g'(q)\right)\;\;{\rm and}\;\;\rho(q)=\frac{(1-q)(g'(q)+\Gamma)}{(g'(1)-g'(q))q}\;.
\ee
The remaining optimal parameters are then given by
\begin{align}
v_*&=\rho(q_0^*)\sqrt{\frac{q_0^*}{f'(q_0^*)\alpha(q_0^*)}}\;,\label{param_2_app}\\
  m_0^*&=\frac{1}{1-q_0^*}\sqrt{\frac{q_0^* \rho(q_0^*)}{f'(q_0^*)}}\left(\sqrt{\frac{\alpha(q_0^*)}{\rho(q_0^*)}}-\sqrt{\frac{\rho(q_0^*)}{\alpha(q_0^*)}}\right)\;.\label{param_3_app}
\end{align}

For a magnetic field 
\be
\Gamma\geq\Gamma_t=\frac{2g''(1)^2}{2g''(1)+g^{(3)}(1)}-g'(1)\;,
\ee
the transition in the CGF from the RS to the RSB regime occurs for $s=s_t(\Gamma)=s_{\rm RSB}(\Gamma)$, i.e. $q_0^*=1$, as a result of the loss of stability of the RS solution. For a magnetic field $\Gamma<\Gamma_t$ instead, it occurs for a value $s=s_t(\Gamma)>s_{\rm RSB}(\Gamma)$ such that the CGF coincides in the two different schemes, i.e. $\phi_{\rm RS}(s_t)=\phi_{\rm 1RSB}(s_t)$. This value of the Laplace parameter is given by
\be
s_t(\Gamma)=\frac{1}{\sqrt{q_t f'(q_t)}}\left(\sqrt{\frac{f'(1)-f'(q_t)}{(1-q_t)f'(1)}}-\sqrt{\frac{(1-q_t)f'(1)}{f'(1)-f'(q_t)}}\right)>s_{\rm RSB}(\Gamma)\;,\;\;\Gamma<\Gamma_t\;,
\ee
with $q_t<1$ obtained as the solution  of 
\be
\mu(q_t)=f'(q_t)\frac{f'(q_t)-q_t f'(1)-q_t f'(1)\ln f'(q_t)/(q_t f'(1))}{(f'(q_t)-q_t f'(1))^2}\;.
\ee
For $\Gamma\geq \Gamma_t$, the only solution is $q_t=1$, and we thus set for convenience $q_t(\Gamma\geq \Gamma_t)=1$ such that $s_t(\Gamma\geq \Gamma_t)=s_{\rm RSB}$.
Note that as $\Gamma\to \Gamma_t$, $q_t\to 1$ and $s_t\to s_{\rm RSB}$. In the opposite limit $\Gamma\to 0$, one has instead that $q_t\approx \Gamma/(\rho_{\rm typ}g'(1)-2g_2)$ and $s_t\to (1-\rho_{\rm typ})/\sqrt{\rho_{\rm typ}g'(1)}\geq 0$, where $\rho_{\rm typ}$ is the solution of the equation
\be
\frac{g(1)}{g'(1)}=\rho_{\rm typ}\frac{\rho_{\rm typ}-1-\ln \rho_{\rm typ}}{(1-\rho_{\rm typ})^2}\;.
\ee

The 1RSB expression for the CGF is then given, using the parametrisation (\ref{param_1_app}-\ref{param_2_app}-\ref{param_3_app}) for Laplace parameter $s\in (-\infty,s_t)]$ as
\begin{align}
\phi_{\rm 1RSB}(s)=&\frac{1}{4} \left(1+\frac{f(q_0^*)-q_0^*f(1)}{q_0^*(1-q_0^*)f'(q_0^*)}+2\frac{f(q_0^*)-f(1)-(q_0^*-1)f'(1)}{q_0^*(f'(1)-f'(q_0^*))}\right)q_0^* f'(q_0^*) s^2\\
   &+\frac{1}{4} \left(-1+2\frac{f(q_0^*)-f(1)-(q_0^*-1)f'(1)}{q_0^*(f'(1)-f'(q_0^*))}+\frac{f(1)-f(q_0^*)}{(1-q_0^*)f'(q_0^*)}\right)s\sqrt{q_0^* f'(q_0^*)(4+q_0^* f'(q_0^*) s^2)}\nn\\
   &+\ln\left(\frac{s\sqrt{q_0^* f'(q_0^*)}+\sqrt{4+q_0^* f'(q_0^*) s^2}}{2}\right)\nn\;.
\end{align}
The expression of the second derivative $\phi_{\rm 1RSB}''(s)$ of the CGF is rather long and cumbersome. We are not able to show explicitly that this function is positive but it can be seen numerically on specific examples, and in particular for the $p$-spin model. The energy $e_{*,{\rm 1RSB}}(s)$ associated to a given value $s$ of the Laplace parameter reads
\be
e_{*,{\rm 1RSB}}(s)=-\phi_{\rm 1RSB}'(s)=\frac{(1-q_0^*)\mu(q_0^*)f'(1)-f(1)+[f(1)-\mu(q_0^*)f'(1)-(1-\mu(q_0^*))f'(q_0^*)]\alpha(q_0^*)}{\sqrt{q_0^* f'(q_0^*)\alpha(q_0^*)}}\;,
\ee
and lie within the range $e\in[e_t,e_c]$, where
\be
e_{*,{\rm 1RSB}}(s_{t})=e_{t}=-\frac{\left((1-q_t) f'(1)-f(1)\right) f'(q_t)+f(1) q_t f'(1)}{\sqrt{q_t f'(1) f'(q_t)(1-q_t) 
   \left(f'(1)-f'(q_t)\right)}}\;,
\ee
corresponding respectively to the parameters $q_0^*=q_t,\alpha(q_t)=(1-q_t)f'(1)/(f'(1)-f'(q_t))$, and 
\be
e_c=e_{*,{\rm 1RSB}}(s\to s(q_0^*=0))=-\frac{1}{\sqrt{g'(1)}}\left(\sqrt{\rho_{\rm typ}}(g'(1)-g(1))+\frac{g(1)}{\sqrt{\rho_{\rm typ}}}\right)\;,
\ee
with $s(q_0^*=0)\to -\infty$ for $\Gamma>0$ and $s(q_0^*=0)=s_t=(1-\rho_{\rm typ})/\sqrt{\rho_{\rm typ}g'(1)}\geq 0$. Note that $e_c=e_{\rm typ}(\Gamma=0)=e_t(\Gamma=0)$ as can be seen by comparing with Eq. \eqref{e_typ_G0} and using the fact that $q_t\approx \Gamma/(\rho_{\rm typ}g'(1)-2g_2)$ as $\Gamma\to 0$. The range of energies thus described collapses to a single point. For a magnetic field $0\leq \Gamma\leq \Gamma_{\rm RSB}$, the typical energy belongs to the interval $[e_t,e_c]$, with
\be
e_{*,{\rm 1RSB}}(s=0)=-\frac{q_{\rm typ} (f(1)-f(q_{\rm typ})) f'(1)+\left((1-q_{\rm typ})^2f'(1)-f(1)+f(q_{\rm typ})\right)f'(q_{\rm typ})}{(1-q_{\rm typ})
   \sqrt{q_{\rm typ} f'(q_{\rm typ})} \left(f'(1)-f'(q_{\rm typ})\right)}\;,
\ee
corresponding to $q_0^*=q_{\rm typ}$ satisfying Eq. \eqref{q_typ_1rsb}. In addition, the expression of the second derivative $\phi_{\rm 1RSB}''(s)$ simplifies considerably for $s=0$, allowing to obtain an explicit expression for the rescaled variance
\be
{\cal V}_{\min}(\Gamma)=\lim_{N\to \infty} N\mbox{Var}(e_{\min})=\phi_{\rm 1RSB}''(0)=g(q_{\rm typ})+\frac{\Gamma-g'(q_{\rm typ})}{2}>0 \;,\;\;0<\Gamma\leq \Gamma_{\rm RSB}\;.
\ee
In the limit $\Gamma\to 0$, using that $q_{\rm typ}\approx q_t\approx \Gamma/(\rho_{\rm typ}g'(1)-2g_2)$, the rescaled variance reads
\be
{\cal V}_{\min}(\Gamma)=\frac{\Gamma^2}{2(\rho_{\rm typ}g'(1)-2g_2)}+O(\Gamma^3)\;,
\ee
and it thus vanishes for $\Gamma=0$. 

Finally, in the special case $\Gamma=0$, the CGF is given by a linear function for any $s\leq s_t$, i.e.
\be
\phi(s)=-s\, e_c\;,\;\;s\leq s_t=\frac{1}{\sqrt{g'(1)}}\left(\frac{1}{\sqrt{\rho_{\rm typ}}}-\sqrt{\rho_{\rm typ}}\right)\;.
\ee

\section{Analysis of the different phases}\label{app:analysis}

Let us now consider the expression of the large-deviation function and the stability in the different phases, namely replica-symmetric (RS), one-step replica symmetry broken (1RSB) and fully replica-symmetry broken phases (FRSB).

\subsection{Replica-symmetric phase}

In the replica-symmetric (RS) phase, the function $z(q)=s$ for any $q\in[0,1]$, which is equivalent to taking the limit $q_0\to 1$. This yields a particularly simple expression for the functional to optimise
\be
\Phi_{\rm RS}(s)=\frac{s}{2}(s f(1)+v f'(1))+\frac{1}{2}\ln\left(1+\frac{s}{v}\right)\;.\label{phi_RS}
\ee
Only the parameter $v$ is to be optimised, such that it satisfies the equation
\be
\left.\partial_{v}\Phi_{\rm RS}\right|_{v=v_*}=\frac{s}{2}\left(f'(1)-\frac{1}{v_*(s+v_*)}\right)\;,\label{sp_eq_RS}
\ee
in agreement with Eq. \eqref{sp_eq_v} for the choice $z_*(q)=s$ for all $q\in[0,1]$. From this expression, we obtain
\be
v=\frac{1}{2}\left(-s+\sqrt{\frac{4}{f'(1)}+s^2}\right)\;,
\ee
which matches exactly the expression in Eq. \eqref{sol_q} in the RS regime where $q_0^*=1$. The cumulant generating function reads in the RS phase
\be
\phi_{\rm RS}(s)=\frac{s^2}{4}(2f(1)-f'(1))+\frac{s f'(1)}{4}\sqrt{\frac{4}{f'(1)}+s^2 }+\ln\left[\frac{\sqrt{f'(1)}}{2}\left(s+\sqrt{\frac{4}{f'(1)}+s^2 }\right)\right]\;.
\ee
An analysis of the eigenvalues of the quadratic form $A_{(ab),(cd)}^{\rm RS}=\left.\frac{\delta^2 \Phi_n(Q)}{\delta Q_{ab}\delta Q_{cd}}\right|_{Q=Q_{\rm RS}}$ defined in Eq. \eqref{quad_form} reveals that the replica symmetric phase is stable for (see Appendix \ref{app:stab} for details)
\be
\Lambda^{\rm RS}=f''(1)-\frac{1}{v_*^2}<0\;,\label{Lamb_RS}
\ee
which gives a lower bound on the value of $s$ for the validity of the RS expression $\phi_{\rm RS}(s)$ of the CGF
\be
s\geq s_{\rm RSB}=\frac{1}{\sqrt{f'(1)}}\left(\sqrt{\frac{f''(1)}{f'(1)}}-\sqrt{\frac{f'(1)}{f''(1)}}\right)\;.\label{s_RSB}
\ee
One can check that the third derivative of $\phi_{\rm RS}(s)$ is positive 
\be
\phi_{\rm RS}^{(3)}(s)=2\left(\frac{4}{f'(1)}+s^2\right)^{3/2}>0\;,
\ee
such that its second derivative is a growing function of $s$. For $s=s_{\rm RSB}$, it reaches the value
\be
\phi_{\rm RS}''(s_{\rm RSB})=f(1)-\frac{f'(1)^2}{f'(1)+f''(1)}\;.
\ee
In order to show that this quantity is non-negative, one can use the nice identity (specialising to the case $q=1$)
\begin{align}
   &\frac{1}{2q}\left.(x\partial_x-y\partial_y)^2\right|_{x=y=q}f(x)f(y)=f(q)(qf''(q)+f'(q))-q f'(q)^2\nn\\
   &=\frac{1}{2}\sum_{r,p=2}^{\infty}g_r g_p (r-p)^2 q^{r+p-1}+\Gamma\sum_{r=2}^{\infty}g_r(r-1)^2\,q^{r}\geq 0\;,\;\;\,g_r\geq 0\;. \label{nice_id}
\end{align}
Thus its value for $s=s_{\rm RSB}$ gives a positive lower bound $\phi_{\rm RS}''(s)\geq\phi_{\rm RS}''(s_{\rm RSB})\geq 0$. Note that only in the case of the pure $p$-spin without magnetic field does one have $\phi_{\rm RS}''(s_{\rm RSB})= 0$, this quantity is strictly positive for any other model. The function $\phi_{\rm RS}(s)$ is thus convex throughout its domain of stability $s\geq s_{\rm RSB}$. For fixed value of $s$, one can compute the associated energy $e_{*,{\rm RS}}\equiv e_{*,{\rm RS}}(s)$ given by 
\be
e_{*,{\rm RS}}(s)=-\phi_{\rm RS}'(s)=-sf(1)-\frac{f'(1)}{2}\left(-s+\sqrt{\frac{4}{f'(1)}+s^2}\right)\;,\;\;s\geq s_{\rm RSB}\;.\label{e_s_RS}
\ee
In particular, if $s_{\rm RSB}\leq 0$, which using the expression $f(q)=g(q)+\Gamma q$ in Eq. \eqref{cov}, is equivalent to having a strong enough magnetic field 
\be
\Gamma\geq\Gamma_{\rm RSB}=g''(1)-g'(1)=\sum_{r=2}^{\infty}r(r-2)\,g_r\;,
\ee
the typical and minimum energy can be recovered as
\be
\overline{e_{\min}}=e_{\rm typ}=e_{*,{\rm RS}}(0)=-\sqrt{g'(1)+\Gamma}\;,\;\;\Gamma\geq \Gamma_{\rm RSB}\;.
\ee
If that condition is met, one can compute higher-order cumulants from $\phi_{\rm RS}(s)$ and in particular the rescaled variance
\be
{\cal V}_{\min}(\Gamma)=\lim_{N\to \infty} N\mbox{Var}(e_{\min})=\phi_{\rm RS}''(0)=g(1)+\frac{\Gamma-g'(1)}{2}\;,\;\;\Gamma\geq \Gamma_{\rm RSB}\;.\label{var_RS}
\ee
Note that $\Gamma_{\rm RSB}\geq g'(1)-2g(1)$ and the variance is thus always positive for $\Gamma\geq \Gamma_{\rm RSB}$.

Let us now consider the LDF ${\cal L}(e)$ in the RS regime. The function ${\cal L}_{\rm RS}(e)$ is given by the Legendre transform of the function $\phi_{\rm RS}(s)$. As the function $e_{*,{\rm RS}}(s)$ is a strictly increasing function of $s$ (unless for the pure $p$-spin that will be considered separately), the identity in Eq. \eqref{e_s_RS} can be inverted exactly to express $s_*\equiv s_*(e)$ in terms of $e$ as
\be
   s_{*,{\rm RS}}(e)=-{\cal L}_{\rm RS}'(e)=\frac{1}{2(f'(1)-f(1))}\left[2e+\frac{f'(1)}{f(1)}\left(-e+\sqrt{e^2-4f(1)\left(1-\frac{f(1)}{f'(1)}\right)}\right)\right]\;,\;\;e\leq e_{\rm RSB}\;,
\ee
where the condition of stability given by the lower bound $s_*\geq s_{\rm RSB}$ translates in terms of $e$ into the upper bound
\be
e\leq e_{\rm RSB}=-\frac{1}{\sqrt{f''(1)}}\left[f'(1)+f(1)\left(\frac{f''(1)}{f'(1)}-1\right)\right]< 0\;.
\ee
The value $e_{\rm RSB}$ is a decreasing function of $\Gamma$ and is negative for $\Gamma=0$,
\be
e_{\rm RSB}(\Gamma=0)=-\frac{1}{g'(1)\sqrt{g''(1)}}\sum_{k,p\geq 2}g_k g_p \left[k p+k(k-2)\right]<0\;,\;\;\partial_\Gamma e_{\rm RSB}=-\frac{\sqrt{g''(1)}}{(\Gamma+g'(1))^2}\sum_{k\geq 2}g_k (k-1)\;.\label{e_RSB_proof}
\ee
One can additionally show that 
\be
e_{\rm RSB}\leq e_{c,{\rm RS}}=-2\sqrt{f(1)\left(1-\frac{f(1)}{f'(1)}\right)}< 0\;,
\ee
by using the identity
\be
e_{\rm RSB}^2-e_{c,{\rm RS}}^2=\frac{1}{f''(1)}\left[f'(1)-f(1)\left(\frac{f''(1)}{f'(1)}+1\right)\right]^2\geq 0\;.
\ee
Only for the pure $p$-spin model does one have that $e_{\rm RSB}=e_{c,{\rm RS}}$. Inserting the expression of $s_*$ into that of $v_*$, one similarly obtains
\be
   v_*=-\frac{e+\sqrt{e^2-e_{c,{\rm RS}}^2}}{2(f'(1)-f(1))}\;.
\ee
Finally, the LDF is obtained in the regime of stability of the RS solution $e\leq e_{\rm RSB}\leq e_{c,{\rm RS}}$ as
\begin{align}
    {\cal L}_{\rm RS}(e)=&-\min_{s\in \mathbb{R}}\left[s e+\phi_{\rm RS}(s)\right]=s_* \phi_{\rm RS}'(s_*)-\phi_{\rm RS}(s_*)=\frac{1}{2}\left(s_*^2 f(1)+\frac{s_*}{s_*+v_*}-\ln\left(1+\frac{s_*}{v_*}\right)\right)\nn\\
    =&\frac{s_*^2}{4} \left(2f(1)-f'(1)\right)+\frac{s_* f'(1)}{4}\sqrt{\frac{4}{f'(1)}+s_*^2}-\ln\left[\frac{\sqrt{f'(1)}}{2}\left(s_*+\sqrt{\frac{4}{f'(1)}+s_*^2}\right)\right]\label{L_rs}\\
    =&-\frac{2 f(1)}{f'(1)}\frac{e^2}{e_{c,{\rm RS}}^2}+\frac{e}{e_{c,{\rm RS}}^2}\left(e-\sqrt{e^2-e_{c,{\rm RS}}^2}\right)-\ln\left(\frac{e-\sqrt{e^2-e_{c,{\rm RS}}^2}}{e_{c,{\rm RS}}}\right)+\frac{1}{2}\ln\frac{f(1)}{f'(1)-f(1)}\label{L_RS_e_text}\;.
\end{align}
This expression only depends on the values of the covariance $f(1)$ and its derivative $f'(1)$ at coinciding points.

For a sufficiently strong magnetic field $\Gamma\geq \Gamma_{\rm RSB}$, the typical energy belongs to the RS phase and one can expand the LDF around its minimum as
\be
{\cal L}_{\rm RS}(e)=\frac{\left(e+\sqrt{f'(1)}\right)^2}{2f(1)-f'(1)}+o\left(e+\sqrt{f'(1)}\right)^2\;,\;\;\Gamma\geq \Gamma_{\rm RSB}\;.
\ee
The expression above can be re-written as
\be
N{\cal L}_{\rm RS}(e)\approx \frac{\left(e-\overline{e_{\min}}\right)^2}{2{\cal V}_{\min}(\Gamma)}\;,\;\;|e-\overline{e_{\min}}|\ll 1\;,
\ee
which matches perfectly the Gaussian tail of the distribution describing the typical fluctuations of the ground state energy. Indeed, the fluctuations of the ground state $e_{\min}$ of a spherical spin glass in the presence of a finite magnetic field have been shown to satisfy a central limit theorem \cite{ChenSen2017}.

As shown in Appendix \ref{app:comp_min}, the expression of the large deviation function in the RS phase ${\cal L}_{\rm RS}(e)$ matches exactly the opposite of the annealed complexity of minima at fixed energy per degree of freedom $e$, defined as 
\be
\Sigma_{\min}(e)=\lim_{N\to \infty}\frac{1}{N}\ln \overline{n_{\min}(e)}=-{\cal L}_{\rm RS}(e)\;,\label{id_comp_LDF}
\ee
where $n_{\min}(e)$ is the density of minima of the Hamiltonian $H({\bf x})$ for a fixed value $e=H({\bf x})/N$. In the replica-symmetric phase, as $\Sigma_{\min}(e)<0$ for any $e\neq e_{\rm typ}$, the probability to find a minimum of the Hamiltonian at an energy $e\neq e_{\rm typ}$ is exponentially small and the LDF describing its occurrence is given by $-\Sigma_{\min}(e)$. As there is typically a single minimum of the Hamiltonian in the RS phase, if there is a minimum at energy $e\leq e_{\rm RSB}$, it must be the global minimum of the Hamiltonian. Hence the LDF describing its occurrence is ${\cal L}(e)$ and the identity in Eq. \eqref{id_comp_LDF} must hold.

For $\Gamma=0$, and if $e_{\rm typ}$ is given within the 1RSB scheme, one can additionally show that 
\be
{\cal L}_{\rm RS}(e_{\rm typ})=-\Sigma_{\min}(e_{\rm typ})=0\;.
\ee
Indeed, inserting the expression given by Eq. \eqref{e_typ_1RSB}
\be
e_{\rm typ}=\frac{e_{c,{\rm RS}}}{2}\left(\sqrt{\frac{(g'(1)-g(1))\rho_{\rm typ}}{g(1)}}+\sqrt{\frac{g(1)}{(g'(1)-g(1))\rho_{\rm typ}}}\right)\;,\;\;\frac{g(1)}{g'(1)}=\rho_{\rm typ}\frac{\rho_{\rm typ}-1-\ln \rho_{\rm typ}}{(1-\rho_{\rm typ})^2}\;,
\ee
into Eq. \eqref{L_RS_e_text}, one obtains
\begin{align}
{\cal L}_{\rm RS}(e_{\rm typ})=\frac{1}{2}\left[\ln\rho_{\rm typ}+(1-\rho_{\rm typ})+\frac{g(1)}{g'(1)}\frac{\left(1-
\rho_{\rm typ}\right)^2}{\rho_{\rm typ}}\right]=0\;.
\end{align}
This identity can be understood using the identification ${\cal L}_{\rm RS}(e)=-\Sigma_{\min}(e)$. As $\Sigma_{\min}(e)<0$ for any $e<e_{\rm typ}$, there is an exponentially small probability to find any minimum with energy $e<e_{\rm typ}$ and if such minimum exists, it would naturally be the global minimum $e=e_{\min}$. At the typical energy $e=e_{\rm typ}$ instead, one expects that there is a sub-exponential number of minima (typically one), such that $\Sigma_{\min}(e_{\rm typ})=0$. Finally, at energy $e>e_{\rm typ}$, one expects an exponentially large number of minima such that $\Sigma_{\min}(e)>0$ but these minima are not expected to be the global minimum $e>e_{\min}$. For this description to be fully rigorous, one would need to consider the quenched complexity, which describes the typical configurations of the energy landscapes instead of the annealed complexity used here. As the identity described above only holds in the case where $\Gamma=0$, one might suspect that for finite $\Gamma>0$ the quenched and annealed complexity do not change sign at the same value of energy and thus do not coincide.

Let us now consider the replica-symmetry broken phase by first considering the full replica-symmetry broken phase.

\subsection{Full replica-symmetry broken phase}\label{FRSB_app}

In the FRSB phase, there is no simplification and the expression of $\Phi(s)$ is directly given in Eq. \eqref{phi_z}. The saddle-point equations are given by Eqs. (\ref{sp_eq_q_0}-\ref{sp_eq_v}) complemented with a third equation, obtained by computing the functional derivative
\be
    \left.\frac{\delta \Phi(s)}{\delta z(q)}\right|_{\rm s.p.}=\frac{s}{2}\left(f'(q)-\int_{q_0^*}^{q}\frac{dr}{\displaystyle \left(v_*+\int_{r}^{1}dt\,z_*(t)\right)^2}-\frac{q_0^*}{\left(v_*+\int_{q_0^*}^1 dq\, z_*(q)\right)\left(s q_0^*+v_*+\int_{q_0^*}^1 dq\, z_*(q)\right)}\right)=0\;,\label{sp_eq_z}
\ee
and which must be valid for any $q\in(q_0^*,1)$. One can check that taking the limit $q\to q_0^*$ in the above equation yields a result consistent with the first saddle-point equation \eqref{sp_eq_q_0} provided $s<z(q_0^*)$, and similarly taking the limit $q\to 1$ in the equation above, the equation matches exactly the second saddle-point equation \eqref{sp_eq_v}. In the FRSB regime, the function $z(q)$ is continuous and strictly increasing in the range $q\in(q_0,1)$. The solution can be determined by taking the derivative of Eq. \eqref{sp_eq_z} with respect to $q$, yielding
\be
f''(q)=\frac{1}{\displaystyle \left(v_*+\int_{q}^1 dr\,z_*(r)\right)^2}\;,\;\;q\in (q_0^*,1]\;,
\ee
which can simply be re-expressed as
\be
v_*+\int_{q}^1 dr\,z_*(r)=\frac{1}{\sqrt{f''(q)}}\;,\;\;q\in (q_0^*,1]\;.\label{int_eq}
\ee
Taking an additional derivative, one obtains an explicit expression for the function $z(q)$, 
\be
z_*(q)=\frac{f^{(3)}(q)}{2 f''(q)^{3/2}}\;,\;\;q\in (q_0^*,1]\;,
\ee
which matches exactly its expression when solving for the typical energy of the spherical spin-glass model. Note that for that solution to be consistent with the hypothesis that $z(q)$ is an increasing function, one must ensure that
\be
z_*'(q)=\frac{{\cal S}\left[f'\right](q)}{2\sqrt{f''(q)}}=\frac{1}{2f''(q)^{5/2}}\left[f''(q)f^{(4)}(q)-\frac{3}{2}f^{(3)}(q)^2\right]\geq 0\;,\;\;q\in (q_0^*,1]\;.\label{schwarz}
\ee
Plugging the expression of $z_*(q)$ into Eq. \eqref{int_eq}, it yields the expression of $v_*$, which reads
\be
v_*=\frac{1}{\sqrt{f''(1)}}\;,
\ee
and similarly takes the same value as for the typical energy in the FRSB phase. The parameter $q_0^*$ is thus the only parameter depending explicitly on $s$ in the FRSB phase.

Inserting the expressions of $z^*(q)$ and $v$ into the first saddle-point equation \eqref{sp_eq_q_0}, one obtains
\be
\left(s-\frac{f^{(3)}(q_0^*)}{2 f''(q_0^*)^{3/2}}\right)\left(f'(q_0^*)-\frac{q_0^*f''(q_0^*)}{s\sqrt{f''(q_0^*)} q_0^*+1}\right)=0\;.
\ee
There are thus two possible solutions
\be
s_1(q_0^*)=\frac{f^{(3)}(q_0^*)}{2 f''(q_0^*)^{3/2}}\;\;{\rm or}\;\;s_2(q_0^*)=\frac{1}{\sqrt{q_0^* f'(q_0^*)}}\left(\sqrt{\frac{q_0^* f''(q_0^*)}{f'(q_0^*)}}-\sqrt{\frac{f'(q_0^*)}{q_0^* f''(q_0^*)}}\right)\;,\label{s_FRSB}
\ee
corresponding respectively to Eq. \eqref{sol_q_not_ok} and \eqref{sol_q}. As we now argue, the correct solution is always given by $s=s_2(q_0^*)$. This solution satisfies 
\be
s_2'(q_0^*)=\left(\frac{g''(q_0^*)}{g'(q_0^*)+\Gamma}+\frac{1}{q_0^*}\right)(z_*(q_0^*)-s_2(q_0^*))=\frac{(\Gamma+g'(q_0^*)+q_0^* g''(q_0^*))\left(\Gamma-\Gamma_0(q_0^*)\right)}{2g''(q_0^*){q_0^*}^2(\Gamma+g'(q_0^*))^2(2g''(q_0^*)+q_0^* g^{(3)}(q_0^*))}\;,\label{ds_2}
\ee
where 
\be
\Gamma_0(q_0^*)=-g'(q_0^*)+\frac{2q_0^* g''(q_0^*)}{2g''(q_0^*)+g^{(3)}(q_0^*))}\;.
\ee
One can show that $\Gamma_0(0)=0$, $\Gamma_0(1)=\Gamma_t$ and 
\be
\Gamma_0'(q_0^*)=-\frac{2{q_0^*}^2 g''(q_0^*)}{(2g''(q_0^*)+q_0^* g^{(3)}(q_0^*))^2}{\cal S}[g'](q_0^*)\leq 0\;.
\ee
Thus, one obtains that for any $q_0^*\in [0,1]$, one has $\Gamma_0(q_0^*)\leq 0$ which ensures that for any $\Gamma\geq 0$ the solution $s_2'(q_0^*)\geq 0$ and $s_2(q_0^*)\leq z_*(q_0^*)$.
The solution $s_2(q_0^*)$ can take either sign and satisfies $s_2(q_{\rm typ})=0$, thus we expect that this is the correct solution.

Using the expressions for $v_*$ and $z_*(q)$, the CGF  $\phi(s)$ reads in the FRSB phase
\begin{align}
    \phi_{\rm FRSB}(s)=&\frac{s}{2}\left[s f(q_0^*)+\frac{f'(q_0^*)}{\sqrt{f''(q_0^*)}}
    +2\int_{q_0^*}^{1} dq\,\sqrt{f''(q)}\right]+\frac{1}{2}\ln\left(1+s\,q_0^*\,\sqrt{f''(q_0^*)}\right)\;,\\
    =&\frac{q_0^* f''(q_0^*)-f'(q_0^*)}{2q_0^* f'(q_0^*)\sqrt{f''(q_0^*)}}\left[\frac{f(q_0^*)\left(q_0^* f''(q_0^*)-f'(q_0^*)\right)+q_0^* f'(q_0^*)^2}{q_0^* f'(q_0^*)\sqrt{f''(q_0^*)}}
    +2\int_{q_0^*}^{1} dq\,\sqrt{f''(q)}\right]+\frac{1}{2}\ln\left(\frac{q_0^* f''(q_0^*)}{f'(q_0^*)}\right)\;.
\end{align}
where $q_0^*$ is the inverse function of $s=s_2(q_0^*)\equiv s(q_0^*)$ as given in Eq. \eqref{s_FRSB}. This CGF is convex as can be shown by taking the second derivative of this expression with respect to $s$,
\be
\phi_{\rm FRSB}''(s)=f(q_0^*)-\frac{q_0^* f'(q_0^*)^2}{f'(q_0^*)+q_0^* f''(q_0^*)}\geq 0\;,\label{phi_FRSB_sec}
\ee
which can be shown to be positive using Eq. \eqref{nice_id}. The first derivative of the CGF gives the associated value of the energy as
\be
e_{*,{\rm FRSB}}(s)=-\phi_{\rm FRSB}'(s)=-\left[\int_{q_0^*}^1 dq\, \sqrt{f''(q)}+\frac{1}{\sqrt{f''(q_0^*)}}\left(f'(q_0^*)+f(q_0^*)\left(\frac{f''(q_0^*)}{f'(q_0^*)}-\frac{1}{q_0^*}\right)\right)\right]\;.\label{e_FRSB}
\ee
One can obviously show that $\partial_s e_{*,{\rm FRSB}}(s)=-\phi_{\rm FRSB}''(s)\leq 0$ from Eq. \eqref{phi_FRSB_sec}, while 
$\partial_{q_0^*} e_{*,{\rm FRSB}}(s)=-\phi_{\rm FRSB}''(s)\times s_2'(q_0^*)\leq 0$ using that we have shown above that $s_2'(q_0^*)\geq 0$. For $s=0$ (corresponding to $q_0^*=q_{\rm typ}$ with $q_{\rm typ} f''(q_{\rm typ})=f'(q_{\rm typ})$) goes to the typical energy
\be
e_{\rm typ}=e_{*,{\rm FRSB}}(s=0)=-\left[\int_{q_{\rm typ}}^1 dq\, \sqrt{f''(q)}+q_{\rm typ}\sqrt{f''(q_{\rm typ})}\right]\;.
\ee
For $s=s_{\rm RSB}$, it goes to $e_{\rm RSB}$ (corresponding to $q_0^*=1$). For $\Gamma>0$ it goes to a finite value 
\be
e_c=\lim_{s\to -\infty}e_{*,{\rm FRSB}}(s)=-\int_{0}^1 dq\, \sqrt{f''(q)}\;,
\ee
as $s\to -\infty$ (corresponding to $q_0^*=0$). Note that the value $e_c$ coincides with the typical energy in the absence of magnetic field.
Using Eq. \eqref{phi_FRSB_sec} as well as the identity $q_{\rm typ} f''(q_{\rm typ})=f'(q_{\rm typ})$, one can further obtain the variance for finite magnetic field as
\be
{\cal V}_{\min}(\Gamma)=\lim_{N\to \infty} N\mbox{Var}(e_{\min})=\phi_{\rm FRSB}''(0)=g(q_{\rm typ})+\frac{q_{\rm typ} (\Gamma-g'(q_{\rm typ}))}{2}\;.\label{var_FRSB}
\ee
If the covariance function $f(q)=g(q)+\Gamma q$ scales as $q\to 0$ as
\be
f(q)=\Gamma q+g_2 q^2+g_p q^r+o(q^r)\;,\;\;
\ee
with $r>2$ the exponent of the leading order correction to the quadratic behaviour, one expects that 
\be
q_{\rm typ}^{*}\approx \left(\frac{\Gamma}{r(r-2)g_r}\right)^{\frac{1}{r-1}}\;,\;\;\Gamma\to 0\;,
\ee
and the variance scales as
\be
{\cal V}_{\min}(\Gamma)=\lim_{N\to \infty} N\mbox{Var}(e_{\min})=\frac{g_r}{2}(r-1)(r-2)\left(\frac{\Gamma}{g_r r(r-2)}\right)^{\frac{r}{r-1}}\;,\;\;\Gamma\to 0\;.
\ee
For zero magnetic field, the ground-state energy "superconcentrates" and the variance of $e_{\min}$ is $o(N^{-1})$. 

Let us now consider the LDF in this regime. In the FRSB regime it is expressed parametrically by fixing $e$ and determining $q_0^*$ from Eq. \eqref{e_FRSB} and inserting its expression in the LDF given by
\begin{align}
{\cal L}_{\rm FRSB}(e)=&s_* \phi_{\rm FRSB}'(s_*)-\phi_{\rm FRSB}(s_*)=\frac{1}{2}\left[s_*^2 f(q_0^*)+1-\frac{1}{1+s_*\,q_0^*\,\sqrt{f''(q_0^*)}}-\ln\left(1+s_*\,q_0^*\,\sqrt{f''(q_0^*)}\right)\right]\\
    =&\frac{1}{2}\left[\frac{f(q_0^*)}{f''(q_0^*)}\left(\frac{f''(q_0^*)}{f'(q_0^*)}-\frac{1}{q_0^*}\right)^2+1-\frac{f'(q_0^*)}{q_0^* f''(q_0^*)}-\ln\left(\frac{q_0^* f''(q_0^*)}{f'(q_0^*)}\right)\right]\;. \label{L_FRSB}  
\end{align}
One can check that for $e=e_{\rm typ}$, i.e. $q_0^*=q_{\rm typ}$ satisfying $q_{\rm typ} f''(q_{\rm typ})=f'(q_{\rm typ})$, the LDF above vanishes. This value corresponds to the unique minimum of ${\cal L}_{\rm FRSB}(e)$, as can be shown using that ${\cal L}_{\rm FRSB}'(e)=-s_{*,\rm FRSB}(e)$ vanishes only for $e=e_{\rm typ}$, i.e. $q_0^*=q_{\rm typ}$, and that ${\cal L}_{\rm FRSB}''(e)=\phi_{\rm FRSB}''(s_{*,\rm FRSB}(e))^{-1}>0$. This yields in particular that ${\cal L}_{\rm FRSB}(e_{\rm RSB})>0$ for $\Gamma<\Gamma_{\rm RSB}$. There are three cases.
\\

{\bf Non zero magnetic field $\Gamma>0$}. Then one must have that $q_{\rm typ}>0$, and expanding the LDF around its minimum, one obtains 
\be
{\cal L}_{\rm FRSB}(e)=\frac{(e-e_{\rm typ})^2}{2{\cal V}_{\min}(\Gamma)}+o(e-e_{\rm typ})^2\;,\;\;\Gamma>0\;.\label{Gaussian_FRSB}
\ee
The LDF diverges logarithmically for finite magnetic field as $e\to e_c$ (corresponding to $q_0^*\to 0$) with
\be
{\cal L}_{\rm FRSB}(e)= -\frac{1}{2}\ln\frac{2\sqrt{2g_2}(e_c-e)}{\Gamma}-\frac{3}{4}+O(e_c-e)\;,\;\;\Gamma>0\;.
\ee
\\

{\bf Zero magnetic field $\Gamma=0$}. For zero magnetic field $\Gamma=0$, the behaviour of the LDF near its minimum $e=e_{\rm typ}=e_{c}$, corresponding to $q_0^*=q_{\rm typ}=0$, is modified. There are two main cases. 

(i) Supposing that $g_3=3! f^{(3)}(0)>0$ as well as $g_4=4! f^{(4)}(0)>0$ (ensuring that the Schwarzian derivative appearing in Eq. \eqref{schwarz} is positive), one can expand the expression of $e$ in Eq. \eqref{e_FRSB} and \eqref{L_FRSB} for $q_0^{*}\to 0$. One obtains that both $e_c-e$ and ${\cal L}_{\rm FRSB}(e)$ scale as ${q_0^*}^4$, and one thus obtains the following linear behaviour
\be
{\cal L}_{\rm FRSB}(e)=z_*(0)(e_c-e)+o(e_c-e)\;,\;\;\Gamma=0\;,\;\;e<e_c\;,
\ee
where we remind that $z_*(0)=\frac{f^{(3)}(0)}{2 f''(0)^{3/2}}$. In that case, one expects that the CGF is given in the regime $s\in(-\infty,z_*(0)]$ by the linear function
\be
\phi_{\rm FRSB}(s)=-s\,e_c\;.
\ee
It displays a fifth order phase transition for $s=z_*(0)$, namely the function $\phi_{\rm FRSB}(s)$ and its fourth first derivatives are continuous for $s=z_*(0)$, while its fifth derivative is discontinuous
\be
\phi_{\rm FRSB}^{(5)}(s=z_*(0)_-)=0\neq \phi_{\rm FRSB}^{(5)}(s=z_*(0)_+)=\frac{9 f''(0)^{3/2}f^{(3)}(0)}{160 S[g'](0)^3}>0\;.
\ee

(ii) For models for which $f^{(3)}(0)=0$, the scaling exponent of the LDF near its minimum becomes non-trivial: supposing the minimum value of $r>2$ such that $f^{(r)}(0)>0$ the LDF scales close to its minimum as  
\be
{\cal L}_{\rm FRSB}(e)=\xi_r (e_c-e)^{\eta}+o(e_c-e)^{\eta}\;,\;\;\eta=\frac{3(r-2)}{2r-3}\;,\label{pow_law}
\ee
where
\be
\xi_r=2^{\frac{r-12}{2(2r-3)}}\left(\frac{r(2r-3)}{r-2}\right)^{\frac{3(r-2)}{2r-3}}\frac{(r-3)^{-\frac{r-3}{2r-3}}\left(r (r-2)g_r\right)^{\frac{3}{2r-3}}}{3 r g_2^{\frac{3r}{2(2r-3)}}}\;.\label{xi_p}
\ee
This large deviation result is quite remarkable. If we assume the simplest possible
matching between the large deviation regime $e_c-e=O(1)$ and the regime of
typical fluctuations, it leads to the conjecture that the
{\it typical} fluctuations of the ground state energy will be of order $e_c - e = O(N^{-1/\eta})$
with $\eta = \frac{3(r-2)}{2 r-3}$. More precisely we conjecture that 
the PDF of the ground-state energy ${\cal P}(e)$ 
takes the following scaling form in the regime of typical fluctuations
\be 
{\cal P}(e)=N^{1/\eta}\, p\left(N^{1/\eta}(e-e_c)\right) \;,\;\;{\rm where}\;\; p(\delta) \sim e^{- \xi_r|\delta|^{\eta}}\;,\;\;\delta\to -\infty\;.
\ee 
Note that the emergence of a non-trivial exponent $\eta$ is not dissimilar to the occurrence of the exponent $3/2$, describing the tail of the Tracy-Widom distribution for the spherical $2$-spin model, although here it appears in a full RSB regime. In fact, in the limit where there is no correction to the quadratic behaviour, i.e. $r\to \infty$, the exponent $3/2$ and correct prefactor $\xi_{\infty}=2/3\times 2^{3/2}$ match smoothly the TW tail. \\

{\bf Scaling regime $\Gamma \to 0$}. As $\Gamma\to 0$, the typical ground state energy approaches $e_c$ as
\be
e_{\rm typ}-e_c\approx -\frac{\Gamma}{2\sqrt{2g_2}}+o(\Gamma)\;.\label{en_dif_FRSB}
\ee
In the limit $\Gamma\to 0$, for energies in the range $e\in[e_c,e_{\rm typ}]$, the LDF takes the scaling form
\be
{\cal L}(e)\approx G\left(\frac{\sqrt{2g_2}(e-e_c)}{\Gamma}\right)\;,\;\;G(z)=
\displaystyle -\frac{1}{4}\left(3+8z+4z^2+2\ln(-2z)\right)\;,\;\;z\in [z_{\rm typ},0]\;,\label{wall_FRSB}
\ee
with $z_{\rm typ}=-1/2$. This scaling function is exactly the same as for the spherical $2$-spin model, obtained above. In contrast, in this scaling limit, the scaling function $G(z)=0$ for $z<-1/2$.

Let now similarly estimate the typical width of fluctuations of energy around $e_{\rm typ}$ allowing to smoothly interpolate between the quadratic behaviour for $\Gamma>0$ and the non-trivial behaviour in Eq. \eqref{pow_law}
\be
(e_{\rm typ}(\Gamma=0)-e)^{\frac{3(r-2)}{2r-3}}\sim \frac{(e_{\rm typ}(\Gamma)-e)^2}{{\cal V}_{\min}(\Gamma)}\Rightarrow e_{\rm typ}(\Gamma)-e\sim \Gamma^{\frac{2r-3}{r-1}}\;.
\ee
The typical energy gap between $e_{\rm typ}$ and $e_c$ in the limit $\Gamma\to 0$ in Eq. \eqref{en_dif_FRSB} is larger than the scale describing this interpolation as $2r-3>r-1$ for $r>2$. Hence there is a well defined scaling regime which describes 
the vicinity of $e_{\rm typ}$ for $\Gamma\to 0$, where the LDF takes the scaling form 
\be
{\cal L}(e)=\Gamma^{\frac{3(r-2)}{r-1}}L_{\rm FRSB}\left(\frac{e-e_{\rm typ}(\Gamma)}{\Gamma^{\frac{2 r - 3}{r-1}}}\right)\;,
\ee
where the function $L_{\rm FRSB}(u)$ is determined parametrically as 
\begin{align}
  L_{\rm FRSB}(u)&=\left(g_r r(r-2)\right)^{\frac{3}{r-1}}\frac{ \left(1-x(u)^{r-1}\right)^2 \left((r-3)
   x(u)^{r-1}+2 r\right)}{96r g_2^3 x(u)^3}\label{scal_FRSB_typ_1}\\
   -u&=\frac{\left(r (r-2)g_r\right)^{\frac{1}{r-1}}}{8\sqrt{2} g_2^{3/2}x(u)}\left(\frac{(r-3)(r-2)}{r(2r-3)} x(u)^{2(r-1)}+\frac{r(r-1)-4}{r(r-2)} x(u)^{r-1}+\frac{(r-1)^2}{(2r-3)(r-2)} x(u)-2\right)\;,\label{scal_FRSB_typ_2}
\end{align}
where $x(u)=q_0^*/q_{\rm typ}$ in that limit $\Gamma\to 0$.

In particular, the function $L_{\rm FRSB}(u)$ has the asymptotic behaviours 
\be
L_{\rm FRSB}(u)\approx \begin{cases}
\displaystyle 2^{\frac{r-12}{2(2r-3)}}\left(\frac{r(2r-3)}{r-2}\right)^{\frac{3(r-2)}{2r-3}}\frac{(r-3)^{-\frac{r-3}{2r-3}}\left(r (r-2)g_r\right)^{\frac{3}{2r-3}}}{3 r g_2^{\frac{3r}{2(2r-3)}}}|u|^{\frac{3(r-2)}{2r-3}}\;,\;\;u\to -\infty\;,\\
&\\
\displaystyle \frac{g_2^{3/2}r\left(r (r-2)g_r\right)^{\frac{1}{r-1}}}{r-1}u^2\;,\;\;u\to 0\;,\\
&\\
\displaystyle \frac{8\sqrt{2}}{3}g_2^{3/2} u^3\;,\;\;u\to +\infty\;,
\end{cases}\label{prefactors_scal_typ}
\ee
One can check that the limit $u \to -\infty$ (first line) matches the $\Gamma=0$ limit given in \eqref{pow_law},
and that the limit $u \to +\infty$ (last line) matches the result \eqref{wall_FRSB} and the behaviour of the 
function $G(z) \simeq \frac{4}{3} (z+\frac{1}{2})^3$ when $z \to z_{\rm typ}^+= - \frac{1}{2}^+$. Finally, the second line matches smoothly the quadratic behaviour of the LDF in Eq. \eqref{Gaussian_FRSB} as $e\to e_{\rm typ}$ for $\Gamma>0$.

Expanding the energy difference $e-e_{\rm RSB}$ and ${\cal L}_{\rm FRSB}(e)$ for $q_0^*\to 1$, one can obtain the Taylor series of ${\cal L}_{\rm FRSB}(e)$ as $e\to e_{\rm RSB}$. Comparing with that of ${\cal L}_{\rm RS}(e)$, one can show that the LDF displays a third order phase transition as $e\to e_{\rm RSB}$ with
\be
{\cal L}_{\rm FRSB}(e)-{\cal L}_{\rm RS}(e)=-\frac{f'(1)^3 f''(1)^{5/2} \left(f'(1)-f''(1)\right)(e-e_{\rm RSB})^3}{3 \left(f(1)
   \left(f''(1)+f'(1)\right)-f'(1)^2\right)^3 \left(f'(1) \left(f^{(3)}(1)+2
   f''(1)\right)-2 f''(1)^2\right)}+o(e-e_{\rm RSB})^3\;.
\ee

For any model such that the Schwarzian derivative ${\cal S}[f'](q)$ of the derivative of the covariance function $f'(q)$ is positive for all $q\in [0,1]$, only the RS and FRSB regimes are required to describe the large deviation function of the ground-state at speed $N$. Note that there exist models for which the Schwarzian derivative ${\cal S}[f'](q)$ changes sign over the interval $q\in[0,1]$ and for which the RSB regime in the LDF should be described by an ansatz combining FRSB and 1RSB features. These models are beyond the scope of this article and will not be considered here. An example of model for which the RSB phase is described within the FRSB framework is
\be
f(q)=\Gamma q+g(q)\;,\;\;g(q)=\frac{q^2}{2}+g_3 q^3+g_4 q^4\;,\;\;0\leq g_4\leq \frac{1}{24}\;,\;\;0\leq g_3\leq \frac{2}{3}\left(\sqrt{g_4(1-8 g_4)}-4g_4\right)\;.\label{f_FRSB}
\ee
In Fig. \ref{fig:LDF_FRSB}, we plot the LDF ${\cal L}(e)$ for one example of the covariance $g(q)$ in Eq. \eqref{f_FRSB} ($g_4=1/48$, $g_3=1/36$) with three different values of magnetic field $\Gamma=0,1/16,3/4$. For the first two values of $\Gamma$, the typical energy $e_{\rm typ}$, where ${\cal L}(e)=0$, belongs to the FRSB regime while for the third it belongs to the RS regime. In each case, a transition from the RS to the FRSB regime occurs for $e=e_{\rm RSB}$. For zero magnetic field the LDF vanishes for $e_c=e_{\rm typ}$ while for positive magnetic field it diverges logarithmically for $e=e_c$.

In the next section, we will consider models for which the RSB phase is not of FRSB type but of 1RSB type, for which ${\cal S}\left[f'\right](q)<0$ for $q\in(0,1)$.

\subsection{One-step replica-symmetry broken phase}\label{1RSB_app}

In the one-step replica-symmetry broken (1RSB) phase, the function $z(q)$ is piece-wise constant with
\be
z(q)=\begin{cases}
s\;,&\;\;0\leq q< q_0\;,\\
m_0\;,&\;\;q_0\leq q<1\;,
\end{cases}
\ee
where $m_0\geq s$ and the value $0<q_0<1$. To obtain the 1RSB expression of the CGF, we need to use Eq. \eqref{ext_phi} and optimise the functional
\be
\Phi_{\rm 1RSB}(s)=\frac{1}{2}\left[m_0 s f(1)+v s f'(1)+s(s-m_0)f(q_0)+\frac{s}{m_0}\ln\left(1+\frac{m_0(1-q_0)}{v}\right)+\ln\left(1+\frac{sq_0}{v+m_0(1-q_0)}\right)\right]\;,\label{phi_1RSB}
\ee
and there are three parameters to optimise over, namely $m_0\geq s$, $v\geq 0$ and $q_0\in [0,1]$. 
The saddle-point equations read
\begin{align}
    \left.\partial_{v}\Phi_{\rm 1RSB}\right|_{\rm s.p.}&=\frac{s}{2}\left(f'(1)-\frac{1-q_0^*}{v(v+m_0^*(1-q_0^*))}-\frac{q_0^*}{(v+m_0^*(1-q_0^*))(v+m_0^*(1-q_0^*)+s q_0^*)}\right)=0\;,\label{Phi_1RSB_v}\\
    \left.\partial_{q_0}\Phi_{\rm 1RSB}\right|_{\rm s.p.}&=\frac{s(s-m_0^*)}{2}\left(f'(q_0^*)-\frac{q_0^*}{(v+m_0^*(1-q_0^*))(v+m_0^*(1-q_0^*)+s q_0^*)}\right)=0\;,\label{Phi_1RSB_q}\\
    \left.\partial_{m_0}\Phi_{\rm 1RSB}\right|_{\rm s.p.}&=\frac{s}{2}\left(f(1)-f(q_0^*)-\frac{1}{{m_0^*}^2}\ln\left(1+\frac{m_0^*(1-q_0^*)}{v}\right)+\frac{(1-q_0^*)(v+s q_0^*+m_0^*(1-2q_0^*))}{m_0^*(v+m_0^*(1-q_0^*))(v+m_0^*(1-q_0^*)+s q_0^*)}\right)=0\;.\label{Phi_1RSB_m}
\end{align}
Note that the first two saddle-point equations (\ref{Phi_1RSB_v},\ref{Phi_1RSB_q}) are recovered by inserting the 1RSB ansatz for $z(q)$ into Eqs. (\ref{sp_eq_q_0}-\ref{sp_eq_v}). The transition to the RS phase occurs either as the optimal value $q_0^*\to 1$ or $m_0^*\to s$. In each of these two cases, one can check that the saddle point equations (\ref{Phi_1RSB_v}-\ref{Phi_1RSB_q}-\ref{Phi_1RSB_m}) and the expression of $\Phi_{\rm 1RSB}(s)$ \eqref{phi_1RSB} coincide respectively with the RS  saddle-point equation \eqref{sp_eq_RS} and the expression of $\Phi_{\rm RS}(s)$ in \eqref{phi_RS}. In order to obtain an expression that differs from the RS expression, we suppose that $m_0^*\neq s$. As mentioned for general level of replica-symmetry, the saddle-point equation \eqref{Phi_1RSB_q} thus yields the condition in \eqref{sol_q}
\be
v_*+\int_{q_0^*}^1 dq\, z_*(q)=v_*+(1-q_0^*)m_0^*=\frac{q_0^*}{2}\left(-s+\sqrt{\frac{4}{q_0^* f'(q_0^*)}+s^2}\right)\;.   \label{nice_eq_1RSB}
\ee
The saddle-point equation \eqref{Phi_1RSB_v} yields on the other hand the simple identity \eqref{id_var}
\be
\int_{q_0^*}^1\frac{dq}{\left(v_*+\int_{q}^1 dr\, z_*(r)\right)^2}=\frac{1-q_0^*}{v_*(v_*+(1-q_0^*)m_0^*)}=f'(1)-f'(q_0^*)\;.
\ee
In order to express our results, it will be convenient to express the CGF parameter $s$ in terms of $q_0^*$ rather than the inverse. We therefore introduce the  parametrisation
\begin{align}
  s&=\frac{1}{\sqrt{q_0^* f'(q_0^*)}}\left(\frac{1}{\sqrt{\alpha(q_0^*)}}-\sqrt{\alpha(q_0^*)}\right)\;,\label{s_1RSB}\\
  v_*&=\rho(q_0^*)\sqrt{\frac{q_0^*}{f'(q_0^*)\alpha(q_0^*)}}\;,\\
  m_0^*&=\frac{1}{1-q_0^*}\sqrt{\frac{q_0^* \rho(q_0^*)}{f'(q_0^*)}}\left(\sqrt{\frac{\alpha(q_0^*)}{\rho(q_0^*)}}-\sqrt{\frac{\rho(q_0^*)}{\alpha(q_0^*)}}\right)\;,
\end{align}
where $\alpha(q_0^*)$ will be determined by the last saddle-point equation \eqref{Phi_1RSB_m} and the function $\rho(q)$, introduced when presenting the results on the typical energy, reads 
\be
\rho(q)=\frac{(1-q)f'(q)}{(f'(1)-f'(q))q}\;.
\ee
The parametrisation above ensures that Eqs. (\ref{Phi_1RSB_v}-\ref{Phi_1RSB_q}) are fulfilled while the remaining equation \eqref{Phi_1RSB_m} yields the following relation between $\alpha$ and $q_0^*$,
\be
\mu(q_0^*)=\rho(q_0^*)\frac{\rho(q_0^*)-\alpha(q_0^*)-\alpha(q_0^*)\ln (\rho(q_0^*)/\alpha(q_0^*))}{(\alpha(q_0^*)-\rho(q_0^*))^2}\;,\label{mu_alph}
\ee
where the function
\be
\mu(q)=\frac{1}{f'(1)-f'(q)}\left(\frac{f(1)-f(q)}{1-q}-f'(q)\right)\;.
\ee
Note that the function $\mu(q)$ is completely independent of the magnetic field $\Gamma$ as can be checked by invariance of the above definition when replacing $f(q)\to f(q)+\Gamma q$. As mentioned when considering the typical energy, the function $\mu(q)$ is an increasing function of $q$ as can be shown by computing its derivative
\be
\mu'(q)=\frac{(f'(1)-f'(q))(f(1)-f(q)-(1-q)f'(q))+(1-q)((1-q)f'(1)-f(1)+f(q))f''(q)}{(1-q)^2(f'(1)-f'(q))^2}\;,
\ee
and using the identities
\begin{align}
    &f(1)-f(q)-(1-q)f'(q)=(1-q)\sum_{p=2}^{\infty}g_p\sum_{k=0}^{p-1}(q^k-q^{p-1})\geq 0\;,\\
    &(1-q)f'(1)-f(1)+f(q)=(1-q)\sum_{p=2}^{\infty}g_p\sum_{k=0}^{p-1}(1-q^k)\geq 0\;.
\end{align}
The function $\mu(q)$ is monotonously increasing apart from the singular case of the 2-spin, where $\mu(q)=1/2$ for any value of $q$, and this will play a role in the analysis below. Thus for any $q\in [0,1]$, one has
\be
0<\mu(0)=\frac{g(1)}{g'(1)}\leq \mu(q)\leq \mu(1)=\frac{1}{2}\;. 
\ee
Note that $g'(1)-2g(1)=\sum_{r\geq 2}(r-2)g_r\geq 0$, such that $g(1)/g'(1)\leq 1/2$. 

Let us now consider the function $\rho(q)$ appearing in Eq. \eqref{mu_alph}. For a magnetic field
\be
\Gamma\geq \Gamma_t=\frac{2g''(1)^2}{2g''(1)+g^{(3)}(1)}-g'(1)\;,\;\;{\rm where}\;\;\Gamma_{\rm RSB}>\Gamma_{t}\;,
\ee
the function $\rho(q)$ is monotonously decreasing in the interval $q\in [0,1]$, such that its minimum over this interval is reached in $q_m(\Gamma\geq \Gamma_t)=1$. For a smaller magnetic field $\Gamma< \Gamma_t$, the function $\rho(q)$ is decreasing for $q$ smaller than its unique minimum $0\leq q_m(\Gamma<\Gamma_t)<1$ and is increasing for $q\geq q_m(\Gamma)$. For any positive magnetic field, it diverges at the origin with $q\rho(q)\to \Gamma/g'(1)$ and reaches a finite value $\rho(1)=f'(1)/f''(1)$ for $q\to 1$. 

Finally, the last function $\alpha(q)$ appearing in Eq. \eqref{mu_alph} can be shown to satisfy $0\leq \rho(q_0^*)\leq \alpha(q_0^*)$ for any value of $q_0^* \in[0,1]$ in order to ensure that $g(1)/g'(1)\leq \mu(q_0^*)\leq 1/2$. Using the same equation, one can also compute explicitly 
\be
\frac{\alpha'(q_0^*)}{\alpha(q_0^*)}=\frac{\rho'(q_0^*)}{\rho(q_0^*)}-\frac{(\alpha(q_0^*)-\rho(q_0^*))^3\mu'(q_0^*)}{\alpha(q_0^*)\rho(q_0^*)[2(\rho(q_0^*)-\alpha(q_0^*))-(\alpha(q_0^*)+\rho(q_0^*))\ln(\rho(q_0^*)/\alpha(q_0^*))]}\leq \frac{\rho'(q_0^*)}{\rho(q_0^*)}\;,
\ee
where we have used that $\mu'(q_0^*)\geq 0$ and $0\leq \rho(q_0^*)\leq \alpha(q_0^*)$. This yields that the function $\alpha(q_0^*)$ is a decreasing function of $q_0^*$, at least in the interval $[0,q_m(\Gamma)]$. In the following, we will show that it will be sufficient to consider the 1RSB solution only in this regime $0\leq q_0^*\leq q_m(\Gamma)$.

 Using the parametrisation introduced above, the CGF is expressed within the 1RSB framework as
\begin{align}
\phi_{\rm 1RSB}(s)=&\frac{1}{2}\left[m_0^* s f(1)+v_* s f'(1)+s(s-m_0^*)f(q_0^*)+\frac{s}{m_0^*}\ln\left(1+\frac{m_0^*(1-q_0^*)}{v_*}\right)+\ln\left(1+\frac{sq_0^*}{v_*+m_0^*(1-q_0^*)}\right)\right]\;,\nn\\
=&\frac{1}{4} \left(1+\frac{f(q_0^*)-q_0^*f(1)}{q_0^*(1-q_0^*)f'(q_0^*)}+2\frac{f(q_0^*)-f(1)-(q_0^*-1)f'(1)}{q_0^*(f'(1)-f'(q_0^*))}\right)q_0^* f'(q_0^*) s^2\nn\\
   &+\frac{1}{4} \left(-1+2\frac{f(q_0^*)-f(1)-(q_0^*-1)f'(1)}{q_0^*(f'(1)-f'(q_0^*))}+\frac{f(1)-f(q_0^*)}{(1-q_0^*)f'(q_0^*)}\right)s\sqrt{q_0^* f'(q_0^*)(4+q_0^* f'(q_0^*) s^2)}\nn\\
   &+\ln\left(\frac{s\sqrt{q_0^* f'(q_0^*)}+\sqrt{4+q_0^* f'(q_0^*) s^2}}{2}\right)\;,
\end{align}
where the value of $q_0^*$ for a fixed $s$ is obtained by inverting Eq. \eqref{s_1RSB} using the expression of $\alpha(q_0^*)$ obtained from Eq. \eqref{mu_alph}.

Computing its first derivative allows to express the associated value of the energy for fixed $s$,
\be
e_{*,{\rm 1RSB}}(s)=-\phi_{\rm 1RSB}'(s)=\frac{(1-q_0^*)\mu(q_0^*)f'(1)-f(1)+[f(1)-\mu(q_0^*)f'(1)-(1-\mu(q_0^*))f'(q_0^*)]\alpha(q_0^*)}{\sqrt{q_0^* f'(q_0^*)\alpha(q_0^*)}}\;.\label{e_1rsb}
\ee
Similarly, one can compute explicitly the second derivative of the CGF but its general expression is rather cumbersome and not given here. Setting instead $s=0$, corresponding to $q_0^*=q_{\rm typ}$, one obtains the rescaled variance, which takes the very simple expression
\be
{\cal V}_{\min}(\Gamma)=\lim_{N\to \infty} N\mbox{Var}(e_{\min})=\phi_{\rm 1RSB}''(0)=g(q_{\rm typ})+\frac{q_{\rm typ} (\Gamma-g'(q_{\rm typ}))}{2}\;.\label{var_1RSB}
\ee
It is quite interesting to note, as already obtained in \eqref{var_gen}, that the expression of the variance in the RS \eqref{var_RS}, FRSB \eqref{var_FRSB} and 1RSB \eqref{var_1RSB} schemes takes exactly the same form as a function of $q_{\rm typ}$. The typical overlap $q_{\rm typ}$ is however obtained in a distinct manner in each of the three schemes: $q_{\rm typ}=1$ in the RS scheme, $q_{\rm typ}$ is the solution of $q_{\rm typ}f''(q_{\rm typ})=f'(q_{\rm typ})$ in the FRSB scheme, and $q_{\rm typ}$ is the solution of Eq. \eqref{mu_alph}, i.e. $q_0^* \to q_{\rm typ}$, satisfying $\alpha(q_{\rm typ})=1$ in the 1RSB scheme.

In the limit $\Gamma\to 0$, the typical overlap $q_{\rm typ}\approx\Gamma/(\rho_{\rm typ}g'(1)-2g_2)$, where $\rho_{\rm typ}=\rho(q_{\rm typ})$ is the solution of Eq. \eqref{mu_alph} satisfying $\alpha(q_{\rm typ})=1$ in the limit $q_{\rm typ}\to 0$, and the variance scales as
\be
{\cal V}_{\min}(\Gamma)=\frac{\Gamma^2}{2(\rho_{\rm typ}g'(1)-2g_2)}+O(\Gamma^3)\;.
\ee

Using Eq. \eqref{e_1rsb} above, one can express $e$ in terms of $q_0^*$ and $\alpha(q_0^*)$, where $\alpha(q_0^*)$ is obtained from Eq. \eqref{mu_alph}, and compute explicitly the LDF ${\cal L}_{\rm 1RSB}(e)$ within the 1RSB framework as
\begin{align}
    {\cal L}_{\rm 1RSB}(e)&=-\min_{s\leq s_{\rm RSB}}\left[s e+\phi_{\rm 1RSB}(s)\right]=s_* \phi_{\rm 1RSB}'(s_*)-\phi_{\rm 1RSB}(s_*)\\
    &=\frac{s_*^2}{4} \left(2f(q_0^*)-q_0^*f'(q_0^*)\right)+\frac{s_* q_0^*f'(q_0^*)}{4}\sqrt{\frac{4}{q_0^*f'(q_0^*)}+s_*^2}-\ln\left[\frac{\sqrt{q_0^*f'(q_0^*)}}{2}\left(s_*+\sqrt{\frac{4}{q_0^*f'(q_0^*)}+s_*^2}\right)\right]\;,\nn\\
    &=\frac{1}{2}\left(1-\alpha(q_0^*)+\ln\alpha(q_0^*)+\frac{f(q_0^*)}{q_0^* f'(q_0^*)}\frac{(1-\alpha(q_0^*))^2}{\alpha(q_0^*)}\right)\;.
\end{align}
For any $\Gamma>0$, in the limit $e\to e_c$, corresponding to $q_0^*\to 0$, the LDF diverges logarithmically as
\be
{\cal L}_{\rm 1RSB}(e)\approx\frac{1}{2}\ln\frac{2\sqrt{\rho_{\rm typ}g'(1)}(e_c-e)}{\Gamma}-\frac{1}{2}-\frac{g_2}{2g'(1)\rho_{\rm typ}}\;,\;e\to e_c\;.\label{log_div_1RSB}
\ee
For $\Gamma=0$ instead, as we will see in more details below, the 1RSB regime is limited to the single point $e=e_c=e_{\rm typ}(\Gamma=0)$, where the LDF vanishes.

Let us now discuss in more details the transition from the RS expression to the 1RSB expressions for $\phi(s)$ and ${\cal L}(e)$ and the domain of validity of the above expressions in terms of values of $q_0^*$. Let us recall that there are two distinct mechanisms prompting a transition from the RSB regime to the RS regime, the parameter $q_0^*$ can reach the value $q_0^*=1$ or the parameter $m_0^*$ can reach the value $m_0^*=s$. The nature of the mechanism at the transition is dictated by the value of the magnetic field $\Gamma$.

\subsubsection{Large magnetic field $\Gamma>\Gamma_t$}

For large magnetic field
\be
\Gamma\geq \Gamma_t=\frac{2g''(1)^2}{2g''(1)+g^{(3)}(1)}-g'(1)\;,
\ee
the transition from the RS to the RSB phase occurs when the parameter $q_0^*$ reaches the value $q_0^*=1$ at the transition. In terms of values of $s$ when considering $\phi(s)$ and of $e$ when considering ${\cal L}(e)$, these transitions thus occur respectively for $s=s_{\rm RSB}$ and $e=e_{\rm RSB}$
(as already predicted within the RS framework), which take the value
\be
s_{\rm RSB}=\frac{1}{\sqrt{f'(1)}}\left(\sqrt{\frac{f''(1)}{f'(1)}}-\sqrt{\frac{f'(1)}{f''(1)}}\right)\;,\;\;e_{\rm RSB}=-\frac{1}{\sqrt{f''(1)}}\left[f'(1)-f(1)+f(1)\frac{f''(1)}{f'(1)}\right]\;.
\ee
These formulae can easily be shown to coincide with the expressions of Eqs. (\ref{s_1RSB}-\ref{mu_alph}-\ref{e_1rsb}) for $q_0^*=1$, where $\mu(1)=1/2$ and $\rho(1)=\alpha(1)=f'(1)/f''(1)$. Expanding the value of $q_0^*$ in the regime $s\to s_{\rm RSB}$, one finds
\be
q_0^*=1-\frac{\left(g'(1)+\Gamma\right)^2 \left(g'(1)+\Gamma_t\right)(s_{\rm RSB}-s)}{\sqrt{g''(1)} (\Gamma-\Gamma_t)
   \left(g''(1)+g'(1)+\Gamma\right)}+O(s-s_{\rm RSB})^2\;.
\ee
Expanding the expressions of $\phi_{\rm 1RSB}(s)$ and $\phi_{\rm RS}(s)$ about $s=s_{\rm RSB}$, one can show that the transition of the CGF occurring for $s=s_{\rm RSB}$ is of third order with
\be
\phi_{\rm 1RSB}(s)-\phi_{\rm RS}(s)=\frac{\sqrt{g''(1)} \left(\Gamma +g'(1)\right)^3 \left(\Gamma _t+g'(1)\right)
   \left(\Gamma_{\rm RSB}-\Gamma \right)}{6 \left(\Gamma-\Gamma_t\right)
   \left(\Gamma +g''(1)+g'(1)\right)^3} (s-s_{\rm RSB})^3+O(s-s_{\rm RSB})^4\;.
\ee
Expanding similarly the functions ${\cal L}_{\rm 1RSB}(e)$ and ${\cal L}_{\rm RS}(e)$, one can similarly obtain that the transition is of third order with 
\be
{\cal L}_{\rm 1RSB}(e)-{\cal L}_{\rm RS}(e)=\frac{\sqrt{g''(1)} \left(\Gamma +g'(1)\right)^3
   \left(g'(1)+\Gamma_t\right) \left(\Gamma -\Gamma_{\rm RSB}\right)}{6
   (\Gamma_t-\Gamma ) \left((\Gamma +g(1)) g''(1)+\left(g(1)-g'(1)\right)
   \left(\Gamma +g'(1)\right)\right)^3}(e-e_{\rm RSB})^3+O(e-e_{\rm RSB})^4\;.
\ee
The divergence of the prefactor in both expressions for $\Gamma=\Gamma_t$ is related to the change in the order of the transition that will be considered in more details in the next section. Exactly for $\Gamma_{\rm RSB}$, the transition is instead of fourth order. Let us now consider the nature of the transition for $\Gamma<\Gamma_t$.

\subsubsection{Low magnetic field}

In the regime 
\be
\Gamma_t=\frac{2g''(1)^2}{2g''(1)+g^{(3)}(1)}-g'(1)>\Gamma>0\;,
\ee
the transition from the 1RSB to the RS transition occurs when the parameter $m_0^*$ reaches $s$. At this transition, the value of $q_0^*=q_t(\Gamma)<1$. Equating the expressions of $s$ and $m_0^*$, one can express explicitly $\alpha$ as a function of $q_t(\Gamma)$ at the transition as
\be
\alpha(q_t)=\frac{(1-q_t)f'(1)}{f'(1)-f'(q_t)}=\frac{q_t f'(1)\rho(q_t)}{f'(q_t)}\;.
\ee
Inserting this expression for $\alpha(q_t)$ into Eq. \eqref{mu_alph}, the value of $q_t$ can then be obtained by solving the self-consistent equation
\begin{align}
\mu(q_t)&=\frac{1}{f'(1)-f'(q_t)}\left(\frac{f(1)-f(q_t)}{1-q_t}-f'(q_t)\right)\\
&=f'(q_t)\frac{f'(q_t)-q_t f'(1)-q_t f'(1)\ln [f'(q_t)/(q_t f'(1))]}{(f'(q_t)-q_t f'(1))^2}\;.  \nn 
\end{align}
Clearly, $q_t=1$ is always a solution of the above equation and will be the correct solution for $\Gamma\geq \Gamma_t$. For $0<\Gamma<\Gamma_t$, an additional solution exists, satisfying $0<q_t<1$. Expanding the value of $q_t$ as $\Gamma\to \Gamma_t$, one obtains
\be
q_t(\Gamma)=1-\frac{2 \left(g^{(3)}(1)+2 g''(1)\right)^2}{g''(1) \left(3
   g^{(3)}(1)^2-2 g^{(4)}(1) g''(1)\right)}(\Gamma_t-\Gamma)+O(\Gamma-\Gamma_t)^2\;.
\ee
Note that the denominator in that expression is proportional to $-{\cal S}[g'](1)$, which is positive for models displaying a 1RSB phase (it would be negative for models displaying FRSB). A similar expansion of the value $q_m(\Gamma)$ where the function $\rho(q)$ reaches its minimum yields instead
\be
q_m(\Gamma)=1-\frac{3 \left(g^{(3)}(1)+2 g''(1)\right)^2}{2g''(1) \left(3
   g^{(3)}(1)^2-2 g^{(4)}(1) g''(1)\right)}(\Gamma_t-\Gamma)+O(\Gamma-\Gamma_t)^2\;,
\ee
such that in the vicinity of $\Gamma=\Gamma_t$ at least, $q_t(\Gamma)\leq q_m(\Gamma)$.

In the opposite limit $\Gamma\to 0$, one must have that 
\be
q_t(\Gamma)=\frac{\Gamma}{g'(1)\rho_{\rm typ}-2g_2}+O(\Gamma^2)\;.\label{q_t_small_G}
\ee
Assuming that the function $g(q)\approx g_2 q^2+g_r q^r$ with $r>2$, one obtains instead
\be
q_m(\Gamma)\approx \frac{\Gamma^{\frac{1}{r-1}}}{r(r-2)g_r}+O(\Gamma^2)\;,
\ee
such that as $\Gamma\to 0$, one has $q_m(\Gamma)\geq q_t(\Gamma)$ as well. We expect that this inequality holds for any value of $\Gamma$.

For $\Gamma\leq \Gamma_t$, the transition of the CGF does not occur at the value $s_{\rm RSB}$ but at the value
\be
s_t=\frac{1}{\sqrt{q_t f'(q_t)}}\left(\sqrt{\frac{f'(1)-f'(q_t)}{(1-q_t)f'(1)}}-\sqrt{\frac{(1-q_t)f'(1)}{f'(1)-f'(q_t)}}\right)\geq s_{\rm RSB}\;.
\ee
Note that for $\Gamma\geq \Gamma_t$, as $q_t=1$ the expression above coincides with $s_{\rm RSB}$. At the transition, the parameter $v=v_t$ is continuous
\be
v_t=\frac{1}{2}\left(-s_t+\sqrt{\frac{4}{f'(1)}+s_t^2}\right)=\rho(q_t)\sqrt{\frac{q_t}{f'(q_t)\alpha(q_t)}}=\sqrt{\frac{\rho(q_t)}{f'(1)}}=\sqrt{\frac{(1-q_t)f'(q_t)}{(f'(1)-f'(q_t))q_t f'(1)}}\;.
\ee
Using that, at the transition $m_0^*=s_t$, one can thus show that the CGF is continuous
\begin{align}
    \phi_{\rm 1RSB}(s_t)=\frac{1}{2}\left[s_t^2 f(1)+v_t s_t f'(1)+\ln\left(\frac{s_t+v_t}{v_t}\right)\right]=\phi_{\rm RS}(s_t)\;.
\end{align}
The first derivative $\phi'(s)$ is also continuous at the value of $s_t$ with
\begin{align}
e_t&=e_{*,{\rm RS}}(s_t)=-\phi_{\rm RS}'(s_t)=-\left(v_t(f'(1)-f(1))+\frac{f(1)}{v_t f'(1)}\right)\\
&=e_{*,{\rm 1RSB}}(s_t)=-\phi_{\rm 1RSB}'(s_t)=-\frac{f(1)(q_t f'(1)-f'(q_t)+(1-q_t)f'(q_t)f'(1)}{\sqrt{q_t f'(q_t)f'(1)(f'(1)-f'(q_t))(1-q_t)}}\;.    
\end{align}
Thus, one can similarly show that the LDF and its first derivative is continuous as well at the transition 
\be
{\cal L}_{\rm RS}(e_t)=s_t\phi_{\rm RS}'(s_t)-\phi_{\rm RS}(s_t)={\cal L}_{\rm 1RSB}(e_t)=s_t\phi_{\rm 1RSB}'(s_t)-\phi_{\rm 1RSB}(s_t)\;.
\ee
The transition is of second order in this regime and occurs for $s=s_t$, i.e. $q=q_t$ with 
\begin{align}
&\phi_{\rm 1RSB}''(s_t)=\frac{f'(1) f'(q_t) \left(2 (q_t-1) \mu(q_t) f'(1)
   (3 q_t \mu(q_t)-2 q_t-\mu(q_t)+1)+f(1) (q_t (q_t+4 \mu(q_t)-2)-2 \mu(q_t)+2)\right)}{-2 q_t \mu(q_t) f'(1)^2+f'(1) (q_t (q_t+4 \mu(q_t)-2)-2 \mu(q_t)+2)
   f'(q_t)-q_t f'(q_t)^2}\nn\\
&+\frac{-2 q_t \mu(q_t) f'(1)^2 \left((q_t-1) \mu(q_t) f'(1)+f(1)\right)-f'(q_t)^2 \left((q_t-1) f'(1) \left(q_t
   (1-2 \mu(q_t))^2-2 (\mu(q_t)-1)^2\right)+f(1) q_t\right)}{-2 q_t \mu(q_t) f'(1)^2+f'(1) (q_t (q_t+4 \mu(q_t)-2)-2 \mu(q_t)+2)
   f'(q_t)-q_t f'(q_t)^2}\;,
\end{align}
which can be shown numerically to be positive on different examples. While this expression is quite cumbersome, one can check explicitly that for $q_t\to 1$, $\mu(q_t)\to 1/2$ and this expression coincides with $\phi_{\rm RS}''(s_t=s_{\rm 1RSB})=f(1)-f'(1)^2/(f'(1)+f''(1))$. Note that ${\cal L}_{\rm 1RSB}''(e_t)=1/\phi_{\rm 1RSB}''(s_t)$. On the other hand, one can compute explicitly
\be
{\cal L}_{\rm RS}''(e_t)=\frac{2}{e_{c,{\rm RS}}}\left(1-\frac{2f(1)}{f'(1)}-\frac{e_t}{\sqrt{e_t^2-e_{c,{\rm RS}}^2}}\right)\;.
\ee

\subsubsection{Zero magnetic field $\Gamma=0$}

For zero magnetic field $\Gamma\to 0$, we have that $q_t\to 0$ as in Eq. \eqref{q_t_small_G} and 
\be
\lim_{\Gamma\to 0}\rho(q_t)=\lim_{\Gamma\to 0}\frac{f'(q_t)}{q_t f'(1)}=\rho_{\rm typ}\;,\;\;\lim_{\Gamma\to 0}\left(\frac{\alpha(q_t)-1}{\Gamma}\right)=-\frac{1-\rho_{\rm typ}}{\rho_{\rm typ}g'(1)-2g_2}\;,
\ee
where we remind that $\rho_{\rm typ}$ is the solution of
\be
\frac{g(1)}{g'(1)}=\rho_{\rm typ}\frac{\rho_{\rm typ}-1-\ln \rho_{\rm typ}}{(1-\rho_{\rm typ})^2}\;.
\ee
Using these identities, one can now check that
\be
e_t=-\frac{1}{\sqrt{g'(1)}}\left(\sqrt{\rho_{\rm typ}}(g'(1)-g(1))+\frac{g(1)}{\sqrt{\rho_{\rm typ}}}\right)=e_{\rm typ}(\Gamma=0)=e_c\;.
\ee
However, the Laplace parameter $s$ reaches a finite value at the transition
\be
s_t=\frac{1}{\sqrt{g'(1)}}\left(\frac{1}{\sqrt{\rho_{\rm typ}}}-\sqrt{\rho_{\rm typ}}\right)\;.
\ee
Since $q_t(\Gamma=0)=0$, one thus naturally expects that $q_0^*$ freezes to the value $q_0^*=0$ in the whole 1RSB regime, i.e. for $s\in(-\infty,s_t]$ and thus the CGF takes in that interval a simple linear form 
\be
\phi_{\rm 1RSB}(s)=-s\,e_{\rm typ}\; \quad , \quad s\in(-\infty,s_t].
\ee
Upon Legendre transform, one can check that the 1RSB expression of the LDF associated to this formula only describes the single point $e=e_{\rm typ}$.

The transition in the CGF is again of second order, namely 
\be
\phi_{\rm 1RSB}(s)-\phi_{\rm RS}(s)=-\rho_{\rm typ}g'(1)\left[\frac{\rho_{\rm typ}-1-\ln \rho_{\rm typ}}{(1-\rho_{\rm typ})^2}-\frac{1}{1+\rho_{\rm typ}}\right](s-s_t)^2+O(s-s_t)^3\;.
\ee
As mentioned in the section on the RS regime, the LDF ${\cal L}_{\rm RS}(e)$ vanishes linearly at the value $e_{\rm typ}$ with ${\cal L}_{\rm RS}'(e_{\rm typ})=-s_t$ (see Eq. \eqref{lin_beh}). One expects that the probability of events $e\geq e_{\rm typ}$ are characterised by a large deviation with speed larger than $N$ and are out of the scope of the present article. Thus, for zero magnetic field  $\Gamma=0$, the LDF is expected to be characterised only by the RS regime ${\cal L}(e)={\cal L}_{\rm RS}(e)$ for all $e\leq e_{\rm typ}$. In addition, it vanishes linearly at the typical value instead of the quadratic behaviour observed for positive magnetic field $\Gamma>0$. This linear behaviour is consistent with a distribution of typical fluctuations displaying an exponential tail. As discussed in the main text, the existence of an exponential tail was proven rigorously for the pure $p$-spin model \cite{subag2017extremal}.

\subsubsection{Scaling regime: $\Gamma \to 0$}

In the regime $\Gamma\to 0$, the differences between the characteristic energies scale (see bottom left panel in Fig. \ref{fig:energies_1RSB}) as
\begin{align}
   e_{\rm typ}-e_c&=-\frac{\Gamma}{2\sqrt{\rho_{\rm typ}g'(1)}}+K_{\rm typ}\Gamma^2+O(\Gamma)^3\;,\\
e_{t}-e_{\rm typ}&=u_t\Gamma^2+O(\Gamma)^3\;,\;\;u_t=-\frac{1-\rho_{\rm typ}}{2 \sqrt{\rho_{\rm typ} g'(1)} \left(\rho_{\rm typ} g'(1)-2g_2\right)}\leq 0\;.\label{u_t}
\end{align}
In the limit $\Gamma\to 0$, $e\to e_{\rm typ}$, there is a scaling function
\begin{align}
  {\cal L}(e)&\approx\Gamma^2 L_{\rm 1RSB}\left(\frac{e-e_{\rm typ}}{\Gamma^2}\right)\\
 L_{\rm 1RSB}\left(u\right)&=\begin{cases}
\displaystyle \frac{(\rho_{\rm typ}-1)^2}{4 \rho_{\rm typ} g'(1) \left(\rho_{\rm typ} g'(1)-2g_2\right)}-\frac{1}{\sqrt{g'(1)}}\left(\frac{1}{\sqrt{\rho_{\rm typ}}}-\sqrt{\rho_{\rm typ}}\right)u&\;,\;\;u\leq u_t<0\;,\\
&\\
\displaystyle (\rho_{\rm typ}g'(1)-2g_2)u^2&\;,\;\;u>u_t\;.
\end{cases}  \label{scal_1RSB_typ}
\end{align}
The first line in this expression is obtained from the RS solution and corresponds to energy $e\leq e_{t}$ while the second line is obtained from ${\cal L}_{\rm 1RSB}(e)$ and describes energies $e\geq e_t$. This scaling function displays a second order transition for $u=K_t-K_{\rm typ}$ in the sense that it is continuous and its first derivative as well at this point. Its asymptotic behaviours as $u\to -\infty$ and $u\to 0$ allow to respectively match the linear behaviour of the LDF ${\cal L}(e)\approx -s_t (e-e_{\rm typ}(\Gamma=0))$ for $\Gamma=0$ and the quadratic behaviour ${\cal L}(e)\approx (e-e_{\rm typ}(\Gamma))^2/(2{\cal V}_{\min}(\Gamma))$ for finite but small $\Gamma$.

Close to the wall, i.e. for $e\to e_c$, the LDF takes instead a different scaling form 
\begin{align}
 {\cal L}(e)&\approx G_{\rm 1RSB}\left(\frac{\sqrt{\rho_{\rm typ}g'(1)}(e-e_c)}{\Gamma};\frac{g_2}{\rho_{\rm typ}g'(1)}\right)\;,\\
 G_{\rm 1RSB}(z;w)&=-\frac{1}{2}\left((1+2z)\left(1+w(1+2z)\right)+\ln(-2z)\right)\;,\;\;z\in[ z_{\rm typ},0]\;, 
\end{align}
with $z_{\rm typ}=z_t=-1/2$. For $z<-1/2$ one has $G_{\rm 1RSB}(z;w)=0$.
Note that in the special case $w=1/2$, this scaling function is the same as for the FRSB case. This scaling function behaves asymptotically as
\be
G_{\rm 1RSB}(z;w)=\begin{cases}
\displaystyle -\frac{1}{2}\left(\ln(-2z)+1+w\right)+O(z)&\;,\;\;z\to 0\;,\\
&\\
\displaystyle \left(1-2w\right)\left(z-z_{\rm typ}\right)^2+\frac{4}{3}\left(z-z_{\rm typ}\right)^3+O(z-z_{\rm typ})^4&\;,\;\;z\to z_{\rm typ}\;.
\end{cases}
\ee
In particular, the first line matches the logarithmic divergence of the LDF in Eq. \eqref{log_div_1RSB} as $e\to e_c$ for $\Gamma>0$. The second line instead, matches the behaviour of the scaling function $L_{\rm 1RSB}(u)$ for $u\to +\infty$.
\\

{\bf Remark on stability of the 1RSB solution:} To ensure the stability of the 1RSB phase, an analysis of the quadratic form (see Appendix \ref{app:stab} for details) reveals that both the following eigenvalues must be negative
\begin{align}
   \Lambda_0^{\rm 1RSB}&=f''(q_0^*)-\frac{1}{(v_*+m_0^*(1-q_0^*))^2}\;,\\
   \Lambda_1^{\rm 1RSB}&=f''(1)-\frac{1}{v_*^2}\;.
\end{align}
In the limit $q_0^*\to 1$, both these eigenvalues coincide and match the expression of $\Lambda^{\rm RS}$ in Eq. \eqref{Lamb_RS}. Ensuring negativity of the above eigenvalues yields the conditions 
\be
\frac{g'(1)-g'(q_0^*)}{(1-q_0^*)g''(1)}\geq \frac{\rho(q_0^*)}{\alpha(q_0^*)}\geq \frac{(1-q_0^*)g''(q_0^*)}{g'(1)-g'(q_0^*)}\;,
\ee
which can be shown to hold numerically on several important examples (e.g. the $p$-spin model).

\section{Stability of the RS and 1RSB phases}\label{app:stab}

To determine the stability of the different phases, one needs to consider the eigenvalues of the quadratic variation $A$ around the saddle-point solution $Q$
\begin{align}
  A_{(ab)(cd)}=&\frac{\delta \Phi_n(q_0^*)}{\delta Q_{ab}\delta Q_{cd}}=\beta^2 f''(Q_{ab})\delta_{(ab)(cd)}-\left(Q^{-1}\right)_{ad}\left(Q^{-1}\right)_{bc}-\left(Q^{-1}\right)_{ac}\left(Q^{-1}\right)_{bd}\;. 
\end{align}
To ensure stability, this qudratic form must have negative eigenvalues. 
\subsection{RS phase}

In the RS phase, 
\be
\left(Q^{-1}\right)_{ab}=\delta_{ab}(p_d-p_0)+p_0=\frac{1}{1-q}\left(\delta_{ab}-\frac{q}{1+(n-1)q}\right)\;.
\ee
The equations for the eigenvalues of the quadratic form read
\begin{align}
    \Lambda_* q_{ab}=&\beta^2 f''(q)q_{ab}-p_0^2\sum_{c\neq d}q_{cd}-p_0(p_d-p_0)\sum_{c}\left(q_{ a c}(1-\delta_{ac})+q_{ bc}(1-\delta_{bc})\right)-(p_d-p_0)^2 q_{ab}\;,\;\;a\neq b\;.\label{stab_RS}
\end{align}
Let us first sum this equation over $a$ and $b\neq a$ to obtain the first eigenvalue, denoted $\Lambda_1^{\rm RS}$,
\begin{align}
    \Lambda_1^{\rm RS} \sum_{a\neq b}q_{ab}=&\left[\beta^2 f''(q)-n(n-1)p_0^2-2p_0(p_d-p_0)(n-1)-(p_d-p_0)^2\right] \sum_{a\neq b}q_{ab}\;.
\end{align}
The degeneracy for the eigenvalue is given by the dimension of space for the corresponding eigenvector and here $d_1^{\rm RS}=n-1$. Let us now suppose that $\sum_{a\neq b}q_{ab}=0$ and sum Eq. \eqref{stab_RS} over $a$, thus obtaining the second eigenvalue $\Lambda_2^{\rm RS}$
\begin{align}
    \Lambda_2^{\rm RS} \sum_{b} q_{ab}(1-\delta_{ab})=&\left[\beta^2 f''(q)-(n-2)p_0(p_d-p_0)-(p_d-p_0)^2\right]\sum_{b} q_{ab}(1-\delta_{ab})\;.
\end{align}
The degenaracy for this eigenvalue is $d_2^{\rm RS}=n-1$ as there are $n$  parameters $\sum_{b} q_{ab}(1-\delta_{ab})$ for $a=1,\cdots,n$ and one constraint $\sum_{a\neq b}q_{ab}=0$. Finally, supposing that $\sum_{b} q_{ab}(1-\delta_{ab})=0$ for $a=1,\cdots,n$, we obtain from Eq. \eqref{stab_RS} the third and last eigenvalue $\Lambda_3^{\rm RS}$
\begin{align}
    \Lambda_3^{\rm RS} q_{ab}=&\left[\beta^2 f''(q)-(p_d-p_0)^2\right] q_{ab}\;.
\end{align}
For this eigenvalue, there are $n(n-1)/2$ independent parameters $q_{ab}=q_{ba}$ for $a,b=1,\cdots,n$ with $q_{aa}=1$ and $n$ constraints $\sum_{b} q_{ab}(1-\delta_{ab})=0$ for $a=1,\cdots,n$, the degeneracy is thus $d_3^{\rm RS}=n(n-3)/2$. In the regime $n\to 0$ and $\beta\to \infty$ with $s=n\beta=O(1)$ and $v=\beta(1-q)$, the eigenvalues for the RS solution read
\begin{align}
    \frac{\Lambda_1^{\rm RS}}{\beta^3}&\to \lambda_1^{\rm RS}=-\frac{2v+s}{v^2(v+s)^2}<0\;,\\
    \frac{\Lambda_2^{\rm RS}}{\beta^3}&\to \lambda_2^{\rm RS}=\frac{2}{v^2(v+s)}<0\;,\\
    \frac{\Lambda_3^{\rm RS}}{\beta^2}&\to \lambda_3^{\rm RS}=f''(1)-\frac{1}{v^2}\;.
\end{align}
Note the different scaling for the last eigenvalue. In particular, the last eigenvalue $\lambda_3^{\rm RS}$ becomes positive for 
\be
v=\frac{1}{2}\left(-s+\sqrt{\frac{4}{f'(1)}+s^2}\right)>\frac{1}{\sqrt{f''(1)}}\;.\label{RSB_crit}
\ee
This criterion can be rewritten in terms of $s$ as
\be
s<s_{\rm RSB}=\frac{1}{\sqrt{f''(1)}}\left(\frac{f''(1)}{f'(1)}-1\right)\;.
\ee

\subsection{1RSB phase}

In the 1RSB phase, 
\begin{align}
  \left(Q^{-1}\right)_{ab}&=(r_d-r_1)\delta_{ab}+(r_1-r_0)P_{ab}+r_0\;,\\
  r_d&=\frac{n-m}{m n (1-q_1+m ({q_1}-{q_0}))}+\frac{1}{n (1-q_1+m ({q_1}-{q_0})+n {q_0})}-\frac{1-m}{m (1-{q_1})}\;,\\
  r_1&=\frac{n-m}{m n (1-q_1+m ({q_1}-{q_0}))}+\frac{1}{n (1-q_1+m ({q_1}-{q_0})+n {q_0})}-\frac{1}{m (1-{q_1})}\\
   r_0&=-\frac{{q_0}}{(m ({q_1}-{q_0})+1-{q_1}) (1-{q_1}+m ({q_1}-{q_0})+n
   {q_0})}\;,
\end{align}
where the matrix $P_{ab}$ is $1$ if $a,b$ belong to the same block and zero otherwise. Using a similar approach as in \cite{fyodorov2007classical}, the equation determining the eigenvalues of the quadratic form reads
\begin{align}
    \Lambda_* q_{ab}=&\beta^2 f''(Q_{ab})q_{ab}-r_0^2\sum_{c\neq d}q_{cd}-(r_1-r_0)^2 m^2 q_{\bar a \bar b}-(r_d-r_1)^2 q_{ab}-m (r_d-r_1)(r_1-r_0)(q_{\bar a  b}+q_{a\bar b})\\
    &-m r_0(r_1-r_0)\sum_{c}(q_{\bar a c}+q_{c \bar b})-r_0(r_d-r_1)\sum_{c}\left( q_{ a c}(1-\delta_{ac})+q_{c  b}(1-\delta_{bc})\right)\;,\;\;a\neq b\;,\\
    q_{a \bar b}&=\frac{1}{m}(q P)_{ab}\;,\;\;q_{\bar a \bar b}=\frac{1}{m^2}(P q P)_{ab}\;,\;\;q_{\bar a \bar a}=\frac{1}{m}(P q P)_{aa}\;.
\end{align}
Using the identities
\begin{align}
&\sum_{b}q_{a \bar b}(1-\delta_{ab})=\frac{1}{m}\sum_{c}q_{ac}(1-\delta_{ac})\sum_{b}P_{bc}(1-\delta_{ab})=\sum_{c}q_{ac}(1-\delta_{ac})-q_{\bar a a}\;,\\
&\sum_{b}q_{\bar a b}(1-\delta_{ab})=\sum_{b}q_{\bar a b}-q_{\bar a a}\;,\\
&\sum_{b}q_{\bar a \bar b}(1-\delta_{ab})=\frac{1}{m^2}\sum_{c\neq d}q_{cd}P_{ad}\sum_{b}P_{bc}(1-\delta_{ab})=\frac{1}{m}\sum_{c\neq d}q_{cd}P_{ad}-\frac{1}{m^2}\sum_{c\neq d}q_{cd}P_{ad}P_{ac}=\sum_{b}q_{\bar a b}-\frac{q_{\bar a \bar a}}{m}\;,\\
&\sum_{b}f''(Q_{ab})q_{ab}(1-\delta_{ab})=f''(q_0)\sum_{b}q_{ab}(1-\delta_{ab})+\left(f''(q_0)-f''(q_1)\right)q_{\bar a a}\;,
\end{align}
let us compute
\begin{align}
    \Lambda_* \sum_{b}q_{ab}(1-\delta_{ab})=&\beta^2 f''(q_0)\sum_{b}q_{ab}(1-\delta_{ab})+\beta^2\left[f''(q_1)-f''(q_0)\right]q_{\bar a a}-r_0^2(n-1)\sum_{c\neq d}q_{cd}\\
    &-(r_1-r_0)^2 m^2\left[\sum_{b}q_{\bar a b}-\frac{q_{\bar a \bar a}}{m}\right]-(r_d-r_1)^2 \sum_{b}q_{ab}(1-\delta_{ab})\nn\\
    &-m (r_d-r_1)(r_1-r_0)\left[\sum_{b}q_{\bar a b}+\sum_{b}q_{ab}(1-\delta_{ab})-2q_{\bar a a}\right]\nn\\
    &-m r_0(r_1-r_0)\left[(n-2)\sum_{b}q_{\bar a b}+\sum_{b\neq c}q_{bc}\right]-r_0(r_d-r_1)\left[(n-2)\sum_{b}q_{ a b}(1-\delta_{ab})+\sum_{b\neq c} q_{bc}\right]\;,\nn\\
    \Lambda_* q_{\bar aa}=&\beta^2 f''(q_1)q_{\bar aa}-\frac{m-1}{m}r_0^2\sum_{c\neq d}q_{cd}-(r_1-r_0)^2(m-1) q_{\bar a \bar a}\nn\\
    &-(r_d-r_1)(r_1-r_0)\left((m-2)q_{\bar a  a}+q_{\bar a \bar a}\right)-(r_d-r_1)^2 q_{\bar aa}\nn\\
    &-2(m-1)r_0(r_1-r_0)\sum_{c}q_{\bar a c}-r_0(r_d-r_1)\sum_{c}\left(q_{\bar a c}+\frac{m-2}{m}q_{ac }(1-\delta_{ac})\right)\;.
\end{align}
Next, using the identity
\begin{align}
&\sum_{a}\sum_{b}q_{\bar a b}=\frac{1}{m}\sum_{b\neq c}q_{bc}\sum_{a}P_{ac}=\sum_{a\neq b}q_{ac}\;,\\
&\sum_{a} q_{\bar a \bar a}=\frac{1}{m}\sum_{a}\sum_{b\neq c}q_{bc}P_{ab}P_{ac}=\frac{1}{m}\sum_{b\neq c}q_{bc}P_{bc}\sum_{a} P_{ab}=\sum_{b\neq c}q_{bc}P_{bc}=m\sum_{a}q_{\bar a a}\;,    
\end{align}
we can obtain the first two eigenvalues $\Lambda_1^{\rm 1RSB},\Lambda_2^{\rm 1RSB}$, satisfying the system of equations
\begin{align}
    \Lambda_* \sum_{a\neq b}q_{ab}=&\left[\beta^2 f''(q_0)-\left(m(r_1-r_0)+r_d-r_1\right)^2-2(n-1)r_0(m(r_1-r_0)+r_d-r_1)\right]\sum_{a\neq b} q_{ab}\\
    &+\left[\beta^2(f''(q_1)-f''(q_0))+m(r_1-r_0)^2+2(r_1-r_0)(r_d-r_1)\right]\sum_{a}q_{\bar a\bar a}\;,\nn\\
    \Lambda_* \sum_{a}q_{\bar a\bar a}=&\left[\beta^2 f''(q_1)-(m-1)(m(r_1-r_0)^2+2(r_d-r_1)(r_1-r_0))-(r_d-r_1)^2\right]\sum_{a} q_{\bar a\bar a}\\
    &-(m-1)\left[n r_0^2+2r_0(m(r_1-r_0)+r_d-r_1)\right]\sum_{a\neq b}q_{ab}\;.\nn
\end{align}
The eigenvalues are obtained by ensuring a determinant equal to zero. Next, we suppose that $\sum_{a\neq b}q_{ab}=0$ and $\sum_{a}q_{\bar a\bar a}=0$ and compute a new pair of eigenvalues $\Lambda_2^{\rm 1RSB},\Lambda_3^{\rm 1RSB}$
\begin{align}
    \Lambda_* \sum_{b}q_{\bar ab}=&\left[\beta^2 f''(q_0)-\left(m(r_1-r_0)+r_d-r_1\right)^2-(n-2)r_0(m(r_1-r_0)+r_d-r_1)\right]\sum_{b}q_{\bar ab}\\
    &+\left[\beta^2(f''(q_1)-f''(q_0))+m(r_1-r_0)^2+2(r_1-r_0)(r_d-r_1)\right]q_{\bar a\bar a}\;,\nn\\
    \Lambda_* q_{\bar a\bar a}=&\left[\beta^2 f''(q_1)-(m-1)(m(r_1-r_0)^2+2(r_d-r_1)(r_1-r_0))-(r_d-r_1)^2\right] q_{\bar a\bar a}\\
    &-(m-1)\left[2r_0(m(r_1-r_0)+r_d-r_1)\right]\sum_{b}q_{\bar ab}\;.\nn
\end{align}
Supposing then that $\sum_{b}q_{\bar ab}=0$ and $q_{\bar a \bar a}=0$, one can obtain a simple equation for yet another eigenvalue
\be
\Lambda_{5}^{\rm 1RSB} q_{\bar a \bar b}=\left[\beta^2f''(q_0)-\left(m(r_1-r_0)+r_d-r_1\right)^2\right]q_{\bar a \bar b}\;,\;\;P_{ ab}=0\;.
\ee
Supposing next that $q_{\bar a \bar b}=0$ for $a,b=1,\cdots,n$ (ensuring thus that $\sum_{b} q_{\bar a b}=0$), one obtains a system for the eigenvalues $\Lambda_6^{\rm 1RSB},\Lambda_7^{\rm 1RSB}$
\begin{align}
    \Lambda_* \sum_{b}q_{ab}(1-\delta_{ab})=&\left[\beta^2 f''(q_0)-(r_d-r_1)^2-m (r_d-r_1)(r_1-r_0)-r_0(n-2)(r_d-r_1)\right]\sum_{b}q_{ab}(1-\delta_{ab})\nn\\
    &+m\left[\beta^2( f''(q_1)-f''(q_0))+2(r_d-r_1)(r_1-r_0)\right]q_{\bar a a}\;,\\
    \Lambda_* q_{\bar a a}=&\left[\beta^2 f''(q_1)-(m-2)(r_d-r_1)(r_1-r_0)-(r_d-r_1)^2\right] q_{\bar a a}\\
    &-\frac{m-2}{m}r_0(r_d-r_1)\sum_{b}q_{ab}(1-\delta_{ab})\;.\nn
\end{align}
To obtain the next eigenvalue, we suppose that $q_{\bar a a}=0$ for $a=1,\cdots,n$, $q_{\bar a \bar b}=0$  for $a,b=1,\cdots,n$ and $\sum_{b} q_{ab}=0$  for $a=1,\cdots,n$, which yields
\begin{align}
\Lambda_8^{\rm 1RSB} q_{\bar a b}&=\left[\beta^2f''(q_0)-\left(r_d-r_1\right)\left(m(r_1-r_0)+r_d-r_1\right)\right]q_{\bar a b}\;,\;\;P_{ ab}=0\;.
\end{align}
Finally supposing that $q_{\bar a b}=0$ for $a,b=1,\cdots,n$ (ensuring thus $\sum_{a}q_{ab}=0$) one obtains the two final eigenvalues $\Lambda_9^{\rm 1RSB},\Lambda_{10}^{\rm 1RSB}$ that correspond respectively to $P_{ab}=1$ and $P_{ab}=0$
\begin{align}
\Lambda_9^{\rm 1RSB} q_{a b}&=\left[\beta^2f''(q_1)-\left(r_d-r_1\right)^2\right]q_{a b}\;,\;\;P_{ab}=1\;,\\
\Lambda_{10}^{\rm 1RSB} q_{a b}&=\left[\beta^2f''(q_0)-\left(r_d-r_1\right)^2\right]q_{a b}\;,\;\;P_{ab}=0\;.
\end{align}


\section{Annealed complexity of minima at fixed energy}\label{app:comp_min}

As shown in the main text, the opposite of the annealed complexity of minima at fixed energy $e$ provides a lower bound for the large-deviation function ${\cal L}_{\rm RS}(e)$. Let us compute this quantity explicitly, by first computing the average density of minima at energy $e$, given by
\begin{align}
    \overline{n_{\min}(e)}=\int_{\sqrt{N}{\cal S}_{N-1}} d{\bf x}\; \overline{\delta(\nabla H({\bf x}))|\det \nabla^2 H({\bf x})|\Theta[\nabla^2 H({\bf x})]\delta(e-H({\bf x})/N)}\;,
\end{align}
where $\Theta[A]=1$ if all the eigenvalues of $A$ are positive and zero otherwise. The Hamiltonian, its gradient and the Hessian have Gaussian zero average statistics which satisfy
\begin{align}
    \overline{H({\bf x})^2}&=N f(1)\;,\\
    \overline{(\nabla H)_i({\bf x}) H({\bf x})}&=0\;,\\
    \overline{(\nabla H)_i({\bf x}) (\nabla H)_j({\bf x})}=-\overline{(\nabla^2 H)_{ij}({\bf x})H ({\bf x})}&=f'(1)\delta_{ij}\;,\\
    \overline{(\nabla^2 H)_{ij}({\bf x})(\nabla H)_k ({\bf x})}&=0\;,\\
    \overline{(\nabla^2 H)_{ij}({\bf x})(\nabla^2 H)_{kl} ({\bf x})}&=\frac{1}{N}\left[(\delta_{ik}\delta_{jl}+\delta_{il}\delta_{jk})f''(1)+\delta_{ij}\delta_{kl}(f''(1)+f'(1))\right]\;.
\end{align}
The Hessian matrix conditioned on zero gradient and a fixed value of the energy has a non-zero average given by
\be
\mathbb{E}\left[(\nabla^2 H)_{ij}({\bf x})|\nabla H({\bf x})={\bf 0},H({\bf x})=Ne\right]=\frac{\overline{(\nabla^2 H)_{ij}({\bf x})H ({\bf x})}}{\overline{H({\bf x})^2}}N e=-\frac{f'(1)}{f(1)}e\;,
\ee
and a modified covariance
\begin{align}
&{\rm Cov}\left[(\nabla^2 H)_{ij}({\bf x}),(\nabla^2 H)_{kl} ({\bf x})|\nabla H({\bf x})={\bf 0},H({\bf x})=Ne\right]\\
&=\overline{(\nabla^2 H)_{ij}({\bf x})(\nabla^2 H)_{kl} ({\bf x})}-\frac{\overline{(\nabla^2 H)_{ij}({\bf x})H ({\bf x})}\times\overline{(\nabla^2 H)_{kl}({\bf x})H ({\bf x})}}{\overline{H({\bf x})^2}}\nn\\
&=\frac{1}{N}\left[(\delta_{ik}\delta_{jl}+\delta_{il}\delta_{jk})f''(1)+\delta_{ij}\delta_{kl}\left(f''(1)+f'(1)-\frac{f'(1)^2}{f(1)}\right)\right]\;.\nn
\end{align}
Using these expressions, one can express
\begin{align}
    \overline{n_{\min}(e)}&=\frac{N^{(N-1)/2}S_{N-1}}{(2\pi f'(1))^{(N-1)/2}}\int_{-\infty}^{\infty} \frac{dx}{\sqrt{2\pi}}e^{-\frac{N}{2}\left(x^2+\frac{e^2}{f(1)}\right)}\mathbb{E}\left[|\det {\cal H}|\Theta[{\cal H}]\right]_{X\in {\rm GOE}}\;,\\
    {\cal H}&=\sqrt{f''(1)}X+\left(\sqrt{f''(1)+f'(1)-\frac{f'(1)^2}{f(1)}}\,x-\frac{f'(1)}{f(1)}e\right)\mathbb{I}\;.
\end{align}
At this stage, it is important to notice that
\be
f(1)(f''(1)+f'(1))-f'(1)^2=\frac{1}{2}\sum_{r,p=2}^{\infty}g_r g_p\,(r-p)^2+\Gamma\sum_{r=2}^{\infty}(r-1)^2\;,\label{exp_1}
\ee
is strictly positive apart in the case of the pure $p$-spin where this quantity is zero. 

Let us first discuss the pure $p$-spin model.
Using that $f'(1)/f(1)=p$, the annealed complexity at fixed energy is readily obtained as
\begin{align}
 \Sigma_{p,\min}(e)&=\lim_{N\to \infty}\frac{1}{N}\ln \overline{n_{p,\min}(e)}\\
 &=-\frac{p e^2}{2}+\frac{1}{2}+\frac{1}{2}\ln(p-1)+\int_{-2}^2 \frac{d\lambda}{2\pi}\,\sqrt{4-\lambda^2}\,\ln\left|\lambda-\frac{p}{\sqrt{p-1}} e\right|\;.\nn 
\end{align}
Using the identity
\be
\int_{-2}^2 \frac{d\lambda}{2\pi}\,\sqrt{4-\lambda^2}\,\ln\left|\lambda+u\right|=-\frac{1}{2}+\frac{u^2}{4} -\Theta(u-2)\left(\frac{u}{4}\sqrt{u^2-4}-\ln\left(\frac{u+\sqrt{u^2-4}}{2}\right)\right)\;,\label{id_LD_GOE}
\ee
it yields \cite{crisanti1995TAP,AufBenArusCerny2013}
\begin{align}
    \Sigma_{p,\min}(e)&=\lim_{N\to \infty}\frac{1}{N}\ln \overline{n_{p,\min}(e)}\\
    &=\frac{1}{2}\ln(p-1)-\frac{p-2}{p}\frac{e^2}{e_{c,\rm RS}^2}+\Theta\left(e_{c,\rm RS}-e\right)\left[\frac{e\sqrt{e^2-e_{c,\rm RS}^2}}{e_{c,\rm RS}^2}+\ln\left(\frac{e-\sqrt{e^2-e_{c,\rm RS}^2}}{e_{c,\rm RS}}\right)\right]\;,\\
    e_{c,\rm RS}&=e_{\rm RSB}=-\frac{2}{p}\sqrt{p-1}\;.
\end{align}
The annealed complexity is negative for any energy $e\leq e_{\rm typ}$, and positive in the complementary range $e> e_{\rm typ}$. 
For this model we find that the annealed complexity satisfies 
\be
\Sigma_{p,\min}(e)=-{\cal L}_{p,{\rm RS}}(e)\;,\;\;e\leq e_{\rm typ}\;.
\ee

Let us now consider more general spherical models for which Eq. \eqref{exp_1} is positive. The associated annealed complexity is obtained as
\begin{align}
 \Sigma_{\min}(e)&=\lim_{N\to \infty}\frac{1}{N}\ln \overline{n_{\min}(e)}\\
 &=\max_{u>2}\left[-\frac{(e f'(1)+u f(1) \sqrt{f''(1)})^2}{2f(1)^2\left(f''(1)+f'(1)-\frac{f'(1)^2}{f(1)}\right)}-\frac{e^2}{2f(1)}+\frac{1}{2}-\frac{1}{2}\ln \frac{f'(1)}{f''(1)}+\int_{-2}^2 \frac{d\lambda}{2\pi}\,\sqrt{4-\lambda^2}\,\ln\left|\lambda+u\right|\right]\;.\nn 
\end{align}
One can then use the identity \eqref{id_LD_GOE} in order to express explicitly the last term of this equation. Recasting the optimisation problem in terms of a new variable $v\in[0,1/\sqrt{f''(1)}]$ such that
\be
u=\frac{1}{v\sqrt{f''(1)}}+v\sqrt{f''(1)}\;,
\ee
one obtains
\begin{align}
 \Sigma_{\min}(e)&=\lim_{N\to \infty}\frac{1}{N}\ln \overline{n_{\min}(e)}\label{comp}\\
 &=\max_{v\in[0,1/\sqrt{f''(1)}]}\left[-\frac{\left(e v f'(1)+f(1) v^2 f''(1)+f(1)\right)^2}{2v^2 f(1)\left(f(1)
   \left(f''(1)+f'(1)\right)-f'(1)^2\right)}-\frac{e^2}{2f(1)}+\frac{1}{2}\left(1+v^2 f''(1)\right)-\ln\left(v\sqrt{f'(1)}\right)\right]\;.\nn 
\end{align}
The saddle-point equation yields
\be
v_*=-\frac{e+\sqrt{e^2-e_{c,{\rm RS}}^2}}{2(f'(1)-f(1))}\;,\;\;e_{c,{\rm RS}}=-2\sqrt{f(1)\left(1-\frac{f(1)}{f'(1)}\right)}\;,
\ee
which lies within the interval $[0,1/\sqrt{f''(1)}]$ for 
\be
e\leq e_{\rm RSB}=-\frac{1}{\sqrt{f''(1)}}\left[f'(1)+f(1)\left(\frac{f''(1)}{f'(1)}-1\right)\right]\;.
\ee
At the saddle-point, the second derivative of the expression to maximise reads
\be
-\frac{(1-v_*^2 f''(1))}{v_*^2}\left(1+\frac{f(1)(1-v_*^2 f''(1))}{v_*^2(f(1)(f'(1)+f''(1))-f'(1)^2)}\right)\leq 0\;,
\ee
thus showing that $v_*$ is indeed the maximum over the interval $[0,1/\sqrt{f''(1)}]$.

Inserting this value in Eq. \eqref{comp}, one can check that for any energy $e\leq e_t$, corresponding to the RS regime,
\begin{align}
\Sigma_{\min}(e)=&-{\cal L}(e)\\
=&\frac{2 f(1)}{f'(1)}\frac{e^2}{e_{c,{\rm RS}}^2}-\frac{e}{e_{c,{\rm RS}}^2}\left(e-\sqrt{e^2-e_{c,{\rm RS}}^2}\right)+\ln\left(\frac{e-\sqrt{e^2-e_{c,{\rm RS}}^2}}{e_{c,{\rm RS}}}\right)-\frac{1}{2}\ln\frac{f(1)}{f'(1)-f(1)}\;.    \label{sig_e_small}
\end{align}
While the identification with ${\cal L}(e)$ is not exact for $e\geq e_t$, the expression of $\Sigma_{\min}(e)$ remains the same for any $e\leq e_{\rm RSB}$.
This identity is not expected to hold within the replica-symmetry broken phase as one naturally expects that $\Sigma_{\min}(e)>0$ in this regime. 

For energies $e>e_{\rm RSB}$, the parameter $v$ is pinned at the value $1/\sqrt{f''(1)}$ and the annealed complexity reads
\be
\Sigma_{\min}(e)=-\frac{4f(1)f''(1)+4 e f'(1)\sqrt{f''(1)}+e^2(f'(1)+f''(1))}{2(f(1)(f'(1)+f''(1))-f'(1)^2)}+1-\frac{1}{2}\ln\frac{f'(1)}{f''(1)}\;,\;\;e\geq e_{\rm RSB}\;.\label{sig_e_large}
\ee
Note that all the expressions for the annealed complexity and the bounds are universal in the sense that they only depend on the covariance function $f(q)$ through its value and the values of its first two derivative for $q=1$. The annealed complexity $\Sigma_{\min}(e)$ is generically positive in a finite band of energy $e\in[{\cal E}_{\min},{\cal E}_{\max}]$ (possibly reduced to a single point in the case ${\cal E}_{\min}={\cal E}_{\max}$). The expression of the annealed complexity in Eq. \eqref{sig_e_large} has two zeros for $\Gamma<\Gamma_{\rm RSB}$,
\be
e_{0,\pm}=-\frac{2f'(1)\sqrt{f''(1)}}{f'(1)+f''(1)}\pm\sqrt{\left(\frac{2(f'(1)-f''(1))}{f'(1)+f''(1)}+\ln\frac{f''(1)}{f'(1)}\right)\left(f(1)-\frac{f'(1)^2}{f'(1)+f''(1)}\right)}\;.
\ee
One can show in particular that in this regime $e_{0,+}\geq e_{\rm RSB}$, thus showing that ${\cal E}_{\max}=e_{0,+}$. The parameters $e_{0,-}$ and $e_{\rm RSB}$ cross for a non-trivial value of the magnetic field $\Gamma=\Gamma_{\rm cr}$ obtained from solving the equation
\be
x(y)=y\frac{y-1-\ln y}{(1-y^2)}\;,\;\;0\leq x=\frac{f(1)}{f'(1)}=\frac{\Gamma_{\rm cr}+g(1)}{\Gamma_{\rm cr}+g'(1)}\leq 1\;,\;\;0\leq y=\frac{f'(1)}{f''(1)}=\frac{\Gamma_{\rm cr}+g'(1)}{g''(1)}<1\;.\label{eq_G_crossing}
\ee
Note that this equation is exactly the same as the equation for $\mu_{\rm typ}$ as a function $\rho_{\rm typ}$, allowing to obtain the value of $e_{\rm typ}$ in the case of zero magnetic field. For $\Gamma=\Gamma_{\rm RSB}$, one has $f'(1)=f''(1)$ and the two zeros $e_{0,\pm}$ coalesce and are equal to $e_{0,\pm}=e_{\rm typ}=-\sqrt{f'(1)}$, and for larger magnetic field $\Gamma>\Gamma_{\rm RSB}$ the expression Eq. \eqref{sig_e_large} has no zero. The expression for the annealed complexity in Eq. \eqref{sig_e_small} has instead a single zero $e_{0,\rm RS}$, which takes a non-trivial value for $0<\Gamma<g'(1)-2g(1)$, i.e. $2f(1)<f'(1)$, while in the converse case $e_{0,\rm RS}=-\sqrt{f'(1)}$. Solving for $e_{0,\rm RS}=e_{\rm RSB}$, one recovers exactly the same equation as in \eqref{eq_G_crossing} and thus the three different values $e_{0,\rm RS}$, $e_{0,-}$ and $e_{\rm RSB}$ cross for $\Gamma=\Gamma_{\rm cr}$. To summarise, we have thus obtained that 
\be
\Sigma_{\min}(e)\geq 0\;,\;\;e\in[{\cal E}_{\min},{\cal E}_{\rm max}]\;,\;\;{\cal E}_{\min}=\begin{cases}
e_{0,{\rm RS}}\;,\;\;0\leq \Gamma\leq \Gamma_{\rm cr}\\
e_{0,-}\;,\;\;\Gamma_{\rm cr}\leq \Gamma\leq \Gamma_{\rm RSB}\\
e_{\rm typ}\;,\;\;\Gamma\geq \Gamma_{\rm RSB}
\end{cases}\;,\;\;{\cal E}_{\max}=\begin{cases}
e_{0,+}\;,\;\;0\leq \Gamma\leq \Gamma_{\rm RSB}\\
e_{\rm typ}\;,\;\;\Gamma\geq \Gamma_{\rm RSB}
\end{cases}\;.
\ee
One naturally expects that for any value of $\Gamma$, the annealed complexity is positive at the typical value of the energy $\Sigma_{\rm min}(e_{\rm typ})\geq 0$ such that $e_{\rm typ}\in[{\cal E}_{\min},{\cal E}_{\rm max}]$. Note that for models for which $e_{\rm typ}$ belongs to the 1RSB phase, $e_{0,{\rm RS}}=e_{\rm typ}$ for $\Gamma=0$.

\bibliography{refs}

\begin{thebibliography}{64}%
\makeatletter
\providecommand \@ifxundefined [1]{%
 \@ifx{#1\undefined}
}%
\providecommand \@ifnum [1]{%
 \ifnum #1\expandafter \@firstoftwo
 \else \expandafter \@secondoftwo
 \fi
}%
\providecommand \@ifx [1]{%
 \ifx #1\expandafter \@firstoftwo
 \else \expandafter \@secondoftwo
 \fi
}%
\providecommand \natexlab [1]{#1}%
\providecommand \enquote  [1]{``#1''}%
\providecommand \bibnamefont  [1]{#1}%
\providecommand \bibfnamefont [1]{#1}%
\providecommand \citenamefont [1]{#1}%
\providecommand \href@noop [0]{\@secondoftwo}%
\providecommand \href [0]{\begingroup \@sanitize@url \@href}%
\providecommand \@href[1]{\@@startlink{#1}\@@href}%
\providecommand \@@href[1]{\endgroup#1\@@endlink}%
\providecommand \@sanitize@url [0]{\catcode `\\12\catcode `\$12\catcode
  `\&12\catcode `\#12\catcode `\^12\catcode `\_12\catcode `\%12\relax}%
\providecommand \@@startlink[1]{}%
\providecommand \@@endlink[0]{}%
\providecommand \url  [0]{\begingroup\@sanitize@url \@url }%
\providecommand \@url [1]{\endgroup\@href {#1}{\urlprefix }}%
\providecommand \urlprefix  [0]{URL }%
\providecommand \Eprint [0]{\href }%
\providecommand \doibase [0]{https://doi.org/}%
\providecommand \selectlanguage [0]{\@gobble}%
\providecommand \bibinfo  [0]{\@secondoftwo}%
\providecommand \bibfield  [0]{\@secondoftwo}%
\providecommand \translation [1]{[#1]}%
\providecommand \BibitemOpen [0]{}%
\providecommand \bibitemStop [0]{}%
\providecommand \bibitemNoStop [0]{.\EOS\space}%
\providecommand \EOS [0]{\spacefactor3000\relax}%
\providecommand \BibitemShut  [1]{\csname bibitem#1\endcsname}%
\let\auto@bib@innerbib\@empty
\bibitem [{\citenamefont {Fyodorov}\ and\ \citenamefont
  {Le~Doussal}(2014)}]{fyodorov2014topology}%
  \BibitemOpen
  \bibfield  {author} {\bibinfo {author} {\bibfnamefont {Y.~V.}\ \bibnamefont
  {Fyodorov}}\ and\ \bibinfo {author} {\bibfnamefont {P.}~\bibnamefont
  {Le~Doussal}},\ }\bibfield  {title} {\bibinfo {title} {Topology
  trivialization and large deviations for the minimum in the simplest random
  optimization},\ }\href@noop {} {\bibfield  {journal} {\bibinfo  {journal}
  {Journal of Statistical Physics}\ }\textbf {\bibinfo {volume} {154}},\
  \bibinfo {pages} {466} (\bibinfo {year} {2014})}\BibitemShut {NoStop}%
\bibitem [{\citenamefont {Dembo}\ and\ \citenamefont
  {Zeitouni}(2015)}]{dembo2015matrix}%
  \BibitemOpen
  \bibfield  {author} {\bibinfo {author} {\bibfnamefont {A.}~\bibnamefont
  {Dembo}}\ and\ \bibinfo {author} {\bibfnamefont {O.}~\bibnamefont
  {Zeitouni}},\ }\bibfield  {title} {\bibinfo {title} {Matrix optimization
  under random external fields},\ }\href@noop {} {\bibfield  {journal}
  {\bibinfo  {journal} {Journal of Statistical Physics}\ }\textbf {\bibinfo
  {volume} {159}},\ \bibinfo {pages} {1306} (\bibinfo {year}
  {2015})}\BibitemShut {NoStop}%
\bibitem [{\citenamefont {Mezard}\ and\ \citenamefont
  {Montanari}(2009)}]{MM2009book}%
  \BibitemOpen
  \bibfield  {author} {\bibinfo {author} {\bibfnamefont {M.}~\bibnamefont
  {Mezard}}\ and\ \bibinfo {author} {\bibfnamefont {A.}~\bibnamefont
  {Montanari}},\ }\href@noop {} {\emph {\bibinfo {title} {Information, physics,
  and computation}}},\ Vol.\ \bibinfo {volume} {Oxford Graduate Texts}\
  (\bibinfo  {publisher} {Oxford University Press},\ \bibinfo {year}
  {2009})\BibitemShut {NoStop}%
\bibitem [{\citenamefont {Choromanska}\ \emph {et~al.}(2015)\citenamefont
  {Choromanska}, \citenamefont {Henaff}, \citenamefont {Mathieu}, \citenamefont
  {BenArous},\ and\ \citenamefont {Y.}}]{Choromanska2015}%
  \BibitemOpen
  \bibfield  {author} {\bibinfo {author} {\bibfnamefont {A.}~\bibnamefont
  {Choromanska}}, \bibinfo {author} {\bibfnamefont {M.}~\bibnamefont {Henaff}},
  \bibinfo {author} {\bibfnamefont {M.}~\bibnamefont {Mathieu}}, \bibinfo
  {author} {\bibfnamefont {G.}~\bibnamefont {BenArous}},\ and\ \bibinfo
  {author} {\bibfnamefont {Y.~L.}\ \bibnamefont {Y.}},\ }\bibfield  {title}
  {\bibinfo {title} {The loss surfaces of multilayer networks},\ }\href@noop {}
  {\bibfield  {journal} {\bibinfo  {journal} {In: Proceedings of the 18th
  International Conference on Artificial Intelligence and Statistics
  (AISTATS)}\ ,\ \bibinfo {pages} {192–}} (\bibinfo {year}
  {2015})}\BibitemShut {NoStop}%
\bibitem [{\citenamefont {Ros}\ and\ \citenamefont
  {Fyodorov}(2022)}]{RosFyoReview}%
  \BibitemOpen
  \bibfield  {author} {\bibinfo {author} {\bibfnamefont {V.}~\bibnamefont
  {Ros}}\ and\ \bibinfo {author} {\bibfnamefont {Y.~V.}\ \bibnamefont
  {Fyodorov}},\ }\bibfield  {title} {\bibinfo {title} {The high-d landscapes
  paradigm: spin-glasses, and beyond},\ }\href@noop {} {\bibfield  {journal}
  {\bibinfo  {journal} {arXiv:2209.07975}\ } (\bibinfo {year}
  {2022})}\BibitemShut {NoStop}%
\bibitem [{\citenamefont {Auffinger}\ \emph {et~al.}(2022)\citenamefont
  {Auffinger}, \citenamefont {Montanari},\ and\ \citenamefont
  {Subag}}]{AufMonSubReview}%
  \BibitemOpen
  \bibfield  {author} {\bibinfo {author} {\bibfnamefont {A.}~\bibnamefont
  {Auffinger}}, \bibinfo {author} {\bibfnamefont {A.}~\bibnamefont
  {Montanari}},\ and\ \bibinfo {author} {\bibfnamefont {E.}~\bibnamefont
  {Subag}},\ }\bibfield  {title} {\bibinfo {title} {Optimization of random
  high-dimensional functions: Structure and algorithms},\ }\href@noop {}
  {\bibfield  {journal} {\bibinfo  {journal} {arXiv:2206.10217}\ } (\bibinfo
  {year} {2022})}\BibitemShut {NoStop}%
\bibitem [{\citenamefont {Sherrington}\ and\ \citenamefont
  {Kirkpatrick}(1975)}]{SK}%
  \BibitemOpen
  \bibfield  {author} {\bibinfo {author} {\bibfnamefont {D.}~\bibnamefont
  {Sherrington}}\ and\ \bibinfo {author} {\bibfnamefont {S.}~\bibnamefont
  {Kirkpatrick}},\ }\bibfield  {title} {\bibinfo {title} {Solvable model of a
  spin-glass},\ }\href@noop {} {\bibfield  {journal} {\bibinfo  {journal}
  {Physical review letters}\ }\textbf {\bibinfo {volume} {35}},\ \bibinfo
  {pages} {1792} (\bibinfo {year} {1975})}\BibitemShut {NoStop}%
\bibitem [{\citenamefont {Montanari}(2021)}]{Montanari_SK}%
  \BibitemOpen
  \bibfield  {author} {\bibinfo {author} {\bibfnamefont {A.}~\bibnamefont
  {Montanari}},\ }\bibfield  {title} {\bibinfo {title} {Optimization of the
  sherrington--kirkpatrick hamiltonian},\ }\href@noop {} {\bibfield  {journal}
  {\bibinfo  {journal} {SIAM Journal of Computing}\ } (\bibinfo {year}
  {2021})}\BibitemShut {NoStop}%
\bibitem [{\citenamefont {Parisi}(1980)}]{ParisiFirst}%
  \BibitemOpen
  \bibfield  {author} {\bibinfo {author} {\bibfnamefont {G.}~\bibnamefont
  {Parisi}},\ }\bibfield  {title} {\bibinfo {title} {A sequence of approximated
  solutions to the sk model for spin glasses},\ }\href@noop {} {\bibfield
  {journal} {\bibinfo  {journal} {J. Phys. A: Math. Gen.}\ }\textbf {\bibinfo
  {volume} {13}},\ \bibinfo {pages} {L115} (\bibinfo {year}
  {1980})}\BibitemShut {NoStop}%
\bibitem [{\citenamefont {M\'ezard}\ \emph {et~al.}(1987)\citenamefont
  {M\'ezard}, \citenamefont {Parisi},\ and\ \citenamefont
  {Virasoro}}]{MPV1987}%
  \BibitemOpen
  \bibfield  {author} {\bibinfo {author} {\bibfnamefont {M.}~\bibnamefont
  {M\'ezard}}, \bibinfo {author} {\bibfnamefont {G.}~\bibnamefont {Parisi}},\
  and\ \bibinfo {author} {\bibfnamefont {M.~A.}\ \bibnamefont {Virasoro}},\
  }\href@noop {} {\emph {\bibinfo {title} {Spin glass theory and beyond: An
  Introduction to the Replica Method and Its Applications}}},\ Vol.\ \bibinfo
  {volume} {Lecture Notes in Physics, vol. 9}\ (\bibinfo  {publisher} {World
  Scientific Publishing Company},\ \bibinfo {year} {1987})\BibitemShut
  {NoStop}%
\bibitem [{\citenamefont {Castellani}\ and\ \citenamefont
  {Cavagna}(2005)}]{CastellaniCavagna2005pedestrian}%
  \BibitemOpen
  \bibfield  {author} {\bibinfo {author} {\bibfnamefont {T.}~\bibnamefont
  {Castellani}}\ and\ \bibinfo {author} {\bibfnamefont {A.}~\bibnamefont
  {Cavagna}},\ }\bibfield  {title} {\bibinfo {title} {Spin-glass theory for
  pedestrians},\ }\href@noop {} {\bibfield  {journal} {\bibinfo  {journal}
  {Journal of Statistical Mechanics: Theory and Experiment}\ }\textbf {\bibinfo
  {volume} {2005}},\ \bibinfo {pages} {P05012} (\bibinfo {year}
  {2005})}\BibitemShut {NoStop}%
\bibitem [{\citenamefont {Rizzo}(2009)}]{rizzo2009chaos}%
  \BibitemOpen
  \bibfield  {author} {\bibinfo {author} {\bibfnamefont {T.}~\bibnamefont
  {Rizzo}},\ }\bibfield  {title} {\bibinfo {title} {Chaos in mean-field
  spin-glass models},\ }in\ \href@noop {} {\emph {\bibinfo {booktitle} {Spin
  Glasses: Statics and Dynamics: Summer School, Paris 2007}}}\ (\bibinfo
  {organization} {Springer},\ \bibinfo {year} {2009})\ pp.\ \bibinfo {pages}
  {143--157}\BibitemShut {NoStop}%
\bibitem [{\citenamefont {Guerra}(2003)}]{guerra2003broken}%
  \BibitemOpen
  \bibfield  {author} {\bibinfo {author} {\bibfnamefont {F.}~\bibnamefont
  {Guerra}},\ }\bibfield  {title} {\bibinfo {title} {Broken replica symmetry
  bounds in the mean field spin glass model},\ }\href@noop {} {\bibfield
  {journal} {\bibinfo  {journal} {Communications in mathematical physics}\
  }\textbf {\bibinfo {volume} {233}},\ \bibinfo {pages} {1} (\bibinfo {year}
  {2003})}\BibitemShut {NoStop}%
\bibitem [{\citenamefont
  {Talagrand}(2006{\natexlab{a}})}]{talagrand2006parisi}%
  \BibitemOpen
  \bibfield  {author} {\bibinfo {author} {\bibfnamefont {M.}~\bibnamefont
  {Talagrand}},\ }\bibfield  {title} {\bibinfo {title} {The parisi formula},\
  }\href@noop {} {\bibfield  {journal} {\bibinfo  {journal} {Annals of
  mathematics}\ ,\ \bibinfo {pages} {221}} (\bibinfo {year}
  {2006}{\natexlab{a}})}\BibitemShut {NoStop}%
\bibitem [{\citenamefont {Panchenko}(2013)}]{Panchenko2013}%
  \BibitemOpen
  \bibfield  {author} {\bibinfo {author} {\bibfnamefont {D.}~\bibnamefont
  {Panchenko}},\ }\href@noop {} {\emph {\bibinfo {title} {The
  Sherrington-Kirkpatrick model}}}\ (\bibinfo  {publisher} {Springer Science \&
  Business Media, New York},\ \bibinfo {year} {2013})\BibitemShut {NoStop}%
\bibitem [{\citenamefont {Auffinger}\ \emph {et~al.}(2020)\citenamefont
  {Auffinger}, \citenamefont {Chen},\ and\ \citenamefont
  {Zeng}}]{auffinger2020sk}%
  \BibitemOpen
  \bibfield  {author} {\bibinfo {author} {\bibfnamefont {A.}~\bibnamefont
  {Auffinger}}, \bibinfo {author} {\bibfnamefont {W.-K.}\ \bibnamefont
  {Chen}},\ and\ \bibinfo {author} {\bibfnamefont {Q.}~\bibnamefont {Zeng}},\
  }\bibfield  {title} {\bibinfo {title} {The sk model is infinite step replica
  symmetry breaking at zero temperature},\ }\href@noop {} {\bibfield  {journal}
  {\bibinfo  {journal} {Communications on Pure and Applied Mathematics}\
  }\textbf {\bibinfo {volume} {73}} (\bibinfo {year} {2020})}\BibitemShut
  {NoStop}%
\bibitem [{\citenamefont {Crisanti}\ and\ \citenamefont
  {Sommers}(1992)}]{CrisantiSommers1992}%
  \BibitemOpen
  \bibfield  {author} {\bibinfo {author} {\bibfnamefont {A.}~\bibnamefont
  {Crisanti}}\ and\ \bibinfo {author} {\bibfnamefont {H.-J.}\ \bibnamefont
  {Sommers}},\ }\bibfield  {title} {\bibinfo {title} {The spherical p-spin
  interaction spin glass model: the statics},\ }\href@noop {} {\bibfield
  {journal} {\bibinfo  {journal} {Zeitschrift f{\"u}r Physik B Condensed
  Matter}\ }\textbf {\bibinfo {volume} {87}},\ \bibinfo {pages} {341} (\bibinfo
  {year} {1992})}\BibitemShut {NoStop}%
\bibitem [{\citenamefont
  {Talagrand}(2006{\natexlab{b}})}]{Talagrand_spherical}%
  \BibitemOpen
  \bibfield  {author} {\bibinfo {author} {\bibfnamefont {M.}~\bibnamefont
  {Talagrand}},\ }\bibfield  {title} {\bibinfo {title} {Free energy of the
  spherical mean field model},\ }\href@noop {} {\bibfield  {journal} {\bibinfo
  {journal} {Probability theory and related fields}\ }\textbf {\bibinfo
  {volume} {134}},\ \bibinfo {pages} {339} (\bibinfo {year}
  {2006}{\natexlab{b}})}\BibitemShut {NoStop}%
\bibitem [{\citenamefont {Antonio~Auffinger}\ and\ \citenamefont
  {Chen}(2017)}]{AufChen2017}%
  \BibitemOpen
  \bibfield  {author} {\bibinfo {author} {\bibfnamefont {A.}~\bibnamefont
  {Antonio~Auffinger}}\ and\ \bibinfo {author} {\bibfnamefont {W.-K.}\
  \bibnamefont {Chen}},\ }\bibfield  {title} {\bibinfo {title} {Parisi formula
  for the ground state energy in the mixed p-spin model},\ }\href@noop {}
  {\bibfield  {journal} {\bibinfo  {journal} {Ann. Probab.}\ }\textbf {\bibinfo
  {volume} {45}},\ \bibinfo {pages} {4617} (\bibinfo {year}
  {2017})}\BibitemShut {NoStop}%
\bibitem [{\citenamefont {Chen}\ and\ \citenamefont {Sen}(2017)}]{ChenSen2017}%
  \BibitemOpen
  \bibfield  {author} {\bibinfo {author} {\bibfnamefont {W.-K.}\ \bibnamefont
  {Chen}}\ and\ \bibinfo {author} {\bibfnamefont {A.}~\bibnamefont {Sen}},\
  }\bibfield  {title} {\bibinfo {title} {Parisi formula, disorder chaos and
  fluctuation for the ground state energy in the spherical mixed p-spin
  models},\ }\href@noop {} {\bibfield  {journal} {\bibinfo  {journal}
  {Communications in Mathematical Physics}\ }\textbf {\bibinfo {volume}
  {350}},\ \bibinfo {pages} {129} (\bibinfo {year} {2017})}\BibitemShut
  {NoStop}%
\bibitem [{\citenamefont {Jagannath}\ and\ \citenamefont
  {Tobasco}(2017)}]{JagTab2017_spherical}%
  \BibitemOpen
  \bibfield  {author} {\bibinfo {author} {\bibfnamefont {A.}~\bibnamefont
  {Jagannath}}\ and\ \bibinfo {author} {\bibfnamefont {I.}~\bibnamefont
  {Tobasco}},\ }\bibfield  {title} {\bibinfo {title} {Low temperature
  asymptotics of spherical mean field spin glasses},\ }\href@noop {} {\bibfield
   {journal} {\bibinfo  {journal} {Commun. Math. Physics}\ }\textbf {\bibinfo
  {volume} {352}},\ \bibinfo {pages} {979–} (\bibinfo {year}
  {2017})}\BibitemShut {NoStop}%
\bibitem [{\citenamefont {Subag}(2018)}]{Subag_spherical_ground}%
  \BibitemOpen
  \bibfield  {author} {\bibinfo {author} {\bibfnamefont {E.}~\bibnamefont
  {Subag}},\ }\bibfield  {title} {\bibinfo {title} {Following the ground-states
  of full-rsb spherical spin glasses},\ }\href@noop {} {\bibfield  {journal}
  {\bibinfo  {journal} {arXiv 1812.04588}\ } (\bibinfo {year}
  {2018})}\BibitemShut {NoStop}%
\bibitem [{\citenamefont {Auffinger}\ and\ \citenamefont
  {Zeng}(2019)}]{AufZEng2019}%
  \BibitemOpen
  \bibfield  {author} {\bibinfo {author} {\bibfnamefont {A.}~\bibnamefont
  {Auffinger}}\ and\ \bibinfo {author} {\bibfnamefont {Q.}~\bibnamefont
  {Zeng}},\ }\bibfield  {title} {\bibinfo {title} {Existence of two-step
  replica symmetry breaking for the spherical mixed p-spin glass at zero
  temperature},\ }\href@noop {} {\bibfield  {journal} {\bibinfo  {journal}
  {Communications in Mathematical Physics}\ }\textbf {\bibinfo {volume}
  {370}},\ \bibinfo {pages} {377} (\bibinfo {year} {2019})}\BibitemShut
  {NoStop}%
\bibitem [{\citenamefont {Auffinger}\ and\ \citenamefont
  {Zhou}(2022)}]{AufZhou2022}%
  \BibitemOpen
  \bibfield  {author} {\bibinfo {author} {\bibfnamefont {A.}~\bibnamefont
  {Auffinger}}\ and\ \bibinfo {author} {\bibfnamefont {Y.}~\bibnamefont
  {Zhou}},\ }\bibfield  {title} {\bibinfo {title} {The spherical $p+s$ spin
  glass at zero temperature},\ }\href@noop {} {\bibfield  {journal} {\bibinfo
  {journal} {arXiv:2209.03866}\ } (\bibinfo {year} {2022})}\BibitemShut
  {NoStop}%
\bibitem [{\citenamefont {Crisanti}\ and\ \citenamefont
  {Leuzzi}(2004)}]{crisantileuzzi2004}%
  \BibitemOpen
  \bibfield  {author} {\bibinfo {author} {\bibfnamefont {A.}~\bibnamefont
  {Crisanti}}\ and\ \bibinfo {author} {\bibfnamefont {L.}~\bibnamefont
  {Leuzzi}},\ }\bibfield  {title} {\bibinfo {title} {Spherical 2+ p spin-glass
  model: An exactly solvable model for glass to spin-glass transition},\
  }\href@noop {} {\bibfield  {journal} {\bibinfo  {journal} {Physical review
  letters}\ }\textbf {\bibinfo {volume} {93}},\ \bibinfo {pages} {217203}
  (\bibinfo {year} {2004})}\BibitemShut {NoStop}%
\bibitem [{\citenamefont {Crisanti}\ and\ \citenamefont
  {Leuzzi}(2006)}]{crisantileuzzi2006}%
  \BibitemOpen
  \bibfield  {author} {\bibinfo {author} {\bibfnamefont {A.}~\bibnamefont
  {Crisanti}}\ and\ \bibinfo {author} {\bibfnamefont {L.}~\bibnamefont
  {Leuzzi}},\ }\bibfield  {title} {\bibinfo {title} {Spherical 2+ p spin-glass
  model: An analytically solvable model with a glass-to-glass transition},\
  }\href@noop {} {\bibfield  {journal} {\bibinfo  {journal} {Physical Review
  B}\ }\textbf {\bibinfo {volume} {73}},\ \bibinfo {pages} {014412} (\bibinfo
  {year} {2006})}\BibitemShut {NoStop}%
\bibitem [{\citenamefont {Fyodorov}(2004)}]{Fyodo2004complexity}%
  \BibitemOpen
  \bibfield  {author} {\bibinfo {author} {\bibfnamefont {Y.~V.}\ \bibnamefont
  {Fyodorov}},\ }\bibfield  {title} {\bibinfo {title} {Complexity of random
  energy landscapes, glass transition, and absolute value of the spectral
  determinant of random matrices},\ }\href@noop {} {\bibfield  {journal}
  {\bibinfo  {journal} {Physical Review Letters}\ }\textbf {\bibinfo {volume}
  {92}},\ \bibinfo {pages} {240601} (\bibinfo {year} {2004})}\BibitemShut
  {NoStop}%
\bibitem [{\citenamefont {Fyodorov}\ and\ \citenamefont
  {Williams}(2007)}]{FyodWilliams2007}%
  \BibitemOpen
  \bibfield  {author} {\bibinfo {author} {\bibfnamefont {Y.~V.}\ \bibnamefont
  {Fyodorov}}\ and\ \bibinfo {author} {\bibfnamefont {I.}~\bibnamefont
  {Williams}},\ }\bibfield  {title} {\bibinfo {title} {Replica symmetry
  breaking condition exposed by random matrix calculation of landscape
  complexity},\ }\href@noop {} {\bibfield  {journal} {\bibinfo  {journal}
  {Journal of Statistical Physics}\ }\textbf {\bibinfo {volume} {129}},\
  \bibinfo {pages} {1081} (\bibinfo {year} {2007})}\BibitemShut {NoStop}%
\bibitem [{\citenamefont {Fyodorov}(2015)}]{FyodRev2015}%
  \BibitemOpen
  \bibfield  {author} {\bibinfo {author} {\bibfnamefont {Y.~V.}\ \bibnamefont
  {Fyodorov}},\ }\bibfield  {title} {\bibinfo {title} {High-dimensional random
  fields and random matrix theory},\ }\href@noop {} {\bibfield  {journal}
  {\bibinfo  {journal} {Markov Process. Relat. Fields}\ }\textbf {\bibinfo
  {volume} {21}},\ \bibinfo {pages} {483} (\bibinfo {year} {2015})}\BibitemShut
  {NoStop}%
\bibitem [{\citenamefont {Auffinger}\ \emph {et~al.}(2013)\citenamefont
  {Auffinger}, \citenamefont {Arous},\ and\ \citenamefont
  {{\v{C}}ern{\`y}}}]{AufBenArusCerny2013}%
  \BibitemOpen
  \bibfield  {author} {\bibinfo {author} {\bibfnamefont {A.}~\bibnamefont
  {Auffinger}}, \bibinfo {author} {\bibfnamefont {G.~B.}\ \bibnamefont
  {Arous}},\ and\ \bibinfo {author} {\bibfnamefont {J.}~\bibnamefont
  {{\v{C}}ern{\`y}}},\ }\bibfield  {title} {\bibinfo {title} {Random matrices
  and complexity of spin glasses},\ }\href@noop {} {\bibfield  {journal}
  {\bibinfo  {journal} {Communications on Pure and Applied Mathematics}\
  }\textbf {\bibinfo {volume} {66}},\ \bibinfo {pages} {165} (\bibinfo {year}
  {2013})}\BibitemShut {NoStop}%
\bibitem [{\citenamefont {Auffinger}\ and\ \citenamefont
  {Ben~Arous}(2013)}]{AufBenArous2013}%
  \BibitemOpen
  \bibfield  {author} {\bibinfo {author} {\bibfnamefont {A.}~\bibnamefont
  {Auffinger}}\ and\ \bibinfo {author} {\bibfnamefont {G.}~\bibnamefont
  {Ben~Arous}},\ }\bibfield  {title} {\bibinfo {title} {Complexity of random
  smooth functions on the high-dimensional sphere},\ }\href@noop {} {\bibfield
  {journal} {\bibinfo  {journal} {Ann. Probab.}\ }\textbf {\bibinfo {volume}
  {41}},\ \bibinfo {pages} {4214} (\bibinfo {year} {2013})}\BibitemShut
  {NoStop}%
\bibitem [{\citenamefont {Subag}(2017{\natexlab{a}})}]{subag2017geometry}%
  \BibitemOpen
  \bibfield  {author} {\bibinfo {author} {\bibfnamefont {E.}~\bibnamefont
  {Subag}},\ }\bibfield  {title} {\bibinfo {title} {The geometry of the gibbs
  measure of pure spherical spin glasses},\ }\href@noop {} {\bibfield
  {journal} {\bibinfo  {journal} {Inventiones mathematicae}\ }\textbf {\bibinfo
  {volume} {210}},\ \bibinfo {pages} {135} (\bibinfo {year}
  {2017}{\natexlab{a}})}\BibitemShut {NoStop}%
\bibitem [{\citenamefont {Crisanti}\ and\ \citenamefont
  {Sommers}(1995)}]{crisanti1995TAP}%
  \BibitemOpen
  \bibfield  {author} {\bibinfo {author} {\bibfnamefont {A.}~\bibnamefont
  {Crisanti}}\ and\ \bibinfo {author} {\bibfnamefont {H.-J.}\ \bibnamefont
  {Sommers}},\ }\bibfield  {title} {\bibinfo {title} {Thouless-anderson-palmer
  approach to the spherical p-spin spin glass model},\ }\href@noop {}
  {\bibfield  {journal} {\bibinfo  {journal} {Journal de Physique I}\ }\textbf
  {\bibinfo {volume} {5}},\ \bibinfo {pages} {805} (\bibinfo {year}
  {1995})}\BibitemShut {NoStop}%
\bibitem [{\citenamefont {Subag}(2021)}]{subag2018TAP}%
  \BibitemOpen
  \bibfield  {author} {\bibinfo {author} {\bibfnamefont {E.}~\bibnamefont
  {Subag}},\ }\bibfield  {title} {\bibinfo {title} {The free energy of
  spherical pure $ p $-spin models--computation from the tap approach},\
  }\href@noop {} {\bibfield  {journal} {\bibinfo  {journal} {arXiv:2101.04352}\
  } (\bibinfo {year} {2021})}\BibitemShut {NoStop}%
\bibitem [{\citenamefont {Belius}(2022)}]{belius2022TAP}%
  \BibitemOpen
  \bibfield  {author} {\bibinfo {author} {\bibfnamefont {D.}~\bibnamefont
  {Belius}},\ }\bibfield  {title} {\bibinfo {title} {High temperature tap upper
  bound for the free energy of mean field spin glasses},\ }\href@noop {}
  {\bibfield  {journal} {\bibinfo  {journal} {arXiv preprint arXiv:2204.00681}\
  } (\bibinfo {year} {2022})}\BibitemShut {NoStop}%
\bibitem [{\citenamefont {Gradenigo}\ \emph {et~al.}(2020)\citenamefont
  {Gradenigo}, \citenamefont {Angelini}, \citenamefont {Leuzzi},\ and\
  \citenamefont {Ricci-Tersenghi}}]{gradenigo2020cavity}%
  \BibitemOpen
  \bibfield  {author} {\bibinfo {author} {\bibfnamefont {G.}~\bibnamefont
  {Gradenigo}}, \bibinfo {author} {\bibfnamefont {M.~C.}\ \bibnamefont
  {Angelini}}, \bibinfo {author} {\bibfnamefont {L.}~\bibnamefont {Leuzzi}},\
  and\ \bibinfo {author} {\bibfnamefont {F.}~\bibnamefont {Ricci-Tersenghi}},\
  }\bibfield  {title} {\bibinfo {title} {Solving the spherical p-spin model
  with the cavity method: equivalence with the replica results},\ }\href@noop
  {} {\bibfield  {journal} {\bibinfo  {journal} {Journal of Statistical
  Mechanics: Theory and Experiment}\ }\textbf {\bibinfo {volume} {2020}},\
  \bibinfo {pages} {113302} (\bibinfo {year} {2020})}\BibitemShut {NoStop}%
\bibitem [{\citenamefont {Kosterlitz}\ \emph {et~al.}(1976)\citenamefont
  {Kosterlitz}, \citenamefont {Thouless},\ and\ \citenamefont
  {Jones}}]{kosterlitz1976spherical}%
  \BibitemOpen
  \bibfield  {author} {\bibinfo {author} {\bibfnamefont {J.~M.}\ \bibnamefont
  {Kosterlitz}}, \bibinfo {author} {\bibfnamefont {D.~J.}\ \bibnamefont
  {Thouless}},\ and\ \bibinfo {author} {\bibfnamefont {R.~C.}\ \bibnamefont
  {Jones}},\ }\bibfield  {title} {\bibinfo {title} {Spherical model of a
  spin-glass},\ }\href@noop {} {\bibfield  {journal} {\bibinfo  {journal}
  {Physical Review Letters}\ }\textbf {\bibinfo {volume} {36}},\ \bibinfo
  {pages} {1217} (\bibinfo {year} {1976})}\BibitemShut {NoStop}%
\bibitem [{\citenamefont {Tracy}\ and\ \citenamefont
  {Widom}(1996)}]{TracyWidom1994GOE}%
  \BibitemOpen
  \bibfield  {author} {\bibinfo {author} {\bibfnamefont {C.~A.}\ \bibnamefont
  {Tracy}}\ and\ \bibinfo {author} {\bibfnamefont {H.}~\bibnamefont {Widom}},\
  }\bibfield  {title} {\bibinfo {title} {On orthogonal and symplectic matrix
  ensembles},\ }\href@noop {} {\bibfield  {journal} {\bibinfo  {journal}
  {Communications in Mathematical Physics}\ }\textbf {\bibinfo {volume}
  {177}},\ \bibinfo {pages} {727} (\bibinfo {year} {1996})}\BibitemShut
  {NoStop}%
\bibitem [{\citenamefont {Arous}\ \emph {et~al.}(2001)\citenamefont {Arous},
  \citenamefont {Dembo},\ and\ \citenamefont {Guionnet}}]{arous2001aging}%
  \BibitemOpen
  \bibfield  {author} {\bibinfo {author} {\bibfnamefont {G.~B.}\ \bibnamefont
  {Arous}}, \bibinfo {author} {\bibfnamefont {A.}~\bibnamefont {Dembo}},\ and\
  \bibinfo {author} {\bibfnamefont {A.}~\bibnamefont {Guionnet}},\ }\bibfield
  {title} {\bibinfo {title} {Aging of spherical spin glasses},\ }\href@noop {}
  {\bibfield  {journal} {\bibinfo  {journal} {Probability theory and related
  fields}\ }\textbf {\bibinfo {volume} {120}},\ \bibinfo {pages} {1} (\bibinfo
  {year} {2001})}\BibitemShut {NoStop}%
\bibitem [{\citenamefont {Dean}\ and\ \citenamefont
  {Majumdar}(2006)}]{deanmajumdar2006LDF}%
  \BibitemOpen
  \bibfield  {author} {\bibinfo {author} {\bibfnamefont {D.~S.}\ \bibnamefont
  {Dean}}\ and\ \bibinfo {author} {\bibfnamefont {S.~N.}\ \bibnamefont
  {Majumdar}},\ }\bibfield  {title} {\bibinfo {title} {Large deviations of
  extreme eigenvalues of random matrices},\ }\href@noop {} {\bibfield
  {journal} {\bibinfo  {journal} {Physical review letters}\ }\textbf {\bibinfo
  {volume} {97}},\ \bibinfo {pages} {160201} (\bibinfo {year}
  {2006})}\BibitemShut {NoStop}%
\bibitem [{\citenamefont {Majumdar}\ and\ \citenamefont
  {Schehr}(2014)}]{majumdar2014top}%
  \BibitemOpen
  \bibfield  {author} {\bibinfo {author} {\bibfnamefont {S.~N.}\ \bibnamefont
  {Majumdar}}\ and\ \bibinfo {author} {\bibfnamefont {G.}~\bibnamefont
  {Schehr}},\ }\bibfield  {title} {\bibinfo {title} {Top eigenvalue of a random
  matrix: large deviations and third order phase transition},\ }\href@noop {}
  {\bibfield  {journal} {\bibinfo  {journal} {Journal of Statistical Mechanics:
  Theory and Experiment}\ }\textbf {\bibinfo {volume} {2014}},\ \bibinfo
  {pages} {P01012} (\bibinfo {year} {2014})}\BibitemShut {NoStop}%
\bibitem [{\citenamefont {Baik}\ \emph {et~al.}(2021)\citenamefont {Baik},
  \citenamefont {Collins-Woodfin}, \citenamefont {Le~Doussal},\ and\
  \citenamefont {Wu}}]{baik2021spherical}%
  \BibitemOpen
  \bibfield  {author} {\bibinfo {author} {\bibfnamefont {J.}~\bibnamefont
  {Baik}}, \bibinfo {author} {\bibfnamefont {E.}~\bibnamefont
  {Collins-Woodfin}}, \bibinfo {author} {\bibfnamefont {P.}~\bibnamefont
  {Le~Doussal}},\ and\ \bibinfo {author} {\bibfnamefont {H.}~\bibnamefont
  {Wu}},\ }\bibfield  {title} {\bibinfo {title} {Spherical spin glass model
  with external field},\ }\href@noop {} {\bibfield  {journal} {\bibinfo
  {journal} {Journal of Statistical Physics}\ }\textbf {\bibinfo {volume}
  {183}},\ \bibinfo {pages} {31} (\bibinfo {year} {2021})}\BibitemShut
  {NoStop}%
\bibitem [{\citenamefont {Cugliandolo}\ and\ \citenamefont
  {Dean}(1995)}]{cugliandolo1995dynamics}%
  \BibitemOpen
  \bibfield  {author} {\bibinfo {author} {\bibfnamefont {L.}~\bibnamefont
  {Cugliandolo}}\ and\ \bibinfo {author} {\bibfnamefont {D.}~\bibnamefont
  {Dean}},\ }\bibfield  {title} {\bibinfo {title} {On the dynamics of a
  spherical spin-glass in a magnetic field},\ }\href@noop {} {\bibfield
  {journal} {\bibinfo  {journal} {Journal of Physics A: Mathematical and
  General}\ }\textbf {\bibinfo {volume} {28}},\ \bibinfo {pages} {L453}
  (\bibinfo {year} {1995})}\BibitemShut {NoStop}%
\bibitem [{\citenamefont {Mehta}\ \emph {et~al.}(2015)\citenamefont {Mehta},
  \citenamefont {Hauenstein}, \citenamefont {Niemerg}, \citenamefont {Simm},\
  and\ \citenamefont {Stariolo}}]{mehta2015spherical}%
  \BibitemOpen
  \bibfield  {author} {\bibinfo {author} {\bibfnamefont {D.}~\bibnamefont
  {Mehta}}, \bibinfo {author} {\bibfnamefont {J.~D.}\ \bibnamefont
  {Hauenstein}}, \bibinfo {author} {\bibfnamefont {M.}~\bibnamefont {Niemerg}},
  \bibinfo {author} {\bibfnamefont {N.~J.}\ \bibnamefont {Simm}},\ and\
  \bibinfo {author} {\bibfnamefont {D.~A.}\ \bibnamefont {Stariolo}},\
  }\bibfield  {title} {\bibinfo {title} {Energy landscape of the finite-size
  mean-field 2-spin spherical model and topology trivialization},\ }\href@noop
  {} {\bibfield  {journal} {\bibinfo  {journal} {Physical Review E}\ }\textbf
  {\bibinfo {volume} {91}},\ \bibinfo {pages} {022133} (\bibinfo {year}
  {2015})}\BibitemShut {NoStop}%
\bibitem [{\citenamefont {Landon}\ and\ \citenamefont
  {Sosoe}(2022)}]{landon2022fluctuations}%
  \BibitemOpen
  \bibfield  {author} {\bibinfo {author} {\bibfnamefont {B.}~\bibnamefont
  {Landon}}\ and\ \bibinfo {author} {\bibfnamefont {P.}~\bibnamefont {Sosoe}},\
  }\bibfield  {title} {\bibinfo {title} {Fluctuations of the overlap at low
  temperature in the 2-spin spherical sk model},\ }in\ \href@noop {} {\emph
  {\bibinfo {booktitle} {Annales de l'Institut Henri Poincare (B) Probabilites
  et statistiques}}},\ Vol.~\bibinfo {volume} {58}\ (\bibinfo {organization}
  {Institut Henri Poincar{\'e}},\ \bibinfo {year} {2022})\ pp.\ \bibinfo
  {pages} {1426--1459}\BibitemShut {NoStop}%
\bibitem [{\citenamefont {Kivimae}(2019)}]{kivimae2019critical}%
  \BibitemOpen
  \bibfield  {author} {\bibinfo {author} {\bibfnamefont {P.}~\bibnamefont
  {Kivimae}},\ }\bibfield  {title} {\bibinfo {title} {Critical fluctuations for
  the spherical sherrington-kirkpatrick model in an external field},\
  }\href@noop {} {\bibfield  {journal} {\bibinfo  {journal} {arXiv preprint
  arXiv:1908.07512}\ } (\bibinfo {year} {2019})}\BibitemShut {NoStop}%
\bibitem [{\citenamefont {Landon}\ and\ \citenamefont
  {Sosoe}(2020)}]{landon2020fluctuations}%
  \BibitemOpen
  \bibfield  {author} {\bibinfo {author} {\bibfnamefont {B.}~\bibnamefont
  {Landon}}\ and\ \bibinfo {author} {\bibfnamefont {P.}~\bibnamefont {Sosoe}},\
  }\bibfield  {title} {\bibinfo {title} {Fluctuations of the 2-spin ssk model
  with magnetic field},\ }\href@noop {} {\bibfield  {journal} {\bibinfo
  {journal} {arXiv preprint arXiv:2009.12514}\ } (\bibinfo {year}
  {2020})}\BibitemShut {NoStop}%
\bibitem [{\citenamefont {Belius}\ \emph {et~al.}(2022)\citenamefont {Belius},
  \citenamefont {{\v{C}}ern{\`y}}, \citenamefont {Nakajima},\ and\
  \citenamefont {Schmidt}}]{belius2022triviality}%
  \BibitemOpen
  \bibfield  {author} {\bibinfo {author} {\bibfnamefont {D.}~\bibnamefont
  {Belius}}, \bibinfo {author} {\bibfnamefont {J.}~\bibnamefont
  {{\v{C}}ern{\`y}}}, \bibinfo {author} {\bibfnamefont {S.}~\bibnamefont
  {Nakajima}},\ and\ \bibinfo {author} {\bibfnamefont {M.~A.}\ \bibnamefont
  {Schmidt}},\ }\bibfield  {title} {\bibinfo {title} {Triviality of the
  geometry of mixed p-spin spherical hamiltonians with external field},\
  }\href@noop {} {\bibfield  {journal} {\bibinfo  {journal} {Journal of
  Statistical Physics}\ }\textbf {\bibinfo {volume} {186}},\ \bibinfo {pages}
  {12} (\bibinfo {year} {2022})}\BibitemShut {NoStop}%
\bibitem [{Note1()}]{Note1}%
  \BibitemOpen
  \bibinfo {note} {See a nice historical exposition of this and related
  identities in \cite {pastore2019large}.}\BibitemShut {Stop}%
\bibitem [{\citenamefont {Pastore}\ \emph {et~al.}(2019)\citenamefont
  {Pastore}, \citenamefont {Di~Gioacchino},\ and\ \citenamefont
  {Rotondo}}]{pastore2019large}%
  \BibitemOpen
  \bibfield  {author} {\bibinfo {author} {\bibfnamefont {M.}~\bibnamefont
  {Pastore}}, \bibinfo {author} {\bibfnamefont {A.}~\bibnamefont
  {Di~Gioacchino}},\ and\ \bibinfo {author} {\bibfnamefont {P.}~\bibnamefont
  {Rotondo}},\ }\bibfield  {title} {\bibinfo {title} {Large deviations of the
  free energy in the p-spin glass spherical model},\ }\href@noop {} {\bibfield
  {journal} {\bibinfo  {journal} {Physical Review Research}\ }\textbf {\bibinfo
  {volume} {1}},\ \bibinfo {pages} {033116} (\bibinfo {year}
  {2019})}\BibitemShut {NoStop}%
\bibitem [{\citenamefont {Rammal}(1981)}]{Rammal_PhD}%
  \BibitemOpen
  \bibfield  {author} {\bibinfo {author} {\bibfnamefont {R.}~\bibnamefont
  {Rammal}},\ }\bibfield  {title} {\bibinfo {title} {Phd thesis, grenoble
  university},\ }\href@noop {} {\bibfield  {journal} {\bibinfo  {journal}
  {unpublished}\ } (\bibinfo {year} {1981})}\BibitemShut {NoStop}%
\bibitem [{\citenamefont {Andreanov}\ \emph {et~al.}(2004)\citenamefont
  {Andreanov}, \citenamefont {Barbieri},\ and\ \citenamefont
  {Martin}}]{andreanov2004large}%
  \BibitemOpen
  \bibfield  {author} {\bibinfo {author} {\bibfnamefont {A.}~\bibnamefont
  {Andreanov}}, \bibinfo {author} {\bibfnamefont {F.}~\bibnamefont
  {Barbieri}},\ and\ \bibinfo {author} {\bibfnamefont {O.~C.}\ \bibnamefont
  {Martin}},\ }\bibfield  {title} {\bibinfo {title} {Large deviations in
  spin-glass ground-state energies},\ }\href@noop {} {\bibfield  {journal}
  {\bibinfo  {journal} {The European Physical Journal B-Condensed Matter and
  Complex Systems}\ }\textbf {\bibinfo {volume} {41}},\ \bibinfo {pages} {365}
  (\bibinfo {year} {2004})}\BibitemShut {NoStop}%
\bibitem [{\citenamefont {Rivoire}(2005)}]{rivoire2005cavity}%
  \BibitemOpen
  \bibfield  {author} {\bibinfo {author} {\bibfnamefont {O.}~\bibnamefont
  {Rivoire}},\ }\bibfield  {title} {\bibinfo {title} {The cavity method for
  large deviations},\ }\href@noop {} {\bibfield  {journal} {\bibinfo  {journal}
  {Journal of Statistical Mechanics: Theory and Experiment}\ }\textbf {\bibinfo
  {volume} {2005}},\ \bibinfo {pages} {P07004} (\bibinfo {year}
  {2005})}\BibitemShut {NoStop}%
\bibitem [{\citenamefont {Parisi}\ and\ \citenamefont
  {Rizzo}(2008)}]{parisi2008large}%
  \BibitemOpen
  \bibfield  {author} {\bibinfo {author} {\bibfnamefont {G.}~\bibnamefont
  {Parisi}}\ and\ \bibinfo {author} {\bibfnamefont {T.}~\bibnamefont {Rizzo}},\
  }\bibfield  {title} {\bibinfo {title} {Large deviations in the free energy of
  mean-field spin glasses},\ }\href@noop {} {\bibfield  {journal} {\bibinfo
  {journal} {Physical review letters}\ }\textbf {\bibinfo {volume} {101}},\
  \bibinfo {pages} {117205} (\bibinfo {year} {2008})}\BibitemShut {NoStop}%
\bibitem [{\citenamefont {Parisi}\ and\ \citenamefont
  {Rizzo}(2009)}]{parisi2009phase}%
  \BibitemOpen
  \bibfield  {author} {\bibinfo {author} {\bibfnamefont {G.}~\bibnamefont
  {Parisi}}\ and\ \bibinfo {author} {\bibfnamefont {T.}~\bibnamefont {Rizzo}},\
  }\bibfield  {title} {\bibinfo {title} {Phase diagram and large deviations in
  the free energy of mean-field spin glasses},\ }\href@noop {} {\bibfield
  {journal} {\bibinfo  {journal} {Physical Review B}\ }\textbf {\bibinfo
  {volume} {79}},\ \bibinfo {pages} {134205} (\bibinfo {year}
  {2009})}\BibitemShut {NoStop}%
\bibitem [{\citenamefont {Parisi}\ and\ \citenamefont
  {Rizzo}(2010)}]{parisi2010universality}%
  \BibitemOpen
  \bibfield  {author} {\bibinfo {author} {\bibfnamefont {G.}~\bibnamefont
  {Parisi}}\ and\ \bibinfo {author} {\bibfnamefont {T.}~\bibnamefont {Rizzo}},\
  }\bibfield  {title} {\bibinfo {title} {Universality and deviations in
  disordered systems},\ }\href@noop {} {\bibfield  {journal} {\bibinfo
  {journal} {Physical Review B}\ }\textbf {\bibinfo {volume} {81}},\ \bibinfo
  {pages} {094201} (\bibinfo {year} {2010})}\BibitemShut {NoStop}%
\bibitem [{\citenamefont {Monthus}\ and\ \citenamefont
  {Garel}(2010)}]{monthus2010matching}%
  \BibitemOpen
  \bibfield  {author} {\bibinfo {author} {\bibfnamefont {C.}~\bibnamefont
  {Monthus}}\ and\ \bibinfo {author} {\bibfnamefont {T.}~\bibnamefont
  {Garel}},\ }\bibfield  {title} {\bibinfo {title} {Matching between typical
  fluctuations and large deviations in disordered systems: application to the
  statistics of the ground state energy in the sk spin-glass model},\
  }\href@noop {} {\bibfield  {journal} {\bibinfo  {journal} {Journal of
  Statistical Mechanics: Theory and Experiment}\ }\textbf {\bibinfo {volume}
  {2010}},\ \bibinfo {pages} {P02023} (\bibinfo {year} {2010})}\BibitemShut
  {NoStop}%
\bibitem [{\citenamefont {Malatesta}\ \emph {et~al.}(2019)\citenamefont
  {Malatesta}, \citenamefont {Parisi},\ and\ \citenamefont
  {Sicuro}}]{malatesta2019fluctuations}%
  \BibitemOpen
  \bibfield  {author} {\bibinfo {author} {\bibfnamefont {E.~M.}\ \bibnamefont
  {Malatesta}}, \bibinfo {author} {\bibfnamefont {G.}~\bibnamefont {Parisi}},\
  and\ \bibinfo {author} {\bibfnamefont {G.}~\bibnamefont {Sicuro}},\
  }\bibfield  {title} {\bibinfo {title} {Fluctuations in the random-link
  matching problem},\ }\href@noop {} {\bibfield  {journal} {\bibinfo  {journal}
  {Physical Review E}\ }\textbf {\bibinfo {volume} {100}},\ \bibinfo {pages}
  {032102} (\bibinfo {year} {2019})}\BibitemShut {NoStop}%
\bibitem [{Note2()}]{Note2}%
  \BibitemOpen
  \bibinfo {note} {Note that Dembo and Zeitouni use the notation 'quenched' for
  a fixed realisation of $J$ and 'annealed' for results averaged over the
  distribution of $J$ but these notations do not exactly cover the standard use
  of these terms in the disordered system community.}\BibitemShut {Stop}%
\bibitem [{\citenamefont {Baik}\ and\ \citenamefont
  {Lee}(2016)}]{baik2016fluctuations}%
  \BibitemOpen
  \bibfield  {author} {\bibinfo {author} {\bibfnamefont {J.}~\bibnamefont
  {Baik}}\ and\ \bibinfo {author} {\bibfnamefont {J.~O.}\ \bibnamefont {Lee}},\
  }\bibfield  {title} {\bibinfo {title} {Fluctuations of the free energy of the
  spherical sherrington--kirkpatrick model},\ }\href@noop {} {\bibfield
  {journal} {\bibinfo  {journal} {Journal of Statistical Physics}\ }\textbf
  {\bibinfo {volume} {165}},\ \bibinfo {pages} {185} (\bibinfo {year}
  {2016})}\BibitemShut {NoStop}%
\bibitem [{\citenamefont {Subag}(2017{\natexlab{b}})}]{subag2017complexity}%
  \BibitemOpen
  \bibfield  {author} {\bibinfo {author} {\bibfnamefont {E.}~\bibnamefont
  {Subag}},\ }\bibfield  {title} {\bibinfo {title} {The complexity of spherical
  $ p $-spin models—a second moment approach},\ }\href@noop {} {\bibfield
  {journal} {\bibinfo  {journal} {The Annals of Probability}\ }\textbf
  {\bibinfo {volume} {45}},\ \bibinfo {pages} {3385} (\bibinfo {year}
  {2017}{\natexlab{b}})}\BibitemShut {NoStop}%
\bibitem [{\citenamefont {Subag}\ and\ \citenamefont
  {Zeitouni}(2017)}]{subag2017extremal}%
  \BibitemOpen
  \bibfield  {author} {\bibinfo {author} {\bibfnamefont {E.}~\bibnamefont
  {Subag}}\ and\ \bibinfo {author} {\bibfnamefont {O.}~\bibnamefont
  {Zeitouni}},\ }\bibfield  {title} {\bibinfo {title} {The extremal process of
  critical points of the pure p-spin spherical spin glass model},\ }\href@noop
  {} {\bibfield  {journal} {\bibinfo  {journal} {Probability theory and related
  fields}\ }\textbf {\bibinfo {volume} {168}},\ \bibinfo {pages} {773}
  (\bibinfo {year} {2017})}\BibitemShut {NoStop}%
\bibitem [{\citenamefont {Fyodorov}\ and\ \citenamefont
  {Sommers}(2007)}]{fyodorov2007classical}%
  \BibitemOpen
  \bibfield  {author} {\bibinfo {author} {\bibfnamefont {Y.~V.}\ \bibnamefont
  {Fyodorov}}\ and\ \bibinfo {author} {\bibfnamefont {H.-J.}\ \bibnamefont
  {Sommers}},\ }\bibfield  {title} {\bibinfo {title} {Classical particle in a
  box with random potential: exploiting rotational symmetry of replicated
  hamiltonian},\ }\href@noop {} {\bibfield  {journal} {\bibinfo  {journal}
  {Nuclear Physics B}\ }\textbf {\bibinfo {volume} {764}},\ \bibinfo {pages}
  {128} (\bibinfo {year} {2007})}\BibitemShut {NoStop}%
\bibitem [{\citenamefont {Zhou}(2023)}]{zhou2023spherical}%
  \BibitemOpen
  \bibfield  {author} {\bibinfo {author} {\bibfnamefont {Y.}~\bibnamefont
  {Zhou}},\ }\bibfield  {title} {\bibinfo {title} {The spherical mixed $ p
  $-spin glass at zero temperature},\ }\href@noop {} {\bibfield  {journal}
  {\bibinfo  {journal} {arXiv preprint arXiv:2303.04943}\ } (\bibinfo {year}
  {2023})}\BibitemShut {NoStop}%
\end{thebibliography}%

\end{document}